%% file: gut.tex
\documentclass[a4paper,11pt]{article}
\usepackage{amsmath}        
\usepackage{amssymb}        
\usepackage{cite}
\usepackage{cancel}
\usepackage{fancybox}
\usepackage{ulem}
\usepackage{pifont}
\usepackage{diagbox}
\usepackage{colortbl}
\usepackage{multicol}
\usepackage{lscape}
\usepackage{colortbl}
\usepackage{multicol}
\usepackage[all,knot]{xy}
\usepackage{longtable}

\makeatletter
 
 \@addtoreset{equation}{section}
\makeatother

\makeatletter
\newcommand{\subsubsubsection}{\@startsection{paragraph}{4}{\z@}%
{1.5\baselineskip \@plus.5\dp0 \@minus.2\dp0}%
{.5\baselineskip \@plus2.3\dp0}%
{\reset@font\normalsize\bfseries}
}
\newcommand{\subsubsubsubsection}{\@startsection{subparagraph}{5}{\z@}%
{1.5\baselineskip \@plus.5\dp0 \@minus.2\dp0}%
{.5\baselineskip \@plus2.3\dp0}%
{\reset@font\normalsize\itshape}
}
\makeatother
\begin{document}

\title{Finite-Dimensional Lie Algebras and Their Representations
for Unified Model Building}

\author{Naoki Yamatsu
\footnote{Electronic address: yamatsu.naoki@phys.kyushu-u.ac.jp}
\\
{\it\small
Department of Physics, Kyushu University, Fukuoka 819-0395, Japan}
}
\date{\today}

\maketitle

\begin{abstract}
 We give information about finite-dimensional Lie algebras and their
 representations for model building in 4 and 5 dimensions; 
 e.g., conjugacy classes, types of  representations, Weyl dimension
 formulas, Dynkin indices, quadratic Casimir invariants, anomaly
 coefficients, projection matrices, and branching rules of Lie algebras
 and their subalgebras up to rank-20.
 We show what kind of  Lie algebras can be applied for grand
 unified theories in 4 and 5 dimensions.
\end{abstract}

\begin{multicols}{2}
\tableofcontents
\end{multicols}

\begin{multicols}{2}
\listoftables
\end{multicols}

\newpage

\section{Introduction}
\label{Sec:Introduction}

We will discuss finite-dimensional Lie algebras and their
representations for unified model building. 
There is already a good report Ref.~\cite{Slansky:1981yr} of Lie
algebras and their representations for particle physicists with the
title ``Group Theory for Unified Model Building'' written by R.~Slansky. 
The paper contains almost all knowledge for usual model building to
construct grand unified models in four-dimensional spacetime with the
finite degrees of freedom of internal space or matter content. 
However, in modern sense, several exceptional cases have emerged.
In this paper, we find missing information for further
unified model building, but it does not contain definitions of Lie
algebras, their theorems and lemmas and their proofs, completely.
They can be confirmed in Dynkin's original papers
Refs.~\cite{Dynkin:1950um,Dynkin:1957um,Dynkin:1957ek,Dynkin:2000}, a
Dynkin's paper's brief review ``Semi-Simple Lie Algebras and
Their Representations'' written by R.~N.~Cahn Ref.~\cite{Cahn:1985wk},
or books about Lie algebras and Lie groups, e.g.,
Refs.~\cite{Weyl:1946,Gilmore:102082,Fulton:1991}. 
Introductory-level knowledge about Lie algebras and groups is given
in Ref.~\cite{Georgi:1982jb}.

At present, the irreducible representations of simple Lie algebras,
whose rank is not exceeding 8, are summarized in Ref.~\cite{McKay:1981}
written by W.~G.~McKay and J.~Patera with the title ``Tables of
Dimensions, Indices, and Branching Rules for Representations of Simple
Lie Algebras.'' The Table 1 of the book includes tables of
dimensions \cite{Weyl:1946,Dynkin:1957um}, 
second order and forth order Dynkin indices
\cite{Patera:1976yd,Okubo:1981td}, and 
type of representations, i.e. complex, self-conjugate, real, and
pseudo-real representations \cite{McKay:1981}.
For each simple Lie algebra of rank $2\leq n\leq 8$ two
pages are devoted, where two pages are used for $E_7$ and $E_8$.

It is known that branching rules of Lie algebras and their Lie
subalgebras can be calculated by using their corresponding projection
matrices. The projection matrices of Lie algebras and their maximal
regular and special Lie subalgebras are listed by W.~Mckay et al. 
in Ref.~\cite{Mckay:1977} up to rank 8, where they do not contain
$\mathfrak{u}_1$ charges. Also, several generic projection matrices of
rank-$n$ classical Lie algebras are derived by using Weyl group orbits
in
Refs.~\cite{Hakova:2008ax,Larouche:2009ax,Larouche:2011bs,Larouche:2011cy}. 

The Table 2 in Ref.~\cite{McKay:1981} is devoted to the tables of the
branching rules of up to 5,000-dimensional representations for the
classical Lie algebras $A_n$, $B_n$, $C_n$, and $D_n$, 
and up to 10,000-dimensional representations for the five exceptional
algebras $E_6$, $E_7$, $E_8$, $F_4$, and $G_2$ by using 
projection matrices in Ref.~\cite{Mckay:1977}.
Its tables do not contain $\mathfrak{u}_1$ charges because
$\mathfrak{u}_1$ is not a semi-simple. R.~Slansky gives us its
information in Ref.~\cite{Slansky:1981yr}, but it is very limited. 

There are useful public codes to calculate some features of irreducible
representations of Lie algebras such as 
dimensions \cite{Weyl:1946,Dynkin:1957um},
conjugacy classes \cite{Dynkin:1957um,Lemire:1979qd},
(second order) Dynkin indices \cite{Dynkin:1957um}, 
quadratic Casimir invariants, 
and anomaly coefficients \cite{Banks1976,Okubo:1977sc,Patera:1981sc},
and also tensor products, 
branching rules and some projection matrices \cite{Mckay:1977}.
For example, 
Susyno program \cite{Fonseca:2011sy} is a Mathematica package,
LieART \cite{Feger:2012bs,Feger:2019tvk} is also a Mathematica package,
and LiE \cite{Leeuwen:1992} is a C program.

Before we finish the introduction, we give some examples when we need 
Lie algebras, their representations, etc. in physics.
For example, to construct a consistent chiral $SU(n)$ gauge theory
\cite{Ball:1988xg} in four dimension, we must consider its $SU(n)$
chiral gauge anomaly. 
(See e.g., Ref.~\cite{Fujikawa:2004xa}.) 
We can easily check whether a matter content is anomaly-free or not
by using anomaly coefficients. 
To calculate the renormalization group equation for gauge coupling
constant, we need (second order) Dynkin indices of irreducible 
representations. (See e.g., Refs.~\cite{Politzer:1974fr,Machacek:1983tz,Machacek:1983fi,Machacek:1984zw}.)
The notion of types of representations (complex, self-conjugate, real,
and pseudo-real representations) is important to find special Lie
subalgebras. The notion of conjugacy classes is also useful to classify
irreducible representations and to calculate tensor products.

Tensor products of Lie algebras are also important for model building
to write down invariant action under certain symmetry transformation. 
There are several calculation techniques by using e.g., Dynkin labels
and Dynkin diagrams in Refs.~\cite{McKay:1981,Cahn:1985wk} and 
Weyl group orbits in Refs.~\cite{Bremner1983,Klimyk:2006ax}.

Branching rules are essential not only to construct grand unified
theories but also to consider models including explicit or spontaneous
broken symmetries. For example, in an $SU(5)$ grand unified theory
discussed in Ref.~\cite{Georgi:1974sy}, 
we have to know how to decompose representations of $SU(5)$ in its
matter content into representations of 
$G_{\rm SM}:=SU(3)\times SU(2)\times U(1)$.
E.g., the gauge bosons with the adjoint representation ${\bf 24}$ of
$SU(5)$ are decomposed into the SM gauge bosons with 
$({\bf 8,1})(0)\oplus({\bf 1,3})(0)\oplus({\bf 1,1})(0)$ of $G_{\rm SM}$
and the so-called $X$ and $Y$ gauge bosons with 
$({\bf 3,2})(5)\oplus({\bf \overline{3},2})(-5)$ of $G_{\rm SM}$,
where we take a normalization of $U(1)$ charges as all the $U(1)$
charges are integers.

A purpose of this paper is to provide basic and useful information about
some properties of irreducible representations of Lie algebras,
branching rules of Lie algebras and their subalgebras, and tensor
products, where the properties of irreducible representations of Lie
algebras include dimensions, quadratic Casimir invariants, Dynkin
indices, anomaly coefficients, conjugacy classes, and types of
representations. 
The branching rules contain not only semi-simple subalgebra but
also $\mathfrak{u}_1$ charges because the information is important for
model building. 
Another purpose is to inform one of calculation methods to obtain the
above things. By using the above information, we will check what kind of
Lie algebras can be applied for grand unification in 4 and 5 dimensions.

This paper is organized as follows. 
In Sec.~\ref{Sec:Basics}, we check basics of Lie algebras and
their subalgebras such as Dynkin diagrams, Cartan matrices, 
and notion of regular, special, and maximal subalgebras.
In Sec.~\ref{Sec:Representations-basics}, we see several features of
representations of Lie algebras such as conjugacy classes, types of
representations, Weyl dimension formulas, Dynkin indices, Casimir
invariants, and anomaly coefficients.
In Sec.~\ref{Sec:Representations-subalgebras}, we consider how to
decompose irreducible representations of Lie algebras into irreducible
representations of their Lie subalgebras.
In Sec.~\ref{Sec:Tensor-product-basics}, we discuss a method to
calculate tensor products of two irreducible representations of a Lie
algebra mainly by using its Dynkin diagram. 
In Sec.~\ref{Sec:Summary-representations}, several features of rank-$n$
classical Lie algebras and the five exceptional Lie algebras are
summarized. 
This section includes several new results about generic projection
matrices of rank-$n$ Lie algebras and their Lie subalgebras.
In Sec.~\ref{Sec:Model-Building}, we show which Lie algebras can be
applied for grand unification in general by using the above discussion.
Section~\ref{Sec:Summary-discussion} is devoted to a summary and
discussion.
Appendix~\ref{Sec:Representations} contains tables of 
representations of classical Lie algebras $A_n$,
$B_n$, $C_n$, $D_n$ $(n=1,2,\cdots,20)$ and
the exceptional Lie algebras $E_6$, $E_7$, $E_8$, $F_4$, and $G_2$, 
which include dimensions, conjugacy classes, Dynkin indices, quadratic
Casimir invariants, anomaly coefficients, and types of representations.
The tables are partially calculated by Susyno program
\cite{Fonseca:2011sy}.
Appendix~\ref{Sec:Positive-roots} contains tables of
positive roots of some classical Lie algebras and the exceptional Lie
algebras.
Appendix~\ref{Sec:Weight-diagrams} contains tables of
weight diagrams of several representations of some classical Lie
algebras and the exceptional Lie algebras.
Appendix~\ref{Sec:Subalgebras} contains a summary table of
maximal regular and special subalgebras of Lie algebra up to rank-20.
Each table of maximal subalgebras of classical Lie algebras $A_n$,
$B_n$, $C_n$, $D_n$ $(n=1,2,\cdots,20)$ and
the exceptional Lie algebras $E_6$, $E_7$, $E_8$, $F_4$, and $G_2$
has one branching rule to determine the corresponding projection matrix.
Appendix~\ref{Sec:Projection-matrices} contains tables of
all projection matrices of classical Lie algebras $A_n$,
$B_n$, $C_n$, $D_n$ $(n=1,2,\cdots,20)$ and
the exceptional Lie algebras $E_6$, $E_7$, $E_8$, $F_4$, and $G_2$ 
and their maximal regular and special subalgebras.
Appendix~\ref{Sec:Branching-rules} contains tables of
branching rules of classical Lie algebras $A_n$,
$B_n$, $C_n$, $D_n$ $(n=1,2,\cdots,20)$ and
the exceptional Lie algebras $E_6$, $E_7$, $E_8$, $F_4$, and $G_2$ 
and their maximal regular and special subalgebras.
The tables are also obtained by Susyno program
\cite{Fonseca:2011sy} by using projection matrices shown in
Appendix~\ref{Sec:Projection-matrices}.
Appendix~\ref{Sec:Teneor-products} contains tables of
tensor products of classical Lie algebras $A_n$,
$B_n$, $C_n$, $D_n$ $(n=1,2,\cdots,20)$, and
the exceptional Lie algebras $E_6$, $E_7$, $E_8$, $F_4$, and $G_2$.
The tables are also calculated by Susyno program
\cite{Fonseca:2011sy}.

\section{Lie algebras and their subalgebras}
\label{Sec:Basics}

First, we list up some technical terms about Lie algebras and subalgebras.
(For more detail, see e.g., Refs.~\cite{Cahn:1985wk,Fuchs:1992}.)
\begin{itemize}
\item A Lie algebra $\mathfrak{g}$ is an algebra such that its
      map $[\ ,\ ]$: $\mathfrak{g}\otimes\mathfrak{g}\to\mathfrak{g}$
      satisfies the following properties:
\begin{itemize}
\item[(1)] $[x,y]=-[y,x]$ for ${}^\forall x,y\in\mathfrak{g}$
      (antisymmetry);
\item[(2)] $[x,[y,z]]+[y,[z,x]]+[z,[x,y]]=0$
       for ${}^\forall x,y,z\in\mathfrak{g}$ (Jacobi identity).
\end{itemize}
\item A Lie subalgebra $\mathfrak{h}$
      ($\mathfrak{h}\subseteq\mathfrak{g}$ ) of the Lie algebra
      $\mathfrak{g}$ itself is a Lie algebra.
\item A proper Lie subalgebra $\mathfrak{h}$ is a Lie subalgebra if 
      $\mathfrak{h}\not=\mathfrak{g}$; i.e., 
      $\mathfrak{h}\subset\mathfrak{g}$;
\item An ideal $\mathfrak{h}$ of the Lie algebra $\mathfrak{g}$ is a
      subalgebra that satisfies the property 
      $[\mathfrak{g},\mathfrak{h}]\subseteq\mathfrak{h}$.
\item An Abelian Lie algebra is a Lie algebra that satisfies
      $[\mathfrak{g},\mathfrak{g}]=0$.
\item A simple Lie algebra is a Lie algebra that does not contain proper
      ideals and that is not Abelian algebra $\mathfrak{u}_1$. 
\item A semi-simple Lie algebra is an algebra that is the direct sum of
      simple Lie algebras.
\item A non-semi-simple Lie algebra is an algebra that is the direct sum
      of a semi-simple and an Abelian Lie algebra.
\end{itemize}

Let us classify a finite-dimensional simple Lie algebra $\mathfrak{g}$ by
using a Cartan matrix $A_{ij}(\mathfrak{g})$,
\begin{align}
A_{ij}(\mathfrak{g}):=2\frac{(\alpha_i,\alpha_j)}{(\alpha_j,\alpha_j)},
\end{align}
where $\alpha_i (i=1,2,\cdots,n)$ are the simple roots of rank-$n$
algebra $\mathfrak{g}$, $(*,*)$ is a scalar product on the root space.
Here we define a Cartan matrix of a simple Lie algebra $\mathfrak{g}$
as it satisfying the following conditions:
\begin{align}
(C0)\ & A_{ij}(\mathfrak{g})\in \mathbb{Z},\nonumber\\
(C1)\ & A_{ii}(\mathfrak{g})=2,\nonumber\\
(C2)\ & A_{ij}(\mathfrak{g})=0\Leftrightarrow 
        A_{ji}(\mathfrak{g})=0,\nonumber\\
(C3)\ & A_{ij}(\mathfrak{g})\in \mathbb{Z}_{\leq 0}\ \mbox{for}\
 i\not=j,\nonumber\\ 
(C4)\ & \mbox{det}A_{ij}(\mathfrak{g})>0,\nonumber\\
(C5)\ & \mbox{irreducible}.
\label{Eq:Finite-Lie-algebra}
\end{align}
(See e.g., Refs.~\cite{Cahn:1985wk,Fuchs:1992}.)

We check what kind of Cartan matrices $A_{ij}(\mathfrak{g})$ satisfy the
above conditions. 
For rank-2, we can write a Cartan matrix $A_{ij}(\mathfrak{g})$ as 
\begin{align}
A(\mathfrak{g})=
\left(
\begin{array}{cc}
2&a_{12}\\
a_{21}&2\\
\end{array}
\right).
\end{align}
Its determinant $\mbox{det}A_{ij}(\mathfrak{g})$ must satisfy the
following condition: 
\begin{align}
\mbox{det}A_{ij}(\mathfrak{g})=4-a_{12}a_{21}>0.
\end{align}
Thus, the possible sets of $(a_{12},a_{21})$ are 
\begin{align}
(a_{12},a_{21})=
(0,0), (-1,-1), (-1,-2), (-1,-3), (-2,-1), (-3,-1),
\end{align}
where $(0,0)$ does not give us a simple Lie algebra, but 
$D_2\simeq A_1\oplus A_1$.
For usual notations, they correspond to 
$D_2$, $A_2$, $C_2(\simeq B_2)$, $G_2$, $B_2$, $G_2$, respectively.
(Note that for rank-2 Lie algebras, a Cartan matrix is the same as its
transpose one, where it can be understood e.g., by using their Dynkin
diagrams discussed later; thus, the matrix with
$(a_{12},a_{21})=(-1,-2)$ is the same as that with  
$(a_{12},a_{21})=(-2,-1)$;
the matrix with $(a_{12},a_{21})=(-1,-3)$ is the same as that with 
$(a_{12},a_{21})=(-3,-1)$.)
The explicit matrices of $A_2$, $B_2(\simeq C_2)$, $C_2$, 
$D_2(\simeq A_1\oplus A_1)$,
and $G_2$ are
\begin{align}
&A(A_2)=
\left(
\begin{array}{cc}
2&-1\\
-1&2\\
\end{array}
\right),\ \ \
A(B_2)=
\left(
\begin{array}{cc}
2&-2\\
-1&2\\
\end{array}
\right),\ \ \
A(C_2)=
\left(
\begin{array}{cc}
2&-1\\
-2&2\\
\end{array}
\right),\nonumber\\
&A(D_2)=
\left(
\begin{array}{cc}
2&0\\
0&2\\
\end{array}
\right),\ \ \
A(G_2)=
\left(
\begin{array}{cc}
2&-3\\
-1&2\\
\end{array}
\right).
\end{align}
Notice that for 
$(a_{12},a_{21})=(-2,-2), (-1,-4)$, the Cartan matrices are 
\begin{align}
&A(A_1^{(1)})=
\left(
\begin{array}{cc}
2&-2\\
-2&2\\
\end{array}
\right),\ \ \
A(A_1^{(2)})=
\left(
\begin{array}{cc}
2&-4\\
-1&2\\
\end{array}
\right).
\end{align}
They lead to $\mbox{det}A(\mathfrak{g})=0$, where the superscript of the
algebra $(r)$, e.g., $(1)$ of $A_1^{(1)}$, is corresponding to so-called
Coxeter label. 
Thus, they are not finite-dimensional Lie algebras,
but Affine Lie algebras. 
The class of the former matrix gives us
important information about a so-called extended Dynkin diagram. 
(See e.g., \cite{Fuchs:1992,Kac:1983} for Affine Lie algebras and
Kac-Moody algebras.)

For rank-3, we can write a Cartan matrix as 
\begin{align}
A_{ij}(\mathfrak{g})=
\left(
\begin{array}{ccc}
2&a_{12}&a_{13}\\
a_{21}&2&a_{23}\\
a_{31}&a_{32}&2\\
\end{array}
\right).
\end{align}
Its determinant $\mbox{det}A_{ij}(\mathfrak{g})$ must satisfy the
following condition: 
\begin{align}
\mbox{det}A_{ij}(\mathfrak{g})=8+a_{12}a_{23}a_{31}+a_{21}a_{32}a_{13}
-2\left(a_{12}a_{21}+a_{13}a_{31}+a_{23}a_{32}\right)
>0.
\end{align}
Thus, the possible sets of 
$(a_{12},a_{21};a_{13},a_{31};a_{23},a_{32})$ are 
\begin{align}
(a_{12},a_{21};a_{13},a_{31};a_{23},a_{32})=&
(-1,-1;0,0;-1,-1),\
(-1,-1;0,0;-2,-1),\nonumber\\
&(-1,-1;0,0;-1,-2),\ 
(-1,-1;-1,-1;0,0).
\end{align}
For usual notations, they correspond to 
$A_3$, $B_3$, $C_3$, and $D_3(\simeq A_3)$, respectively.
(Note that for rank-3 Lie algebras, the Cartan matrix of $A_3$ is the
same as that of $D_3$, where it can be understood e.g., by using
their Dynkin diagrams discussed later.)
The explicit matrices of $A_3$, $B_3$, $C_3$, $D_3(\simeq A_3)$,
and $G_2$ are
\begin{align}
&A(A_3)=
\left(
\begin{array}{ccc}
2&-1&0\\
-1&2&-1\\
0&-1&2\\
\end{array}
\right),\ \ \
A(B_3)=
\left(
\begin{array}{ccc}
2&-1&0\\
-1&2&-2\\
0&-1&2\\
\end{array}
\right),\nonumber\\
&A(C_3)=
\left(
\begin{array}{ccc}
2&-1&0\\
-1&2&-1\\
0&-2&2\\
\end{array}
\right),\ \ \
A(D_3)=
\left(
\begin{array}{ccc}
2&-1&-1\\
-1&2&0\\
-1&0&2\\
\end{array}
\right).
\end{align}
Notice that for appropriate sets
$(a_{12},a_{21};a_{13},a_{31};a_{23},a_{32})$, the Cartan matrices 
with $\mbox{det}A(\mathfrak{g})=0$ are 
\begin{align}
&A(A_2^{(1)})=
\left(
\begin{array}{ccc}
2&-1&-1\\
-1&2&-1\\
-1&-1&2\\
\end{array}
\right),\ \ \
A(D_5^{(2)})=
\left(
\begin{array}{ccc}
2&-1&0\\
-2&2&-2\\
0&-1&2\\
\end{array}
\right),\nonumber\\
&A(A_4^{(2)})=
\left(
\begin{array}{ccc}
2&-2&0\\
-1&2&-2\\
0&-1&2\\
\end{array}
\right),\ \ \
A(C_2^{(1)})=
\left(
\begin{array}{ccc}
2&-2&0\\
-1&2&-1\\
0&-2&2\\
\end{array}
\right),\nonumber\\
&A(G_2^{(1)})=
\left(
\begin{array}{ccc}
2&-1&0\\
-1&2&-3\\
0&-1&2\\
\end{array}
\right),\ \ \
A(D_4^{(3)})=
\left(
\begin{array}{ccc}
2&-3&0\\
-1&2&-1\\
0&-1&2\\
\end{array}
\right).
\end{align}

For rank-4, algebras $A_4$, $B_4$, $C_4$, $D_4$, and $F_4$ satisfy
the condition $\mbox{det}A(\mathfrak{g})>0$.
The explicit matrices of $A_4$, $B_4$, $C_4$, $D_4$, and $F_4$ are
\begin{align}
&A(A_4)=
\left(
\begin{array}{cccc}
 2&-1& 0& 0\\
-1& 2&-1& 0\\
 0&-1& 2&-1\\
 0& 0&-1& 2\\
\end{array}
\right),\ \ \
A(B_4)=
\left(
\begin{array}{cccc}
 2&-1& 0& 0\\
-1& 2&-1& 0\\
 0&-1& 2&-2\\
 0& 0&-1& 2\\
\end{array}
\right),\nonumber\\
&A(C_4)=
\left(
\begin{array}{cccc}
 2&-1& 0& 0\\
-1& 2&-1& 0\\
 0&-1& 2&-1\\
 0& 0&-2& 2\\
\end{array}
\right),\ \ \
A(D_4)=
\left(
\begin{array}{cccc}
 2&-1& 0& 0\\
-1& 2&-1&-1\\
 0&-1& 2& 0\\
 0&-1& 0& 2\\
\end{array}
\right),\nonumber\\
&A(F_4)=
\left(
\begin{array}{cccc}
 2&-1& 0& 0\\
-1& 2&-2& 0\\
 0&-1& 2&-1\\
 0& 0&-1& 2\\
\end{array}
\right).
\end{align}
The algebras $A_3^{(1)}$, $B_3^{(1)}$, $C_3^{(1)}$, $A_5^{(2)}$, 
$A_6^{(2)}$, and $D_6^{(2)}$ satisfy
the condition $\mbox{det}A(\mathfrak{g})=0$.
The explicit matrices of them are
\begin{align}
&A(A_3^{(1)})=
\left(
\begin{array}{cccc}
 2&-1& 0&-1\\
-1& 2&-1& 0\\
 0&-1& 2&-1\\
-1& 0&-1& 2\\
\end{array}
\right),\ \ \
A(B_3^{(1)})=
\left(
\begin{array}{cccc}
 2& 0&-1& 0\\
 0& 2&-1& 0\\
-1&-1& 2&-2\\
 0& 0&-1& 2\\
\end{array}
\right),\nonumber\\
&A(A_5^{(2)})=
\left(
\begin{array}{cccc}
 2& 0&-1& 0\\
 0& 2&-1& 0\\
-1&-1& 2&-2\\
 0& 0&-1& 2\\
\end{array}
\right),\ \ \
A(C_3^{(1)})=
\left(
\begin{array}{cccc}
 2&-2& 0& 0\\
-1& 2&-1& 0\\
 0&-1& 2&-1\\
 0& 0&-2& 2\\
\end{array}
\right),\nonumber\\
&A(D_6^{(2)})=
\left(
\begin{array}{cccc}
 2& 0&-1& 0\\
 0& 2&-1& 0\\
-1&-1& 2&-1\\
 0& 0&-2& 2\\
\end{array}
\right),\ \ \
A(A_6^{(2)})=
\left(
\begin{array}{cccc}
 2&-2& 0& 0\\
-1& 2&-1& 0\\
 0&-1& 2&-2\\
 0& 0&-1& 2\\
\end{array}
\right).
\end{align}

For rank-5, algebras $A_5$, $B_5$, $C_5$, and $D_5$ satisfy
the condition $\mbox{det}A(\mathfrak{g})>0$.
The explicit matrices of $A_5$, $B_5$, $C_5$, and $D_5$ are
\begin{align}
&A(A_5)=
\left(
\begin{array}{ccccc}
 2&-1& 0& 0& 0\\
-1& 2&-1& 0& 0\\
 0&-1& 2&-1& 0\\
 0& 0&-1& 2&-1\\
 0& 0& 0&-1& 2\\
\end{array}
\right),\ \ \
A(B_5)=
\left(
\begin{array}{ccccc}
 2&-1& 0& 0& 0\\
-1& 2&-1& 0& 0\\
 0&-1& 2&-1& 0\\
 0& 0&-1& 2&-2\\
 0& 0& 0&-1& 2\\
\end{array}
\right),\nonumber\\
&A(C_5)=
\left(
\begin{array}{ccccc}
 2&-1& 0& 0& 0\\
-1& 2&-1& 0& 0\\
 0&-1& 2&-1& 0\\
 0& 0&-1& 2&-1\\
 0& 0& 0&-2& 2\\
\end{array}
\right),\ \ \
A(D_5)=
\left(
\begin{array}{ccccc}
 2&-1& 0& 0& 0\\
-1& 2&-1& 0& 0\\
 0&-1& 2&-1&-1\\
 0& 0&-1& 2& 0\\
 0& 0&-1& 0& 0\\
\end{array}
\right).
\end{align}
The algebras $A_4^{(1)}$, $A_3^{(1)}$, $A_3^{(1)}$, $F_4^{(1)}$, 
$A_7^{(2)}$, $A_8^{(2)}$, $D_7^{(2)}$, $F_4^{(1)}$, and $E_6^{(2)}$
satisfy the condition $\mbox{det}A(\mathfrak{g})=0$.
The explicit matrices of them are
\begin{align}
&A(A_4^{(1)})=
\left(
\begin{array}{ccccc}
 2&-1& 0& 0&-1\\
-1& 2&-1& 0& 0\\
 0&-1& 2&-1& 0\\
 0& 0&-1& 2&-1\\
-1& 0& 0&-1& 2\\
\end{array}
\right),\ \ \
A(B_4^{(1)})=
\left(
\begin{array}{ccccc}
 2& 0&-1& 0& 0\\
 0& 2&-1& 0& 0\\
-1&-1& 2&-1& 0\\
 0& 0&-1& 2&-2\\
 0& 0& 0&-1& 2\\
\end{array}
\right),\nonumber\\
&A(C_4^{(1)})=
\left(
\begin{array}{ccccc}
 2&-2& 0& 0& 0\\
-1& 2&-1& 0& 0\\
 0&-1& 2&-1& 0\\
 0& 0&-1& 2&-1\\
 0& 0& 0&-2& 2\\
\end{array}
\right),\ \ \
A(D_4^{(1)})=
\left(
\begin{array}{ccccc}
 2&-1& 0& 0& 0\\
-1& 2&-1&-1&-1\\
 0&-1& 2& 0& 0\\
 0&-1& 0& 2& 0\\
 0&-1& 0& 0& 2\\
\end{array}
\right),\nonumber\\
&A(F_4^{(1)})=
\left(
\begin{array}{ccccc}
 2&-1& 0& 0& 0\\
-1& 2&-2& 0& 0\\
 0&-1& 2&-1& 0\\
 0& 0&-1& 2&-1\\
 0& 0& 0&-1& 2\\
\end{array}
\right).
\end{align}

We can also check rank-$n$ $(n\geq 4)$ algebras by using the same
technique. 
The condition $(C4)$ in Eq.~(\ref{Eq:Finite-Lie-algebra}) 
$\mbox{det}A_{ij}>0$ does not constrain the rank of the classical
algebras $A_n$, $B_n$, $C_n$, and $D_n$, 
but it strongly constrains the rank of the exceptional algebras $E_n$,
$F_n$, and $G_n$.
In other words, the rank of the classical algebras is unlimited for
large $n$, but the rank of the exceptional algebras $E_n$, $F_n$, and
$G_n$ is limited by $n=8$, $n=4$, and $n=2$, respectively.
By using words of the Affine Lie algebras, 
$E_9=E_8^{(1)}$, $F_5=F_4^{(1)}$, and $G_3=G_2^{(1)}$.

Note that 
$\mbox{det}A(A_n)=n+1$ ($n\geq 1$),
$\mbox{det}A(B_n)=2$ ($n\geq 1$),
$\mbox{det}A(C_n)=2$ ($n\geq 1$),
$\mbox{det}A(D_n)=4$ ($n\geq 3$),
$\mbox{det}A(E_n)=9-n$ ($n\geq 6$),
$\mbox{det}A(F_4)=1$, and $\mbox{det}A(G_2)=1$.

\subsection{(Extended) Dynkin diagrams and Cartan matrices}

Finite-dimensional Lie algebras are classified into 
$A_n$, $B_n$, $C_n$, $D_n$ $(n\geq 1)$, 
$E_n$ $(n=6,7,8)$, $F_4$, and $G_2$. 
For their associate groups, we only consider compact groups.
We sometimes denote the Lie algebras 
$A_n=\mathfrak{su}_{n+1}$, 
$B_n=\mathfrak{so}_{2n+1}$,
$C_n=\mathfrak{usp}_{2n}$, and
$D_n=\mathfrak{so}_{2n}$.
Also, their associate compact groups of $A_n$, $B_n$, $C_n$, $D_n$,
$E_n$, $F_4$, and $G_2$ are 
$SU(n+1)$, $SO(2n+1)$, $USp(2n)$, $SO(2n)$,
$E_n$, $F_4$, and $G_2$, respectively.

The simple Lie algebras, their associate compact groups, and
their extended Dynkin diagrams are summarized in the following table.
\begin{center}
\begin{longtable}{cccc}
\caption{Extended Dynkin diagrams}
\label{Table:Dynkin-diagrams}\\
\hline\hline
Algebra&Group&Rank&Extended Dynkin diagram\\\hline\hline
\endfirsthead
\multicolumn{4}{c}{Table~\ref{Table:Dynkin-diagrams} (continued)}\\\hline\hline
Algebra&Group&Rank&Extended Dynkin diagram\\\hline\hline
\endhead
\hline
\endfoot
$A_n=\mathfrak{su}_{n+1}$&$SU(n+1)$ &${}^\forall n$&
$\xygraph{
    \circ ([]!{+(0,-.3)} {{}_1}) (
        - []!{+(2.5,+1.0)}  \circ ([]!{+(0,-.3)} {{}_x})
        - []!{+(2.5,-1.0)}  \circ ([]!{+(0,-.3)} {{}}))
        - [r]
    \circ ([]!{+(0,-.3)} {{}_2}) 
        - [r]
    \circ ([]!{+(0,-.3)} {{}_3}) 
        - [r] \cdots - [r]
    \circ ([]!{+(0,-.3)} {{}_{n-1}}) - [r]
    \circ ([]!{+(0,-.3)} {{}_n})}
$\\
$B_n=\mathfrak{so}_{2n+1}$&$SO(2n+1)$ &${}^\forall n$&
$\xygraph{!~:{@{=}|@{}}
    \circ ([]!{+(0,-.3)} {{}_2}) (
        - []!{+(-1,.5)}  \circ ([]!{+(0,-.3)} {{}_1}),
        - []!{+(-1,-.5)} \circ ([]!{+(0,-.3)} {{}_x}))
        - [r]
    \circ ([]!{+(0,-.3)} {{}_3}) 
        - [r] \cdots - [r]
    \circ   ([]!{+(0,-.3)} {{}_{n-1}}) : [r]
    \bullet ([]!{+(0,-.3)} {{}_n})}
$\\
$C_n=\mathfrak{usp}_{2n}$&$USp(2n)$ &${}^\forall n$&
$\xygraph{!~:{@{=}|@{}}
    \circ   ([]!{+(0,-.3)} {{}_x}) : [r]
    \bullet ([]!{+(0,-.3)} {{}_1}) - [r]
    \bullet ([]!{+(0,-.3)} {{}_2}) - [r] \cdots - [r]
    \bullet ([]!{+(0,-.3)} {{}_{n-1}}) : [r]
    \circ   ([]!{+(0,-.3)} {{}_n})}
$\\
$D_n=\mathfrak{so}_{2n}$&$SO(2n)$ &${}^\forall n$&
$\xygraph{
    \circ ([]!{+(0,-.3)} {{}_2}) (
        - []!{+(-1,.5)}  \circ ([]!{+(0,-.3)} {{}_1}),
        - []!{+(-1,-.5)} \circ ([]!{+(0,-.3)} {{}_x}))
        - [r] 
    \circ ([]!{+(0,-.3)} {{}_3})
        - [r] \cdots - [r]
    \circ ([]!{+(0,-.3)} {{}_{n-2}}) (
        - []!{+(1,.5)}  \circ ([]!{+(0,-.3)} {{}_{n-1}}),
        - []!{+(1,-.5)} \circ ([]!{+(0,-.3)} {{}_n})
)}
$\\
$E_6$&$E_6$ &$6$     &
$\xygraph{
    \circ ([]!{+(0,-.3)}  {{}_1}) - [r]
    \circ ([]!{+(0,-.3)}  {{}_2}) - [r]
    \circ ([]!{+(.3,-.3)} {\hspace{-2em}{}_3}) (
        - [u] \circ ([]!{+(.3,0)}  {{}_6}) (
        - [u] \circ ([]!{+(.3,0)}  {{}_x})
),
        - [r] \circ ([]!{+(0,-.3)} {{}_4})
        - [r] \circ ([]!{+(0,-.3)} {{}_5})
)}
$\\
$E_7$&$E_7$ &$7$     &
$\xygraph{
    \circ ([]!{+(0,-.3)}  {{}_x}) - [r]
    \circ ([]!{+(0,-.3)}  {{}_1}) - [r]
    \circ ([]!{+(0,-.3)}  {{}_2}) - [r]
    \circ ([]!{+(.3,-.3)} {\hspace{-2em}{}_3}) (
        - [u] \circ ([]!{+(.3,0)}  {{}_7}),
        - [r] \circ ([]!{+(0,-.3)} {{}_4})
        - [r] \circ ([]!{+(0,-.3)} {{}_5})
        - [r] \circ ([]!{+(0,-.3)} {{}_6})
)}
$\\
$E_8$&$E_8$ &$8$     &
$\xygraph{
    \circ ([]!{+(0,-.3)}  {{}_1}) - [r]
    \circ ([]!{+(0,-.3)}  {{}_2}) - [r]
    \circ ([]!{+(.3,-.3)} {\hspace{-2em}{}_3}) (
        - [u] \circ ([]!{+(.3,0)}  {{}_8}),
        - [r] \circ ([]!{+(0,-.3)} {{}_4})
        - [r] \circ ([]!{+(0,-.3)} {{}_5})
        - [r] \circ ([]!{+(0,-.3)} {{}_6})
        - [r] \circ ([]!{+(0,-.3)} {{}_7})
        - [r] \circ ([]!{+(0,-.3)} {{}_x}) 
)}
$\\
$F_4$&$F_4$ &$4$         &
$\xygraph{!~:{@{=}|@{}}
    \circ   ([]!{+(0,-.3)} {{}_x}) - [r]
    \circ   ([]!{+(0,-.3)} {{}_1}) - [r]
    \circ   ([]!{+(0,-.3)} {{}_2}) : [r]
    \bullet ([]!{+(0,-.3)} {{}_3}) - [r]
    \bullet ([]!{+(0,-.3)} {{}_4})}
$\\
$G_2$&$G_2$ &$2$         &
$
\xygraph{!~:{@3{-}|@{}}
    \circ   ([]!{+(0,-.3)} {{}_x}) - [r]
    \circ   ([]!{+(0,-.3)} {{}_1}) : [r]
    \bullet ([]!{+(0,-.3)} {{}_2})}
$\\
\end{longtable}
\end{center}
where Group stands for a compact group.
Since $D_2\simeq A_1\oplus A_1$ 
($\mathfrak{so}_4\simeq\mathfrak{su}_2\oplus\mathfrak{su}_2$), 
$D_2=\mathfrak{so}_4$ is not a simple Lie algebra. 
Note that we took the same notation in 
Refs.~\cite{Slansky:1981yr,Georgi:1982jb} to write the (extended) Dynkin
diagrams. 
The Dynkin diagram of a simple algebra is connected, and the Dynkin
diagram of a non-simple algebra is disconnected. Each simple algebra has
simple roots of one or two different lengths. The black circles denote
the shorter roots.

The Dynkin diagrams have the same information of their corresponding
Cartan matrices $A(\mathfrak{g})$ of Lie algebras $\mathfrak{g}$. (See, e.g.,
Ref.~\cite{Dynkin:1957um,Slansky:1981yr,Gilmore:102082} 
in detail.) 
The Cartan matrices of simple Lie algebras are listed in the following
table.
\begin{center}
\begin{longtable}{cccc}
\caption{Cartan matrices}
\label{Table:Cartan-matrices}\\
\hline\hline
Algebra&Group&Rank&Cartan matrix\\\hline\hline
\endfirsthead
\multicolumn{4}{c}{Table~\ref{Table:Cartan-matrices} (continued)}\\\hline\hline
Algebra&Group&Rank&Cartan matrix\\\hline\hline
\endhead
\hline
\endfoot
$A_n=\mathfrak{su}_{n+1}$&$SU(n+1)$ &${}^\forall n$&
$A(A_n)=
\left(
\begin{array}{ccccccc}
2&-1&0&\cdots&0&0&0\\
-1&2&-1&\cdots&0&0&0\\
0&-1&2&\cdots&0&0&0\\
\vdots&\vdots&\vdots&\ddots&\vdots&\vdots&\vdots\\
0&0&0&\cdots&2&-1&0\\
0&0&0&\cdots&-1&2&-1\\
0&0&0&\cdots&0&-1&2\\
\end{array}
\right)
$\\
$B_n=\mathfrak{so}_{2n+1}$&$SO(2n+1)$ &${}^\forall n$&
$A(B_n)=
\left(
\begin{array}{ccccccc}
2&-1&0&\cdots&0&0&0\\
-1&2&-1&\cdots&0&0&0\\
0&-1&2&\cdots&0&0&0\\
\vdots&\vdots&\vdots&\ddots&\vdots&\vdots&\vdots\\
0&0&0&\cdots&2&-1&0\\
0&0&0&\cdots&-1&2&-2\\
0&0&0&\cdots&0&-1&2\\
\end{array}
\right)
$\\
$C_n=\mathfrak{usp}_{2n}$&$USp(2n)$ &${}^\forall n$&
$A(C_n)=
\left(
\begin{array}{ccccccc}
2&-1&0&\cdots&0&0&0\\
-1&2&-1&\cdots&0&0&0\\
0&-1&2&\cdots&0&0&0\\
\vdots&\vdots&\vdots&\ddots&\vdots&\vdots&\vdots\\
0&0&0&\cdots&2&-1&0\\
0&0&0&\cdots&-1&2&-1\\
0&0&0&\cdots&0&-2&2\\
\end{array}
\right)
$\\
$D_n=\mathfrak{so}_{2n}$&$SO(2n)$ &${}^\forall n$&
$A(D_n)=
\left(
\begin{array}{ccccccc}
2&-1&0&\cdots&0&0&0\\
-1&2&-1&\cdots&0&0&0\\
0&-1&2&\cdots&0&0&0\\
\vdots&\vdots&\vdots&\ddots&\vdots&\vdots&\vdots\\
0&0&0&\cdots&2&-1&-1\\
0&0&0&\cdots&-1&2&0\\
0&0&0&\cdots&-1&0&2\\
\end{array}
\right)
$\\
$E_6$&$E_6$ &$6$     &
$A(E_6)=
\left(
\begin{array}{cccccc}
 2&-1& 0& 0& 0& 0\\
-1& 2&-1& 0& 0& 0\\
 0&-1& 2&-1& 0&-1\\
 0& 0&-1& 2&-1& 0\\
 0& 0& 0&-1& 2& 0\\
 0& 0&-1& 0& 0& 2\\
\end{array}
\right)
$\\
$E_7$&$E_7$ &$7$     &
$A(E_7)=
\left(
\begin{array}{ccccccc}
 2&-1& 0& 0& 0& 0& 0\\
-1& 2&-1& 0& 0& 0& 0\\
 0&-1& 2&-1& 0& 0&-1\\
 0& 0&-1& 2&-1& 0& 0\\
 0& 0& 0&-1& 2&-1& 0\\
 0& 0& 0& 0&-1& 2& 0\\
 0& 0&-1& 0& 0& 0& 2\\
\end{array}
\right)
$\\
$E_8$&$E_8$ &$8$     &
$A(E_8)=
\left(
\begin{array}{cccccccc}
 2&-1& 0& 0& 0& 0& 0& 0\\
-1& 2&-1& 0& 0& 0& 0& 0\\
 0&-1& 2&-1& 0& 0& 0&-1\\
 0& 0&-1& 2&-1& 0& 0& 0\\
 0& 0& 0&-1& 2&-1& 0& 0\\
 0& 0& 0& 0&-1& 2&-1& 0\\
 0& 0& 0& 0& 0&-1& 2& 0\\
 0& 0&-1& 0& 0& 0& 0& 2\\
\end{array}
\right)
$\\
$F_4$&$F_4$ &$4$         &
$A(F_4)=
\left(
\begin{array}{cccc}
 2&-1& 0& 0\\
-1& 2&-2& 0\\
 0&-1& 2&-1\\
 0& 0&-1& 2\\
\end{array}
\right)
$\\
$G_2$&$G_2$ &$2$         &
$A(G_2)=
\left(
\begin{array}{cc}
 2&-3\\
-1& 2\\
\end{array}
\right)
$\\
\end{longtable}
\end{center}

The inverse Cartan matrices of simple Lie algebras $\mathfrak{g}$ in the
following table are defined by 
\begin{align}
G(\mathfrak{g})_{ij}:=\left(A(\mathfrak{g})^{-1}\right)_{ij}
\frac{(\alpha_j,\alpha_j)}{2},
\label{Eq::Cartan-matrices-inverse}
\end{align}
The matrices are useful when we calculate the Weyl dimension formulas,
second order Casimir invariants, etc. 
\begin{center}
\begin{longtable}{cc}
\caption{Inverse Cartan matrices}
\label{Table:Cartan-matrices-inverse}\\
\hline\hline
Algebra&Inverse Cartan matrix\\\hline\hline
\endfirsthead
\multicolumn{2}{c}{Table~\ref{Table:Cartan-matrices-inverse} (continued)}\\\hline\hline
Algebra&Inverse Cartan matrix\\\hline\hline
\endhead
\hline
\endfoot
$A_n$&
$G(A_n)=\frac{1}{n+1}
\left(
\begin{array}{cccccc}
1\cdot n    &1\cdot (n-1)&1\cdot (n-2)&\cdots&1\cdot 2    &1\cdot 1\\
1\cdot (n-1)&2\cdot (n-1)&2\cdot (n-2)&\cdots&2\cdot 2    &2\cdot 1\\
1\cdot (n-2)&2\cdot (n-2)&3\cdot (n-2)&\cdots&3\cdot 2    &3\cdot 1\\
\vdots      &\vdots      &\vdots      &\ddots&\vdots      &\vdots  \\
1\cdot 2    &2\cdot 2    &3\cdot 2    &\cdots&(n-1)\cdot 2&(n-1)\cdot 1\\
1\cdot 1    &2\cdot 1    &3\cdot 1    &\cdots&(n-1)\cdot 1&n\cdot 1\\
\end{array}
\right)
$\\
$B_n$&
$G(B_n)=\frac{1}{2}
\left(
\begin{array}{cccccc}
2&2&2&\cdots&2&1\\
2&4&4&\cdots&4&2\\
2&4&6&\cdots&6&3\\
\vdots&\vdots&\vdots&\ddots&\vdots&\vdots  \\
2&4&6&\cdots&2(n-1)&n-1\\
1&2&3&\cdots&n-1&n/2\\
\end{array}
\right)
$\\
$C_n$&
$G(C_n)=\frac{1}{2}
\left(
\begin{array}{cccccc}
1&1&1&\cdots&1&1\\
1&2&2&\cdots&2&2\\
1&2&3&\cdots&3&3\\
\vdots&\vdots&\vdots&\ddots&\vdots&\vdots  \\
1&2&3&\cdots&n-1&n-1\\
1&2&3&\cdots&n-1&n\\
\end{array}
\right)
$\\
$D_n$&
$G(D_n)=\frac{1}{2}
\left(
\begin{array}{ccccccc}
2&2&2&\cdots&2&1&1\\
2&4&4&\cdots&4&2&2\\
2&4&6&\cdots&6&3&3\\
\vdots&\vdots&\vdots&\ddots&\vdots&\vdots&\vdots  \\
2&4&6&\cdots&2(n-2)&n-2&n-2\\
1&2&3&\cdots&n-2&n/2&(n-2)/2\\
1&2&3&\cdots&n-2&(n-2)/2&n/2\\
\end{array}
\right)
$\\
$E_6$&
$G(E_6)=\frac{1}{3}
\left(
\begin{array}{cccccc}
4&5&6&4&2&3\\
5&10&12&8&4&6\\
6&12&18&12&6&9\\
4&8&12&10&5&6\\
2&4&6&5&4&3\\
3&6&9&6&3&6\\
\end{array}
\right)
$\\
$E_7$&
$G(E_7)=\frac{1}{2}
\left(
\begin{array}{ccccccc}
4&6&8&6&4&2&4\\
6&12&16&12&8&4&8\\
8&16&24&18&12&6&12\\
6&12&18&15&10&5&9\\
4&8&12&10&8&4&6\\
2&4&6&5&4&3&3\\
4&8&12&9&6&3&7\\
\end{array}
\right)
$\\
$E_8$&
$G(E_8)=
\left(
\begin{array}{cccccccc}
4&7&10&8&6&4&2&5\\
7&14&20&16&12&8&4&10\\
10&20&30&24&18&12&6&15\\
8&16&24&20&15&10&5&12\\
6&12&18&15&12&8&4&9\\
4&8&12&10&8&6&3&6\\
2&4&6&5&4&3&2&3\\
5&10&15&12&9&6&3&8\\
\end{array}
\right)
$\\
$F_4$&
$G(F_4)=
\left(
\begin{array}{cccc}
2&3&2&1\\
3&6&4&2\\
2&4&3&3/2\\
1&2&3/2&1\\
\end{array}
\right)
$\\
$G_2$&
$G(G_2)=\frac{1}{3}
\left(
\begin{array}{cc}
6&3\\
3&2\\
\end{array}
\right)
$\\
\end{longtable}
\end{center}

Here we list up isomorphisms for classical algebras:
\begin{align}
&A_1\simeq B_1\simeq C_1\simeq D_1,\ \ \ B_2\simeq C_2,\ \ \ 
D_2\simeq A_1\oplus A_1,\ \ \
D_3\simeq A_3.
\end{align}
In some literature, we can also find additional isomorphisms
for exceptional algebras
\begin{align}
E_5\simeq D_5,\ \ \ E_4\simeq A_4,\ \ \ 
E_3\simeq A_1\oplus A_2,\ \ \ E_2\simeq A_1\oplus A_1\ \ \
E_1\simeq A_1.
\end{align}
If we restrict the rank of the classical algebras, we do not need to
define isomorphisms. This is because when we can define $A_n$ $(n\geq 1)$,
$B_n$ $(n\geq 3)$, $C_n$ $(n\geq 2)$, $D_n$ $(n\geq 4)$, there is no
isomorphism.

\subsection{Subalgebras}
\label{Sec:Subalgebras}

Let us first introduce several important notions of Lie subalgebras.
(To understand this part, one may need to know basics of
representations discussed in Sec.~\ref{Sec:Representations-basics}.)

There are two types of subalgebras of simple Lie algebras defined by
E.B.~Dynkin in Refs.~\cite{Dynkin:1957um,Dynkin:1957ek}.
One is an $R$-subalgebra; the other is an $S$-subalgebra.
An $R$-subalgebra is a subalgebra that contains several regular
subalgebras, where all $R$-subalgebras is obtained by deleting dots from
(extended) Dynkin diagrams.
An $S$-subalgebra is a subalgebra that is not any $R$-subalgebra.
It can be found by using dimensions and type of irreducible
representations of a Lie algebra and its subalgebra. 
(Some $S$-subalgebras can be found by using (extended) Dynkin diagrams
but it is difficult to find almost all $S$-subalgebras. The rank of an
$S$-subalgebra is always smaller than the rank of its original Lie
algebra.) 

Another important concept is a maximal subalgebra; a maximal subalgebra
$\mathfrak{h}$ of a subalgebra $\mathfrak{g}$ is a subalgebra such that
there in no any subalgebra $\mathfrak{k}$ satisfies 
$\mathfrak{h}\subset\mathfrak{k}\subset\mathfrak{g}$.
Here, we only discuss maximal Lie subalgebras because non-maximal
subalgebras can be obtained by using maximal subalgebras; 
e.g., two maximal subalgebras 
$D_5=\mathfrak{so}_{10}\supset A_4\oplus\mathfrak{u}_1=
\mathfrak{su}_5\oplus\mathfrak{u}_1$ 
and $A_4=\mathfrak{su}_{5}\supset A_2\oplus A_1\oplus\mathfrak{u}_1
=\mathfrak{su}_3\oplus\mathfrak{su}_2\oplus\mathfrak{u}_1$
lead to non-maximal subalgebra
$D_5\supset A_2\oplus A_1\oplus\mathfrak{u}_1\oplus\mathfrak{u}_1=
\mathfrak{su}_3\oplus\mathfrak{su}_2\oplus\mathfrak{u}_1\oplus\mathfrak{u}_1$.

Here we summarize how to obtain the maximal subalgebras $\mathfrak{h}$
of the simple Lie algebra $\mathfrak{g}$:
\begin{enumerate}
\item Maximal $R$-subalgebras, which are maximal and regular
      subalgebras, can be found by using extended Dynkin diagrams in
      Table~\ref{Table:Dynkin-diagrams} and the following rules:
\begin{itemize}
\item[(R-1)] Extended Dynkin diagrams whose one node is dropped are
	     maximal semi-simple regular subalgebras,
	     (Note that for $A_n$, the removed extended Dynkin diagram
	     $A_n$ whose one node is dropped is the same as the Dynkin
	     diagram of $A_n$, so we cannot obtain useful information.)
\item[(R-2)] Extended Dynkin diagrams of $A_n=\mathfrak{su}_{n+1}$ whose
	   two nodes are dropped are maximal non-semi-simple regular
	   subalgebras. In other words,
	     Dynkin diagrams of $A_n=\mathfrak{su}_{n+1}$ whose
	   one node is dropped are maximal non-semi-simple regular
	   subalgebras. More explicitly,
	     $A_n\supset A_{m}\oplus A_{n-m-1}\oplus\mathfrak{u}_1$ 
	     $(m\leq [n/2])$. Its corresponding Dynkin diagram is 
\begin{align}
&\underset{A_n}
{\xygraph{
    \circ ([]!{+(0,-.3)} {{}_1}) - [r]
    \circ ([]!{+(0,-.3)} {{}_2}) - [r]
    \cdots - [r]
    \circ ([]!{+(0,-.3)} {{}_{m}}) - [r] 
    \circ ([]!{+(0,-.3)} {{}_{m+1}}) - [r] 
    \circ ([]!{+(0,-.3)} {{}_{m+2}}) - [r] 
    \cdots - [r]
    \circ ([]!{+(0,-.3)} {{}_n})}
}\nonumber\\
\Rightarrow
&
\underset{A_{m}\oplus A_{n-m-1}\oplus\mathfrak{u}_1}
{\xygraph{
    \circ ([]!{+(0,-.3)} {{}_1}) - [r]
    \circ ([]!{+(0,-.3)} {{}_2}) - [r]
    \cdots - [r]
    \circ ([]!{+(0,-.3)} {{}_{m}}) - [r] 
    {\circ\hspace{-0.645em}\times} ([]!{+(0,-.3)} {{}_{m+1}}) - [r] 
    \circ ([]!{+(0,-.3)} {{}_{m+2}}) - [r] 
    \cdots - [r]
    \circ ([]!{+(0,-.3)} {{}_n})}
},
\end{align}
	     where the symbol $\times$ on the node $\circ$ stands for
	     removing $\circ$ in the Dynkin diagram.
\item[(R-3)] Extended Dynkin diagrams of
	     $B_n=\mathfrak{so}_{2n+1}$,
	     $C_n=\mathfrak{usp}_{2n}$,
	     $D_n=\mathfrak{so}_{2n}$,
	     $E_6$, and $E_7$ 
	     whose appropriate two nodes are dropped are maximal
	     non-semi-simple regular subalgebras.
	     More explicitly, 
	     $B_n\supset B_{n-1}\oplus\mathfrak{u}_1$;
	     $C_n\supset A_{n-1}\oplus\mathfrak{u}_1$;
	     $D_n\supset A_{n-1}\oplus\mathfrak{u}_1$;
	     $D_n\supset D_{n-1}\oplus\mathfrak{u}_1$;
	     $E_6\supset D_5\oplus\mathfrak{u}_1$; and 
	     $E_7\supset E_6\oplus\mathfrak{u}_1$, respectively.
	     They are related with a parity symmetry of their
	     extended Dynkin diagrams.
	     Their corresponding Dynkin diagrams are given by
\begin{align}
&\underset{B_n}
{\xygraph{!~:{@{=}|@{}}
    \circ ([]!{+(0,-.3)} {{}_2}) (
        - []!{+(-1,.5)}  \circ ([]!{+(0,-.3)} {{}_1}),
        - []!{+(-1,-.5)} \circ ([]!{+(0,-.3)} {{}_x}))
        - [r]
    \circ ([]!{+(0,-.3)} {{}_3}) 
        - [r] \cdots - [r]
    \circ   ([]!{+(0,-.3)} {{}_{n-1}}) : [r]
    \bullet ([]!{+(0,-.3)} {{}_n})}
}
\Rightarrow
\underset{B_{n-1}\oplus\mathfrak{u}_1}
{\xygraph{!~:{@{=}|@{}}
    \circ ([]!{+(0,-.3)} {{}_2}) (
        - []!{+(-1,.5)}  {\circ\hspace{-0.645em}\times} ([]!{+(0,-.3)} {{}_1}),
        - []!{+(-1,-.5)} {\circ\hspace{-0.645em}\times} ([]!{+(0,-.3)} {{}_x}))
        - [r]
    \circ ([]!{+(0,-.3)} {{}_3}) 
        - [r] \cdots - [r]
    \circ   ([]!{+(0,-.3)} {{}_{n-1}}) : [r]
    \bullet ([]!{+(0,-.3)} {{}_n})}
},
\end{align}

\begin{align}
&\underset{C_n}
{\xygraph{!~:{@{=}|@{}}
    \circ   ([]!{+(0,-.3)} {{}_x}) : [r]
    \bullet ([]!{+(0,-.3)} {{}_1}) - [r]
    \bullet ([]!{+(0,-.3)} {{}_2}) - [r] \cdots - [r]
    \bullet ([]!{+(0,-.3)} {{}_{n-1}}) : [r]
    \circ   ([]!{+(0,-.3)} {{}_n})}
}
\Rightarrow
\underset{A_{n-1}\oplus\mathfrak{u}_1}
{\xygraph{!~:{@{=}|@{}}
    {\circ\hspace{-0.645em}\times}   ([]!{+(0,-.3)} {{}_x}) : [r]
    \bullet ([]!{+(0,-.3)} {{}_1}) - [r]
    \bullet ([]!{+(0,-.3)} {{}_2}) - [r] \cdots - [r]
    \bullet ([]!{+(0,-.3)} {{}_{n-1}}) : [r]
    {\circ\hspace{-0.645em}\times}   ([]!{+(0,-.3)} {{}_n})}
},
\end{align}

\begin{align}
&\underset{D_n}
{\xygraph{
    \circ ([]!{+(0,-.3)} {{}_2}) (
        - []!{+(-1,.5)}  \circ ([]!{+(0,-.3)} {{}_1}),
        - []!{+(-1,-.5)} \circ ([]!{+(0,-.3)} {{}_x}))
        - [r] \cdots - [r]
    \circ ([]!{+(0,-.3)} {{}_{n-2}}) (
        - []!{+(1,.5)}  \circ ([]!{+(0,-.3)} {{}_{n-1}}),
        - []!{+(1,-.5)} \circ ([]!{+(0,-.3)} {{}_n})
)}
}
\Rightarrow
\left\{
\begin{array}{c}
\underset{D_{n-1}\oplus\mathfrak{u}_1}
{\xygraph{
    \circ ([]!{+(0,-.3)} {{}_2}) (
        - []!{+(-1,.5)}  {\circ\hspace{-0.645em}\times} ([]!{+(0,-.3)} {{}_1}),
        - []!{+(-1,-.5)} {\circ\hspace{-0.645em}\times} ([]!{+(0,-.3)} {{}_x}))
        - [r] \cdots - [r]
    \circ ([]!{+(0,-.3)} {{}_{n-2}}) (
        - []!{+(1,.5)}  \circ ([]!{+(0,-.3)} {{}_{n-1}}),
        - []!{+(1,-.5)} \circ ([]!{+(0,-.3)} {{}_n})
)}
}\\
\underset{A_{n-1}\oplus\mathfrak{u}_1}
{\xygraph{
    \circ ([]!{+(0,-.3)} {{}_2}) (
        - []!{+(-1,.5)}  \circ ([]!{+(0,-.3)} {{}_1}),
        - []!{+(-1,-.5)} {\circ\hspace{-0.645em}\times} ([]!{+(0,-.3)} {{}_x}))
        - [r] \cdots - [r]
    \circ ([]!{+(0,-.3)} {{}_{n-2}}) (
        - []!{+(1,.5)}  \circ ([]!{+(0,-.3)} {{}_{n-1}}),
        - []!{+(1,-.5)} {\circ\hspace{-0.645em}\times} ([]!{+(0,-.3)} {{}_n})
)}
}\\
\end{array}
\right.,
\end{align}

\begin{align}
&\underset{E_6}
{\xygraph{
     ([]!{+(0,-.3)}  {{}_1}) - [r]
    \circ ([]!{+(0,-.3)}  {{}_2}) - [r]
    \circ ([]!{+(.3,-.3)} {\hspace{-2em}{}_3}) (
        - [u] \circ ([]!{+(.3,0)}  {{}_6}) (
        - [u] \circ ([]!{+(.3,0)}  {{}_x})
),
        - [r] \circ ([]!{+(0,-.3)} {{}_4})
        - [r] \circ ([]!{+(0,-.3)} {{}_5})
)}
}
\Rightarrow
\underset{D_5\oplus\mathfrak{u}_1}
{\xygraph{
    {\circ\hspace{-0.645em}\times} ([]!{+(0,-.3)}  {{}_1}) - [r]
    \circ ([]!{+(0,-.3)}  {{}_2}) - [r]
    \circ ([]!{+(.3,-.3)} {\hspace{-2em}{}_3}) (
        - [u] \circ ([]!{+(.3,0)}  {{}_6}) (
        - [u] {\circ\hspace{-0.645em}\times} ([]!{+(.3,0)}  {{}_x})
),
        - [r] \circ ([]!{+(0,-.3)} {{}_4})
        - [r] \circ ([]!{+(0,-.3)} {{}_5})
)}
},
\end{align}

\begin{align}
&\underset{E_7}
{\xygraph{
    \circ ([]!{+(0,-.3)}  {{}_x}) - [r]
    \circ ([]!{+(0,-.3)}  {{}_1}) - [r]
    \circ ([]!{+(0,-.3)}  {{}_2}) - [r]
    \circ ([]!{+(.3,-.3)} {\hspace{-2em}{}_3}) (
        - [u] \circ ([]!{+(.3,0)}  {{}_7}),
        - [r] \circ ([]!{+(0,-.3)} {{}_4})
        - [r] \circ ([]!{+(0,-.3)} {{}_5})
        - [r] \circ ([]!{+(0,-.3)} {{}_6})
)}
}
\nonumber\\
\Rightarrow
&
\underset{E_6\oplus\mathfrak{u}_1}
{\xygraph{
    {\circ\hspace{-0.645em}\times} ([]!{+(0,-.3)}  {{}_x}) - [r]
    \circ ([]!{+(0,-.3)}  {{}_1}) - [r]
    \circ ([]!{+(0,-.3)}  {{}_2}) - [r]
    \circ ([]!{+(.3,-.3)} {\hspace{-2em}{}_3}) (
        - [u] \circ ([]!{+(.3,0)}  {{}_7}),
        - [r] \circ ([]!{+(0,-.3)} {{}_4})
        - [r] \circ ([]!{+(0,-.3)} {{}_5})
        - [r] {\circ\hspace{-0.645em}\times} ([]!{+(0,-.3)} {{}_6})
)}
}.
\end{align}
\end{itemize}
\item Maximal $S$-subalgebras, which are maximal and special
      subalgebras, can be found by the following rules: 
\begin{itemize}
\item[(S-1)] For classical Lie algebras $\mathfrak{g}$,
	     several maximal non-simple $S$-subalgebras $\mathfrak{h}$ of
	     $\mathfrak{g}$ are listed below:
\begin{center}
\begin{longtable}{ccllc}
\caption{Maximal $S$-subalgebras of classical algebras (1)}
\label{Table:Maximal-S-sub-classical-1}\\
\hline\hline
Rank&Algebra $\mathfrak{g}$&\multicolumn{2}{l}{Maximal $S$-Subalgebra $\mathfrak{h}$}&Type\\\hline\hline
\endfirsthead
\multicolumn{4}{c}{Table~\ref{Table:Maximal-S-sub-classical-1} (continued)}\\\hline\hline
Rank&Algebra $\mathfrak{g}$&\multicolumn{2}{l}{Maximal $S$-Subalgebra $\mathfrak{h}$}&Type\\\hline\hline
\endhead
\hline
\endfoot
$mn-1$ &$\mathfrak{su}_{mn}$&$\mathfrak{su}_m\oplus\mathfrak{su}_n$
&$(m,n\geq 2)$&$(S)$\\
$\left[\frac{mn}{2}\right]$
&$\mathfrak{so}_{mn}$&$\mathfrak{so}_m\oplus\mathfrak{so}_n$
&$(m,n\geq 3; m,n\not=4)$ &$(S)$\\
$2mn$ &$\mathfrak{so}_{4mn}$&$\mathfrak{usp}_{2m}\oplus\mathfrak{usp}_{2n}$
&$(m,n\geq 1)$&$(S)$\\
$mn$ &$\mathfrak{usp}_{2mn}$&$\mathfrak{so}_m\oplus\mathfrak{usp}_{2n}$
&$(m\geq 3,m\not=4, n\geq 1\ \mbox{or}\ m=4, n=1)$ &$(S)$\\
\end{longtable}
\end{center}
	     where $\mathfrak{usp}_{2}\simeq\mathfrak{su}_{2}$,
	     $\mathfrak{so}_{3}\simeq\mathfrak{su}_{2}$,
	     $\mathfrak{so}_{4}\simeq\mathfrak{usp}_{2}\oplus\mathfrak{usp}_{2}\simeq\mathfrak{su}_{2}\oplus\mathfrak{su}_{2}$, and
	     $\mathfrak{so}_{5}\simeq\mathfrak{usp}_{4}$.
	     (See Theorems 1.3 and 1.4 in
	     Ref.~\cite{Dynkin:1957ek} and their proof.)
\item[(S-2)] For classical Lie algebras $\mathfrak{g}$,
	     maximal $S$-subalgebras $\mathfrak{h}$ of
	     $\mathfrak{g}$ can be found as below.
	     If a Lie algebra $\mathfrak{h}$ has an $n$ dimensional
	     complex representation, it is a maximal subalgebra of
	     a Lie algebra $A_{n}=\mathfrak{su}_{n+1}$.
	     If a Lie algebra $\mathfrak{h}$ has an $n$ dimensional
	     real representation, it is a maximal subalgebra of
	     $B_{k}=\mathfrak{so}_{2k+1}$ for $n=2k+1(k\in\mathbb{Z})$
	     and
	     $D_{k}=\mathfrak{so}_{2k}$ for  $n=2k (k\in\mathbb{Z})$. 
	     If a Lie algebra $\mathfrak{h}$ has a
	     $2k (k\in\mathbb{Z})$
	     dimensional pseudo-real representation, it is a maximal
	     subalgebra of 
	     $C_{k}=\mathfrak{usp}_{2k}$,
	     where real, pseudo-real, and complex representations are
	     listed in
	     Tables~\ref{Table:Complex-representations-1} and
	     \ref{Table:Self-conjugate-representations}.
	     (For its proof, see e.g.,Ref.~\cite{Dynkin:1957ek}.)
	     Note that there are several exceptions.
	     The above examples are summarized in the following table:
\begin{center}
\begin{longtable}{ccllc}
\caption{Maximal $S$-subalgebras of classical algebras (2)}
\label{Table:Maximal-S-sub-classical-2}\\
\hline\hline
Rank&Algebra $\mathfrak{g}$&\multicolumn{2}{l}{Maximal $S$-subalgebra $\mathfrak{h}$}&Type\\\hline\hline
\endfirsthead
\multicolumn{5}{c}{Table~\ref{Table:Maximal-S-sub-classical-2} (continued)}\\\hline\hline
Rank&Algebra $\mathfrak{g}$&\multicolumn{2}{l}{Maximal $S$-subalgebra
 $\mathfrak{h}$}&Type\\\hline\hline
\endhead
\hline
\endfoot
$2n$   &$A_{2n}$  &$B_n$ &$(n\geq 2)$&$(S)$\\
$2n-1$ &$A_{2n-1}$&$C_n;D_n$ &$(n\geq 2)$&$(S)$\\
$\frac{(n-1)(n+2)}{2}$ 
&$A_{\frac{(n-1)(n+2)}{2}}$  &$A_n$ &$(n\geq 3)$&$(S)$\\
$\frac{n(n+3)}{2}$&$A_{\frac{n(n+3)}{2}}$  &$A_n$ &$(n\geq 2)$&$(S)$\\
$15$  &$A_{15}$  &$D_5$ &&$(S)$\\
$26$  &$A_{26}$  &$E_6$ &&$(S)$\\
$3$   &$B_{3}$   &$G_2$ &&$(S)$\\
$7$   &$C_{7}$   &$C_3$ &&$(S)$\\
$10$  &$C_{10}$  &$A_5$ &&$(S)$\\
$16$  &$C_{16}$  &$D_6$ &&$(S)$\\
$28$  &$C_{28}$  &$E_7$ &&$(S)$\\
$n+1$ &$D_{n+1}$ &$B_n$ &$(n\geq 3)$&$(S)$\\
$m+n+1$&${D}_{m+n+1}$&${B}_m\oplus{B}_n$ &$(m+n\geq 4)$&$(S)$\\
$7$   &$D_7$     &$G_2;C_3$ &&$(S)$\\
$8$   &$D_8$     &$B_4$ &&$(S)$\\
$13$  &$D_{13}$  &$F_4$ &&$(S)$\\
\end{longtable}
\end{center}
	     where $m,n\in\mathbb{Z}_{\geq 1}$, and 
	     the table includes several duplications.
	     This table is a part of Table 5 in
	     Ref.~\cite{Dynkin:1957ek}.
\item[(S-3)] All the maximal $S$-subalgebras $\mathfrak{h}$ of the
	     exceptional simple Lie algebras $\mathfrak{g}$ are listed
	     below. It is shown by
	     E.B.~Dynkin in Ref.~\cite{Dynkin:1957um}.
\begin{center}
\begin{longtable}{cclc}
\caption{Maximal $S$-subalgebras of exceptional algebras}
\label{Table:Maximal-S-sub-exceptional}\\
\hline\hline
Rank&Algebra $\mathfrak{g}$&Maximal $S$-Subalgebra $\mathfrak{h}$&Type\\\hline\hline
\endfirsthead
\multicolumn{4}{c}{Table~\ref{Table:Maximal-S-sub-exceptional} (continued)}\\\hline\hline
Rank&Algebra $\mathfrak{g}$&Maximal $S$-Subalgebra $\mathfrak{h}$&Type\\\hline\hline
\endhead
\hline
\endfoot
$6$    &$E_6$&$A_1^9;\ G_2^3;\ C_4^1;\ G_2^1\oplus A_2^2{}'';\ F_4^1$ &$(S)$\\
$7$    &$E_7$&$A_1^{399};\ A_1^{231};\ A_2^{21};\ G_2^1\oplus C_3^1{}'';\
	      F_4^1\oplus A_1^3{}'';\ G_2^2\oplus A_1^7;\
              A_1^{24}\oplus A_1^{15}$ &$(S)$\\
$8$    &$E_8$&$A_1^{1240};\ A_1^{760};\ A_1^{520};\ 
               G_2^1\oplus F_4^1;\ A_2^6{}'\oplus A_1^{16};\ B_2^{12}$ 
              &$(S)$\\
$4$    &$F_4$&$A_1^{156};\ G_2^1\oplus A_1^8$ &$(S)$\\
$2$    &$G_2$&$A_1^{28}$&$(S)$\\
\end{longtable}
\end{center}
	     where the superscript e.g., $9$ of $A_1^9$ stands for the
	     ratio of the Dynkin indices of an algebra $\mathfrak{g}$
	     and its subalgebra $\mathfrak{h}$. It is called the
	     embedding index of $\mathfrak{h}\subset\mathfrak{g}$
	     denoted as $j_{\mathfrak{h}\subset\mathfrak{g}}$.
	     (For its explanation, see e.g.,
	     Refs.~\cite{Cahn:1985wk,Fuchs:1992}).
\end{itemize}
\end{enumerate}

Here we summarize maximal subalgebras of classical and exceptional
simple Lie algebras, where their rank is up to 16 and Type $(R)$ and
$(S)$ stand for $R$- and $S$-subalgebras, respectively. 
(A part of the following table is given in Table 14 and 15 in
Ref.~\cite{Slansky:1981yr} up to Rank-8.)
\begin{center}
\begin{longtable}{crcp{10cm}c}
\caption{Maximal subalgebras}
\label{Table:Maximal-subalgebra}\\
\hline\hline
Rank&\multicolumn{2}{l}{Algebra $\mathfrak{g}$}&Maximal subalgebras $\mathfrak{h}$&Type\\\hline\hline
\endfirsthead
\multicolumn{4}{c}{Table~\ref{Table:Maximal-subalgebra} (continued)}\\
\hline
Rank&\multicolumn{2}{l}{Algebra $\mathfrak{g}$}&Maximal subalgebras $\mathfrak{h}$&Type\\\hline\hline
\endhead
\hline
\endfoot
$1$ &$\mathfrak{su}_2$&$\supset$&$\mathfrak{u}_1$&$(R)$\\
    &\multicolumn{4}{l}{($\mathfrak{su}_2,\mathfrak{so}_3,\mathfrak{usp}_2$ are all isomorphic.)}\\
\hline
$2$ &$\mathfrak{su}_3$&$\supset$&$\mathfrak{su}_2\oplus\mathfrak{u}_1$&$(R)$\\
    &       &$\supset$&$\mathfrak{su}_2$&$(S)$\\
    &$\mathfrak{usp}_4$&$\supset$&$\mathfrak{su}_2\oplus\mathfrak{su}_2;\
     \mathfrak{su}_2\oplus\mathfrak{u}_1$&$(R)$\\
    &       &$\supset$&$\mathfrak{su}_2$&$(S)$\\
    &\multicolumn{4}{l}
{($\mathfrak{so}_5$ is isomorphic to $\mathfrak{usp}_4$. 
$\mathfrak{so}_4\simeq\mathfrak{su}_2\oplus\mathfrak{su}_2$.)}\\
    &$G_2$  &$\supset$&$\mathfrak{su}_3;\
    \mathfrak{su}_2\oplus\mathfrak{su}_2$&$(R)$\\
\end{longtable}
\end{center}
(Note that Table~\ref{Table:Maximal-subalgebra} in the first version of
this paper had some mistakes. There were some non-maximal
$S$-subalgebras, e.g., 
$\mathfrak{so}_{4n}\supset\mathfrak{su}_{2}\oplus\mathfrak{su}_2\oplus\mathfrak{so}_{n}$,
which is not maximal because 
$\mathfrak{so}_{4n}\supset\mathfrak{su}_{2}\oplus\mathfrak{usp}_{2n}
\supset\mathfrak{su}_{2}\oplus\mathfrak{su}_2\oplus\mathfrak{so}_{n}
\simeq \mathfrak{so}_{4}\oplus\mathfrak{so}_{n}$
$(n\in\mathbb{Z}_{\geq 3})$,
where Refs.~\cite{Mckay:1977,McKay:1981,Slansky:1981yr} also contain a
non-maximal $S$-subalgebra
$\mathfrak{so}_{12}\supset\mathfrak{su}_{2}\oplus\mathfrak{su}_{2}\oplus\mathfrak{su}_{2}$.
In addition, Table~\ref{Table:Maximal-subalgebra} in the first version of
this paper have not contained some maximal $S$-subalgebras, e.g.,
$\mathfrak{usp}_{2n}\supset\mathfrak{su}_{2}\oplus\mathfrak{so}_{n}$
$(n\in\mathbb{Z}_{\geq 3})$,
$\mathfrak{usp}_{14}\supset\mathfrak{usp}_{6}$.
Refs.~\cite{Mckay:1977,McKay:1981,Slansky:1981yr} also do not
contain a maximal $S$-subalgebra
$\mathfrak{usp}_{14}\supset\mathfrak{usp}_{6}$, which
is listed in Ref.~\cite{Dynkin:1957ek}.)

Let us check the above table by using several examples.
First, we consider a Lie algebra $A_4=\mathfrak{su}_5$.
From the above rule (R-2), by using Dynkin diagrams, we find its
$R$-subalgebras $\mathfrak{su}_4\oplus\mathfrak{u}_1$ and 
$\mathfrak{su}_3\oplus\mathfrak{su}_2\oplus\mathfrak{u}_1$:
\begin{align}
\underset{\mathfrak{su}_5}
{\xygraph{
    \circ ([]!{+(0,-.3)}  {{}_1}) - [r]
    \circ ([]!{+(0,-.3)}  {{}_2}) - [r]
    \circ ([]!{+(0,-.3)}  {{}_3}) - [r] 
    \circ ([]!{+(0,-.3)}  {{}_4})
}}
&\Rightarrow
\left\{
\begin{array}{l}
\underset{\mathfrak{su}_4\oplus\mathfrak{u}_1}
{\xygraph{
    \circ ([]!{+(0,-.3)}  {{}_1}) - [r]
    \circ ([]!{+(0,-.3)}  {{}_2}) - [r]
    \circ ([]!{+(0,-.3)}  {{}_3}) - [r] 
    {\circ\hspace{-0.645em}\times} ([]!{+(0,-.3)}  {{}_4})
}}\\
\underset{\mathfrak{su}_3\oplus\mathfrak{su}_2\oplus\mathfrak{u}_1}
{\xygraph{
    \circ ([]!{+(0,-.3)}  {{}_1}) - [r]
    \circ ([]!{+(0,-.3)}  {{}_2}) - [r]
    {\circ\hspace{-0.645em}\times} ([]!{+(0,-.3)}  {{}_3}) - [r] 
    \circ ([]!{+(0,-.3)}  {{}_4})
}}
\end{array}
\right\}.
\end{align}
In addition, from the rule (S-2), we find its $S$-subalgebra
$A_4=\mathfrak{su}_5\supset B_2=\mathfrak{so}_5\simeq\mathfrak{usp}_4$.
Second, we consider a Lie algebra $B_4=\mathfrak{so}_9$.
From the above rule (R-2), by using Dynkin diagrams, we find its
$R$-subalgebras $\mathfrak{so}_8$, 
$\mathfrak{su}_2\oplus\mathfrak{su}_2\oplus\mathfrak{usp}_4$,
$\mathfrak{su}_4\oplus\mathfrak{su}_2$, and
$\mathfrak{so}_7\oplus\mathfrak{u}_1$:
\begin{align}
\underset{\mathfrak{so}_9}
{\xygraph{!~:{@{=}|@{}}
    \circ ([]!{+(0,-.3)} {{}_2}) (
        - []!{+(-1,.5)}  \circ ([]!{+(0,-.3)} {{}_1}),
        - []!{+(-1,-.5)} \circ ([]!{+(0,-.3)} {{}_x}))
        - [r]
    \circ   ([]!{+(0,-.3)} {{}_3}) : [r]
    \bullet ([]!{+(0,-.3)} {{}_4})}
}
&\Rightarrow
\left\{
\begin{array}{c}
\underset{\mathfrak{so}_8}
{\xygraph{!~:{@{=}|@{}}
    \circ ([]!{+(0,-.3)} {{}_2}) (
        - []!{+(-1,.5)}  \circ ([]!{+(0,-.3)} {{}_1}),
        - []!{+(-1,-.5)} \circ ([]!{+(0,-.3)} {{}_x}))
        - [r]
    \circ   ([]!{+(0,-.3)} {{}_3}) : [r]
    {\bullet\hspace{-0.645em}\times} ([]!{+(0,-.3)} {{}_4})}
}\\
\underset{\mathfrak{su}_2\oplus\mathfrak{su}_2\oplus\mathfrak{usp}_4}
{\xygraph{!~:{@{=}|@{}}
    {\circ\hspace{-0.645em}\times} ([]!{+(0,-.3)} {{}_2}) (
        - []!{+(-1,.5)}  \circ ([]!{+(0,-.3)} {{}_1}),
        - []!{+(-1,-.5)} \circ ([]!{+(0,-.3)} {{}_x}))
        - [r]
    \circ   ([]!{+(0,-.3)} {{}_3}) : [r]
    \bullet ([]!{+(0,-.3)} {{}_4})}
}\\
\underset{\mathfrak{su}_4\oplus\mathfrak{su}_2}
{\xygraph{!~:{@{=}|@{}}
    \circ ([]!{+(0,-.3)} {{}_2}) (
        - []!{+(-1,.5)}  \circ ([]!{+(0,-.3)} {{}_1}),
        - []!{+(-1,-.5)} \circ ([]!{+(0,-.3)} {{}_x}))
        - [r]
    {\circ\hspace{-0.645em}\times}   ([]!{+(0,-.3)} {{}_3}) : [r]
    \bullet ([]!{+(0,-.3)} {{}_4})}
}\\
\underset{\mathfrak{so}_7\oplus\mathfrak{u}_1}
{\xygraph{!~:{@{=}|@{}}
    \circ ([]!{+(0,-.3)} {{}_2}) (
        - []!{+(-1,.5)}  {\circ\hspace{-0.645em}\times} ([]!{+(0,-.3)} {{}_1}),
        - []!{+(-1,-.5)} {\circ\hspace{-0.645em}\times} ([]!{+(0,-.3)} {{}_x}))
        - [r]
    \circ   ([]!{+(0,-.3)} {{}_3}) : [r]
    \bullet ([]!{+(0,-.3)} {{}_4})}
}\\
\end{array}
\right\}.
\end{align}
From the rule (S-1), we find its $S$-subalgebras
$\mathfrak{su}_2\oplus\mathfrak{su}_2$ and
$\mathfrak{su}_2$.
Third, we consider a Lie algebra $C_4=\mathfrak{usp}_8$.
From the above rule (R-2), by using Dynkin diagrams, we find its
$R$-subalgebras
$\mathfrak{su}_4\oplus\mathfrak{u}_1$,
$\mathfrak{su}_2\oplus\mathfrak{usp}_6$, and  
$\mathfrak{usp}_4\oplus\mathfrak{usp}_4$:
\begin{align}
\underset{\mathfrak{usp}_8}
{\xygraph{!~:{@{=}|@{}}
    \circ   ([]!{+(0,-.3)} {{}_x}) : [r]
    \bullet ([]!{+(0,-.3)} {{}_1}) - [r]
    \bullet ([]!{+(0,-.3)} {{}_2}) - [r] 
    \bullet ([]!{+(0,-.3)} {{}_3}) : [r]
    \circ   ([]!{+(0,-.3)} {{}_4})}
}
&\Rightarrow
\left\{
\begin{array}{c}
\underset{\mathfrak{su}_4\oplus\mathfrak{u}_1}
{\xygraph{!~:{@{=}|@{}}
    {\circ\hspace{-0.645em}\times}    ([]!{+(0,-.3)} {{}_x}) : [r]
    \bullet ([]!{+(0,-.3)} {{}_1}) - [r]
    \bullet ([]!{+(0,-.3)} {{}_2}) - [r] 
    \bullet ([]!{+(0,-.3)} {{}_3}) : [r]
    {\circ\hspace{-0.645em}\times}    ([]!{+(0,-.3)} {{}_4})}
}\\
\underset{\mathfrak{su}_2\oplus\mathfrak{usp}_6}
{\xygraph{!~:{@{=}|@{}}
    \circ   ([]!{+(0,-.3)} {{}_x}) : [r]
    {\bullet\hspace{-0.645em}\times} ([]!{+(0,-.3)} {{}_1}) - [r]
    \bullet ([]!{+(0,-.3)} {{}_2}) - [r] 
    \bullet ([]!{+(0,-.3)} {{}_3}) : [r]
    \circ   ([]!{+(0,-.3)} {{}_4})}
}\\
\underset{\mathfrak{usp}_4\oplus\mathfrak{usp}_4}
{\xygraph{!~:{@{=}|@{}}
    \circ   ([]!{+(0,-.3)} {{}_x}) : [r]
    \bullet ([]!{+(0,-.3)} {{}_1}) - [r]
    {\bullet\hspace{-0.645em}\times} ([]!{+(0,-.3)} {{}_2}) - [r] 
    \bullet ([]!{+(0,-.3)} {{}_3}) : [r]
    \circ   ([]!{+(0,-.3)} {{}_4})}
}\\
\end{array}
\right\}.
\end{align}
From the rule (S-1), we find its $S$-subalgebras
$\mathfrak{su}_2$ and
$\mathfrak{su}_2\oplus\mathfrak{su}_2\oplus\mathfrak{su}_2$.
Forth, we consider a Lie algebra $D_4=\mathfrak{so}_8$.
From the above rule (R-2), by using Dynkin diagrams, we find its
$R$-subalgebras
$\mathfrak{su}_2\oplus\mathfrak{su}_2\oplus\mathfrak{su}_2
\oplus\mathfrak{su}_2$ and $\mathfrak{su}_4\oplus\mathfrak{u}_1$:
\begin{align}
\underset{\mathfrak{so}_8}
{\xygraph{
    \circ ([]!{+(0,-.3)} {{}_2}) 
       (
        - []!{+(-1,.5)}  \circ ([]!{+(0,-.3)} {{}_1}),
        - []!{+(-1,-.5)} \circ ([]!{+(0,-.3)} {{}_x}))
       (
        - []!{+(1,.5)}  \circ ([]!{+(0,-.3)} {{}_3}),
        - []!{+(1,-.5)} \circ ([]!{+(0,-.3)} {{}_4})
)}}
&\Rightarrow
\left\{
\begin{array}{c}
\underset{\mathfrak{su}_2\oplus\mathfrak{su}_2\oplus
\mathfrak{su}_2\oplus\mathfrak{su}_2}
{\xygraph{
    {\circ\hspace{-0.645em}\times} ([]!{+(0,-.3)} {{}_2}) 
       (
        - []!{+(-1,.5)}  \circ ([]!{+(0,-.3)} {{}_1}),
        - []!{+(-1,-.5)} \circ ([]!{+(0,-.3)} {{}_x}))
       (
        - []!{+(1,.5)}  \circ ([]!{+(0,-.3)} {{}_3}),
        - []!{+(1,-.5)} \circ ([]!{+(0,-.3)} {{}_4})
)}}\\
\underset{\mathfrak{su}_4\oplus\mathfrak{u}_1}
{\xygraph{
      \circ ([]!{+(0,-.3)} {{}_2}) 
       (
        - []!{+(-1,.5)}  \circ ([]!{+(0,-.3)} {{}_1}),
        - []!{+(-1,-.5)} {\circ\hspace{-0.645em}\times} ([]!{+(0,-.3)} {{}_x}))
       (
        - []!{+(1,.5)}  \circ ([]!{+(0,-.3)} {{}_3}),
        - []!{+(1,-.5)}  {\circ\hspace{-0.645em}\times} ([]!{+(0,-.3)} {{}_4})
)}}\\
\end{array}
\right\}.
\end{align}
From the rules (S-1) and (S-2), we find its $S$-subalgebras
$\mathfrak{su}_3$, 
$\mathfrak{su}_2\oplus\mathfrak{usp}_4$, and 
$\mathfrak{so}_7$.
Finally, we consider a Lie algebra $F_4$.
From the above rule (R-2), by using Dynkin diagrams, we find its
$R$-subalgebras
$\mathfrak{so}_9$,
$\mathfrak{su}_4\oplus\mathfrak{su}_2$,
$\mathfrak{su}_3\oplus\mathfrak{su}_3$, and 
$\mathfrak{su}_2\oplus\mathfrak{usp}_6$:
\begin{align}
\underset{F_4}
{\xygraph{!~:{@{=}|@{}}
    \circ   ([]!{+(0,-.3)} {{}_x}) - [r]
    \circ   ([]!{+(0,-.3)} {{}_1}) - [r]
    \circ   ([]!{+(0,-.3)} {{}_2}) : [r]
    \bullet ([]!{+(0,-.3)} {{}_3}) - [r]
    \bullet ([]!{+(0,-.3)} {{}_4})}
}
&\Rightarrow
\left\{
\begin{array}{c}
\underset{\mathfrak{so}_9}
{\xygraph{!~:{@{=}|@{}}
    \circ   ([]!{+(0,-.3)} {{}_x}) - [r]
    \circ   ([]!{+(0,-.3)} {{}_1}) - [r]
    \circ   ([]!{+(0,-.3)} {{}_2}) : [r]
    \bullet ([]!{+(0,-.3)} {{}_3}) - [r]
    {\bullet\hspace{-0.645em}\times} ([]!{+(0,-.3)} {{}_4})}
}\\
\underset{\mathfrak{su}_4\oplus\mathfrak{su}_2}
{\xygraph{!~:{@{=}|@{}}
    \circ   ([]!{+(0,-.3)} {{}_x}) - [r]
    \circ   ([]!{+(0,-.3)} {{}_1}) - [r]
    \circ   ([]!{+(0,-.3)} {{}_2}) : [r]
    {\bullet\hspace{-0.645em}\times} ([]!{+(0,-.3)} {{}_3}) - [r]
    \bullet ([]!{+(0,-.3)} {{}_4})}
}\\
\underset{\mathfrak{su}_3\oplus\mathfrak{su}_3}
{\xygraph{!~:{@{=}|@{}}
    \circ   ([]!{+(0,-.3)} {{}_x}) - [r]
    \circ   ([]!{+(0,-.3)} {{}_1}) - [r]
    {\circ\hspace{-0.645em}\times}   ([]!{+(0,-.3)} {{}_2}) : [r]
    \bullet ([]!{+(0,-.3)} {{}_3}) - [r]
    \bullet ([]!{+(0,-.3)} {{}_4})}
}\\
\underset{\mathfrak{su}_2\oplus\mathfrak{usp}_6}
{\xygraph{!~:{@{=}|@{}}
    \circ   ([]!{+(0,-.3)} {{}_x}) - [r]
    {\circ\hspace{-0.645em}\times}   ([]!{+(0,-.3)} {{}_1}) - [r]
    \circ   ([]!{+(0,-.3)} {{}_2}) : [r]
    \bullet ([]!{+(0,-.3)} {{}_3}) - [r]
    \bullet ([]!{+(0,-.3)} {{}_4})}
}\\
\end{array}
\right\}.
\end{align}
From the rule (S-3), we find its $S$-subalgebras
$A_1=\mathfrak{su}_2$, and $G_2\oplus A_1=G_2\oplus\mathfrak{su}_2$.

\section{Representations of algebras}
\label{Sec:Representations-basics}

We can classify an irreducible representation of a simple Lie algebra by
a weight vector $\Lambda$ or the Dynkin labels $(a_1a_2\cdots a_n)$ 
\cite{Dynkin:1957um,Slansky:1981yr}.
Each root or a weight vector $\Lambda$ in root space is a linear
combination of the simple roots $\alpha_i$:
\begin{align}
\Lambda=\sum_{i=1}^{n}\lambda_i\frac{2}{(\alpha_i,\alpha_i)}\alpha_i,
\end{align}
where  $\lambda_i$ stands for an $i$th coordinate in the root space, and
$(\alpha_i,\alpha_i)$ is a scalar product of a simple root $\alpha_i$
and itself, i.e., its length of the simple root $\alpha_i$. 
A set of the labels $(\lambda_1\lambda_2\cdots\lambda_n)$ represents a
coordinate in the root space.
Also, the irreducible representation can be rewritten in a different
basis: 
\begin{align}
a_i=2\frac{(\Lambda,\alpha_i)}{(\alpha_i,\alpha_i)}
=\sum_{j=1}^n\lambda_j\frac{2}{(\alpha_j,\alpha_j)}A_{ji}(\mathfrak{g})
,\ \ \
i=1,2,\cdots,n,
\end{align}
where $a_i$ is called a Dynkin label (or a Dynkin coefficient), 
$\Lambda$ is specified by a set of semi-positive integers called Dynkin
labels, and $A_{ji}(\mathfrak{g})$ is the $(ji)$ element of the Cartan
matrix of a Lie algebra $\mathfrak{g}$ given in
Table~\ref{Table:Cartan-matrices}. 
This basis is called the Dynkin basis. 
The relation between the coordinate in the root space $\lambda_i$ and
the Dynkin label $a_i$ is given by
\begin{align}
\lambda_i=\sum_{j=1}^{n}G(\mathfrak{g})_{ij}a_j,
\end{align}
where the matrix $G(\mathfrak{g})$ is defined in
Eq.~(\ref{Eq::Cartan-matrices-inverse}) and its explicit form is given
in Table~\ref{Table:Cartan-matrices-inverse}.
From now on, we will use the Dynkin labels $(a_1a_2\cdots a_n)$
as a representation of $\mathfrak{g}$.

\subsection{Conjugacy class}

Let us introduce conjugacy classes of simple Lie algebras $\mathfrak{g}$
given in Ref.~\cite{Dynkin:1957um}.
As we will see later, it is a useful concept to discuss tensor product
representations. 
We summarize the conjugacy classes of a representation $R$ of
semi-simple Lie algebras $\mathfrak{g}$ \cite{Lemire:1979qd}:
\begin{center}
\begin{longtable}{cl}
\caption{Conjugacy classes}
\label{Table:Conjugacy-classes}\\
\hline\hline
$\mathfrak{g}$:&
Conjugacy classes of classical and exceptional Lie algebras\\\hline\hline
\endfirsthead
\multicolumn{2}{c}{Table~\ref{Table:Conjugacy-classes} (continued)}\\\hline\hline
$\mathfrak{g}$:&
Conjugacy classes of classical and exceptional Lie algebras\\\hline\hline
\endhead
\hline
\endfoot
$A_n:$&
$C_c(R):=a_1+2a_2+3a_3+\cdots\ \ \  (\bmod.\ n+1)$\\
$B_n:$&
$C_c(R):=a_n\ \ \  (\bmod.\ 2)$\\
$C_n:$&
$C_c(R):=a_1+a_3+a_5+\cdots\ \ \  (\bmod.\ 2)$\\
$D_n:$&
$C_{c1}(R):=a_{n-1}+a_n\ \ \  (\bmod.\ 2)$\\
&$C_{c2}(R):=
\left\{
\begin{array}{ll}
2a_1+2a_3+\cdots+2a_{n-2}+(n-2)a_{n-1}+n a_n\ (\bmod.\ 4)
&\mbox{for}\ n=2k+1,\\
2a_1+2a_3+\cdots+2a_{n-3}
+(n-2)a_{n-1}+n a_n\ (\bmod.\ 4)
&\mbox{for}\ n=2k\\
\end{array}
\right.$\\
$E_6:$&
$C_c(R):=a_1-a_2+a_4-a_5\ \ \  (\bmod.\ 3)$\\
$E_7:$&
$C_c(R):=a_4+a_6+a_7\ \ \  (\bmod.\ 2)$\\
$E_8:$&
$C_c(R):=0$\\
$F_4:$&
$C_c(R):=0$\\
$G_2:$&
$C_c(R):=0$\\
\end{longtable}
\end{center}
For $G_2$, $F_4$, and $E_8$, all representations belong to
the same conjugacy class.
It can be shown by using the property of the Cartan matrices of Lie
algebras. 
(For its proof, see, e.g., Ref.~\cite{Dynkin:1957um}.)

We consider which representations belongs to the same conjugacy class.
First, from Table~\ref{Table:Conjugacy-classes}, the conjugacy class
of $A_5=\mathfrak{su}_6$ is given by
\begin{align}
C_c(a_1a_2a_3a_4a_5)=a_1+2a_2+3a_3+4a_4+5a_5\ \ \  (\bmod.\ 6).
\end{align}
Thus, we find that there are six conjugacy classes in
$\mathfrak{su}_6$. We can classify the representations of
$\mathfrak{su}_6$ based on the conjugacy classes:
\begin{align}
&C_c(00000)=C_c(00200)=\cdots=0,\ \ 
 C_c(10000)=C_c(20001)=\cdots=1,\nonumber\\
&C_c(01000)=C_c(20000)=\cdots=2,\ \ 
 C_c(00100)=C_c(11000)=\cdots=3,\nonumber\\
&C_c(00010)=C_c(02000)=\cdots=4,\ \ 
 C_c(00001)=C_c(10010)=\cdots=5.
\end{align}
Second, from Table~\ref{Table:Conjugacy-classes}, the conjugacy class
of $B_4=\mathfrak{so}_9$ is given by
\begin{align}
C_c(a_1a_2a_3a_4)=a_4\ \ \  (\bmod.\ 2).
\end{align}
Thus, there are only two conjugacy classes in $\mathfrak{so}_9$:
\begin{align}
&C_c(1000)=C_c(0100)=C_c(0010)=\cdots=0,\nonumber\\
&C_c(0001)=C_c(1001)=C_c(1111)=\cdots=1.
\end{align}
Third, from Table~\ref{Table:Conjugacy-classes}, the conjugacy class
of $C_5=\mathfrak{usp}_{10}$ is given by
\begin{align}
C_c(a_1a_2a_3a_4a_5)=a_1+a_3+a_5\ \ \  (\bmod.\ 2).
\end{align}
Thus, there are only two conjugacy classes in $\mathfrak{usp}_{10}$:
\begin{align}
&C_c(00000)=C_c(01000)=C_c(00010)=C_c(20000)=\cdots=0,\nonumber\\
&C_c(10000)=C_c(00100)=C_c(00001)=C_c(10101)=\cdots=1. 
\end{align}
Forth, from Table~\ref{Table:Conjugacy-classes}, the conjugacy classes
of $D_5=\mathfrak{so}_{10}$ are given by
\begin{align}
\{C_{c1},C_{c2}\}(a_1a_2a_3a_4a_5)=
\{a_4+a_5\ (\bmod.\ 2),\ 2a_1+2a_3+3a_4+5a_5\ (\bmod.\ 4)\}.
\end{align}
Thus, there are four conjugacy classes in $\mathfrak{so}_{10}$:
\begin{align}
&\{C_{c1},C_{c2}\}(00000)=\{C_{c1},C_{c2}\}(01000)=\cdots=\{0,0\},
\nonumber\\
&\{C_{c1},C_{c2}\}(10000)=\{C_{c1},C_{c2}\}(00100)=\cdots=\{0,2\},
\nonumber\\
&\{C_{c1},C_{c2}\}(00001)=\{C_{c1},C_{c2}\}(10010)=\cdots=\{1,1\},
\nonumber\\
&\{C_{c1},C_{c2}\}(00010)=\{C_{c1},C_{c2}\}(10001)=\cdots=\{1,3\}.
\end{align}
Fifth, from Table~\ref{Table:Conjugacy-classes}, the conjugacy classes
of $D_6=\mathfrak{so}_{12}$ are given by
\begin{align}
\{C_{c1},C_{c2}\}(a_1a_2a_3a_4a_5a_6)=
\{a_5+a_6\ (\bmod.\ 2),\ 2a_1+2a_3+4a_5+6a_6\ (\bmod.\ 4)\}.
\end{align}
Thus, there are four conjugacy classes in $\mathfrak{so}_{12}$:
\begin{align}
&\{C_{c1},C_{c2}\}(000000)=\{C_{c1},C_{c2}\}(010000)=\cdots=\{0,0\},
 \nonumber\\
&\{C_{c1},C_{c2}\}(100000)=\{C_{c1},C_{c2}\}(001000)=\cdots=\{0,2\},
\nonumber\\
&\{C_{c1},C_{c2}\}(000010)=\{C_{c1},C_{c2}\}(100001)=\cdots=\{1,0\},
\nonumber\\
&\{C_{c1},C_{c2}\}(000001)=\{C_{c1},C_{c2}\}(100010)=\cdots=\{1,2\}.
\end{align}
Sixth, from Table~\ref{Table:Conjugacy-classes}, the conjugacy class
of $E_6$ is given by
\begin{align}
C_c(a_1a_2a_3a_4a_5a_6):=a_1-a_2+a_4-a_5\ \ \  (\bmod.\ 3).
\end{align}
Thus, there are three conjugacy classes in $E_6$:
\begin{align}
&C_c(000000)=C_c(001000)=C_c(000001)=\cdots=0,\nonumber\\
&C_c(100000)=C_c(000100)=C_c(101000)=\cdots=1,\nonumber\\
&C_c(010000)=C_c(000010)=C_c(001010)=\cdots=2.
\end{align}
Finally, Table~\ref{Table:Conjugacy-classes}, the conjugacy class
of $E_7$ is given by
\begin{align}
C_c(a_1a_2a_3a_4a_5a_6a_7):=a_4+a_6+a_7\ \ \  (\bmod.\ 2).
\end{align}
Thus, there are two conjugacy classes in $E_7$:
\begin{align}
&C_c(1000000)=C_c(0100000)=C_c(0010000)=C_c(0000100)=\cdots=0,\nonumber\\
&C_c(0001000)=C_c(0000010)=C_c(0000001)=C_c(1000100)=\cdots=1.
\end{align}

\subsection{Complex, self-conjugate, real, and pseudo-real}

Let us check complex and self-conjugate representations and how to
identify them discussed in Ref.~\cite{McKay:1981}.
A complex representation has no invariant bilinear form by
itself, while a self-conjugate (or self-contragredient) representation has
an invariant bilinear form by itself.
Further, self-conjugate representations are classified into real
(orthogonal) representations and pseudo-real (symplectic)
representations; a real representation has an invariant symmetric
bilinear form by itself, while a pseudo-real representation has an
invariant skew symmetric bilinear form by itself.
(As we discuss later, they are necessary e.g., to construct
four-dimensional anomaly-free chiral gauge theories and to find Lie
subalgebras, especially, $S$-subalgebras.)

First, we consider complex representations of Lie algebras.
We define conjugate representations as the followings by using
Dynkin labels; 
for the representations of $A_n=\mathfrak{su}_{n+1}$, 
$\overline{(a_1a_2\cdots a_{n-1}a_n)}=
(a_n a_{n-1}\cdots a_2a_1)$;
for the representations of $D_{2n+1}=\mathfrak{so}_{4n+2}$,
$\overline{(a_1a_2\cdots a_{2n-1}a_{2n}a_{2n+1})}
=(a_1a_2\cdots a_{2n-1}a_{2n+1}a_{2n})$;
for the representations of $E_6$,
$\overline{(a_1a_2a_3a_4a_5a_6)}
=(a_5a_4a_3a_2a_1a_6)$;
for the representations of the other simple Lie algebras,
$\overline{(a_1a_2\cdots a_{n-1}a_{n})}
=(a_1a_2\cdots a_{n-1}a_{n})$.
Second, we can regard a self-conjugate representation as a representation
satisfying 
$\overline{(a_1a_2\cdots a_{n-1}a_n)}=
(a_1a_2\cdots a_{n-1}a_n)$,
while we can regard a complex representation as a representation
satisfying 
$\overline{(a_1a_2\cdots a_{n-1}a_n)}\not=
(a_1a_2\cdots a_{n-1}a_n)$.
Thus, all the Lie algebras except $A_n$, $D_{2n+1}$, and $E_6$, i.e,.
$B_n=\mathfrak{so}_{2n+1}$, $C_n=\mathfrak{usp}_{2n}$, 
$D_{2n}=\mathfrak{so}_{4n}$, $E_7$, $E_8$, $F_4$, and $G_2$,
have only self-conjugate representations.
The Lie algebras $A_n$, $D_{2n+1}$, and $E_6$ have several complex 
representations; for $A_n$, if a representation of
$A_n$ satisfies a 
condition $(a_1a_2\cdots a_{n-1}a_n)
\not=(a_n a_{n-1}\cdots a_2a_1)$, then
the representation is a complex representation of $A_n$;
for $D_{2n+1}=\mathfrak{so}_{4n+2}$, if a representation of $D_{2n+1}$
satisfies a condition  
$(a_1\cdots a_{2n-1}a_{2n}a_{2n+1})\not=
(a_1\cdots a_{2n-1}a_{2n+1}a_{2n})$, 
then the representation is a complex representation of $D_{2n+1}$;
for $E_n$, if a representation of $E_6$ satisfies a condition 
$(a_1a_2a_3a_4a_5a_6)
\not=(a_5a_4a_3a_2a_1a_6)$,
then the representation is a complex representation of $E_6$.

We summarize what kinds of representations are complex in the following
table.
\begin{center}
\begin{longtable}{ccl}
\caption{Complex representations (1)}
\label{Table:Complex-representations-1}\\
\hline\hline
Algebra&Rank&Condition for a complex representation\\\hline\hline
\endfirsthead
\multicolumn{3}{c}{Table~\ref{Table:Complex-representations-1} (continued)}\\\hline\hline
Algebra&Rank&Condition for a complex representation\\\hline\hline
\endhead
\hline
\endfoot
$A_n=\mathfrak{su}_{n+1}$&${}^\forall n$&$(a_1a_2\cdots a_{n-1}a_n)\not=(a_n a_{n-1}\cdots a_2a_1)$\\
$D_{2n+1}=\mathfrak{so}_{4n+2}$&${}^\forall n$&
$(a_1\cdots a_{2n-1}a_{2n}a_{2n+1})\not=
(a_1\cdots a_{2n-1}a_{2n+1}a_{2n})$\\
$E_6$     &$6$        &$(a_1a_2a_3a_4a_5a_6)\not=(a_5a_4a_3a_2a_1a_6)$\\
\end{longtable}
\end{center}
Its corresponding description in terms of Dynkin diagrams is shown in
the following table: 
\begin{center}
\begin{longtable}{cl}
\caption{Complex representations (2)}
\label{Table:Complex-representations-2}\\
\hline\hline
$\mathfrak{g}$:&Condition for a complex representation\\\hline\hline
\endfirsthead
\multicolumn{2}{c}{Table~\ref{Table:Complex-representations-1} (continued)}\\\hline\hline
$\mathfrak{g}$:&Condition for a complex representation\\\hline\hline
\endhead
\hline
\endfoot
$A_n$:&
${\xygraph{
    \circ ([]!{+(0,-.3)} {a_1})  - [r]
    \circ ([]!{+(0,-.3)} {a_2})  - [r]
    \cdots                        - [r]
    \circ ([]!{+(0,-.3)} {a_{n-1}}) - [r]
    \circ ([]!{+(0,-.3)} {a_n})}}
\not=
{\xygraph{
    \circ ([]!{+(0,-.3)} {a_{n}})  - [r]
    \circ ([]!{+(0,-.3)} {a_{n-1}})  - [r]
    \cdots                        - [r]
    \circ ([]!{+(0,-.3)} {a_2}) - [r]
    \circ ([]!{+(0,-.3)} {a_1})}}
$\\
$D_{2n+1}$:&
${\xygraph{
    \circ ([]!{+(0,-.3)} {a_1})  - [r] 
    \circ ([]!{+(0,-.3)} {a_2})  - [r] 
    \cdots                        - [r]
    \circ ([]!{+(0,-.3)} {a_{2n-1}}) (
        - []!{+(1,.5)}  \circ ([]!{+(0,-.3)} {a_{2n}}),
        - []!{+(1,-.5)} \circ ([]!{+(0,-.3)} {a_{2n+1}})
)}}
\not=
{\xygraph{
    \circ ([]!{+(0,-.3)} {a_1})  - [r] 
    \circ ([]!{+(0,-.3)} {a_2})  - [r] 
    \cdots                        - [r]
    \circ ([]!{+(0,-.3)} {a_{2n-1}}) (
        - []!{+(1,.5)}  \circ ([]!{+(0,-.3)} {a_{2n+1}}),
        - []!{+(1,-.5)} \circ ([]!{+(0,-.3)} {a_{2n}})
)}}
$\\
$E_6$: &
${\xygraph{
    \circ ([]!{+(0,-.3)}  {a_1}) - [r]
    \circ ([]!{+(0,-.3)}  {a_2}) - [r]
    \circ ([]!{+(.3,-.3)} {\hspace{-2em}a_3}) (
        - [u] \circ ([]!{+(.3,0)}  {a_6}),
        - [r] \circ ([]!{+(0,-.3)} {a_4})
        - [r] \circ ([]!{+(0,-.3)} {a_5})
)}}
\not=
{\xygraph{
    \circ ([]!{+(0,-.3)}  {a_5}) - [r]
    \circ ([]!{+(0,-.3)}  {a_4}) - [r]
    \circ ([]!{+(.3,-.3)} {\hspace{-2em}a_3}) (
        - [u] \circ ([]!{+(.3,0)}  {a_6}),
        - [r] \circ ([]!{+(0,-.3)} {a_2})
        - [r] \circ ([]!{+(0,-.3)} {a_1})
)}}
$\\
\end{longtable}
\end{center}

Next, self-conjugate representations are summarized in the following
table.
\begin{center}
\begin{longtable}{clp{4cm}l}
\caption{Self-conjugate representations}
\label{Table:Self-conjugate-representations}\\
\hline\hline
Algebra&Rank&Condition for a self-conjugate rep.&Real/PR\\\hline\hline
\endfirsthead
\multicolumn{4}{c}{Table~\ref{Table:Self-conjugate-representations} (continued)}\\\hline\hline
Algebra&Rank&Condition for a self-conjugate rep.&Real/PR\\\hline\hline
\endhead
\hline
\endfoot
$\mathfrak{su}_{n+1}$ &$n=2,3,4;6,7,8;\cdots$
&$(a_1\cdots a_n)=(a_n\cdots a_1)$
&Real only\\\cline{2-4}
&$n=1,5,9,13,\cdots$
&$(a_1\cdots a_n)=(a_n\cdots a_1)$
&Real if $a_{(n+1)/2}$ is even\\
&&
&PR if $a_{(n+1)/2}$ is odd\\
\hline
$\mathfrak{so}_{2n+1}$ 
&$n=3,4;7,8;11,12;\cdots$
&None
&Real only
\\\cline{2-4}
&$n=1,2;5,6;9,10;\cdots$
&None
&Real if $a_n$ is even\\
&
&($\alpha_n$ is the short root.)
&PR if $a_{n}$ is odd\\
\hline
$\mathfrak{usp}_{2n}$ 
&${}^\forall n$
&None
&Real if $\sum_{i {\rm odd}}a_i$ is even\\
&
&($\alpha_n$ is the long root.)
&PR if $\sum_{i {\rm odd}}a_i$ is odd\\
\hline
$\mathfrak{so}_{2n}$ 
&$n=2k+1$
&$a_{n-1}=a_n$
&Real only
\\\cline{2-4}
&$n=2k=4\ell$
&None
&Real only
\\\cline{2-4}
&$n=2k=4\ell+2$
&None
&Real if $a_{n-1}+a_n$ is even\\
&
&($\alpha_n$ and $\alpha_{n-1}$ are the spinor roots.)
&PR if $a_{n-1}+a_n$ is odd\\
\hline
$E_6$
&6
&$(a_1a_2a_3a_4a_5a_6)$ $=(a_5a_4a_3a_2a_1a_6)$
&Real only\\
\hline
$E_7$ 
&7
&None
&Real if $a_4+a_6+a_7$ is even\\
&&
&PR if $a_4+a_6+a_7$ is odd\\
\hline
$E_8$ 
&8
&None
&Real only\\
\hline
$F_4$ 
&4
&None
&Real only\\
\hline
$G_2$ 
&2
&None
&Real only\\
\end{longtable}
\end{center}
where PR stands for a pseudo-real representation, 
and $k,\ell\in\mathbb{Z}_{\geq 1}$.
The table is shown in Table 12 in Ref.~\cite{Slansky:1981yr}.

Here we check several examples of complex and self-conjugate (real and
pseudo-real) representations.
(Note that for self-conjugate representations, we use
Table~\ref{Table:Self-conjugate-representations} to identify real or
pseudo-real representations as well as Dynkin diagrams below.) 
For $A_4=\mathfrak{su}_5$, a representation $(a_1a_2a_3a_4)=(0100)$ is a 
complex representation because
$(a_1a_2a_3a_4)=(0100)\not=(0010)=(a_4a_3a_2a_1)$
or its Dynkin diagram is 
\begin{align}
A_4=\mathfrak{su}_5:\ &
{\xygraph{
    \circ ([]!{+(0,-.3)} {0})  - [r]
    \circ ([]!{+(0,-.3)} {1})  - [r]
    \circ ([]!{+(0,-.3)} {0})  - [r]
    \circ ([]!{+(0,-.3)} {0})}}
\not=
{\xygraph{
    \circ ([]!{+(0,-.3)} {0})  - [r]
    \circ ([]!{+(0,-.3)} {0})  - [r]
    \circ ([]!{+(0,-.3)} {1})  - [r]
    \circ ([]!{+(0,-.3)} {0})}}.
\end{align}
Another representation $(a_1a_2a_3a_4)=(1001)$ is a real representation
because $n=4$ and $(a_1a_2a_3a_4)=(1001)=(a_4a_3a_2a_1)$ or
its Dynkin diagram is 
\begin{align}
A_4=\mathfrak{su}_5:\ &
{\xygraph{
    \circ ([]!{+(0,-.3)} {1})  - [r]
    \circ ([]!{+(0,-.3)} {0})  - [r]
    \circ ([]!{+(0,-.3)} {0})  - [r]
    \circ ([]!{+(0,-.3)} {1})}}
=
{\xygraph{
    \circ ([]!{+(0,-.3)} {1})  - [r]
    \circ ([]!{+(0,-.3)} {0})  - [r]
    \circ ([]!{+(0,-.3)} {0})  - [r]
    \circ ([]!{+(0,-.3)} {1})}}.
\end{align}
For $A_5=\mathfrak{su}_6$, a representation $(a_1a_2a_3a_4a_5)=(01000)$
is a complex representation because
$(a_1a_2a_3a_4a_5)=(01000)\not=(00010)=(a_5a_4a_3a_2a_1)$
or its Dynkin diagram is 
\begin{align}
A_5=\mathfrak{su}_6:\ &
{\xygraph{
    \circ ([]!{+(0,-.3)} {0})  - [r]
    \circ ([]!{+(0,-.3)} {1})  - [r]
    \circ ([]!{+(0,-.3)} {0})  - [r]
    \circ ([]!{+(0,-.3)} {0})  - [r]
    \circ ([]!{+(0,-.3)} {0})}}
\not=
{\xygraph{
    \circ ([]!{+(0,-.3)} {0})  - [r]
    \circ ([]!{+(0,-.3)} {0})  - [r]
    \circ ([]!{+(0,-.3)} {0})  - [r]
    \circ ([]!{+(0,-.3)} {1})  - [r]
    \circ ([]!{+(0,-.3)} {0})}}.
\end{align}
Another representation $(a_1a_2a_3a_4a_5)=(10001)$ is a real
representation 
because $n=5$, $a_{3}=0$ and 
$(a_1a_2a_3a_4a_5)=(10001)=(a_5a_4a_3a_2a_1)$ or
its Dynkin diagram is 
\begin{align}
A_5=\mathfrak{su}_6:\ &
{\xygraph{
    \circ ([]!{+(0,-.3)} {1})  - [r]
    \circ ([]!{+(0,-.3)} {0})  - [r]
    \circ ([]!{+(0,-.3)} {0})  - [r]
    \circ ([]!{+(0,-.3)} {0})  - [r]
    \circ ([]!{+(0,-.3)} {1})}}
=
{\xygraph{
    \circ ([]!{+(0,-.3)} {1})  - [r]
    \circ ([]!{+(0,-.3)} {0})  - [r]
    \circ ([]!{+(0,-.3)} {0})  - [r]
    \circ ([]!{+(0,-.3)} {0})  - [r]
    \circ ([]!{+(0,-.3)} {1})}}.
\end{align}
A representation $(a_1a_2a_3a_4a_5)=(00100)$ is a pseudo-real
representation 
because $n=5$, $a_{3}=1$ and
$(a_1a_2a_3a_4a_5)=(00100)=(a_5a_4a_3a_2a_1)$ or 
its Dynkin diagram is 
\begin{align}
A_5=\mathfrak{su}_6:\ &
{\xygraph{
    \circ ([]!{+(0,-.3)} {0})  - [r]
    \circ ([]!{+(0,-.3)} {0})  - [r]
    \circ ([]!{+(0,-.3)} {1})  - [r]
    \circ ([]!{+(0,-.3)} {0})  - [r]
    \circ ([]!{+(0,-.3)} {0})}}
=
{\xygraph{
    \circ ([]!{+(0,-.3)} {0})  - [r]
    \circ ([]!{+(0,-.3)} {0})  - [r]
    \circ ([]!{+(0,-.3)} {1})  - [r]
    \circ ([]!{+(0,-.3)} {0})  - [r]
    \circ ([]!{+(0,-.3)} {0})}}.
\end{align}
For $B_5=\mathfrak{so}_{11}$, a representation
$(a_1a_2a_3a_4a_5)=(01000)$ 
is a real representation because
$a_5=0$ or its Dynkin diagram is 
\begin{align}
B_{5}=\mathfrak{so}_{11}:\ &
\xygraph{!~:{@{=}|@{}}
    \circ   ([]!{+(0,-.3)} {0}) - [r]
    \circ   ([]!{+(0,-.3)} {1}) - [r]
    \circ   ([]!{+(0,-.3)} {0}) - [r]
    \circ   ([]!{+(0,-.3)} {0}) : [r]
    \bullet ([]!{+(0,-.3)} {0})}.
\end{align}
Another representation
$(a_1a_2a_3a_4a_5)=(00001)$ 
is a pseudo-real representation because
$a_5=1$ or its Dynkin diagram is 
\begin{align}
B_{5}=\mathfrak{so}_{11}:\ &
\xygraph{!~:{@{=}|@{}}
    \circ   ([]!{+(0,-.3)} {0}) - [r]
    \circ   ([]!{+(0,-.3)} {0}) - [r]
    \circ   ([]!{+(0,-.3)} {0}) - [r]
    \circ   ([]!{+(0,-.3)} {0}) : [r]
    \bullet ([]!{+(0,-.3)} {1})}.
\end{align}
For $C_5=\mathfrak{usp}_{10}$, a representation
$(a_1a_2a_3a_4a_5)=(10000)$ 
is a pseudo-real representation because
$a_1+a_3+a_5=1$ or its Dynkin diagram is 
\begin{align}
C_5=\mathfrak{usp}_{10}:\ &
\xygraph{!~:{@{=}|@{}}
    \bullet ([]!{+(0,-.3)} {1}) - [r]
    \bullet ([]!{+(0,-.3)} {0}) - [r]
    \bullet ([]!{+(0,-.3)} {0}) - [r]
    \bullet ([]!{+(0,-.3)} {0}) : [r]
    \circ   ([]!{+(0,-.3)} {0})}.
\end{align}
Another representation
$(a_1a_2a_3a_4a_5)=(10001)$ 
is a pseudo-real representation because
$a_1+a_3+a_5=0 (\bmod.\ 2)$
or its Dynkin diagram is 
\begin{align}
C_5=\mathfrak{usp}_{10}:\ &
\xygraph{!~:{@{=}|@{}}
    \bullet ([]!{+(0,-.3)} {1}) - [r]
    \bullet ([]!{+(0,-.3)} {0}) - [r]
    \bullet ([]!{+(0,-.3)} {0}) - [r]
    \bullet ([]!{+(0,-.3)} {0}) : [r]
    \circ   ([]!{+(0,-.3)} {1})}.
\end{align}
For $D_5=\mathfrak{so}_{10}$, a representation
$(a_1a_2a_3a_4a_5)=(00001)$ is a complex representation because 
$n=5$ and 
$(a_1a_2a_3a_4a_5)=(00001)\not=(00010)=(a_1a_2a_3a_5a_4)$
or its Dynkin diagram is 
\begin{align}
D_5=\mathfrak{so}_{10}:\ &
{\xygraph{
    \circ ([]!{+(0,-.3)} {0})  - [r] 
    \circ ([]!{+(0,-.3)} {0})  - [r] 
    \circ ([]!{+(0,-.3)} {0}) (
        - []!{+(1,.5)}  \circ ([]!{+(0,-.3)} {0}),
        - []!{+(1,-.5)} \circ ([]!{+(0,-.3)} {1})
)}}
\not=
{\xygraph{
    \circ ([]!{+(0,-.3)} {0})  - [r] 
    \circ ([]!{+(0,-.3)} {0})  - [r] 
    \circ ([]!{+(0,-.3)} {0}) (
        - []!{+(1,.5)}  \circ ([]!{+(0,-.3)} {1}),
        - []!{+(1,-.5)} \circ ([]!{+(0,-.3)} {0})
)}}.
\end{align}
Another representation $(a_1a_2a_3a_4a_5)=(01000)$ is a real
representation because 
$n=5$ and 
$(a_1a_2a_3a_4a_5)=(01000)=(a_1a_2a_3a_5a_4)$ or
its Dynkin diagram is 
\begin{align}
D_5=\mathfrak{so}_{10}:\ &
{\xygraph{
    \circ ([]!{+(0,-.3)} {0})  - [r] 
    \circ ([]!{+(0,-.3)} {1})  - [r] 
    \circ ([]!{+(0,-.3)} {0}) (
        - []!{+(1,.5)}  \circ ([]!{+(0,-.3)} {0}),
        - []!{+(1,-.5)} \circ ([]!{+(0,-.3)} {0})
)}}
=
{\xygraph{
    \circ ([]!{+(0,-.3)} {0})  - [r] 
    \circ ([]!{+(0,-.3)} {1})  - [r] 
    \circ ([]!{+(0,-.3)} {0}) (
        - []!{+(1,.5)}  \circ ([]!{+(0,-.3)} {0}),
        - []!{+(1,-.5)} \circ ([]!{+(0,-.3)} {0})
)}}.
\end{align}
For $D_6=\mathfrak{so}_{12}$, a representation
$(a_1a_2a_3a_4a_5a_6)=(000001)$ is a pseudo-real representation because  
$n=6$ and $a_5+a_6=1$
or its Dynkin diagram is 
\begin{align}
D_6=\mathfrak{so}_{12}:\ &
{\xygraph{
    \circ ([]!{+(0,-.3)} {0})  - [r] 
    \circ ([]!{+(0,-.3)} {0})  - [r] 
    \circ ([]!{+(0,-.3)} {0})  - [r] 
    \circ ([]!{+(0,-.3)} {0}) (
        - []!{+(1,.5)}  \circ ([]!{+(0,-.3)} {0}),
        - []!{+(1,-.5)} \circ ([]!{+(0,-.3)} {1})
)}}.
\end{align}
Another representation $(a_1a_2a_3a_4a_5)=(01000)$ is a real
representation because 
$n=6$ and $a_5+a_6=0$ or
its Dynkin diagram is 
\begin{align}
D_6=\mathfrak{so}_{12}:\ &
{\xygraph{
    \circ ([]!{+(0,-.3)} {0})  - [r] 
    \circ ([]!{+(0,-.3)} {1})  - [r] 
    \circ ([]!{+(0,-.3)} {0})  - [r] 
    \circ ([]!{+(0,-.3)} {0}) (
        - []!{+(1,.5)}  \circ ([]!{+(0,-.3)} {0}),
        - []!{+(1,-.5)} \circ ([]!{+(0,-.3)} {0})
)}}.
\end{align}
For $E_6$, a representation $(a_1a_2a_3a_4a_5a_6)=(100000)$ is a complex
representation because 
$(a_1a_2a_3a_4a_5a_6)=(100000)\not=(000010)=(a_5a_4a_3a_2a_1a_6)$
or its Dynkin diagram is 
\begin{align}
E_6:\ &
{\xygraph{
    \circ ([]!{+(0,-.3)}  {1}) - [r]
    \circ ([]!{+(0,-.3)}  {0}) - [r]
    \circ ([]!{+(.3,-.3)} {\hspace{-2em}0}) (
        - [u] \circ ([]!{+(.3,0)}  {0}),
        - [r] \circ ([]!{+(0,-.3)} {0})
        - [r] \circ ([]!{+(0,-.3)} {0})
)}}
\not=
{\xygraph{
    \circ ([]!{+(0,-.3)}  {0}) - [r]
    \circ ([]!{+(0,-.3)}  {0}) - [r]
    \circ ([]!{+(.3,-.3)} {\hspace{-2em}0}) (
        - [u] \circ ([]!{+(.3,0)}  {0}),
        - [r] \circ ([]!{+(0,-.3)} {0})
        - [r] \circ ([]!{+(0,-.3)} {1})
)}}.
\end{align}
Another representation $(a_1a_2a_3a_4a_5a_6)=(001000)$ is a
self-conjugate representation because 
$(a_1a_2a_3a_4a_5a_6)=(001000)=(a_5a_4a_3a_2a_1a_6)$ or
its Dynkin diagram is 
\begin{align}
E_6:\ &
{\xygraph{
    \circ ([]!{+(0,-.3)}  {0}) - [r]
    \circ ([]!{+(0,-.3)}  {0}) - [r]
    \circ ([]!{+(.3,-.3)} {\hspace{-2em}1}) (
        - [u] \circ ([]!{+(.3,0)}  {0}),
        - [r] \circ ([]!{+(0,-.3)} {0})
        - [r] \circ ([]!{+(0,-.3)} {0})
)}}
=
{\xygraph{
    \circ ([]!{+(0,-.3)}  {0}) - [r]
    \circ ([]!{+(0,-.3)}  {0}) - [r]
    \circ ([]!{+(.3,-.3)} {\hspace{-2em}1}) (
        - [u] \circ ([]!{+(.3,0)}  {0}),
        - [r] \circ ([]!{+(0,-.3)} {0})
        - [r] \circ ([]!{+(0,-.3)} {0})
)}}.
\end{align}
For $E_7$, a representation $(a_1a_2a_3a_4a_5a_6a_7)=(0001000)$ is a
pseudo-real representation because 
$a_4+a_6+a_7=1$ or its Dynkin diagram is 
\begin{align}
E_7:\ 
\xygraph{
    \circ ([]!{+(0,-.3)}  {0}) - [r]
    \circ ([]!{+(0,-.3)}  {0}) - [r]
    \circ ([]!{+(.3,-.3)} {\hspace{-2em}0}) (
        - [u] \circ ([]!{+(.3,0)}  {0}),
        - [r] \circ ([]!{+(0,-.3)} {1})
        - [r] \circ ([]!{+(0,-.3)} {0})
        - [r] \circ ([]!{+(0,-.3)} {0})
)}.
\end{align}
Another representation $(a_1a_2a_3a_4a_5a_6a_7)=(0010000)$ is a
real representation because 
$a_4+a_6+a_7=0$ or its Dynkin diagram is 
\begin{align}
E_7:\ 
\xygraph{
    \circ ([]!{+(0,-.3)}  {0}) - [r]
    \circ ([]!{+(0,-.3)}  {0}) - [r]
    \circ ([]!{+(.3,-.3)} {\hspace{-2em}1}) (
        - [u] \circ ([]!{+(.3,0)}  {0}),
        - [r] \circ ([]!{+(0,-.3)} {0})
        - [r] \circ ([]!{+(0,-.3)} {0})
        - [r] \circ ([]!{+(0,-.3)} {0})
)}.
\end{align}

Here we summarize examples of complex and self-conjugate representations
in the following tables.
\begin{center}
\begin{longtable}{ccllc}
\caption{Examples of complex representations}
\label{Table:Complex-representations-examples}\\
\hline\hline
Class&Compact group &\multicolumn{2}{l}{Representation}&Dimension\\\hline\hline
\endfirsthead
\multicolumn{5}{c}{Table~\ref{Table:Complex-representations-examples} (continued)}\\\hline\hline
Class&Compact group &\multicolumn{2}{l}{Representation}&Dimension\\\hline\hline
\endhead
\hline
\endfoot
 $A_2$&$SU(3)$ &(10),    &(01)    &3\\[0.25em]
 $A_4$&$SU(5)$ &(1000),  &(0001)  &5\\
 &        &(0100),  &(0010)  &10\\[0.25em]
 $A_5$&$SU(6)$ &(10000), &(00001) &6\\
 &        &(01000), &(00010) &15\\[0.25em]
 $D_5$&$SO(10)$&(00001), &(00010) &16\\[0.25em]
 $E_6$&$E_6$   &(100000),&(000010)&27\\
 &        &(010000),&(000100)&351\\
\end{longtable}
\end{center}

\begin{center}
\begin{longtable}{cclc}
\caption{Examples of self-conjugate representations}
\label{Table:Self-representations-examples}\\
\hline\hline
Class   &Compact group&Representation    &Dimension\\\hline\hline
\endfirsthead
\multicolumn{4}{c}{Table~\ref{Table:Self-representations-examples} (continued)}\\\hline\hline
Class   &Compact group&Representation    &Dimension\\\hline\hline
\endhead
\hline
\endfoot
 $A_2$&$SU(3)$     &(11)     &8\\[0.25em]
 $A_4$&$SU(5)$     &(1001)   &24\\
 &            &(0110)   &75\\[0.25em]
 $A_5$&$SU(6)$     &(10001)  &35\\
 &            &(01010)  &189\\[0.25em]
 $D_5$&$SO(10)$    &(10000)  &10\\
 &            &(01000)  &45\\[0.25em]
 $E_6$&$E_6$       &(000001) &78\\
\end{longtable}
\end{center}

\subsection{Weyl dimension formula}
\label{Sec:Weyl-dimensional-formula}

The dimension of the representation $R$ of a rank-$n$ algebra 
$\mathfrak{g}$ with the highest weight $\Lambda:=(a_1a_2\cdots a_n)$,
where $a_i (i=1,2,\cdots n)$ is integer, is derived from the so-called
Weyl dimension formula \cite{Weyl:1946,Dynkin:1957um}, 
\begin{align}
d(R)=
\prod_{\alpha\in \Delta_+}
\frac{(\Lambda+g,\alpha)}{(g,\alpha)},
\label{Eq:Weyl-formula}
\end{align}
where $\Delta_+$ is a set of the positive roots of the Lie algebra
$\mathfrak{g}$, $\alpha$ is a positive root, 
and $g:=(\sum_{\alpha\in \Delta_+}\alpha)/2$ in the root space.
(The formula in Eq.~(\ref{Eq:Weyl-formula}) can be derived from the
Weyl character formula \cite{Weyl:1946}. For its derivation, see, e.g.,
Refs.~\cite{Cahn:1985wk,Fulton:1991}.)
From the above Weyl formula in the root space, we obtain the following
expression in the Dynkin basis:
\begin{align}
d(R)=
\prod_{\alpha\in \Delta_+}
\frac{\sum_{i=1}^n k_\alpha^i(a_i+1)(\alpha_i,\alpha_i)}
{\sum_{i=1}^n k_\alpha^i(\alpha_i,\alpha_i)},
\label{Eq:Weyl-formula'}
\end{align}
where we used the Cartan matrix
$A_{ij}(G)=2(\alpha_i,\alpha_j)/(\alpha_j,\alpha_j)$,
$\Lambda=\sum_{i=1}^n2\lambda_i\alpha_i/(\alpha_i,\alpha_i)$
and $a_i=\sum_{j=1}^n2\lambda_j A_{ij}(G)/(\alpha_j,\alpha_j)$,
$\alpha=\sum_{i=1}^n k_\alpha^i\alpha_i$, where $k_\alpha^i$ is a
non-negative integer.

We rewrite the Weyl dimension formula in
Eq.~(\ref{Eq:Weyl-formula'}).  
From the Cartan matrices in Table~\ref{Table:Cartan-matrices},
for the algebras $A_n=\mathfrak{su}_{n+1}$, $D_n=\mathfrak{so}_{2n}$,
$E_6$, $E_7$, and $E_8$,  
the value of $(\alpha_i,\alpha_i)$ is the same for any $i$,
while 
for the algebras $B_n=\mathfrak{so}_{2n+1}$, $C_n=\mathfrak{usp}_{2n}$,
$F_4$, and $G_2$, the value of  
$(\alpha_i,\alpha_i)$ $(i=1,2,\cdots,n)$ is not the same for any $i$:
$(\alpha_i,\alpha_i)=2(\alpha_n,\alpha_n)$ $(i=1,2,\cdots,n-1)$
for $B_n$; 
$(\alpha_i,\alpha_i)=(\alpha_n,\alpha_n)/2$ $(i=1,2,\cdots,n-1)$
for $C_n$;
$(\alpha_1,\alpha_1)=(\alpha_2,\alpha_2)=
2(\alpha_3,\alpha_3)=2(\alpha_4,\alpha_4)$ for $F_4$;
and $(\alpha_1,\alpha_1)=3(\alpha_2,\alpha_2)$ for $G_2$.
First, by using the explicit values of
the scalar product $(\alpha_i,\alpha_i)$ of the Lie algebras, the
formula become 
\begin{align}
A_n:\ &
d(R)=
\prod_{\alpha\in \Delta_+}
\frac{\sum_{i=1}^n k_\alpha^i(a_i+1)}
{\sum_{i=1}^n k_\alpha^i},\\
B_n:\ &
d(R)=
\prod_{\alpha\in \Delta_+}
\frac{\left(\sum_{i=1}^{n-1}2k_\alpha^i(a_i+1)\right)+k_\alpha^n(a_n+1)}
{\left(\sum_{i=1}^n2k_\alpha^i\right)+k_\alpha^n},\\
C_n:\ &
d(R)=
\prod_{\alpha\in \Delta_+}
\frac{\left(\sum_{i=1}^{n-1}k_\alpha^i(a_i+1)\right)+2k_\alpha^n(a_n+1)}
{\left(\sum_{i=1}^n k_\alpha^i\right)+2k_\alpha^n},\\
D_n:\ &
d(R)=
\prod_{\alpha\in \Delta_+}
\frac{\sum_{i=1}^n k_\alpha^i(a_i+1)}
{\sum_{i=1}^n k_\alpha^i},\\
E_n:\ &
d(R)=
\prod_{\alpha\in \Delta_+}
\frac{\sum_{i=1}^n k_\alpha^i(a_i+1)}
{\sum_{i=1}^n k_\alpha^i},\\
F_4:\ &
d(R)=
\prod_{\alpha\in \Delta_+}
\frac{2k_\alpha^1(a_1+1)+2k_\alpha^2(a_2+1)
+k_\alpha^3(a_3+1)+k_\alpha^4(a_4+1)}
{2k_\alpha^1+2k_\alpha^2+k_\alpha^3+k_\alpha^4},\\
G_2:\ &
d(R)=
\prod_{\alpha\in \Delta_+}
\frac{3k_\alpha^1(a_1+1)+k_\alpha^2(a_2+1)}
{3k_\alpha^1+k_\alpha^2}.
\end{align}

Next, by using the positive roots of the Lie algebras,
we can obtain more useful forms of the Weyl dimension formula of each
algebra. The coefficients $k_\alpha^i$ for the classical and exceptional
algebras $A_n (n=1,2,\cdots,6)$, $B_n (n=2,3,4,5)$, $C_n (n=2,3,4,5)$, 
$D_n (n=4,5,6,7)$, $E_n (n=6,7,8)$, $F_4$, and $G_2$ are shown in
Appendix~\ref{Sec:Positive-roots}. Once we know the positive roots of a
Lie algebra, it is easy to write down its explicit Weyl dimension
formula. Here we only give the Weyl dimension formula of 
$\mathfrak{g}=A_n,B_n,C_n,D_n$ as follows (because there is no
information about representations of Lie algebras with more than rank-16
in Appendix~\ref{Sec:Representations}):
\begin{align}
A_n:\ 
d(R)=&
\prod_{k=1}^n\prod_{i=1}^k
\left(\frac{\sum_{j=i}^{k}a_j}{k-i+1}+1\right),
\label{Eq:Weyl-formula-A}\\
B_n:\ 
d(R)=&
\left[
\prod_{k=1}^n\prod_{i=1}^k
\left(\frac{\sum_{j=i}^{k}a_j}{k-i+1}+1\right)
\right]
\prod_{k=0}^{n-2}
\prod_{i=k}^{n-2}
\left(
\frac{\left(\sum_{j=k+1}^{i}a_{n-1-j}\right)
+\left(\sum_{j=0}^{k}2a_{n-1-j}\right)+a_n}{k+i+3}+1\right),
\label{Eq:Weyl-formula-B}\\
C_n:\ 
d(R)=&
\left[
\prod_{k=1}^n\prod_{i=1}^k
\left(\frac{\sum_{j=i}^{k}a_j}{k-i+1}+1\right)
\right]
\prod_{k=0}^{n-2}
\prod_{i=k+1}^{n-1}
\left(
\frac{\left(\sum_{j=k+1}^{i}a_{n-j}\right)
+\left(\sum_{j=0}^{k}2a_{n-j}\right)}{k+i+2}+1\right),
\label{Eq:Weyl-formula-C}\\
D_n:\ 
d(R)=&
\left[
\prod_{k=1}^{n-2}\prod_{i=1}^k
\left(\frac{\sum_{j=i}^{k}a_j}{k-i+1}+1\right)
\right]
\left[
\prod_{k=0}^{n-2}
\prod_{\ell=n-1}^{n}
\left(
\frac{\left(\sum_{j=1}^{k}a_{n-1-j}\right)+a_\ell}{k+1}+1\right)
\right]\nonumber\\
&\times
\left[
\prod_{k=0}^{n-3}
\prod_{i=k+1}^{n-2}
\left(
\frac{\left(\sum_{j=k+1}^{i}a_{n-1-j}\right)
 +\left(\sum_{j=1}^{k}2a_{n-1-j}\right)+a_{n-1}+a_n}{i+k+2}+1\right)
\right].
\label{Eq:Weyl-formula-D}
\end{align}

We check how to use the Weyl dimension formula of the Lie algebras
$\mathfrak{g}$.
First, the Weyl dimension formula of $A_4=\mathfrak{su}_5$ is 
\begin{align}
A_4:\ d(R)=&
\prod_{k=1}^4\prod_{i=1}^k
\left(\frac{\sum_{j=i}^{k}a_j}{k-i+1}+1\right).
\end{align}
E.g., the dimension of a representation $(a_1a_2a_3a_4)=(1000)$
is 
\begin{align}
A_4:\ d(1000)=&
\prod_{k=1}^4\prod_{i=1}^k
\left(\frac{\sum_{j=i}^{k}\delta_{j1}}{k-i+1}+1\right)
=\prod_{k=1}^4\left(\frac{1}{k}+1\right)
=\prod_{k=1}^4\frac{k+1}{k}=5.
\end{align}
Next, the Weyl dimension formula of $B_5=\mathfrak{so}_{10}$ is
\begin{align}
B_5:\ 
d(R)=&
\left[
\prod_{k=1}^5\prod_{i=1}^k
\left(\frac{\sum_{j=i}^{k}a_j}{k-i+1}+1\right)
\right]
\prod_{k=0}^{3}
\prod_{i=k}^{3}
\left(
\frac{\left(\sum_{j=k+1}^{i}a_{4-j}\right)
+\left(\sum_{j=0}^{k}2a_{4-j}\right)+a_5}{k+i+3}+1\right).
\end{align}
E.g., the dimension of a representation $(a_1a_2a_3a_4a_5)=(00001)$
is 
\begin{align}
B_5:\ 
d(00001)=&
\left[
\prod_{k=1}^5\prod_{i=1}^k
\left(\frac{\sum_{j=i}^{k}\delta_{j5}}{k-i+1}+1\right)
\right]
\prod_{k=0}^{3}
\prod_{i=k}^{3}
\left(\frac{1}{k+i+3}+1\right)\nonumber\\
=&
\left[
\prod_{i=1}^5
\left(\frac{7-i}{6-i}\right)
\right]
\prod_{k=0}^{3}
\prod_{i=k}^{3}
\left(\frac{k+i+4}{k+i+3}\right)=32.
\end{align}
Next, the Weyl dimension formula of $C_5=\mathfrak{usp}_{10}$ is
\begin{align}
C_5:\ 
d(R)=&
\left[
\prod_{k=1}^5\prod_{i=1}^k
\left(\frac{\sum_{j=i}^{k}a_j}{k-i+1}+1\right)
\right]
\prod_{k=0}^{3}
\prod_{i=k+1}^{4}
\left(
\frac{\left(\sum_{j=k+1}^{i}a_{5-j}\right)
+\left(\sum_{j=0}^{k}2a_{5-j}\right)}{k+i+2}+1\right).
\end{align}
E.g., the dimension of a representation $(a_1a_2a_3a_4a_5)=(20000)$
is 
\begin{align}
C_5:\ 
d(20000)=&
\left[
\prod_{k=1}^5\prod_{i=1}^k
\left(\frac{\sum_{j=i}^{k}2\delta_{j,1}}{k-i+1}+1\right)
\right]
\prod_{k=0}^{3}
\prod_{i=k+1}^{4}
\left(
\frac{\left(\sum_{j=k+1}^{i}2\delta_{5-j,1}\right)}{k+i+2}+1\right)
\nonumber\\
=&
\left[\prod_{k=1}^5\left(\frac{k+2}{k}\right)\right]
\prod_{k=0}^{3}\left(\frac{k+8}{k+6}\right)=55.
\end{align}
Next, the Weyl dimension formula of $D_5=\mathfrak{so}_{10}$ is
\begin{align}
D_5:\ 
d(R)=&
\left[
\prod_{k=1}^{3}\prod_{i=1}^k
\left(\frac{\sum_{j=i}^{k}a_j}{k-i+1}+1\right)
\right]
\left[
\prod_{k=0}^{3}
\prod_{\ell=4}^{5}
\left(
\frac{\left(\sum_{j=1}^{k}a_{4-j}\right)+a_\ell}{k+1}+1\right)
\right]\nonumber\\
&\times
\left[
\prod_{k=0}^{2}
\prod_{i=k+1}^{3}
\left(
\frac{\left(\sum_{j=k+1}^{i}a_{4-j}\right)
 +\left(\sum_{j=1}^{k}2a_{4-j}\right)+a_{4}+a_5}{i+k+2}+1\right)
\right].
\end{align}
E.g., the dimension of a representation $(a_1a_2a_3a_4a_5)=(00001)$
is 
\begin{align}
D_5:\ 
d(00001)=&
\left[\prod_{k=0}^{3} \left(\frac{k+2}{k+1}\right)\right]
\left[\prod_{k=0}^{2}
\prod_{i=k+1}^{3}\left(\frac{i+k+3}{i+k+2}\right)\right]=16.
\end{align}

\subsection{Dynkin index and Casimir invariant}
\label{Sec:Dynkin-index}

Let us introduce the Dynkin index of a representation $R$ of the simple
Lie algebra $\mathfrak{g}$ with the highest weight $\Lambda$ given
in Ref.~\cite{Dynkin:1957um}:
\begin{align}
T(R)=\frac{d(R)}{d(\mathfrak{g})}C_2(R),
\label{Eq:Dynkin-index-root}
\end{align}
where $d(R)$ is the dimension of a representation $R$, $d(\mathfrak{g})$
is the dimension of the adjoint representation of the Lie algebra
$\mathfrak{g}$, and $C_2(R)$ is its corresponding quadratic Casimir
invariant: 
\begin{align}
C_2(R)= \frac{(\Lambda,\Lambda+2g)}{2}.
\label{Eq:2-Casimir-root}
\end{align}
(Original Dynkin indices $\ell(R)$ defined by E.~Dynkin in
Ref.~\cite{Dynkin:1957um} are integer, but in our convention, some of
the Dynkin indices $T(R)$ are half-integer, which seems to be widely
used in Physics. The relation between two is $\ell(R)=2T(R)$.)
The Dynkin index satisfies the following relations:
\begin{align}
&T(R_i)=T(\overline{R_i}),
\label{Eq:Dynkin-index-R1}\\
&T(R_1\oplus R_2)=T(R_1)+T(R_2),
\label{Eq:Dynkin-index-R2}\\
&T(R_1\otimes R_2)=d(R_2)T(R_1)+d(R_1)T(R_2).
\label{Eq:Dynkin-index-R3}
\end{align}
(For its proof, see, e.g., \cite{Dynkin:1957um}.)

In the Dynkin basis, we can rewrite $(\Lambda,\Lambda+2g)$
in Eq.~(\ref{Eq:2-Casimir-root}) as
\begin{align}
(\Lambda,\Lambda+2g)=
\sum_{i,j=1}^{n}(a_i+2g_i)G(\mathfrak{g})_{ij}a_i,
\end{align}
where $G(\mathfrak{g})$ is the inverse Cartan matrix of the Lie algebra
$\mathfrak{g}$ in Table~\ref{Table:Cartan-matrices-inverse}.

Thus, the second order Casimir invariant and the Dynkin index of Lie
algebras $\mathfrak{g}$ in Eqs.~(\ref{Eq:2-Casimir-root}) and
(\ref{Eq:Dynkin-index-root}) are 
\begin{align}
C_2(R)=&\frac{1}{2}
\sum_{i,j=1}^{n}(a_i+2g_i)G(\mathfrak{g})_{ij}a_j,
\label{Eq:2-Casimir}\\
T(R)=&\frac{d(R)}{2 d(\mathfrak{g})}
\sum_{i,j=1}^{n}(a_i+2g_i)G(\mathfrak{g})_{ij}a_j,
\label{Eq:Dynkin-index}
\end{align}
where the dimensions of representations $R$ and $G$ can be calculated
by using the Weyl dimension formulas e.g., in
Eqs.~(\ref{Eq:Weyl-formula-A})-(\ref{Eq:Weyl-formula-D}).

We see several examples below.
For $A_1=\mathfrak{su}_2$, substituting $G(A_1)_{11}=1/2$ 
in Table~\ref{Table:Cartan-matrices-inverse}
into Eqs.~(\ref{Eq:2-Casimir}) and
(\ref{Eq:Dynkin-index}), we get  the Casimir invariant and the Dynkin
index of a representation $(a_1)$: 
\begin{align}
C_2(a_1)=&\frac{1}{4}a_1(a_1+2),\\
T(a_1)=&\frac{1}{12}a_1(a_1+1)(a_1+2).
\end{align}
E.g., the Casimir invariant and the Dynkin index of 
representations $(1)$ and $(2)$ are 
\begin{align}
&C_2(1)=\frac{3}{4},\ \ \ T(1)=\frac{1}{2};\\
&C_2(2)=2,\ \ \ T(2)=2.
\end{align}
For $A_4=\mathfrak{su}_5$, from Eqs.~(\ref{Eq:2-Casimir}) and
(\ref{Eq:Dynkin-index}), we get the Casimir invariant and the Dynkin
index of a representation $(a_1a_2a_3a_4)$:
\begin{align}
C_2(a_1a_2a_3a_4)=&
2a_1\left(\frac{4a_1+3a_2+2a_3+a_4}{20}+1\right)
+3a_2\left(\frac{3a_1+6a_2+4a_3+2a_4}{30}+1\right)\nonumber\\
&+3a_3\left(\frac{2a_1+4a_2+6a_3+3a_4}{30}+1\right)
+2a_4\left(\frac{a_1+2a_2+3a_3+4a_4}{20}+1\right),\\
T(a_1a_2a_3a_4)=&\frac{1}{24}
d(a_1a_2a_3a_4)C_2(a_1a_2a_3a_4),
\end{align}
where the metric matrix $G(A_4)$ and the dimension $d(a_1a_2a_3a_4)$
derived from Table~\ref{Table:Cartan-matrices-inverse}
and Eq.~(\ref{Eq:Weyl-formula-A}) are 
\begin{align}
G(A_4)&=\frac{1}{5}
\left(
\begin{array}{cccc}
4&3&2&1\\
3&6&4&2\\
2&4&6&3\\
1&2&3&4\\
\end{array}
\right),\\
d(a_1a_2a_3a_4)&=
\prod_{k=1}^4\prod_{i=1}^k
\left(\frac{\sum_{j=i}^{k}a_j}{k-i+1}+1\right).
\end{align}
E.g., the Casimir invariant and the Dynkin index of 
representations $(1000)$, $(0100)$ and $(1001)$ are 
\begin{align}
&C_2(1000)=\frac{12}{5},\ \ \ T(1000)=\frac{1}{2};\\
&C_2(0100)=\frac{18}{5},\ \ \ T(0100)=\frac{3}{2};\\
&C_2(1001)=5,\ \ \ T(1001)=5.
\end{align}
Next, for $B_5=\mathfrak{so}_{11}$, from Eqs.~(\ref{Eq:2-Casimir}) and
(\ref{Eq:Dynkin-index}), we get the Casimir invariant and the Dynkin
index of a representation $(a_1a_2a_3a_4a_5)$:
\begin{align}
C_2(a_1a_2a_3a_4a_5)=&
\frac{9}{2}a_1\left(\frac{2a_1+2a_2+2a_3+2a_4+a_5}{18}+1\right)\nonumber\\
&+8a_2\left(\frac{a_1+2a_2+2a_3+2a_4+a_5}{16}+1\right)\nonumber\\
&+\frac{21}{2}a_3\left(\frac{2a_1+4a_2+4a_3+6a_4+3a_5}{42}+1\right)\nonumber\\
&+12a_4\left(\frac{a_1+2a_2+3a_3+4a_4+2a_5}{24}+1\right)\nonumber\\
&+\frac{25}{4}a_5\left(\frac{2a_1+4a_2+6a_3+8a_4+5a_5}{50}+1\right),\\
T(a_1a_2a_3a_4a_5)=&\frac{d(a_1a_2a_3a_4a_5)}{55}C_2(a_1a_2a_3a_4a_5),
\end{align}
where the metric matrix $G(B_5)$ and the dimension $d(a_1a_2a_3a_4a_5)$ 
derived from Table~\ref{Table:Cartan-matrices-inverse}
and Eq.~(\ref{Eq:Weyl-formula-B}) are 
\begin{align}
G(B_5)&=\frac{1}{2}
\left(
\begin{array}{ccccc}
2&2&2&2&1\\
2&4&4&4&2\\
2&4&6&6&3\\
2&4&6&8&4\\
1&2&3&4&5/2\\
\end{array}
\right),\\
d(a_1a_2a_3a_4a_5)=&
\left[
\prod_{k=1}^5\prod_{i=1}^k
\left(\frac{\sum_{j=i}^{k}a_j}{k-i+1}+1\right)
\right]
\prod_{k=0}^{3}
\prod_{i=k}^{3}
\left(
\frac{\left(\sum_{j=k+1}^{i}a_{4-j}\right)
+\left(\sum_{j=0}^{k}2a_{4-j}\right)+a_5}{k+i+3}+1\right).
\end{align}
E.g., the Casimir invariant and the Dynkin index of 
representations $(10000)$ and $(01000)$
are 
\begin{align}
&C_2(10000)=5,\ \ \ T(10000)=1;\\
&C_2(01000)=9,\ \ \ T(01000)=9.
\end{align}
For $C_4=\mathfrak{usp}_8$, from Eqs.~(\ref{Eq:2-Casimir}) and
(\ref{Eq:Dynkin-index}), we get the Casimir invariant and the Dynkin
index of a representation $(a_1a_2a_3a_4)$:
\begin{align}
C_2(a_1a_2a_3a_4)=&
2a_1\left(\frac{a_1+a_2+a_3+a_4}{8}+1\right)
+\frac{7}{2}a_2\left(\frac{a_1+2a_2+2a_3+2a_4}{14}+1\right)\nonumber\\
&+\frac{9}{2}a_3\left(\frac{a_1+2a_2+3a_3+3a_4}{18}+1\right)
+5a_4\left(\frac{a_1+2a_2+3a_3+4a_4}{20}+1\right),\\
T(a_1a_2a_3a_4)=&\frac{d(a_1a_2a_3a_4)}{36}C_2(a_1a_2a_3a_4),
\end{align}
where the metric matrix $G(C_4)$ and the dimension $d(a_1a_2a_3a_4)$ 
derived from Table~\ref{Table:Cartan-matrices-inverse}
and Eq.~(\ref{Eq:Weyl-formula-C}) are 
\begin{align}
G(C_4)&=\frac{1}{2}
\left(
\begin{array}{cccc}
1&1&1&1\\
1&2&2&2\\
1&2&3&3\\
1&2&3&4\\
\end{array}
\right),\\
d(a_1a_2a_3a_4)=&
\left[
\prod_{k=1}^4\prod_{i=1}^k
\left(\frac{\sum_{j=i}^{k}a_j}{k-i+1}+1\right)
\right]
\prod_{k=0}^{2}
\prod_{i=k+1}^{3}
\left(
\frac{\left(\sum_{j=k+1}^{i}a_{4-j}\right)
+\left(\sum_{j=0}^{k}2a_{4-j}\right)}{k+i+2}+1\right).
\end{align}
E.g., the Casimir invariant and the Dynkin index of representations
$(1000)$, $(2000)$ and $(0100)$ are 
\begin{align}
&C_2(1000)=\frac{9}{4},\ \ \ T(1000)=\frac{1}{2};\\
&C_2(2000)=5,\ \ \ T(2000)=5;\\
&C_2(0100)=4,\ \ \ T(0100)=3.
\end{align}
For $D_4=\mathfrak{so}_8$, from Eqs.~(\ref{Eq:2-Casimir}) and
(\ref{Eq:Dynkin-index}), we get the Casimir invariant and the Dynkin
index of a representation $(a_1a_2a_3a_4)$:
\begin{align}
C_2(a_1a_2a_3a_4)=&
  3a_1\left(\frac{2a_1+2a_+a_3+a_4}{12}+1\right)
 +5a_2\left(\frac{a_1+2a_2+a_3+a_4}{10}+1\right)\nonumber\\
&+3a_3\left(\frac{a_1+2a_2+2a_3+a_4}{12}+1\right)
 +3a_4\left(\frac{a_1+2a_2+a_3+2a_4}{12}+1\right),\\
T(a_1a_2a_3a_4)=&\frac{d(a_1a_2a_3a_4)}{28}C_2(a_1a_2a_3a_4),
\end{align}
where the metric matrix $G(D_4)$ and the dimension $d(a_1a_2a_3a_4)$ 
derived from Table~\ref{Table:Cartan-matrices-inverse}
and Eq.~(\ref{Eq:Weyl-formula-D}) are 
\begin{align}
G(D_4)&=\frac{1}{2}
\left(
\begin{array}{ccccccc}
2&2&1&1\\
2&4&2&2\\
1&2&2&1\\
1&2&1&2\\
\end{array}
\right),\\
d(a_1a_2a_3a_4)=&
\left[
\prod_{k=1}^4\prod_{i=1}^k
\left(\frac{\sum_{j=i}^{k}a_j}{k-i+1}+1\right)
\right]
\prod_{k=0}^{2}
\prod_{i=k}^{2}
\left(
\frac{\left(\sum_{j=k+1}^{i}a_{3-j}\right)
+\left(\sum_{j=0}^{k}2a_{3-j}\right)+a_4}{k+i+3}+1\right).
\end{align}
E.g., the Casimir invariant and the Dynkin index of representations
$(1000)$, $(0100)$, $(0010)$ and $(0001)$ are 
\begin{align}
&C_2(1000)=\frac{7}{2},\ \ \ T(1000)=1;\\
&C_2(0100)=6,\ \ \ T(0100)=6;\\
&C_2(0010)=\frac{7}{2},\ \ \ T(0010)=1;\\
&C_2(0001)=\frac{7}{2},\ \ \ T(0001)=1.
\end{align}

Here we give an example when Casimir invariants and Dynkin indices are
necessary in four-dimensional gauge theories. The renormalization group
equation (RGE) for a gauge coupling constant discussed in e.g.,
Ref.~\cite{Slansky:1981yr}, is given by 
\begin{align}
\mu\frac{dg}{d\mu}=\beta(g),
\label{Eq:RGE-gauge-coupling-4D}
\end{align}
where $g$ is a gauge coupling constant, and 
$\beta(g)$ is a $\beta$ function for a gauge coupling constant. In
a model that includes real vector, Weyl fermion, and real scalar fields,
the formula of the $\beta$ function at one-loop level in
the four-dimensional gauge theory is given by
\begin{align}
\beta^{\rm 1-loop}(g)
=-\frac{g^3}{16\pi^2}\left[
\frac{11}{3}C_2(\mathfrak{g})
-\frac{2}{3}\sum_{\rm Weyl}T(R_F)
-\frac{1}{6}\sum_{\rm Real}T(R_S)\right],
\end{align}
where Vector, Weyl, and Real stand for real vector, Weyl fermion, and
real scalar field in terms of four-dimensional theories, respectively.
Since the vector bosons are gauge bosons, they belong to the
adjoint representation of the Lie algebra $\mathfrak{g}$. 
$C_2(\mathfrak{g})$ is the quadratic Casimir invariant of the adjoint
representation of the Lie algebra $\mathfrak{g}$, and 
$T(R_i)$ is a Dynkin index of the irreducible representation $R_{i}$ of
the Lie algebra $\mathfrak{g}$. (For two-loop level RGEs, see
e.g., Refs.~\cite{Machacek:1983tz,Machacek:1983fi,Machacek:1984zw}.)

\subsection{Anomaly coefficient}

Let us consider anomaly coefficients of the representations of
$A_{n-1}=\mathfrak{su}_n$ $(n\geq 3)$ introduced in
Ref.~\cite{Banks1976}.
It is related with so-called chiral gauge anomaly in four-dimensional
gauge theories. The anomaly can be generated only by complex
representations of Lie algebras, while any self-conjugate representation
does not produce the anomaly. There are several complex
representations of the Lie algebras $A_{n-1}=\mathfrak{su}_n (n\geq 3)$,
$D_{2n+1}=\mathfrak{so}_{4n+2} (n\geq 1)$, and $E_6$. 
However, any complex representation of $D_{2n+1}$ and $E_6$ has no
anomaly. (For its proof, see e.g., Ref.~\cite{Okubo:1977sc}).
Thus, only the complex representations of 
$A_{n-1}=\mathfrak{su}_n (n\geq 3)$ may produce the anomaly discussed
in Ref.~\cite{Patera:1981sc}. 
The general expression of the anomaly of any representation of $A_n$ is 
obtained in Ref.~\cite{Banks1976}. 

The anomaly coefficient of the representation $R$ of $\mathfrak{su}_n$
$(a_1a_2\cdots a_{n-1})$, which is defined as the coefficient of the
symmetric trace of three generators of a representation $R$ of the Lie
algebra $A_{n-1}=\mathfrak{su}_n$, is given in Ref.~\cite{Banks1976}:
\begin{align}
\frac{A(R)}{d(R)}:=
\sum_{i,j,k=1}^{n-1}a_{ijk}(a_i+1)(a_j+1)(a_k+1),
\label{Eq:anomaly-coefficients}
\end{align}
where $a_{ijk}$ is a completely symmetric under exchanging between $i$, 
$j$ and $k$, and for $i\leq j\leq k$,
\begin{align}
a_{ijk}=N_n i(n-2j)(n-k),
\label{Eq:anomaly-coefficients-a}
\end{align}
where $N_n$ is a normalization factor depending on its rank but
independent of $i,j,k$. We take it as satisfying
$A(10\cdots 0)=+1$.  
The anomaly coefficient $A(R)$ satisfies the following properties:
\begin{align}
&A(R_i)=-A(\overline{R_i}),
\label{Eq:anomaly-coefficients-R1}\\
&A(R_1\oplus R_2)=
A(R_1)+A(R_2),
\label{Eq:anomaly-coefficients-R2}\\
&A(R_1\otimes R_2)=
d(R_2)A(R_1)+d(R_1)A(R_2),
\label{Eq:anomaly-coefficients-R3}
\end{align}
where $R_1$ and $R_2$ are representations of $\mathfrak{su}_n$.
They suggest that for any self-conjugate (real or pseudo-real)
representation of $A_{n-1}$, $A(R_i=\overline{R}_i)=0$.
This indicates that all self-conjugate representations do not contribute
to the four-dimensional chiral anomaly.
Notice that any representation of $A_1=\mathfrak{su}_2$ is
a self-conjugate representation satisfying $R_i=\overline{R}_i$, so any
anomaly coefficient is zero: $A(R_i)=0$.
(For its proof, see, e.g.,\cite{Banks1976,Okubo:1977sc,Patera:1981sc}.)

We check several examples about the anomaly coefficients $A(R)$.
From Eq.~(\ref{Eq:anomaly-coefficients}), the anomaly coefficient
$A(R)$ of a representation of $\mathfrak{su}_3$ $(a_1a_2)$ is 
\begin{align}
\frac{A(a_1a_2)}{d(a_1a_2)}
=&\frac{1}{60}(a_1-a_2)(a_1+2a_2+3)(2a_1+a_2+3),
\end{align}
where $N_3=1/60$ by using $A(10)/d(10)=20N_3=+1/3$.
E.g., the anomaly coefficients of the representations $(10)$, $(01)$,
$(11)$, $(20)$, and $(02)$ are 
\begin{align}
A(10)=-A(01)=1,\ \
A(11)=0,\ \
A(20)=-A(02)=7,
\end{align}
where $d(10)=d(01)=3$, $d(11)=8$, $d(20)=d(02)=6$.
Next, from Eq.~(\ref{Eq:anomaly-coefficients}), the anomaly coefficient
$A(R)$ of a representation of $\mathfrak{su}_4$ $(a_1a_2a_3)$ is 
\begin{align}
\frac{A(a_1a_2a_3)}{d(a_1a_2a_3)}=&
\frac{1}{60}(a_1-a_3)(a_1+a_3+2)(a_1+2a_2+a_3+4),
\end{align}
where $N_4=1/60$ by using $A(100)/d(100)=15N_4=+1/4$.
E.g., the anomaly coefficients of the representations $(100)$, $(001)$,
$(010)$, $(101)$, $(200)$, and $(002)$ are 
\begin{align}
A(100)=-A(001)=+1,\ \ 
A(010)=A(101)=0,\ \ 
A(200)=-A(002)=8.
\end{align}
For more information, see the tables in
Appendix~\ref{Sec:Representations}.

Here we check how to construct anomaly-free chiral gauge theories by
using anomaly coefficients of $\mathfrak{su}_n$. By using anomaly
coefficients $A(R)$, four-dimensional gauge anomalies can be expressed
by 
\begin{align}
A^{\rm total}:=\sum_{R_i\in{\rm Left}}A(R_i)-\sum_{R_i\in{\rm Right}}A(R_i),
\end{align}
where Left and Right stand for left-handed and right-handed Weyl
fermions, respectively. Of course, Dirac fermions does not produce any
anomaly because a Dirac fermion contain left-handed and right-handed
Weyl fermions with the same representation.
In addition, the anomaly coefficient of a left-handed or right-handed
Weyl fermion with a self-conjugate representation is zero.
In four-dimensional theories, a right-handed Weyl fermion with a
representation $R$ of a Lie algebra $\mathfrak{g}$ can be 
converted to a left-handed Weyl fermion with its conjugate representation
$\overline{R}$ of the Lie algebra $\mathfrak{g}$. Therefore, we can use 
only left-handed Weyl fermions to express the anomaly in
four-dimensional theories:
\begin{align}
A^{\rm total}=
\sum_{R_i\in{\rm Left}}A(R_i)=A\left(\bigoplus_i R_i\right).
\label{Eq:total-anomaly-coefficients}
\end{align}
The anomaly-free condition $A^{\rm total}=0$ is satisfied if 
$A(\bigoplus_i R_i)=0$. It is easily checked by using anomaly
coefficients in the tables in Appendix~\ref{Sec:Representations}.
As well-known, in four-dimensional $SU(5)$ chiral gauge theories, 
a left-handed Weyl fermion with an $SU(5)$ reducible 
${\bf 10}\oplus{\bf \overline{5}}$ representation has no anomaly
because 
\begin{align}
A({\bf 10}\oplus{\bf \overline{5}})=
A({\bf 10})+A({\bf \overline{5}})=(+1)+(-1)=0,
\end{align}
where ${\bf 10}=(0100)$, ${\bf \overline{5}}=(0001)$,
$A({\bf 10})=+1$, and $A({\bf \overline{5}})=-1$ given
in Table~\ref{Table:Representations-A4} in
Appendix~\ref{Sec:Representations}. 
For four-dimensional $SU(6)$ chiral gauge theories, 
a left-handed Weyl fermion with an $SU(6)$ reducible 
${\bf 15}\oplus{\bf \overline{6}}\oplus{\bf \overline{6}}$
representation has no anomaly because 
\begin{align}
A({\bf 15}\oplus{\bf \overline{6}}\oplus{\bf \overline{6}})=
A({\bf 15})+A({\bf \overline{6}})+A({\bf \overline{6}})=(+2)+(-1)+(-1)=0,
\end{align}
where ${\bf 15}=(01000)$, ${\bf \overline{6}}=(00001)$,
$A({\bf 15})=+2$, and $A({\bf \overline{6}})=-1$
given in Table~\ref{Table:Representations-A5} in
Appendix~\ref{Sec:Representations}. 
For four-dimensional $SU(3)$ chiral gauge theories, 
a left-handed Weyl fermion with an $SU(3)$ reducible 
${\bf 6}\oplus{\bf 3}^{\oplus 7}$ representation has no anomaly because 
\begin{align}
A({\bf 6}\oplus{\bf 3}^{\oplus 7})=
A({\bf 6})+7\times A({\bf 3})=(-7)+7(+1)=0,
\end{align}
where ${\bf 6}=(02)$, ${\bf 3}=(10)$,
$A({\bf 6})=-7$, and $A({\bf 3})=+1$
given in Table~\ref{Table:Representations-A2} in
Appendix~\ref{Sec:Representations}. 

We comment on so-called mixed anomalies in four-dimensional gauge
theories. If a chiral gauge theory in four dimension contains not
only a semi-simple group $G$ but also $U(1)$, we have to take into
account the pure anomaly coefficients for $G$ and $U(1)$, but also
their mixed gauge anomaly coefficients for $G-G-U(1)$.
Note that in four-dimensional theories the pure anomaly coefficients for
gravity and the mixed anomaly coefficients for $G-{\rm grav}.-{\rm grav}$
are 
automatically zero, and if the mixed gauge anomaly coefficients for
$G-G-U(1)$ are zero, the mixed anomaly coefficients for
$U(1)-{\rm grav.}-{\rm grav.}$ are zero. In this paper, an anomaly-free
condition for $G$ means that its total pure gauge anomaly coefficient is
zero.

We also comment on anomaly-free irreducible complex
representations of $A_{n-1}=\mathfrak{su}_n$ $(n\geq 3)$. There is no
anomaly-free irreducible complex representation of $\mathfrak{su}_3$ and
$\mathfrak{su}_4$. 
It is also known that there are anomaly-free irreducible complex
representations of $\mathfrak{su}_5$ discussed in
Ref.~\cite{Eichten1982}. 
We have the general formula of the anomaly coefficients of
$\mathfrak{su}_n$. 
It is easy to find several anomaly-free irreducible complex
representations. For example, $\mathfrak{su}_5$ representations, e.g.,
$(3370)$, $(5181)$, $(7,7,15,1)$, $(5,12,17,0)$, $(15,0,16,8)$,
$(11,3,17,3)$, $(16,4,17,10)$ and their conjugate representations are
anomaly-free complex representations. Unfortunately, their dimension is
huge, so it seems to be difficult to apply them for Physics.

\subsection{Higher order Dynkin indices and Casimir invariants}

Higher order Dynkin indices and Casimir invariants are discussed in
Refs.~\cite{Patera:1976yd,Okubo:1981td,Okubo:1982dt,Okubo:1983sv}. 
The forth order Dynkin indices of semi-simple Lie algebras are listed
in Table 1 of Ref.~\cite{McKay:1981}.
It is known that any rank-$n$ simple Lie algebra has $n$ fundamental
Casimir invariant. (See, e.g., Ref.~\cite{Gilmore:102082}.) 
Here we list degrees of the fundamental $p$th order Casimir invariants
$C_{p}(R)$ of simple Lie algebras:
\begin{center}
\begin{longtable}{ccl}
\caption{Higher order Casimir invariants}
\label{Table:Casimir-invariants-higher}\\
\hline\hline
Class  &Compact group&Degree $p$\\\hline\hline
\endfirsthead
\multicolumn{3}{c}{Table~\ref{Table:Casimir-invariants-higher} (continued)}\\\hline\hline
Class  &Compact group&Degree $p$\\\hline\hline
\endhead
\hline
\endfoot
$A_n=\mathfrak{su}_{n+1}$ &$SU(n+1)$      &$2,3,4,\cdots,n+1$\\
$B_n=\mathfrak{so}_{2n+1}$&$SO(2n+1)$   &$2,4,6\cdots,2n$\\
$C_n=\mathfrak{usp}_{2n}$  &$USp(2n)$     &$2,4,6\cdots,2n$\\
$D_n=\mathfrak{so}_{2n}$  &$SO(2n)$     &$2,4,6\cdots,2n-2,n$\\
$E_6$  &$E_6$        &$2,5,6,8,9,12$\\
$E_7$  &$E_7$        &$2,6,8,10,12,14,18$\\
$E_8$  &$E_8$        &$2,8,12,14,18,20,24,30$\\
$F_4$  &$F_4$        &$2,6,8,12$\\
$G_2$  &$G_2$        &$2,6$\\
\end{longtable}
\end{center}
In this paper, we consider only the second and third order Dynkin 
indices corresponding to ordinary the Dynkin indices and anomaly
coefficients, respectively. However, the higher order Dynkin indices
play an important role for discussing anomaly cancellation in higher
dimensional spacetime. The $(n+1)$th order Dynkin index is related
with gauge anomalies in $2n$-dimensional gauge theories.
Fourth order Dynkin indices are listed in Table 1 in
Ref.~\cite{McKay:1981}.

There are many studies about anomalies; e.g., higher-order anomaly
cancellation in Ref.~\cite{Okubo:1985qk}; anomaly cancellation in
$d$-dimensional gauge and gravity theories e.g., in
Ref.~\cite{Bilal:2008qx}; a global (non-perturbative) anomaly discussed in
Refs.~\cite{Witten:1982fp,Geng:1987fg,Okubo:1987dh,Okubo:1989vn}, known
as the Witten global anomaly; 6-dimensional global anomaly discussed in
Refs.~\cite{Borghini:2001sa,vonGersdorff:2006nt}.
For general review about anomalies, see e.g., Ref.~\cite{Fujikawa:2004xa}.

\subsection{Naming convention for representations}

Here we check our naming convention especially for whether
the name of each complex representation of $A_{n-1}=\mathfrak{su}_n$,
$D_{n=2k+1}=\mathfrak{so}_{2n=4k+2}$, and $E_6$ has a bar or not.

We first consider a representation $R$ of $A_{n-1}=\mathfrak{su}_n$
whose Dynkin label is given by $(a_1a_2\cdots a_{n-1})$.
We determine whether the name of $R$ has a bar or not as below.
\begin{enumerate}
\item[(A-1)] First, we check whether a representation $R$ is the same as 
	     the conjugate representation or not. If the Dynkin label of
	     $R$, $(a_1a_2\cdots a_{n-1})$, 
	     is equal to that of the conjugate representation,
	     $(a_{n-1}\cdots a_2a_1)$, then the 
	     representation is self-conjugate, so the name of $R$ is no
	     bar. If
	     the Dynkin label of $R$, $(a_1a_2\cdots a_{n-1})$, is
	     different from that of the conjugate
	     representation, $(a_{n-1}\cdots a_2a_1)$, then $R$ is a
	     complex representation.
\item[(A-2)] Next, for a complex representation $R$, we check the
	     conjugacy class $C_c(R)$ of $A_{n-1}=\mathfrak{su}_n$,
	     where $C_c(R)=a_1+2a_2+3a_3+\cdots\ (\mbox{mod}.\ n)$
	     for $(a_1a_2a_3\cdots a_{n-1})$. 
	     For a representation $R$ whose the conjugacy class $C_c(R)$
	     is not $0$ or $[n/2]$, if the $C_c(R)$ is less than
	     $[n/2]$, the name has no bar; if the $C_c(R)$ is more than
	     $[n/2]$, the name has a bar. 
\item[(A-3)] Finally, for a complex representation $R$ whose $C_c(R)$ is
	     $0$ or $[n/2]$, if the value of the first Dynkin label
	     $a_1$ is less than that of the last Dynkin label
	     $a_{n-1}$, the name has a bar;
	     if the value of the first Dynkin label
	     $a_1$ is more than that of the last Dynkin label
	     $a_{n-1}$, the name has no bar;
	     if the value of the first Dynkin label $a_1$ is equal to 
	     that of the last Dynkin label $a_{n-1}$, then
	     we compare the value of the second Dynkin label
	     $a_2$ and the $(n-2)$th Dynkin label $a_{n-2}$.
	     If the value of the second Dynkin label $a_2$ is less
	     than that the $(n-2)$th Dynkin label $a_{n-2}$,
	     the name has a bar; if not, the name has no bar.
	     If the value of the second Dynkin label $a_2$ is equal to 
	     that of the $(n-2)$th Dynkin label $a_{n-2}$,
	     then we compare the value of the third Dynkin label
	     $a_3$ and the $(n-3)$th Dynkin label $a_{n-3}$.
	     We continue the process until the value of $a_{k}$
	     $(k=1,2,\cdots,[(n-1)/2])$ is different from $a_{n-k}$.
\end{enumerate}

We consider a representation $R$ of $D_{n=2k+1}=\mathfrak{so}_{2n=4k+2}$
whose Dynkin label is given by $(a_1\cdots a_{n-1}a_{n})$.
We also determine whether the name of $R$ has a bar or not as below.
\begin{enumerate}
\item[(D-1)] First, we check whether a representation $R$ is the same as
	     the conjugate representation or not. If the Dynkin label of
	     $R$, $(a_1\cdots a_{n-1}a_{n})$, is equal to that 
	     of the conjugate representation,
	     $(a_1\cdots a_{n}a_{n-1})$, then the representation is
	     self-conjugate,
	     so the name of $R$ has no bar. If the Dynkin label of $R$,
	     $(a_1\cdots a_{n-1}a_{n})$, is different from that 
	     of the conjugate representation,
	     $(a_1\cdots a_{n}a_{n-1})$, then $R$ is a 
	     complex representation. 
\item[(D-2)] Next, for a complex representation $R$, we check the
	     conjugacy classes 
	     $C_{c1}(R)$ and $C_{c2}(R)$ of
	     $D_{n=2k+1}=\mathfrak{so}_{2n=4n+2}$, where 
	     $C_{c1}(R)=a_{n-1}+a_{n}\ (\mbox{mod}.\ 2)$ and 
	     $C_{c2}(R)=2a_1+2a_2+\cdots+2a_{n-1}+(n-2)a_{n-1}+na_{n}\
	     (\mbox{mod}.\  4)$ for
	     $(a_1a_2\cdots a_{n-2}a_{n-1}a_{n})$.  
	     If the $(C_{c1}(R),C_{c2}(R))$ is $(1,1)$, the name has no
	     bar; if the $(C_{c1}(R),C_{c2}(R))$ is $(1,3)$, the name
	     has a bar.
\item[(D-3)] Finally, for a complex representation $R$ whose
	     $(C_{c1}(R),C_{c2}(R))$ is $(0,0)$ or $(0,2)$,
	     if the value of the Dynkin label $a_{n-1}$ is less than
	     that of the Dynkin label $a_{n}$,
	     the name has a bar;
	     if the value of the Dynkin label $a_{n}$ is less than
	     that of the Dynkin label $a_{n-1}$,
	     the name has no bar.
\end{enumerate}

We consider a representation $R$ of $E_6$ whose Dynkin label is
$(a_1a_2a_3a_4a_5a_6)$.
We determine whether the name of $R$ has a bar or not as below.
\begin{enumerate}
\item[(E-1)] First, we check whether a representation $R$ is the same as
	     the conjugate representation or not. If the Dynkin label of
	     $R$, $(a_1a_2a_3a_4a_5a_6)$ is equal to that 
	     of the conjugate representation,
	     $(a_5a_4a_3a_2a_1a_6)$, then the representation is
	     self-conjugate, so the name of $R$ has no bar. If the Dynkin
	     label of $R$, $(a_1a_2a_3a_4a_5a_6)$, is different from that
	     of the conjugate representation,
	     $(a_5a_4a_3a_2a_1a_6)$, then $R$ is a complex
	     representation.  
\item[(E-2)] Next, for a complex representation $R$, we check the
	     conjugacy classes $C_c(R)$ of $E_6$, where
	     $C_c(R)=a_1-a_2+a_4-a_5\ (\mbox{mod}.\ 3)$
	     for $(a_1a_2a_3a_4a_5a_6)$.
	     If the $C_c(R)$ is $1$, the name has no bar;
	     if the $C_c(R)$ is $2$, the name has a bar.
\item[(E-3)] Finally, for a complex representation $R$ whose $C_c(R)$ is
	     $0$, if the value of the first Dynkin label $a_1$ is less
	     than that of the fifth Dynkin label $a_5$,
	     the name has a bar;
	     if the value of the first Dynkin label $a_1$ is more
	     than that of the fifth Dynkin label $a_5$,
	     the name has no bar;
	     if the value of the first Dynkin label
	     $a_1$ is equal to that of the fifth Dynkin label
	     $a_5$, then we compare the value of the second Dynkin label
	     $a_2$ and the forth Dynkin label $a_4$.
	     If the value of the second Dynkin label $a_2$ is less
	     than that of the forth Dynkin label $a_4$,
	     the name has a bar; if not, the name has no bar.
\end{enumerate}

We check how to determine whether the name of each complex
representation of $A_{n-1}=\mathfrak{su}_n$,
$D_{n=2k+1}=\mathfrak{so}_{2n=4k+2}$ has a bar or not, explicitly by
using the above procedures.

First, we consider representations of $A_{2}=\mathfrak{su}_3$ whose
Dynkin label is given by $(a_1a_2)$. Here we check examples such as
$(10)$, $(01)$, $(20)$, $(02)$, $(11)$, $(30)$, and $(03)$ of
$A_{2}=\mathfrak{su}_3$.
From the above procedure (A-1), 
$(10)$, $(01)$, $(20)$, $(02)$, $(30)$, and $(03)$ of
$A_{2}=\mathfrak{su}_3$ are complex representations, while
$(11)$ of $A_{2}=\mathfrak{su}_3$ is a self-conjugate representation.
So, $(11)$ of $A_{2}=\mathfrak{su}_3$ has no bar.
From (A-2), $C_c(10)=C_c(02)=1$, $C_c(01)=C_c(20)=2$, and
$C_c(30)=C_c(03)=0$.
So, $(10)$ and $(02)$ of $A_{2}=\mathfrak{su}_3$ has no bar, while
$(01)$ and $(20)$ of $A_{2}=\mathfrak{su}_3$ have a bar.
From (A-3), 
$(30)$ of $A_{2}=\mathfrak{su}_3$ has no bar, while
$(03)$ of $A_{2}=\mathfrak{su}_3$ has a bar.
By using Weyl dimension formula discussed in
Sec.~\ref{Sec:Weyl-dimensional-formula}, we can calculate the dimension
of representations $(10)$, $(01)$, $(20)$, $(02)$, $(11)$, $(30)$, and
$(03)$ of $A_{2}=\mathfrak{su}_3$. We can denote them as
their corresponding names:
\begin{align}
 (10)={\bf 3},\ (01)={\bf \overline{3}},\
 (20)={\bf \overline{6}},\ (02)={\bf 6},\
 (11)={\bf 8},\
 (30)={\bf 10},\ (03)={\bf \overline{10}}. 
\end{align}
Note that $(20)$ is denoted as ${\bf 6}$ in Table~23 of
Ref.~\cite{Slansky:1981yr}.

Second, we consider representations of $A_{5}=\mathfrak{su}_6$ whose
Dynkin label is given by $(a_1a_2a_3a_4a_5)$. Here we check examples
such as $(10000)$, $(00001)$, $(30000)$, $(00003)$, $(11001)$,
$(10011)$, $(10020)$, and $(02001)$ of $A_{5}=\mathfrak{su}_6$.
From the above procedure (A-1), 
$(10000)$, $(00001)$, $(30000)$, $(00003)$, $(11001)$,
$(10011)$, $(10020)$, and $(02001)$ of $A_{5}=\mathfrak{su}_6$
are complex representations.
From (A-2), $C_c(10000)=1$,
$C_c(11001)=2$,
$C_c(30000)=C_c(00003)=C_c(10020)=C_c(02001)=3$,
$C_c(10011)=4$, and 
$C_c(00001)=5$.
So, $(10000)$ and $(11001)$ of $A_{5}=\mathfrak{su}_6$ has no bar, while
$(10011)$ and $(00001)$ of $A_{5}=\mathfrak{su}_6$ have a bar.
From (A-3), 
$(30000)$ and $(10020)$ of $A_{5}=\mathfrak{su}_6$ has no bar, while
$(00003)$ and $(02001)$ of $A_{5}=\mathfrak{su}_6$ has a bar.
By using Weyl dimension formula discussed in
Sec.~\ref{Sec:Weyl-dimensional-formula}, we can calculate the dimension
of representations $(10000)$, $(00001)$, $(30000)$, $(00003)$, $(11001)$,
$(10011)$, $(10020)$, and $(02001)$ of $A_{5}=\mathfrak{su}_6$.
We can denote them as their corresponding names:
\begin{align}
 &(10000)={\bf 6},\ (00001)={\bf \overline{6}},\
 (30000)={\bf 56},\ (00003)={\bf \overline{56}},\nonumber\\
 &(11001)={\bf 384},\ (10011)={\bf \overline{384}},\
 (10020)={\bf 560},\ (02001)={\bf \overline{560}}.
\end{align}
Note that $(02001)$ is denoted as ${\bf 560}$ in Table~31 of
Ref.~\cite{Slansky:1981yr}.

Third, we consider representations of $D_{5}=\mathfrak{so}_{10}$ whose
Dynkin label is given by $(a_1a_2a_3a_4a_5)$. Here we check examples
such as $(00001)$, $(00010)$, $(00020)$, $(00002)$, $(10010)$,
$(10001)$, $(10020)$, and $(10002)$ of $D_{5}=\mathfrak{so}_{10}$.
From the above procedure (D-1),
$(00001)$, $(00010)$, $(00020)$, $(00002)$, $(10010)$,
$(10001)$, $(10020)$, and $(10002)$ of $D_{5}=\mathfrak{so}_{10}$
are complex representations.
From (D-2),
$(C_{c1}(00001),C_{c2}(00001)) = (C_{c1}(10010),C_{c2}(10010))=(1,1)$, 
$(C_{c1}(00010),C_{c2}(00010)) = (C_{c1}(10001),C_{c2}(10001))=(1,3)$,
$(C_{c1}(00020),C_{c2}(00020)) = (C_{c1}(00002),C_{c2}(00002))=(0,2)$, and
$(C_{c1}(10020),C_{c2}(10020)) = (C_{c1}(10002),C_{c2}(10002))=(0,0)$.
So, $(00001)$ and $(10010)$ of $D_{5}=\mathfrak{so}_{10}$ has no bar,
while 
$(00010)$ and $(10001)$ of $D_{5}=\mathfrak{so}_{10}$ have a bar.
From (D-3), 
$(00020)$ and $(10020)$ of $D_{5}=\mathfrak{so}_{10}$ has no bar, while
$(00002)$ and $(10002)$ of $D_{5}=\mathfrak{so}_{10}$ has a bar.
By using Weyl dimension formula discussed in
Sec.~\ref{Sec:Weyl-dimensional-formula}, we can calculate the dimension
of representations $(00001)$, $(00010)$, $(00020)$, $(00002)$, $(10010)$,
$(10001)$, $(10020)$, and $(10002)$ of $D_{5}=\mathfrak{so}_{10}$.
We can denote them as their corresponding names:
\begin{align}
 &(00001)={\bf 16},\ (00010)={\bf \overline{16}},\
 (00020)={\bf 126},\ (00002)={\bf \overline{126}},\nonumber\\
 &(10010)={\bf 144},\ (10001)={\bf \overline{144}},\
 (10020)={\bf 1050},\ (10002)={\bf \overline{1050}}.
\end{align}
Note that $(00002)$ and $(10002)$ are denoted as ${\bf 126}$ and
${\bf 1050}$ in Table~41 of Ref.~\cite{Slansky:1981yr}, respectively.

Forth, we consider representations of $D_{7}=\mathfrak{so}_{14}$ whose
Dynkin label is given by $(a_1a_2a_3a_4a_5a_6a_7)$. Here we check examples
such as $(0000010)$, $(0000001)$, $(0000020)$, $(0000002)$, $(1000001)$,
$(1000010)$, $(1000020)$, and $(1000002)$ of $D_{7}=\mathfrak{so}_{14}$.
From the above procedure (D-1),
$(0000001)$, $(0000010)$, $(0000020)$, $(0000002)$, $(1000001)$,
$(1000010)$, $(1000020)$, and $(1000002)$ of $D_{7}=\mathfrak{so}_{14}$
are complex representations.
From (D-2),
$(C_{c1}(0000010),C_{c2}(0000010)) =
(C_{c1}(1000001),C_{c2}(1000001)) = (1,1)$,
$(C_{c1}(0000001),C_{c2}(0000001)) =
(C_{c1}(1000010),C_{c2}(1000010)) = (1,3)$,\\
$(C_{c1}(0000020),C_{c2}(0000020)) =
(C_{c1}(0000002),C_{c2}(0000002)) = (0,2)$, and \\ 
$(C_{c1}(1000020),C_{c2}(1000020)) =
(C_{c1}(1000002),C_{c2}(1000002)) = (0,0)$.
So, $(0000010)$ and $(1000001)$ of $D_{7}=\mathfrak{so}_{14}$ has no bar,
while 
$(0000001)$ and $(1000010)$ of $D_{7}=\mathfrak{so}_{14}$ have a bar.
From (D-3), 
$(0000020)$ and $(1000020)$ of $D_{7}=\mathfrak{so}_{14}$ has no bar, while
$(0000002)$ and $(1000002)$ of $D_{7}=\mathfrak{so}_{14}$ has a bar.
By using Weyl dimension formula discussed in
Sec.~\ref{Sec:Weyl-dimensional-formula}, we can calculate the dimension
of representations $(0000010)$, $(0000001)$, $(0000020)$, $(0000002)$,
$(1000001)$, $(1000010)$, $(1000020)$, and $(1000002)$ of
$D_{7}=\mathfrak{so}_{14}$. 
We can denote them as their corresponding names:
\begin{align}
 &(0000010)={\bf 64},\ (0000001)={\bf \overline{64}},\
  (0000020)={\bf 1716},\ (0000002)={\bf \overline{1716}},\nonumber\\
 &(1000001)={\bf 832},\ (1000010)={\bf \overline{832}},\
  (1000020)={\bf 21021},\ (1000002)={\bf \overline{21021}}.
\end{align}
Note that $(0000001)$ and $(0000002)$ are denoted as ${\bf 64}$ and
${\bf 1716}$ in Table~55 of Ref.~\cite{Slansky:1981yr}, respectively.

Finally, we consider representations of $E_{6}$ whose
Dynkin label is given by $(a_1a_2a_3a_4a_5a_6)$. Here we check examples
such as $(100000)$, $(000010)$, $(000100)$, $(010000)$, $(100001)$,
$(000011)$, $(300000)$, and $(000030)$ of $E_{6}$.
From the above procedure (E-1),
$(100000)$, $(000010)$, $(000100)$, $(010000)$, $(100001)$,
$(000011)$, $(300000)$, and $(000030)$ of $E_{6}$
are complex representations.
From (E-2),
$C_c(100000)=C_c(000100)=C_c(100001)=1$,
$C_c(000010)=C_c(010000)=C_c(000011)=2$, and
$C_c(300000)=C_c(000030)=0$.
So, $(100000)$, $(000100)$, and $(100001)$ of $E_{6}$ has no bar,
while 
$(000010)$, $(010000)$, and $(000011)$ of $E_{6}$ have a bar.
From (E-3), 
$(300000)$ of $E_{6}$ has no bar, while
$(000030)$ of $E_{6}$ has a bar.
By using Weyl dimension formula discussed in
Sec.~\ref{Sec:Weyl-dimensional-formula}, we can calculate the dimension
of representations
$(100000)$, $(000010)$, $(000100)$, $(010000)$, $(100001)$,
$(000011)$, $(300000)$, and $(000030)$ of $E_{6}$.
We can denote them as their corresponding names:
\begin{align}
 &(100000)={\bf 27},\ (000010)={\bf \overline{27}},\
  (000100)={\bf 351},\ (010000)={\bf \overline{351}},\nonumber\\
 &(100001)={\bf 1728},\ (000011)={\bf \overline{1728}},\
  (300000)={\bf 3003},\ (000030)={\bf \overline{3003}}.
\end{align}

From the above discussion, the naming convention for representations in
the paper is somewhat different from that in Ref.~\cite{Slansky:1981yr}.
So, it is better to check Dynkin labels and naming convention for
representations when one uses several references.

\section{Representations of subalgebras}
\label{Sec:Representations-subalgebras}

Let us consider how to decompose irreducible representations $R$ of a
Lie algebras $\mathfrak{g}$ into irreducible representations $R'$s of
their Lie subalgebras $\mathfrak{h}$ discussed in
Sec.~\ref{Sec:Subalgebras}.
It is essential to apply a Lie algebra $\mathfrak{g}$
for unified model building discussed in Sec.~\ref{Sec:Model-Building}.
To achieve it, we need to know the relations between a simple Lie
algebra $\mathfrak{g}$ and its maximal Lie subalgebra $\mathfrak{h}$.
We also check how to obtain non-maximal Lie subalgebras $\mathfrak{h'}$
from the viewpoint of the Lie algebra $\mathfrak{g}$, which
satisfies $\mathfrak{h'}\subset\mathfrak{h}\subset\mathfrak{g}$.

\subsection{Branching rules and projection matrices}

First, we check two technical terms; branching rules and
projection matrices introduced in Ref.~\cite{Mckay:1977}.
A branching rule is a relation between a representation of a Lie algebra
$\mathfrak{g}$ and representations of its Lie subalgebra $\mathfrak{h}$.
E.g., for $\mathfrak{su}_3\supset\mathfrak{su}_2\oplus\mathfrak{u}_1$,
${\bf 3}$ of $\mathfrak{g}=\mathfrak{su}_3$ is decomposed into 
${\bf 2}(+1)$ and ${\bf 1}(-2)$ of
$\mathfrak{h}=\mathfrak{su}_2\oplus\mathfrak{u}_1$; i.e., 
one of branching rules
of $\mathfrak{su}_3\supset\mathfrak{su}_2\oplus\mathfrak{u}_1$ is 
\begin{align}
{\bf 3}={\bf 2}(+1)\oplus{\bf 1}(-2).
\end{align}
Also, branching rules between a Lie algebra $\mathfrak{g}$ and its Lie
subalgebra $\mathfrak{h}$ are determined by single projection matrix.
The projection matrix of a Lie algebra $\mathfrak{g}$ and its subalgebra
$\mathfrak{h}$ is fixed by the weight diagrams of a branching rule of
them, where weight diagrams of a Lie algebra $\mathfrak{g}$ can be
written down by using the Cartan matrix of $\mathfrak{g}$ and Dynkin
labels. (Several examples of weight diagrams are shown in
Appendix~\ref{Sec:Weight-diagrams}. For more information about weight
diagrams, see e.g.,
Refs.~\cite{Slansky:1981yr,Georgi:1982jb,Cahn:1985wk}.) 
E.g., for $\mathfrak{su}_3\supset\mathfrak{su}_2\oplus\mathfrak{u}_1$,
its projection matrix
$P_{\mathfrak{su}_3\supset\mathfrak{su}_2\oplus\mathfrak{u}_1}$ is
determined by a branching rule ${\bf 3}={\bf 2}(+1)\oplus{\bf 1}(-2)$
and the following weight diagram:
\begin{align}
{\bf 3}\Leftrightarrow
\begin{array}{|c|}\hline
(10)\\
(\bar{1}1)\\
(0\bar{1})\\\hline
\end{array}
\hspace{1em}
\underset{P_{\mathfrak{su}_3\supset\mathfrak{su}_2\oplus\mathfrak{u}_1}}
{\Longrightarrow}
\hspace{1em}
\begin{array}{|c|}\hline
(1)(+1)\\
(\bar{1})(+1)\\
(0)(-2)\\\hline
\end{array}
\Leftrightarrow{\bf 2}(+1)\oplus{\bf 1}(-2),
\label{Eq:branching-rule-demo}
\end{align}
where $\bar{1}$ stands for $-1$,
$(10)$, $(\bar{1}1)$, and $(0\bar{1})$ in the first box
are the weights of the representation $(10)$ of a Lie algebra
$\mathfrak{su}_3$,
$(1)(+1)$, $(\bar{1})(+1)$, and $(0)(-2)$ in the second box
are the weights of representations $(1)(+1)$ and $(0)(-2)$ of its
subalgebras $\mathfrak{su}_2\oplus\mathfrak{u}_1$.
The projection matrix is a $2\times 2$ matrix because $\mathfrak{su}_3$
is rank-2 and $\mathfrak{su}_2\oplus\mathfrak{u}_1$ is also
rank-2. 
\begin{align}
P_{\mathfrak{su}_3\supset \mathfrak{su}_2\oplus\mathfrak{u}_1}=
\left(
\begin{array}{cc}
a_{11}&a_{12}\\
a_{21}&a_{22}\\
\end{array}
\right),
\end{align}
where $a_{ij}$ $(i,j=1,2)$ are unknown values.
We introduce an $\mathfrak{su}_3$ weight vector $w_{\mathfrak{su}_3}$,
and an $\mathfrak{su}_2\oplus\mathfrak{u}_1$ weight vector
$v_{\mathfrak{su}_2\oplus\mathfrak{u}_1}$ that satisfy the following
relation, where a weight vector is a row vector:
\begin{align}
v_{\mathfrak{su}_2\oplus\mathfrak{u}_1}^T=
P_{\mathfrak{su}_3\supset\mathfrak{su}_2\oplus\mathfrak{u}_1}
w_{\mathfrak{su}_3}^T,
\end{align}
where for ${\bf 3}={\bf 2}(+1)\oplus{\bf 1}(-2)$, 
$w_{\mathfrak{su}_3}=(1\ 0),(-1\ 1),(0\ -1)$ and 
$v_{\mathfrak{su}_2\oplus\mathfrak{u}_1}=(1\ 1),(-1\ 1),(0\ -2)$.
By using the relation between 
$\{w_{\mathfrak{su}_3},v_{\mathfrak{su}_2\oplus\mathfrak{u}_1}\}
=\{(1\ 0),(1\ 1)\},\{(-1\ 1),(-1\ 1)\},\{(0\ -1),(0\ -2)\}$, 
we find $a_{11}=1$, $a_{12}=0$, $a_{21}=1$, and $a_{22}=2$.
Therefore, we can find the projection matrix as 
\begin{align}
P_{\mathfrak{su}_3\supset\mathfrak{su}_2\oplus\mathfrak{u}_1}
=\left(
\begin{array}{cc}
1&0\\
1&2\\
\end{array}
\right).
\label{Eq:projection-matrix-demo}
\end{align}

Here we check how much projection matrices are useful to obtain
branching rules. E.g., for
$\mathfrak{su}_3\supset\mathfrak{su}_2\oplus\mathfrak{u}_1$,
we decompose an irreducible representation ${\bf 6}$ of
$\mathfrak{su}_3$, whose Dynkin label is $(02)$, 
into some irreducible representations of 
$\mathfrak{su}_2\oplus\mathfrak{u}_1$. 
The weight diagram of ${\bf 6}$ of $\mathfrak{su}_3$ is 
\begin{align}
{\bf 6}\Leftrightarrow
\begin{array}{|c|}\hline
(02)\\
(10)\\
(\bar{1}1)\ (2\bar{2})\\
(0\bar{1})\\
(\bar{2}0)\\\hline
\end{array}.
\end{align}
By using the projection matrix in Eq.~(\ref{Eq:projection-matrix-demo}),
we get the weight diagram of $\mathfrak{su}_2\oplus\mathfrak{u}_1$:
\begin{align}
{\bf 6}\Leftrightarrow
\begin{array}{|c|}\hline
(02)\\
(10)\\
(\bar{1}1)\ (2\bar{2})\\
(0\bar{1})\\
(\bar{2}0)\\
\hline
\end{array}
\hspace{1em}
\underset{P_{\mathfrak{su}_3\supset\mathfrak{su}_2\oplus\mathfrak{u}_1}}
{\Longrightarrow}
\hspace{1em}
\begin{array}{|c|}\hline
(0)(+4)\\
(1)(+1)\\
(\bar{1})(+1)\ (2)(-2)\\
(0)(-2)\\
(\bar{2})(-2)\\
\hline
\end{array}.
\end{align}
By translating the weight diagram of 
$\mathfrak{su}_2\oplus\mathfrak{u}_1$ into irreducible representations
of $\mathfrak{su}_2\oplus\mathfrak{u}_1$, we obtain
\begin{align}
{\bf 6}\Leftrightarrow
\begin{array}{|c|}\hline
(02)\\
(10)\\
(\bar{1}1)\ (2\bar{2})\\
(0\bar{1})\\
(\bar{2}0)\\
\hline
\end{array}
\hspace{1em}
\underset{P_{\mathfrak{su}_3\supset\mathfrak{su}_2\oplus\mathfrak{u}_1}}
{\Longrightarrow}
\hspace{1em}
\begin{array}{|c|}\hline
(0)(+4)\\
(1)(+1)\\
(\bar{1})(+1)\ (2)(-2)\\
(0)(-2)\\
(\bar{2})(-2)\\
\hline
\end{array}
\Leftrightarrow{\bf 3}(-2)\oplus{\bf 2}(+1)\oplus{\bf 1}(+4).
\end{align}
Thus, we find another branching rule of
$\mathfrak{su}_3\supset\mathfrak{su}_2\oplus\mathfrak{u}_1$:
\begin{align}
{\bf 6}={\bf 3}(-2)\oplus{\bf 2}(+1)\oplus{\bf 1}(+4).
\end{align}
The above processes are the same as any branching rule of
$\mathfrak{su}_3\supset\mathfrak{su}_2\oplus\mathfrak{u}_1$.
In principle, we can get any branching rule of 
$\mathfrak{su}_3\supset\mathfrak{su}_2\oplus\mathfrak{u}_1$
by using one projection matrix 
$P_{\mathfrak{su}_3\supset\mathfrak{su}_2\oplus\mathfrak{u}_1}$.
Of course, in reality, it is almost impossible to decompose e.g.,
$10,000$ dimensional representation of $\mathfrak{su}_3$ or
$10,000$ representations of $\mathfrak{su}_3$ by hand.
Fortunately, since the calculation is systematic, there are several
public codes for the calculation. (See e.g.,
Refs.~\cite{Fonseca:2011sy,Feger:2012bs}.)

One may wonder whether each projection matrix of a Lie algebra
$\mathfrak{g}$ and its subalgebra $\mathfrak{h}$ is unique or not.
In general, its projection matrix is not unique. 
We will see its example below. For example, 
to obtain the projection matrix of
$\mathfrak{su}_3\supset\mathfrak{su}_2\oplus\mathfrak{u}_1$ 
in Eq.~(\ref{Eq:projection-matrix-demo}),
we use a branching rule in Eq.~(\ref{Eq:branching-rule-demo}); i.e.,
$\{w_{\mathfrak{su}_3},v_{\mathfrak{su}_2\oplus\mathfrak{u}_1}\}
=\{(1\ 0),(1\ 1)\},\{(-1\ 1),(-1\ 1)\},\{(0\ -1),(0\ -2)\}$.
Instead of it, we can use 
$\{w_{\mathfrak{su}_3},v_{\mathfrak{su}_2\oplus\mathfrak{u}_1}\}
=\{(1\ 0),(1\ 1)\},\{(-1\ 1),(0\ -2)\},\{(0\ -1),(-1\ 1)\}$.
\begin{align}
{\bf 3}\Leftrightarrow
\begin{array}{|c|}\hline
(10)\\
(\bar{1}1)\\
(0\bar{1})\\\hline
\end{array}
\hspace{1em}
\underset{P_{\mathfrak{su}_3\supset\mathfrak{su}_2\oplus\mathfrak{u}_1}}
{\Longrightarrow}
\hspace{1em}
\begin{array}{|c|}\hline
(1)(+1)\\
(0)(-2)\\
(\bar{1})(+1)\\
\hline
\end{array}
\Leftrightarrow{\bf 2}(+1)\oplus{\bf 1}(-2).
\end{align}
It leads to $a_{11}=1$, $a_{12}=1$, $a_{21}=1$, and $a_{22}=-1$.
Thus, we get the projection matrix
\begin{align}
P_{\mathfrak{su}_3\supset\mathfrak{su}_2\oplus\mathfrak{u}_1}
=\left(
\begin{array}{cc}
1&1\\
1&-1\\
\end{array}
\right).
\label{Eq:projection-matrix-demo-2}
\end{align}
To apply the projection matrix for ${\bf 6}$, we get its branching rule
\begin{align}
{\bf 6}\Leftrightarrow
\begin{array}{|c|}\hline
(02)\\
(10)\\
(\bar{1}1)\ (2\bar{2})\\
(0\bar{1})\\
(\bar{2}0)\\
\hline
\end{array}
\hspace{1em}
\underset{P_{\mathfrak{su}_3\supset\mathfrak{su}_2\oplus\mathfrak{u}_1}}
{\Longrightarrow}
\hspace{1em}
\begin{array}{|c|}\hline
(2)(-2)\\
(1)(+1)\\
(0)(-2)\ (0)(+4)\\
(\bar{1})(+1)\\
(\bar{2})(-2)\\
\hline
\end{array}
\Leftrightarrow{\bf 3}(-2)\oplus{\bf 2}(+1)\oplus{\bf 1}(+4).
\end{align}
From the above example, we find that projection matrices derived from a
branching rule are not always unique, but apparently different
projection matrices lead to the same branching rules.

\subsection{Dynkin diagrams}

Let us discuss how to calculate projection matrices by using Dynkin
diagrams below. Here we introduce three notions: removing, folding, and
shrinking. All $R$-subalgebras can be found by using removing; a part of
$S$-subalgebras can be found by using folding or shrinking; in general,
all subalgebras can be found by using a combination of removing,
folding, and shrinking. 

We start to discuss $R$-subalgebras by using (extended) Dynkin diagrams.
An example of removed Dynkin diagrams is 
$A_2=\mathfrak{su}_3\supset\mathfrak{su}_2\oplus\mathfrak{u}_1$.
The Dynkin diagram of
$A_2=\mathfrak{su}_3\supset\mathfrak{su}_2\oplus\mathfrak{u}_1$ is
\begin{align}
\underset{\mathfrak{su}_3}
{\xygraph{
    \circ ([]!{+(0,-.3)}  {{}_1}) - [r]
    \circ ([]!{+(0,-.3)}  {{}_2})
}}
&\underset{\rm Removing}{\Longrightarrow}
\underset{\mathfrak{su}_2\oplus\mathfrak{u}_1}
{\xygraph{
    \circ ([]!{+(0,-.3)}  {{}_1}) - [r]
    {\circ\hspace{-0.645em}\times} ([]!{+(0,-.3)}  {{}_2})
}}.
\end{align}
In this case, we removed the 2nd node of $\mathfrak{su}_3$. The remaining
1st node of $\mathfrak{su}_3$ stands for a subalgebra $\mathfrak{su}_2$.
The removed node from Dynkin diagrams means $\mathfrak{u}_1$.
Thus, we find the subalgebras as $\mathfrak{su}_2\oplus\mathfrak{u}_1$.
We can also identify the 1st node of $\mathfrak{su}_3$ as the node of
the subalgebra $\mathfrak{su}_2$. I.e.,
its corresponding projection matrix is
\begin{align}
P_{\mathfrak{su}_3\supset \mathfrak{su}_2\oplus\mathfrak{u}_1}=
\left(
\begin{array}{cc}
1&0\\
a_{21}&a_{22}\\
\end{array}
\right),
\end{align}
where $a_{21}$ and $a_{22}$ are still undetermined values.
By using the branching rule in Eq.~(\ref{Eq:branching-rule-demo}), 
we have to use the relations of weight vectors 
$\{w_{\mathfrak{su}_3},v_{\mathfrak{su}_2\oplus\mathfrak{u}_1}\}
=\{(1\ 0),(1\ 1)\},\{(-1\ 1),(-1\ 1)\},\{(0\ -1),(0\ -2)\}$
instead of 
$\{w_{\mathfrak{su}_3},v_{\mathfrak{su}_2\oplus\mathfrak{u}_1}\}
=\{(1\ 0),(1\ 1)\},\{(-1\ 1),(0\ -2)\},\{(0\ -1),(-1\ 1)\}$.
Thus, we get the same projection matrix in
Eq.~(\ref{Eq:projection-matrix-demo}):
\begin{align}
P_{\mathfrak{su}_3\supset \mathfrak{su}_2\oplus\mathfrak{u}_1}=
\left(
\begin{array}{cc}
1&0\\
1&2\\
\end{array}
\right).
\end{align}

Another example of removed Dynkin diagrams is 
$A_4=\mathfrak{su}_5\supset
\mathfrak{su}_3\oplus\mathfrak{su}_2\oplus\mathfrak{u}_1$.
The Dynkin diagram is 
\begin{align}
\underset{\mathfrak{su}_5}
{\xygraph{
    \circ ([]!{+(0,-.3)}  {{}_1}) - [r]
    \circ ([]!{+(0,-.3)}  {{}_2}) - [r]
    \circ ([]!{+(0,-.3)}  {{}_3}) - [r] 
    \circ ([]!{+(0,-.3)}  {{}_4})
}}
&\underset{\rm Removing}{\Longrightarrow}
\underset{\mathfrak{su}_3\oplus\mathfrak{su}_2\oplus\mathfrak{u}_1}
{\xygraph{
    \circ ([]!{+(0,-.3)}  {{}_1}) - [r]
    \circ ([]!{+(0,-.3)}  {{}_2}) - [r]
    {\circ\hspace{-0.645em}\times} ([]!{+(0,-.3)}  {{}_3}) - [r] 
    \circ ([]!{+(0,-.3)}  {{}_4})
}}.
\end{align}
When we use an $\mathfrak{su}_5$ weight vector $w_{\mathfrak{su}_5}$
and an ${\mathfrak{su}_3\oplus\mathfrak{su}_2\oplus\mathfrak{u}_1}$
weight vector
$v_{\mathfrak{su}_3\oplus\mathfrak{su}_2\oplus\mathfrak{u}_1}$,
both $\mathfrak{su}_5$ and
${\mathfrak{su}_3\oplus\mathfrak{su}_2\oplus\mathfrak{u}_1}$ are rank-4,
so its projection matrix
$P_{\mathfrak{su}_5\supset\mathfrak{su}_3\oplus\mathfrak{su}_2\oplus\mathfrak{u}_1}$
is a $4\times 4$ matrix. The 1st and 2nd nodes of $\mathfrak{su}_5$ can
be identified to the 1st and 2nd nodes of $\mathfrak{su}_3$, and
the 4th node of $\mathfrak{su}_5$ can
be identified to the node of $\mathfrak{su}_2$. Thus, we can write 
its projection matrix as 
\begin{align}
P_{\mathfrak{su}_5\supset\mathfrak{su}_3\oplus\mathfrak{su}_2\oplus\mathfrak{u}_1}&=
\left(
\begin{array}{cccc}
1&0&0&0\\
0&1&0&0\\
0&0&0&1\\
a_{41}&a_{42}&a_{43}&a_{44}\\
\end{array}
\right),
\end{align}
where $a_{4j}$ $(j=1,2,3,4)$ are undetermined values. Here we use the
following branching rule 
\begin{align}
{\bf 5}\Leftrightarrow
\begin{array}{|c|}\hline
(1000)\\
(\bar{1}100)\\
(0\bar{1}10)\\
(00\bar{1}1)\\
(000\bar{1})\\
\hline
\end{array}
\hspace{1em}
\underset{P_{\mathfrak{su}_5\supset\mathfrak{su}_3\oplus\mathfrak{su}_2\oplus\mathfrak{u}_1}}
{\Longrightarrow}
\hspace{1em}
\begin{array}{|c|}\hline
(10)(0)(+2)\\
(\bar{1}1)(0)(+2)\\
(0\bar{1})(0)(+2)\\
(00)(1)(-3)\\
(00)(\bar{1})(-3)\\
\hline
\end{array}
\Leftrightarrow({\bf 3,1})(+2)\oplus({\bf 1,2})(-3).
\end{align}
When we use the relations
$\{w_{\mathfrak{su}_5},v_{\mathfrak{su}_3\oplus\mathfrak{su}_2\oplus\mathfrak{u}_1}\}
=\{(1\ 0\ 0\ 0),(1\ 0\ 0\ 2)\},
\{(-1\ 1\ 0\ 0),(-1\ 1\ 0\ 2)\},
\{(0\ -1\ 1\ 0),(0\ -1\ 0\ 2)\},
\{(0\ 0\ -1\ 1),(0\ 0\ 1\ -3)\},
\{(0\ 0\ 0\ -1),(0\ 0\ -1\ -3)\}$, 
we get $a_{41}=2$, $a_{42}=4$, $a_{43}=6$, and $a_{44}=3$.
Thus, its projection matrix is
\begin{align}
P_{\mathfrak{su}_5\supset\mathfrak{su}_3\oplus\mathfrak{su}_2\oplus\mathfrak{u}_1}&=
\left(
\begin{array}{cccc}
1&0&0&0\\
0&1&0&0\\
0&0&0&1\\
2&4&6&3\\
\end{array}
\right).
\end{align}

Next, we discuss $S$-subalgebras by using Dynkin diagrams.
An example of folded Dynkin diagrams is 
$D_5=\mathfrak{so}_{10}\supset\mathfrak{so}_{9}$.
The Dynkin diagram is
\begin{align}
\underset{\mathfrak{so}_{10}}
{\xygraph{
    \circ ([]!{+(0,-.3)} {{}_1}) - [r]
    \circ ([]!{+(0,-.3)} {{}_2}) - [r]
    \circ ([]!{+(0,-.3)} {{}_3}) (
        - []!{+(1,.5)}  \circ ([]!{+(0,-.3)} {{}_{4}}),
        - []!{+(1,-.5)} \circ ([]!{+(0,-.3)} {{}_{5}})
)}}
\underset{\rm Folding}{\Longrightarrow}
\underset{\mathfrak{so}_{9}}
{\xygraph{!~:{@{=}|@{}}
    \circ   ([]!{+(0,-.3)} {{}_1}) - [r]
    \circ   ([]!{+(0,-.3)} {{}_2}) - [r]
    \circ   ([]!{+(0,-.3)} {{}_3}) : [r]
    \bullet ([]!{+(0,-.3)} {{}_{4,5}})
}}.
\end{align}
When we use an $\mathfrak{so}_{10}$ weight vector $w_{\mathfrak{so}_{10}}$
and an ${\mathfrak{so}_9}$ weight vector $v_{\mathfrak{so}_9}$,
$\mathfrak{so}_{10}$ is rank-5 and ${\mathfrak{so}_9}$ is rank-4,
so its projection matrix
$P_{\mathfrak{so}_{10}\supset\mathfrak{so}_9}$
is a $4\times 5$ matrix. The 1st, 2nd, and 3rd nodes of
$\mathfrak{so}_{10}$ can be identified to the 1st, 2nd, and 3rd nodes of
$\mathfrak{so}_{9}$, and the 4th and 5th nodes of $\mathfrak{so}_{10}$
can be identified to the 4th node of $\mathfrak{so}_9$. Thus, we can
write its projection matrix as 
\begin{align}
P_{\mathfrak{so}_{10}\supset\mathfrak{so}_9}&=
\left(
\begin{array}{ccccc}
1&0&0&0&0\\
0&1&0&0&0\\
0&0&1&0&0\\
0&0&0&a_{44}&a_{45}\\
\end{array}
\right),
\end{align}
where $a_{44}$ and $a_{45}$ are undetermined values. Here we use the
following branching rule 
\begin{align}
{\bf 10}\Leftrightarrow
\begin{array}{|c|}\hline
(10000)\\
(\bar{1}1000)\\
(0\bar{1}100)\\
(00\bar{1}11)\\
(000\bar{1}1)\ (0001\bar{1})\\
(001\bar{1}\bar{1})\\
(01\bar{1}00)\\
(1\bar{1}000)\\
(\bar{1}0000)\\
\hline
\end{array}
\hspace{1em}
\underset{P_{\mathfrak{so}_{10}\supset\mathfrak{so}_9}}
{\Longrightarrow}
\hspace{1em}
\begin{array}{|c|}\hline
(1000)\\
(\bar{1}100)\\
(0\bar{1}10)\\
(00\bar{1}2)\\
(0000)\ (0000)\\
(001\bar{2})\\
(01\bar{1}0)\\
(1\bar{1}00)\\
(\bar{1}000)\\
\hline
\end{array}
\Leftrightarrow{\bf 9}\oplus{\bf 1}.
\end{align}
When we use the relations
$\{w_{\mathfrak{so}_{10}},v_{\mathfrak{so}_9}\}
=\{(0\ 0\ -1\ 1\ 1),(0\ 0\ -1\ 2)\},
\{(0\ 0\ 0\ -1\ 1),(0\ 0\ 0\ 0)\}$,
we get $a_{44}=a_{45}=1$.
Thus, its projection matrix is
\begin{align}
P_{\mathfrak{so}_{10}\supset\mathfrak{so}_{9}}&=
\left(
\begin{array}{ccccc}
1&0&0&0&0\\
0&1&0&0&0\\
0&0&1&0&0\\
0&0&0&1&1\\
\end{array}
\right).
\end{align}

Another example of folded Dynkin diagrams is 
$B_3=\mathfrak{so}_{7}\supset G_2$.
The Dynkin diagram is
\begin{align}
\underset{\mathfrak{so}_{7}}
{\xygraph{!~:{@{=}|@{}}
    \circ ([]!{+(0,-.3)} {{}_1}) - [r]
    \circ ([]!{+(0,-.3)} {{}_2}) : [r]
    \bullet ([]!{+(0,-.3)} {{}_3})}
}
\underset{\rm Folding}{\Longrightarrow}
\underset{G_2}
{\xygraph{!~:{@3{-}|@{}}
    \circ   ([]!{+(0,-.3)} {{}_2}) : [r]
    \bullet ([]!{+(0,-.3)} {{}_{1,3}})}
}.
\end{align}
When we use an $\mathfrak{so}_{7}$ weight vector $w_{\mathfrak{so}_{7}}$
and a $G_2$ weight vector $v_{G_2}$, $\mathfrak{so}_{7}$ is rank-3 and
$G_2$ is rank-2, so its projection matrix　
$P_{\mathfrak{so}_7\supset G_2}$ is a $2\times 3$ matrix. The 2nd node
of $\mathfrak{so}_7$ can be identified to the 1st node of $G_2$, and the
1st and 3rd nodes of $\mathfrak{so}_{7}$ can be identified to the 2nd
node of $G_2$. Thus, we can write its projection matrix as 
\begin{align}
P_{\mathfrak{so}_7\supset G_2}&=
\left(
\begin{array}{ccc}
0&1&0\\
a_{21}&0&a_{23}\\
\end{array}
\right),
\end{align}
where $a_{21}$ and $a_{23}$ are undetermined values. Here we use the
following branching rule 
\begin{align}
{\bf 7}\Leftrightarrow
\begin{array}{|c|}\hline
(100)\\
(\bar{1}10)\\
(0\bar{1}2)\\
(000)\\
(01\bar{2})\\
(1\bar{1}0)\\
(\bar{1}00)\\
\hline
\end{array}
\hspace{1em}
\underset{P_{\mathfrak{so}_7\supset G_2}}
{\Longrightarrow}
\hspace{1em}
\begin{array}{|c|}\hline
(01)\\
(1\bar{1})\\
(\bar{1}2)\\
(00)\\
(1\bar{2})\\
(\bar{1}1)\\
(0\bar{1})\\
\hline
\end{array}
\Leftrightarrow{\bf 7}.
\end{align}
When we use the relations
$\{w_{\mathfrak{so}_7},v_{G_2}\}
=\{(-1\ 1\ 0),(1\ -1)\},
\{(0\ -1\ 2),(-1\ 2)\}$,
we get $a_{21}=a_{23}=1$.
Thus, its projection matrix is
\begin{align}
P_{\mathfrak{so}_{7}\supset G_2}&=
\left(
\begin{array}{ccc}
0&1&0\\
1&0&1\\
\end{array}
\right).
\end{align}

An  example of shrunk Dynkin diagrams is 
$A_2=\mathfrak{su}_3\supset\mathfrak{su}_2$.
The Dynkin diagram is
\begin{align}
\underset{\mathfrak{su}_{3}}
{\xygraph{
    \circ ([]!{+(0,-.3)} {{}_1}) - [r]
    \circ ([]!{+(0,-.3)} {{}_2})
}}
\underset{\rm Shrinking}{\Longrightarrow}
\underset{\mathfrak{su}_2}
{\xygraph{!~:{@{=}|@{}}
    \circ   ([]!{+(0,-.3)} {{}_{1,2}})
}}.
\end{align}
When we use an $\mathfrak{su}_3$ weight vector $w_{\mathfrak{su}_3}$
and an ${\mathfrak{su}_2}$ weight vector $v_{\mathfrak{su}_2}$,
$\mathfrak{su}_3$ is rank-2 and ${\mathfrak{su}_2}$ is rank-1,
so its projection matrix $P_{\mathfrak{su}_3\supset\mathfrak{su}_2}$ 
is a $1\times 2$ matrix. The 1st and 2nd nodes of $\mathfrak{su}_{3}$
can be identified to the node of $\mathfrak{su}_{2}$. Thus, we can write
its projection matrix as  
\begin{align}
P_{\mathfrak{su}_3\supset\mathfrak{su}_2}&=
\left(
\begin{array}{cc}
a_{11}&a_{12}\\
\end{array}
\right),
\end{align}
where $a_{11}$ and $a_{12}$ are undetermined values. Here we use the
following branching rule 
\begin{align}
{\bf 3}\Leftrightarrow
\begin{array}{|c|}\hline
(10)\\
(\bar{1}1)\\
(0\bar{1})\\
\hline
\end{array}
\hspace{1em}
\underset{P_{\mathfrak{su}_{3}\supset\mathfrak{su}_2}}
{\Longrightarrow}
\hspace{1em}
\begin{array}{|c|}\hline
(2)\\
(0)\\
(\bar{2})\\
\hline
\end{array}
\Leftrightarrow{\bf 3}.
\end{align}
When we use the relations
$\{w_{\mathfrak{su}_{3}},v_{\mathfrak{su}_2}\}
=\{(1\ 0),(2)\}, \{(-1\ 1),(0)\}, \{(0\ -1),(-2)\}$,
we get $a_{11}=a_{12}=2$.
Thus, its projection matrix is
\begin{align}
P_{\mathfrak{su}_3\supset\mathfrak{su}_2}&=
\left(
\begin{array}{cc}
2&2\\
\end{array}
\right).
\end{align}

\subsection{Recipe for calculating branching rules}
\label{Sec:Recipe-branching-rules}

We recall how to obtain the maximal subalgebras $\mathfrak{h}$
of the simple Lie algebra $\mathfrak{g}$ listed in
Sec.~\ref{Sec:Subalgebras}. (See Sec.~\ref{Sec:Subalgebras} in detail).
\begin{enumerate}
\item Maximal $R$-subalgebras can be found by using extended Dynkin
      diagrams in Table~\ref{Table:Dynkin-diagrams} and the
      following rules:
\begin{itemize}
\item[(A:R-1)] Extended Dynkin diagrams whose one node is dropped are
	     maximal semi-simple regular subalgebras.
\item[(A:R-2)] Extended Dynkin diagrams of $A_n=\mathfrak{su}_{n+1}$ whose
	     two nodes are dropped are maximal non-semi-simple regular
	     subalgebras. In other words,
	     Dynkin diagrams of $A_n=\mathfrak{su}_{n+1}$ whose
	     one node is dropped are maximal non-semi-simple regular
	     subalgebras. 
\item[(A:R-3)] Extended Dynkin diagrams of
	     $B_n=\mathfrak{so}_{2n+1}$,
	     $C_n=\mathfrak{so}_{2n+1}$,
	     $D_n=\mathfrak{so}_{2n+1}$,
	     $E_6$, and $E_7$ 
	     whose appropriate two nodes are dropped are maximal
	     non-semi-simple regular subalgebras.
	     They are related with a parity symmetry of their
	     extended Dynkin diagrams.
\end{itemize}
\item Maximal $S$-subalgebras can be found by the following rules:
\begin{itemize}
\item[(A:S-1)] For classical Lie algebras $\mathfrak{g}$,
	     the maximal non-simple $S$-subalgebras $\mathfrak{h}$ of
	     $\mathfrak{g}$ are listed in
	     Table~\ref{Table:Maximal-S-sub-classical-1}.
\item[(A:S-2)] For classical Lie algebras $\mathfrak{g}$,
	     the maximal $S$-subalgebras $\mathfrak{h}$ of
	     $\mathfrak{g}$ can be found as below.
	     If a Lie algebra $\mathfrak{h}$ has an $n$ dimensional
	     complex representation, it is a maximal subalgebra of
	     a Lie algebra $A_{n}=\mathfrak{su}_{n+1}$.
	     If a Lie algebra $\mathfrak{h}$ has an $n$ dimensional
	     real representation, it is a maximal subalgebra of
	     $B_{k}=\mathfrak{so}_{2k+1}$ for $n=2k+1(k\in\mathbb{Z})$
	     and  
	     $D_{k}=\mathfrak{so}_{2k}$ for  $n=2k (k\in\mathbb{Z})$. 
	     If a Lie algebra $\mathfrak{h}$ has a
	     $2k (k\in\mathbb{Z})$
	     dimensional pseudo-real representation, it is a maximal
	     subalgebra of 
	     $C_{k}=\mathfrak{usp}_{2k}$. 
	     (See Table~\ref{Table:Self-conjugate-representations}.)
	     Note that there are several exceptions.
	     The above examples are summarized in
	     Table~\ref{Table:Maximal-S-sub-classical-2}.
\item[(A:S-3)] All the maximal $S$-subalgebras $\mathfrak{h}$ of the
	     exceptional simple Lie algebras $\mathfrak{g}$ are listed
	     in Table~\ref{Table:Maximal-S-sub-exceptional}.
\end{itemize}
\end{enumerate}
Note that one may wonder whether Dynkin diagrams are truly necessary to
find subalgebras or not. The answer is no because we can find any
subalgebra 
by using the similar method as Rule (S-2), where it is a tedious way.
However, Dynkin diagrams are very useful tools especially to find all
$R$-subalgebras and several types of $S$-subalgebras and their branching
rules. This is because for $R$-subalgebras, at least one branching rule
is almost obvious. E.g.,  
for $\mathfrak{su}_{n+1}\supset\mathfrak{su}_{n}\oplus\mathfrak{u}_1$,
${\bf n+1}={\bf n}(+1)\oplus{\bf 1}(-n)$;
for $\mathfrak{so}_{2n+1}\supset\mathfrak{so}_{2n}$,
${\bf 2n+1}={\bf 2n}\oplus{\bf 1}$;
for $\mathfrak{usp}_{2n}\supset\mathfrak{usp}_{2k}\oplus\mathfrak{usp}_{2n-2k}$,
${\bf 2n}={\bf 2k}\oplus{\bf 2n-2k}$;
for $\mathfrak{so}_{2n}\supset\mathfrak{so}_{2k}\oplus\mathfrak{so}_{2n-2k}$,
${\bf 2n}={\bf 2k}\oplus{\bf 2n-2k}$.

We check how to calculate projection matrices of the maximal subalgebras
$\mathfrak{h}$ of the simple Lie algebra $\mathfrak{g}$ listed in the
table in Sec.~\ref{Sec:Subalgebras}. 
\begin{enumerate}
\item The projection matrices between a Lie algebra $\mathfrak{g}$ and
      its maximal $R$-subalgebra $\mathfrak{h}$ 
      can be found by the following rules:
\begin{itemize}
\item[(P:R-1)] The projection matrices of the maximal algebras are 
	     $n\times n$ matrices, where $\mathfrak{g}$ and
	     $\mathfrak{h}$ are rank-$n$.
\item[(P:R-2)] We identify the remaining nodes of the Dynkin diagram of
	     an algebra $\mathfrak{g}$ to the nodes of Dynkin diagrams
	     of its semi-simple subalgebras $\mathfrak{h}$ except the
	     highest root of the extend Dynkin diagram, which is
	     the $x$th node of extended Dynkin diagrams shown in
	     Table~\ref{Table:Dynkin-diagrams}. This is because the
	     relation between the $x$th node of of extend Dynkin
	     diagrams and its projection matrix is not always obvious.
\item[(P:R-3)] We calculate the remaining matrix elements of 
	     the highest root of the extended Dynkin diagram of the
	     algebra $\mathfrak{g}$ or the removed node(s) 
	     by using its branching rule of e.g., the lowest dimensional
	     representation of the algebra $\mathfrak{g}$ and the
	     representations of its subalgebra $\mathfrak{h}$.
\end{itemize}
\item The projection matrices between a Lie algebra $\mathfrak{g}$ and
      their maximal $S$-subalgebras $\mathfrak{h}$ 
      can be found by the following rules:
\begin{itemize}
\item[(P:S-1)] The projection matrices of the maximal algebras are 
	     $n\times m$ matrices, where $\mathfrak{g}$ is rank-$n$
	     and $\mathfrak{h}$ is rank-$m$ $(m<n)$.
\item[(P:S-2)] By using the above rules (A:S-1)-(A:S-3), we find one
	     branching rule of $\mathfrak{g}\supset\mathfrak{h}$.
\item[(P:S-3)] We write down weight diagrams of the representations of
	     $\mathfrak{g}$ and $\mathfrak{h}$.
\item[(P:S-4)] We find an appropriate combination between the
	     representations of $\mathfrak{g}$ and $\mathfrak{h}$.
\item[(P:S-5)] By using them, we calculate the matrix elements of the
	     projection matrix.
\end{itemize}
\end{enumerate}
Note that we discussed how to calculate projection matrices of $R$- and
$S$-subalgebras (regular and special subalgebras). In principle, by
using one branching rule between a Lie algebra $\mathfrak{g}$ and its
$R$- or $S$-subalgebra $\mathfrak{h}$, we can calculate its projection
matrix regardless of $R$- or $S$-subalgebras. 
Section~\ref{Sec:Summary-representations} includes projection matrices
between rank-$n$ Lie algebras and all $R$-subalgebras and also
between rank-$n$ Lie algebras and several $S$-subalgebras.
Appendix~\ref{Sec:Projection-matrices} contains projection matrices
between rank-20 or less Lie algebras
and their maximal $R$- and $S$-subalgebras. In addition,
Appendix~\ref{Sec:Branching-rules} contains their branching rules. 

We check how to get projection matrices of non-maximal subalgebras
$\mathfrak{h}$ of the simple Lie algebra $\mathfrak{g}$. The number of
non-maximal subalgebras is huge and their projection matrices can be
obtained by using several maximal subalgebras. This paper does not
contain explicit forms of them except some projection matrices related
with the standard model and grand unification discussed in
Sec.~\ref{Sec:Model-Building}. If necessary, we can calculate them by
using e.g., the following procedure.
\begin{enumerate}
\item The projection matrices between a Lie algebra $\mathfrak{g}$ and
      its non-maximal subalgebras $\mathfrak{h}$ 
      can be calculated by the following rules:
\begin{itemize}
\item[(P:N-1)] The projection matrices of the non-maximal algebras are 
	     $m\times n$ matrices, where $\mathfrak{g}$ is rank-$n$ and
	     $\mathfrak{h}$ is rank-$m$ $(m\leq n)$.
\item[(P:N-2)] We find Lie subalgebras $\mathfrak{k}^{(a)}$
	     $(a=1,2,\cdots,\ell)$ that satisfy
	     $\mathfrak{h}\subset\mathfrak{k}^{(1)}\subset
	     \cdots\subset\mathfrak{k}^{(\ell)}\subset\mathfrak{g}$,
	     where each
	     $\mathfrak{g}\supset\mathfrak{k}^{(\ell)}$,
	     $\mathfrak{k}^{(\ell)}\supset\mathfrak{k}^{(\ell-1)}$,
	     $\cdots$, 
	     $\mathfrak{k}^{(1)}\supset\mathfrak{h}$ is
	     a pair of a Lie algebra and its maximal subalgebra.
\item[(P:N-3)] We obtain the projection matrix of 
	     $\mathfrak{g}\supset\mathfrak{h}$
	     by multiplying the projection matrices of the intermediate
	     Lie algebras and their maximal subalgebras.
\end{itemize}
\end{enumerate}

We check how to calculate branching rules of the subalgebras
$\mathfrak{h}$ of an algebra $\mathfrak{g}$.
\begin{enumerate}
\item Branching rules of an algebra $\mathfrak{g}$ and its maximal
      subalgebras $\mathfrak{h}$ listed in
      Table~\ref{Table:Maximal-subalgebra} in 
      Sec.~\ref{Sec:Subalgebras} can be found by the following rules:
\begin{itemize}
\item[(B:M-1)] By using the projection matrix of
	     $\mathfrak{g}\supset\mathfrak{h}$ and 
	     an appropriate computer program such as 
	     Susyno program \cite{Fonseca:2011sy} and 
	     LieART \cite{Feger:2012bs},
	     we obtain branching rules of
	     $\mathfrak{g}\supset\mathfrak{h}$. 

	     Note that projection matrices of Lie algebras 
	     $\mathfrak{g}$ up to rank-20 are given in
	     Appendix~\ref{Sec:Projection-matrices} 
	     and several projection matrices of rank-$n$ Lie algebras
	     are also given in Sec.~\ref{Sec:Summary-representations}.
\end{itemize}
\item Non-maximal subalgebras
      Branching rules of an algebra $\mathfrak{g}$ and its non-maximal
      subalgebras $\mathfrak{h}$ can be found by the following rules:
\begin{itemize}
\item[(B:N-1)] As the same as the above maximal subalgebra case,
	     by using the projection matrix derived from the above
	     procedure (P:N-1)-(P:N-3) and 
	     an appropriate computer program,
	     we obtain branching rules of
	     $\mathfrak{g}\supset\mathfrak{h}$. 
\end{itemize}
\end{enumerate}

Before we finish this section, we will see how to obtain projection
matrices of a simple Lie algebra $\mathfrak{g}$ and its non-maximal
subalgebra $\mathfrak{h}$ by using projection matrices of
a simple Lie algebra $\mathfrak{g}'$ and its maximal subalgebra
$\mathfrak{h}'$ given in Appendix~\ref{Sec:Projection-matrices}.
First, we consider non-maximal Lie subalgebras of 
$A_{15}=\mathfrak{su}_{16}$. The projection matrices of 
$\mathfrak{su}_{16}\supset\mathfrak{so}_{10}$,
$\mathfrak{so}_{10}\supset\mathfrak{su}_5\oplus\mathfrak{u}_1$ 
$\mathfrak{su}_5\supset\mathfrak{su}_3\oplus\mathfrak{su}_2\oplus\mathfrak{u}_1$
are given by 
\begin{align}
P_{\mathfrak{su}_{16}\supset\mathfrak{so}_{10}}&=
\left(
\begin{array}{ccccccccccccccc}
0&0&0&0&1&0&1&0&1&0&1&0&0&0&0\\
0&0&1&2&1&1&0&0&0&1&1&2&1&0&0\\
0&1&0&0&0&0&1&2&1&0&0&0&0&1&0\\
0&0&1&0&1&2&1&0&0&0&0&0&0&0&1\\
1&0&0&0&0&0&0&0&1&2&1&0&1&0&0\\
\end{array}
\right),\\
P_{\mathfrak{so}_{10}\supset\mathfrak{su}_5\oplus\mathfrak{u}_1}&=
\left(
\begin{array}{ccccc}
0&0&0&1&0\\
0&0&1&0&0\\
0&1&0&0&0\\
1&0&0&0&0\\
2&4&6&3&5\\
\end{array}
\right),\\
P_{\mathfrak{su}_5\supset\mathfrak{su}_3\oplus\mathfrak{su}_2\oplus\mathfrak{u}_1}&=
\left(
\begin{array}{cccc}
1&0&0&0\\
0&1&0&0\\
0&0&0&1\\
2&4&6&3\\
\end{array}
\right).
\end{align}
By using them, we obtain the projection matrix of 
$\mathfrak{su}_{16}\supset\mathfrak{so}_{10}
\supset\mathfrak{su}_5\oplus\mathfrak{u}_1$:
\begin{align}
P_{\mathfrak{su}_{16}\supset\mathfrak{su}_5\oplus\mathfrak{u}_1}&=
P_{\mathfrak{so}_{10}\supset\mathfrak{su}_5\oplus\mathfrak{u}_1}
P_{\mathfrak{su}_{16}\supset\mathfrak{so}_{10}}
\nonumber\\
&=
\left(
\begin{array}{ccccccccccccccc}
0&0&1&0&1&2&1&0&0&0&0&0&0&0&1\\
0&1&0&0&0&0&1&2&1&0&0&0&0&1&0\\
0&0&1&2&1&1&0&0&0&1&1&2&1&0&0\\
0&0&0&0&1&0&1&0&1&0&1&0&0&0&0\\
5&6&7&8&9&10&11&12&13&14&11&8&9&6&3\\
\end{array}
\right)
\end{align}
and also we obtain the projection matrix of 
$\mathfrak{su}_{16}\supset\mathfrak{so}_{10}
\supset\mathfrak{su}_5\oplus\mathfrak{u}_1
\supset\mathfrak{su}_3\oplus\mathfrak{su}_2\oplus\mathfrak{u}_1\oplus\mathfrak{u}_{1}$:
\begin{align}
P_{\mathfrak{su}_{16}\supset\mathfrak{su}_3\oplus\mathfrak{su}_2\oplus\mathfrak{u}_1\oplus\mathfrak{u}_{1}}&=
\left(
P_{\mathfrak{su}_5\supset\mathfrak{su}_3\oplus\mathfrak{su}_2\oplus\mathfrak{u}_1}
\oplus I_1\right)
P_{\mathfrak{so}_{10}\supset\mathfrak{su}_5\oplus\mathfrak{u}_1}
P_{\mathfrak{su}_{16}\supset\mathfrak{so}_{10}}
\nonumber\\
&=
\left(
\begin{array}{ccccccccccccccc}
0&0&1&0&1&2&1&0&0&0&0&0&0&0&1\\
0&1&0&0&0&0&1&2&1&0&0&0&0&1&0\\
0&0&0&0&1&0&1&0&1&0&1&0&0&0&0\\
0&4&8&12&11&10&9&8&7&6&9&12&6&4&2\\
5&6&7&8&9&10&11&12&13&14&11&8&9&6&3\\
\end{array}
\right).
\end{align}
Second, we consider non-maximal Lie subalgebras of 
$B_{13}=\mathfrak{so}_{27}$. The projection matrices of 
$\mathfrak{so}_{27}\supset\mathfrak{so}_{7}$, 
$\mathfrak{so}_{7}\supset G_2$,
$G_2\supset\mathfrak{su}_3$, and 
$\mathfrak{su}_3\supset\mathfrak{su}_2$ are given by 
\begin{align}
P_{\mathfrak{so}_{27}\supset\mathfrak{so}_{7}}
&=\left(
\begin{array}{ccccccccccccc}
2&2&3&1&2&1&2&1&1&3&2&2&1\\
0&1&0&2&2&2&3&4&2&1&3&2&1\\
0&0&2&2&2&4&2&2&6&6&4&6&3\\
\end{array}
\right),\\
P_{\mathfrak{so}_{7}\supset G_2}
&=\left(
\begin{array}{ccc}
0&1&0\\
1&0&1\\
\end{array}
\right),\\
P_{G_2\supset\mathfrak{su}_3}
&=\left(
\begin{array}{cc}
1&1\\
1&0\\
\end{array}
\right),\\
P_{\mathfrak{su}_{3}\supset\mathfrak{su}_2}
&=\left(
\begin{array}{cc}
2&2\\
\end{array}
\right).
\end{align}
They lead to the projection matrices of 
$\mathfrak{so}_{27}\supset\mathfrak{so}_{7}\supset G_2$,
$\mathfrak{so}_{27}\supset\mathfrak{so}_{7}\supset
G_2\supset\mathfrak{su}_3$, and
$\mathfrak{so}_{27}\supset\mathfrak{so}_{7}\supset
G_2\supset\mathfrak{su}_3\supset\mathfrak{su}_2$:
\begin{align}
P_{\mathfrak{so}_{27}\supset G_2}&=
P_{\mathfrak{so}_{7}\supset G_2}
P_{\mathfrak{so}_{27}\supset\mathfrak{so}_{7}}
\nonumber\\
&=\left(
\begin{array}{ccccccccccccc}
0&1&0&2&2&2&3&4&2&1&3&2&1\\
2&2&5&3&4&5&4&3&7&9&6&8&4\\
\end{array}
\right),\\
P_{\mathfrak{so}_{27}\supset\mathfrak{su}_3}&=
P_{G_2\supset\mathfrak{su}_3}
P_{\mathfrak{so}_{7}\supset G_2}
P_{\mathfrak{so}_{27}\supset\mathfrak{so}_{7}}
\nonumber\\
&=\left(
\begin{array}{ccccccccccccc}
2&3&5&5&6&7&7&7&9&10&9&10&5\\
0&1&0&2&2&2&3&4&2&1&3&2&1\\
\end{array}
\right),\\
P_{\mathfrak{so}_{27}\supset\mathfrak{su}_2}&=
P_{\mathfrak{su}_3\supset\mathfrak{su}_2}
P_{G_2\supset\mathfrak{su}_3}
P_{\mathfrak{so}_{7}\supset G_2}
P_{\mathfrak{so}_{27}\supset\mathfrak{so}_{7}}
\nonumber\\
&=\left(
\begin{array}{ccccccccccccc}
4&8&10&14&16&18&20&22&22&22&24&12\\
\end{array}
\right).
\end{align}
Third, we consider non-maximal Lie subalgebras of 
$C_{10}=\mathfrak{usp}_{20}$. The projection matrices of 
$\mathfrak{usp}_{20}\supset\mathfrak{su}_{6}$, 
$\mathfrak{su}_{6}\supset\mathfrak{usp}_{6}$, and 
$\mathfrak{usp}_{6}\supset\mathfrak{su}_3\oplus\mathfrak{u}_1$ are given
by 
\begin{align}
P_{\mathfrak{usp}_{20}\supset\mathfrak{su}_6}
&=\left(
\begin{array}{cccccccccc}
0&0&1&1&0&1&1&0&1&2\\
0&1&0&1&1&0&1&1&1&0\\
1&0&0&0&0&1&1&2&1&2\\
0&1&2&1&2&1&1&0&0&0\\
0&0&0&1&1&2&1&2&3&2\\
\end{array}
\right),\\
P_{\mathfrak{su}_6\supset\mathfrak{usp}_{6}}
&=\left(
\begin{array}{ccccc}
1&0&0&0&1\\
0&1&0&1&0\\
0&0&1&0&0\\
\end{array}
\right),\\
P_{\mathfrak{usp}_{6}\supset\mathfrak{su}_3\oplus\mathfrak{u}_1}
&=
\left(
\begin{array}{ccc}
1&0&0\\
0&1&0\\
1&2&3\\
\end{array}
\right).
\end{align}
They lead to the projection matrices of 
$\mathfrak{usp}_{20}\supset\mathfrak{su}_6\supset\mathfrak{usp}_{6}$ and 
$\mathfrak{usp}_{20}\supset\mathfrak{su}_6\supset\mathfrak{usp}_{6}
\supset\mathfrak{su}_3\oplus\mathfrak{u}_1$:
\begin{align}
P_{\mathfrak{usp}_{20}\supset\mathfrak{usp}_{6}}
&=P_{\mathfrak{su}_6\supset\mathfrak{usp}_{6}}
P_{\mathfrak{usp}_{20}\supset\mathfrak{su}_6}
\nonumber\\
&=\left(
\begin{array}{cccccccccc}
0&0&1&2&1&3&2&2&4&4\\
0&2&2&2&3&1&2&1&1&0\\
1&0&0&0&0&1&1&2&1&2\\
\end{array}
\right),\\
P_{\mathfrak{usp}_{20}\supset\mathfrak{su}_3\oplus\mathfrak{u}_1}
&=
P_{\mathfrak{usp}_{6}\supset\mathfrak{su}_3\oplus\mathfrak{u}_1}
P_{\mathfrak{su}_6\supset\mathfrak{usp}_{6}}
P_{\mathfrak{usp}_{20}\supset\mathfrak{su}_6}
\nonumber\\
&=\left(
\begin{array}{cccccccccc}
0&0&1&2&1&3&2&2&4&4\\
0&2&2&2&3&1&2&1&1&0\\
3&4&5&6&7&8&9&10&9&10\\
\end{array}
\right).
\end{align}
Finally, we consider non-maximal Lie subalgebras of 
$D_{13}=\mathfrak{so}_{26}$. The projection matrices of 
$\mathfrak{so}_{27}\supset F_4$, 
$F_4\supset\mathfrak{so}_{9}$,
$\mathfrak{so}_9\supset\mathfrak{su}_4\oplus\mathfrak{su}_2$, and 
$\mathfrak{su}_4\supset\mathfrak{su}_3\oplus\mathfrak{u}_1$ 
are given by 
\begin{align}
P_{\mathfrak{so}_{26}\supset F_4}
&=\left(
\begin{array}{ccccccccccccc}
0&0&0&1&0&1&0&1&1&0&0&0&0\\
0&0&1&0&0&0&1&1&0&1&1&0&0\\
0&1&0&1&2&1&0&0&1&1&0&1&1\\
1&0&0&0&0&1&2&1&2&1&3&1&1\\
\end{array}
\right),\\
P_{F_4\supset\mathfrak{so}_{9}}
&=\left(
\begin{array}{cccc}
-2&-3&-2&-1\\
1&0&0&0\\
0&1&0&0\\
0&0&1&0\\
\end{array}
\right),\\
P_{\mathfrak{so}_{9}\supset\mathfrak{su}_4\oplus\mathfrak{su}_2}
&=\left(
\begin{array}{cccc}
1&0&0&0\\
0&1&0&0\\
-1&-2&-2&-1\\
0&0&0&1\\
\end{array}
\right),\\
P_{\mathfrak{su}_4\supset\mathfrak{su}_3\oplus\mathfrak{u}_1}
&=\left(
\begin{array}{ccc}
1&0&0\\
0&1&0\\
1&2&3\\
\end{array}
\right).
\end{align}
They lead to the projection matrices of 
$\mathfrak{so}_{26}\supset F_4\supset\mathfrak{so}_{9}$,
$\mathfrak{so}_{26}\supset F_4\supset\mathfrak{so}_{9}
\supset\mathfrak{su}_4\oplus\mathfrak{su}_2$,
$\mathfrak{so}_{26}\supset F_4\supset\mathfrak{so}_{9}
\supset\mathfrak{su}_4\oplus\mathfrak{su}_2
\supset\mathfrak{su}_3\oplus\mathfrak{u}_1\oplus\mathfrak{su}_2$:
\begin{align}
P_{\mathfrak{so}_{26}\supset\mathfrak{so}_{9}}
&=
P_{F_4\supset\mathfrak{so}_{9}}
P_{\mathfrak{so}_{26}\supset F_4}
\nonumber\\
&=\left(
\begin{array}{ccccccccccccc}
-1&-2&-3&-4&-4&-5&-5&-6&-6&-6&-6&-3&-3\\
0&0&0&1&0&1&0&1&1&0&0&0&0\\
0&0&1&0&0&0&1&1&0&1&1&0&0\\
0&1&0&1&2&1&0&0&1&1&0&1&1\\
\end{array}
\right),\\
P_{\mathfrak{so}_{26}\supset\mathfrak{su}_4\oplus\mathfrak{su}_2}
&=
P_{\mathfrak{so}_{9}\supset\mathfrak{su}_4\oplus\mathfrak{su}_2}
P_{F_4\supset\mathfrak{so}_{9}}
P_{\mathfrak{so}_{26}\supset F_4}
\nonumber\\
&=\left(
\begin{array}{ccccccccccccc}
-1&-2&-3&-4&-4&-5&-5&-6&-6&-6&-6&-3&-3\\
0&0&0&1&0&1&0&1&1&0&0&0&0\\
1&1&1&1&2&2&3&2&3&3&4&2&2\\
0&1&0&1&2&1&0&0&1&1&0&1&1\\
\end{array}
\right),\\
P_{\mathfrak{so}_{26}\supset\mathfrak{su}_3\oplus\mathfrak{u}_1\oplus\mathfrak{su}_2}
&=
\left(P_{\mathfrak{su}_4\supset\mathfrak{su}_3\oplus\mathfrak{u}_1}
\oplus I_1\right)
P_{\mathfrak{so}_{9}\supset\mathfrak{su}_4\oplus\mathfrak{su}_2}
P_{F_4\supset\mathfrak{so}_{9}}
P_{\mathfrak{so}_{26}\supset F_4}
\nonumber\\
&=\left(
\begin{array}{ccccccccccccc}
-1&-2&-3&-4&-4&-5&-5&-6&-6&-6&-6&-3&-3\\
0&0&0&1&0&1&0&1&1&0&0&0&0\\
2&1&0&1&2&3&4&2&5&3&6&3&3\\
0&1&0&1&2&1&0&0&1&1&0&1&1\\
\end{array}
\right).
\end{align}
(By using the above projection matrices, it is easy to calculate their 
branching rules.)

\section{Tensor product}
\label{Sec:Tensor-product-basics}

Let us discuss how to calculate tensor products of two irreducible
representations $R_1$ and $R_2$ of a simple Lie algebra $\mathfrak{g}$,
where $\Lambda_1$ and $\Lambda_2$ are their highest weights.
\begin{align}
R_1\otimes R_2=
R_1'\oplus  R_2'\oplus \cdots,
\end{align}
where $R_1'$, $R_2'$, etc. stand for irreducible representations of the 
Lie algebra $\mathfrak{g}$.
We determine the product representations $R_1'$, $R_2'$, etc. of
$\mathfrak{g}$.
First, the highest weight of one of the product representations of
$\mathfrak{g}$ is $\Lambda_1+\Lambda_2$. To obtain the highest weight of
non-trivial representations of the product representations of 
$\mathfrak{g}$, we introduce three useful techniques; 
Dynkin's theorem for the second highest representation
\cite{Dynkin:1957ek},
conjugacy classes \cite{Dynkin:1957um},
and Dynkin's method of parts
\cite{Dynkin:1957ek}.
(For review, see, e.g., Ref.~\cite{Cahn:1985wk}.)

\subsection{Dynkin's theorem for second highest representation}

Let us consider the Dynkin's theorem for the second
highest representation given in Ref.~\cite{Dynkin:1957ek}. 
By using this, we can get additional one or more irreducible
representations with highest weights of a Lie algebra $\mathfrak{g}$. 
To apply the theorem for our purpose, we have to use the Dynkin diagram
of $\mathfrak{g}$ and add the two irreducible representations
$\Lambda_1$ and $\Lambda_2$ on the Dynkin diagram, and then we identify
the shortest path between $\Lambda_1$ and $\Lambda_2$. 
If the path is $\alpha_i,\alpha_{i+1},\cdots,\alpha_{i+k}$, then 
$\Lambda_1+\Lambda_2-\alpha_i-\alpha_{i+1}-\cdots-\alpha_{i+k}$ is the
highest weight of an irreducible representation within the product
representations. 
(For its proof, see Refs.~\cite{Dynkin:1957ek,Cahn:1985wk}.) 

To understand the technique, let us consider an $\mathfrak{so}_{11}$
algebra and its representations $\Lambda_1=(1,0,0,0,0)$ and
$\Lambda_2=(0,0,0,0,1)$.  
The highest weight of one of the product representations is
$\Lambda_1':=(1,0,0,0,1)$ because of 
$\Lambda_1+\Lambda_2=(1,0,0,0,0)+(0,0,0,0,1)=(1,0,0,0,1)$.
By using the Dynkin's theorem for the second highest representation
discussed in Refs.~\cite{Dynkin:1957ek,Cahn:1985wk}, we get another
product representation. To check the shortest path between
$\Lambda_1=(1,0,0,0,0)$ and $\Lambda_2=(0,0,0,0,1)$, we draw a Dynkin
diagram as the following: 
\begin{align}
&\underset{\mathfrak{so}_{11}}
{\xygraph{!~:{@{=}|@{}}
    \circ ([]!{+(0,-.3)}  {\alpha_1}) (
        - [u] \circ ([]!{+(.3,0)}  {\Lambda_1}),
        - [r] \circ ([]!{+(0,-.3)} {\alpha_2})
        - [r] \circ ([]!{+(0,-.3)} {\alpha_3}) 
        - [r] \circ ([]!{+(0,-.3)} {\alpha_4})
        : [r] \bullet ([]!{+(.3,-.3)} {\hspace{-1.5em}\alpha_5}) (
        - [u] \circ ([]!{+(.3,0)}  {\Lambda_2}),
))}},
\end{align}
where $\alpha_i (i=1,2,\cdots,5)$ are the simple roots of
$\mathfrak{so}_{11}$. 
From the diagram, the shortest path is
$\alpha_1,\alpha_2,\alpha_3,\alpha_4,\alpha_5$.
Thus, the highest weight of one of the product
representations is 
$\Lambda_2':=
\Lambda_1+\Lambda_2-\alpha_1-\alpha_2-\alpha_3-\alpha_4-\alpha_5
=(0,0,0,0,1)$.
By using Weyl dimension formula, we find $d(\Lambda_1)=11$,
$d(\Lambda_2)=32$, $d(\Lambda_1')=320$, and $d(\Lambda_2')=32$.
Thus, $d(\Lambda_1)\times d(\Lambda_2)=d(\Lambda_1')+d(\Lambda_2')$, so
there is no other product representation.

\subsection{Conjugacy class}

Let us check how to use the concept of conjugacy classes of simple Lie
algebras $\mathfrak{g}$ given in Ref.~\cite{Dynkin:1957um}.
First, we consider a tensor product of representations $R_1$ and $R_2$
of $\mathfrak{g}$ with the number of a conjugacy class $C_c(R_1)$ and
$C_c(R_2)$, respectively. Its product representations of $\mathfrak{g}$
have the same value for $C_c(R_1\otimes R_2)=C_c(R_1)+C_c(R_2)$ up to an
appropriate modulus.  
That is, all product representations of $\mathfrak{g}$ must belong to
the same conjugacy class. 
Instead of giving its proof, let us consider how to use it.
(See Ref.~\cite{Dynkin:1957um} for its proof.)

To understand how to use this concept, let us consider an
$\mathfrak{su}_3$ algebra. The concept for $\mathfrak{su}_3$ is known as
triality. 
By using the highest weight representation and Dynkin's theorem for the
second highest representation, the product representations of
$\Lambda_1=(10)$ and $\Lambda_2=(10)$ of $\mathfrak{su}_3$ are
$\Lambda_1'=(20)$ and $\Lambda_2'=(01)$ of $\mathfrak{su}_3$ because one
of the highest representations is  
$\Lambda_1'=\Lambda_1+\Lambda_2=(20)$ and 
$\Lambda_2'=\Lambda_1+\Lambda_2-\alpha_1=(01)$ from the following Dynkin
diagram: 
\begin{align}
&\underset{\mathfrak{su}_3}
{\xygraph{!~:{@{=}|@{}}
    \circ ([]!{+(0,-.3)}  {\alpha_1}) (
        - [u] \circ ([]!{+(.3,0)}  {\hspace{1.5em}\Lambda_1,\Lambda_2}),
        - [r] \circ ([]!{+(0,-.3)} {\alpha_2})
)}}.
\end{align}
For $A_2=\mathfrak{su}_3$, $C_c(\Lambda)=a_1+2a_2\ (\bmod.\ 3)$ for
$\Lambda=(a_1a_2)$. 
By using it, we have $C_c(\Lambda_1=(10))=C_c(\Lambda_2=(10))=1$, and thus
its product representations of $\mathfrak{su}_3$ must have
$C_c(\Lambda_1')=C_c(\Lambda_2')=2$. In fact,
$C_c(\Lambda_1'=(20))=C_c(\Lambda_2'=(01))=2$. Thus, we can check it 
by using the conjugacy class. In general, we cannot determine all
product representations only by using he highest weight representation
and the Dynkin's theorem for the second highest representation. In that
case, it may be useful to determine the product representations.
We can also check this result by using the Dynkin indices and anomaly 
coefficients. 

Another example is a tensor product of a Lie algebra $E_7$.
By using the highest weight representation and Dynkin's theorem for the
second highest representation, the highest and second highest product
representations of $\Lambda_1=(1000000)$ and $\Lambda_2=(0000010)$ are
$\Lambda_1'=(1000010)$ and $\Lambda_2'=(0000001)$, respectively, because
one of the highest representations is  
$\Lambda_1'=\Lambda_1+\Lambda_2=(1000010)$ and 
$\Lambda_2'=\Lambda_1+\Lambda_2
-\alpha_1-\alpha_2-\alpha_3-\alpha_4-\alpha_5-\alpha_6=(0000001)$ from
the following Dynkin diagram: 
\begin{align}
\xygraph{
    \circ ([]!{+(0,-.3)}  {\alpha_1}) (
        - [u] \circ ([]!{+(.3,0)}  {\hspace{1.5em}\Lambda_1}),
        - [r] 
    \circ ([]!{+(0,-.3)}  {\alpha_2}) - [r]
 \circ ([]!{+(.3,-.3)} {\hspace{-2em}\alpha_3}) (
        - [u] \circ ([]!{+(.3,0)}  {\alpha_7}),
        - [r] \circ ([]!{+(0,-.3)} {\alpha_4})
        - [r] \circ ([]!{+(0,-.3)} {\alpha_5})
        - [r] \circ ([]!{+(0,-.3)} {\alpha_6}) (
        - [u] \circ ([]!{+(.3,0)}  {\hspace{1.5em}\Lambda_2})
)))}.
\end{align}
Since the dimension of the product representation of $E_7$ is
$d(\Lambda_1\otimes\Lambda_2)=d(\Lambda_1)d(\Lambda_2)=
133\times 56=7448$ and 
the dimension of the highest and second highest product representations
are $d(\Lambda_1')=6480$ $d(\Lambda_2')=912$, 
$d(\Lambda_1\otimes\Lambda_2)-d(\Lambda_1')-d(\Lambda_2')=56>0$, and
then we continue the calculation.
The conjugacy class of the product representation 
$C_c(\Lambda_1\otimes\Lambda_2)=1$, so we list up the representations
satisfying conditions $d(\Lambda)\leq 56$ and $C_c(\Lambda)=1$.
The conditions are satisfied by only one representation (0000010).
Its dimension is $d(0000010)=56$.
Thus, the tensor product is 
\begin{align}
(1000000)\otimes(0000010)=
(1000010)\oplus(0000001)\oplus(0000010).
\end{align}
We can also check it by using the Dynkin indices.
Since $T(0000010)=6$, $T(1000000)=18$, $T(0000001)=180$,
and $T(1000010)=1620$, 
we get $T((0000010)\otimes(1000000))=
d(0000010)T(1000000)+d(1000000)T(0000010)=1806$, and 
$T(1000010)+T(0000001)+T(0000010)=1806$.
Thus, $T((0000010)\otimes(1000000))=T(1000010)+T(0000001)+T(0000010)$.

\subsection{Dynkin's method of parts}

Let us introduce the Dynkin's method of parts given in
Ref.~\cite{Dynkin:1957ek}.
When we calculate a tensor product of two irreducible representation
$\Lambda_1$ and $\Lambda_2$ of a simple Lie algebra $\mathfrak{g}$, 
we also calculate that of two irreducible representation $\Lambda_1'$
and $\Lambda_2'$ of its subalgebra $\mathfrak{h}$.
The product representations of the original algebra $\mathfrak{g}$ must
include those of its subalgebra $\mathfrak{h}$. 

To understand it, let us consider an example of $\mathfrak{so}_{11}$.
By using the highest weight representation and the Dynkin's theorem for
the second highest representation, the highest and second highest
product representations of $\Lambda_1=(10001)$ and $\Lambda_2=(01000)$
of $\mathfrak{so}_{11}$ are
$\Lambda_1'=(11001)$ and $\Lambda_2'=(00101)$, respectively, because one
of the highest representations of $\mathfrak{so}_{11}$ is  
$\Lambda_1'=\Lambda_1+\Lambda_2=(11001)$ and 
$\Lambda_2'=\Lambda_1+\Lambda_2-\alpha_1-\alpha_2=(00101)$ from the
following Dynkin diagram: 
\begin{align}
&\underset{\mathfrak{so}_{11}}
{\xygraph{!~:{@{=}|@{}}
              \circ ([]!{+(0,-.3)} {\alpha_1}) (
        - [u] \circ ([]!{+(.3,0)}  {\hspace{-4em}\Lambda_1})
        - []!{+(4.0,-1.0)}  \bullet ([]!{+(0,-.3)} {{}})),
        - [r] \circ ([]!{+(0,-.3)} {\alpha_2}) (
        - [u] \circ ([]!{+(.3,0)}  {\Lambda_2}),
        - [r] \circ ([]!{+(0,-.3)} {\alpha_3}) 
        - [r] \circ ([]!{+(0,-.3)} {\alpha_4})
        : [r] \bullet ([]!{+(.3,-.3)} {\hspace{-1.5em}\alpha_5}) 
)}}.
\end{align}
Since the dimension of the product representation of
$\mathfrak{so}_{11}$ is
$d(\Lambda_1\otimes\Lambda_2)=d(\Lambda_1)d(\Lambda_2)=
320\times 55=17600$ and 
the dimension of the highest and second highest product representations
of $\mathfrak{so}_{11}$
are $d(\Lambda_1')=10240$, $d(\Lambda_2')=3520$, 
$d(\Lambda_1\otimes\Lambda_2)-d(\Lambda_1')-d(\Lambda_2')=3840>0$, and
then we continue the calculation.

The conjugacy class of the product representation of $\mathfrak{so}_{11}$
$C_c(\Lambda_1\otimes\Lambda_2)=1$, so we list up the representations
satisfying conditions $d(\Lambda)\leq 3840$ and $C_c(\Lambda)=1$.
The conditions are satisfied by only five representations (00101),
(20001), (01001), (10001), and (00001). Their dimensions are 
$d(00101)=3520$, $d(20001)=1760$, $d(01001)=1408$, $d(10001)=320$,
and $d(00001)=32$.

Here we use the Dynkin's method of parts. 
First, we consider a Lie subalgebra $\mathfrak{h}=\mathfrak{su}_5$ of
the Lie algebra $\mathfrak{so}_{11}$ as  
\begin{align}
&\underset{\mathfrak{so}_{11}\supset\mathfrak{su}_5}
{\xygraph{!~:{@{=}|@{}}
              \circ ([]!{+(0,-.3)} {\alpha_1}) (
        - [u] \circ ([]!{+(.3,0)}  {\hspace{-4em}\Lambda_1})
        - []!{+(4.0,-1.0)}  \bullet ([]!{+(0,-.3)} {{}})),
        - [r] \circ ([]!{+(0,-.3)} {\alpha_2}) (
        - [u] \circ ([]!{+(.3,0)}  {\Lambda_2}),
        - [r] \circ ([]!{+(0,-.3)} {\alpha_3}) 
        - [r] \circ ([]!{+(0,-.3)} {\alpha_4})
        : [r] {\bullet\hspace{-0.645em}\times} ([]!{+(.3,-.3)} {\hspace{-1.5em}\alpha_5}) 
)}},
\end{align}
where the symbol $\times$ on the node $\bullet$ stands for removing
$\bullet$ in the Dynkin diagram. We denote the representations of the
remaining algebra of $\mathfrak{su}_5$ as $\tilde{\Lambda}_1$ and
$\tilde{\Lambda}_2$. 
Since $\Lambda_1=(10001)\ni\tilde{\Lambda}_1=(1000)$ and 
$\Lambda_2=(01000)\ni\tilde{\Lambda}_2=(0100)$, the product
representations of $\mathfrak{su}_{5}$ are $\tilde{\Lambda}_1'=(1100)$
and $\tilde{\Lambda}_2'=(0010)$. 
Unfortunately, this is no addition constraint for the remaining product
representations of $\mathfrak{su}_5$ because 
$\Lambda_1'=(11001)\ni\tilde{\Lambda}_1'=(1100)$ and 
$\Lambda_2'=(00101)\ni\tilde{\Lambda}_2'=(0010)$.
Next, we consider a Lie subalgebra $\mathfrak{so}_{9}$ of the Lie
algebra $\mathfrak{so}_{11}$ as  
\begin{align}
&\underset{\mathfrak{so}_{11}\supset\mathfrak{so}_9}
{\xygraph{!~:{@{=}|@{}}
              {\circ\hspace{-0.645em}\times} ([]!{+(0,-.3)} {\alpha_1}) (
        - [u] \circ ([]!{+(.3,0)}  {\hspace{-4em}\Lambda_1})
        - []!{+(4.0,-1.0)}  \bullet ([]!{+(0,-.3)} {{}})),
        - [r] \circ ([]!{+(0,-.3)} {\alpha_2}) (
        - [u] \circ ([]!{+(.3,0)}  {\Lambda_2}),
        - [r] \circ ([]!{+(0,-.3)} {\alpha_3}) 
        - [r] \circ ([]!{+(0,-.3)} {\alpha_4})
        : [r] \bullet ([]!{+(.3,-.3)} {\hspace{-1.5em}\alpha_5}) 
)}}.
\end{align}
Since $\tilde{\Lambda}_1=(0001)$ and $\tilde{\Lambda}_2=(1000)$ of
$\mathfrak{so}_9$, the product representations are
$\tilde{\Lambda}_1'=(1001)$ and 
$\tilde{\Lambda}_2'=(0001)$. 
$\tilde{\Lambda}_2'=(0001)$ is not a decomposed representation of
$\tilde{\Lambda}_1=(11001)$ nor $\tilde{\Lambda}_2=(00101)$, and then 
at least one of the representations (20001), (10001), or (00001) of
$\mathfrak{so}_{11}$ must be one of the product representations of
$\mathfrak{so}_{11}$. 
Unfortunately, the Dynkin's method of parts is not much useful in this 
case.

Next we use the relations of the Dynkin index give in 
Eqs.~(\ref{Eq:Dynkin-index-R1})-(\ref{Eq:Dynkin-index-R3}).
By using $T(\Lambda_1=(10001))=72$, $T(\Lambda_2=(01000))=9$, 
$T(\Lambda_1'=(11001))=4352$, and $T(\Lambda_2'=(00101))=1304$,
we get $T(\Lambda_1\otimes\Lambda_2)-T(\Lambda_1')-T(\Lambda_2')=1184$.
Also, $T(00101)=1304$, $T(20001)=604$, 
$T(01001)=432$, $T(10001)=72$, and $T(00001)=4$,
and then we can eliminate $(00101)$ because $T(00101)=1304>1184$.
Now we try to fit $d(R_{rem})=3840$ and $T(R_{rem})=1184$ by using a
linear combination of $d(20001)=1760$, $d(01001)=1408$, $d(10001)=320$,
$d(00001)=32$, and 
$T(00101)=1304$, $T(20001)=604$, $T(01001)=432$, $T(10001)=72$, 
$T(00001)=4$, respectively.
We can find that the remaining product representations are 
$(00101)$, $(20001)$, $(01001)$, $(10001)$, $(10001)$, and $(00001)$.
Thus, the tensor product is
\begin{align}
(10001)\otimes(01000)=
(11001)\oplus(00101)\oplus(20001)\oplus(01001)
\oplus(10001)\oplus(10001)\oplus(00001).
\end{align}

\subsection{Recipe for calculating tensor product}
\label{Sec:Recipe-tensor-product}

Let us summarize how to decompose product representations of a tensor 
product of representations $R_1$ and $R_2$ of a Lie algebra
$\mathfrak{g}$ with highest weights
$\Lambda_1$ and $\Lambda_2$ into the irreducible representations.
\begin{itemize}
\item[(T-1)] A product representation $R_1'$ has the highest weight 
      $\Lambda_1'=\Lambda_1+\Lambda_2$.
\item[(T-2)] By using the Dynkin's theorem for second highest
      representation, another product representation $R_2'$ has the 
      highest weight 
      $\Lambda_2'=\Lambda_1+\Lambda_2
      -(\alpha_i+\alpha_{i+1}+\cdots+\alpha_j)$, 
      where $i$ and $j$ are determined by the shortest path of the Dynkin
      diagram between $\Lambda_1$ and $\Lambda_2$.
\item[(T-3)] Check $d(R_1)d(R_2)-d(R_1')-d(R_2')$. If 
      $d(R_1)d(R_2)-d(R_1')-d(R_2')=0$, the calculation is done.
      If $d(R_1)d(R_2)-d(R_1')-d(R_2')\not=0$, 
      list up representations that satisfy two conditions:
      its dimension is less than  $d(R_1)d(R_2)-d(R_1')-d(R_2')$ 
      and their conjugacy class is the same as that of the tensor
      product representation $R_1\otimes R_2$, $C_c(R_1\otimes R_2)$. 
      If the representation is unique, the calculation is done.
\item[(T-4)] Use the Dynkin's method of parts to reduce its candidate
      for product representations.
\item[(T-5)] If all the product representations are not determined,
      use the Dynkin index (and also the anomaly coefficients for
      $\mathfrak{su}_n$).
\end{itemize}

First, it may be better to check whether necessary tensor products are
listed in Appendix~\ref{Sec:Teneor-products} or not. If you can find
them, you do not need to calculate them. Also, one may consult some
computer programs such as Susyno \cite{Fonseca:2011sy}, LieART
\cite{Feger:2012bs}, or LIE \cite{Leeuwen:1992}. 
Even in these cases, it is  
better to verify whether the tensor products are correct or not by using
their dimensions, Dynkin indices, (anomaly coefficients for
$\mathfrak{su}_n$), 
and conjugacy classes given in Appendix~\ref{Sec:Representations}. 
(Appendix~\ref{Sec:Teneor-products} contains several tensor products of
classical Lie algebras up to rank-20 and the exceptional
Lie algebras.) 

For the above calculation, we used only the Dynkin diagram, but 
it is better to use the Young tableau method in e.g.,
Ref.~\cite{Fulton:1997}. 
There is the method not only for $\mathfrak{su}_n$ but also for
$\mathfrak{so}_{2n+1}$, 
$\mathfrak{usp}_{2n}$, and $\mathfrak{so}_{2n}$ given in
Refs.~\cite{Black:1983,Koike:1987aa,Koike:1987bb}.
For its review, see, e.g., Ref.~\cite{Fulton:1997}.
Their discussion is based on the Weyl character formula
\cite{Weyl:1946}.

\section{Summary for representations of Lie algebras and their subalgebras}
\label{Sec:Summary-representations}

In this section, let us summarize several features of Lie algebras
$A_n$, $B_n$, $C_n$, $D_n$, $E_{6}$, $E_{7}$, $E_{8}$, $F_4$, and $G_2$.
Their explicit examples for Rank-20 and lower are
summarized in Appendix~\ref{Sec:Representations}. For usual usage, 
Appendix~\ref{Sec:Representations} contains enough information, but 
the following discussion may be useful if someone consider more than
Rank-20 algebra or too higher dimensional representations.

Let us summarize nontrivial representations of Lie algebras
$\mathfrak{g}$ with the lowest dimension and the adjoint representation
of $\mathfrak{g}$.
Nontrivial representations $R$ of $\mathfrak{g}$ with the lowest
dimension are listed in the following table.
\begin{center}
\begin{longtable}{cclc}
\caption{Lowest dimensional representations}
\label{Table:Representaitons-lowest-dimension}\\
\hline\hline
Algebra $\mathfrak{g}$&Rank     &Nontrivial representation of $\mathfrak{g}$&$d(R)$\\\hline\hline
\endfirsthead
\multicolumn{4}{c}{Table~\ref{Table:Representaitons-lowest-dimension} (continued)}\\\hline\hline
Algebra $\mathfrak{g}$&Rank     &Nontrivial representation of $\mathfrak{g}$&$d(R)$\\\hline\hline
\endhead
\hline
\endfoot
$A_n=\mathfrak{su}_{n+1}$&$n\geq 1$&$(10\cdots00)$,\ $(00\cdots01)$&$n+1$\\
$B_n=\mathfrak{so}_{2n+1}$&$n\geq 3$&$(10\cdots00)$&$2n+1$\\
$C_n=\mathfrak{usp}_{2n}$&$n\geq 2$&$(10\cdots00)$&$2n$\\
$D_4=\mathfrak{so}_{8}$&$4$      &$(1000)$,\ $(0010)$,\ $(0001)$&$8$\\
$D_n=\mathfrak{so}_{2n}$&$n\geq 5$&$(10\cdots00)$&$2n$\\
$E_6$&$6$&$(100000)$,\ $(000010)$&$27$\\
$E_7$&$7$&$(0000010)$&$56$\\
$E_8$&$8$&$(00000010)$&$248$\\
$F_4$&$4$&$(0001)$&$26$\\
$G_2$&$2$&$(01)$&$7$\\
\end{longtable}
\end{center}
The adjoint representations of Lie algebras $\mathfrak{g}$ are listed in
the following table.
\begin{center}
\begin{longtable}{cclc}
\caption{Adjoint representations}
\label{Table:Representaitons-adjoint}\\
\hline\hline
Algebra $\mathfrak{g}$&Rank&Adjoint representation&$d(\mathfrak{g})$ \\\hline\hline
\endfirsthead
\multicolumn{4}{c}{Table~\ref{Table:Representaitons-lowest-dimension} (continued)}\\\hline\hline
Algebra $\mathfrak{g}$&Rank&Adjoint representation&$d(\mathfrak{g})$ \\\hline\hline
\endhead
\hline
\endfoot
$A_1=\mathfrak{su}_2$&$1$&$(2)$&$3$\\
$A_n=\mathfrak{su}_{n+1}$&$n\geq 2$&$(100\cdots01)$&$n(n+2)$\\
$B_n=\mathfrak{so}_{2n+1}$&$n\geq 3$&$(010\cdots00)$&$n(2n+1)$\\
$C_n=\mathfrak{usp}_{2n}$&$n\geq 2$&$(200\cdots00)$&$n(2n+1)$\\
$D_n=\mathfrak{so}_{2n}$&$n\geq 4$&$(010\cdots00)$&$n(2n-1)$\\
$E_6$&$6$&$(000001)$&$78$\\
$E_7$&$7$&$(1000000)$&$133$\\
$E_8$&$8$&$(00000010)$&$248$\\
$F_4$&$4$&$(1000)$&$52$\\
$G_2$&$2$&$(10)$&$14$\\
\end{longtable}
\end{center}
The dimensions of the fundamental representations of the Lie algebras
are given in Ref.~\cite{Dynkin:1957um}, where the fundamental
representations of the Lie algebras are defined as representations whose
Dynkin labels are zero except only one coefficient. (Note that one may
use another definition for a fundamental representation. E.g., the
fundamental representation of $A_n=\mathfrak{su}_{n+1}$ is only an
$(n+1)$-dimensional representation $(10\cdots 00)$, and its conjugate
representation $(00\cdots 01)$ is the anti-fundamental representation.
In the definition, a Lie algebra $A_n$ has only one fundamental
representation, while in our definition, $A_n$ has $n$ fundamental
representations.)
\begin{center}
\begin{longtable}{cl}
\caption{Dimension of fundamental representations}
\label{Table:Representaitons-fundamental-dimension}\\
\hline\hline
Algebra $\mathfrak{g}$&Dynkin diagrams with dimensions\\\hline\hline
\endfirsthead
\multicolumn{2}{c}{Table~\ref{Table:Representaitons-fundamental-dimension} (continued)}\\\hline\hline
Algebra $\mathfrak{g}$&Dynkin diagrams with dimensions\\\hline\hline
\endhead
\hline
\endfoot
$A_n$:&
${\xygraph{
    \circ ([]!{+(0,-.3)} {{}_{n+1}})  - [r]
    \circ ([]!{+(0,-.3)} {{}_{\frac{n(n+1)}{2}}})  - [r]
    \cdots                                      - [r]
    \circ ([]!{+(0,-.3)} {{}_{{}_{n+1}C_{k}}})  - [r] 
    \cdots                        - [r]
    \circ ([]!{+(0,-.3)} {{}_{\frac{n(n+1)}{2}}}) - [r]
    \circ ([]!{+(0,-.3)} {{}_{n+1}})}}
$\\
$B_n$:&
${\xygraph{!~:{@{=}|@{}}
    \circ   ([]!{+(0,-.3)} {{}_{2n+1}}) - [r]
    \circ   ([]!{+(0,-.3)} {{}_{n(2n+1)}})  - [r] 
    \cdots                             - [r]
    \circ   ([]!{+(0,-.3)} {{}_{{}_{2n+1}C_k}})  - [r] 
    \cdots                             - [r]
    \circ   ([]!{+(0,-.3)} {{}_{\frac{(2n+1)!}{(n-1)!(n+2)!}}}) : [r]
    \bullet ([]!{+(0,-.3)} {{}_{2^n}})}}
$\\
$C_n$:&
${\xygraph{!~:{@{=}|@{}}
    \bullet ([]!{+(0,-.3)} {{}_{2n}}) - [r]
    \bullet ([]!{+(0,-.3)} {{}_{(n-1)(2n+1)}}) - [r] 
    \cdots                      - [r]
    \bullet ([]!{+(0,-.3)} {{}_{{}_{2n}C_{k}-{}_{2n}C_{k-2}}}) - [r] 
    \cdots                      - [r]
    \bullet ([]!{+(0,-.3)} {{}_{\frac{3(n+1)(2n)!}{(n-1)!(n+3)!}}}) : [r]
    \circ   ([]!{+(0,-.3)} {\hspace{1.5em}{}_{\frac{3(n+1)(2n)!}{n!(n+2)!}}})}}
$\\
$D_n$:&
${\xygraph{
    \circ ([]!{+(0,-.3)} {{}_{2n}})  - [r] 
    \circ ([]!{+(0,-.3)} {{}_{n(2n-1)}})  - [r] 
    \cdots                        - [r]
    \circ ([]!{+(0,-.3)} {{}_{{}_{2n}C_{k}}})  - [r] 
    \cdots                        - [r]
    \circ ([]!{+(0,-.3)} {{}_{\frac{(2n)!}{(n-2)!(n+2)!}}}) (
        - []!{+(1,.5)}  \circ ([]!{+(0,-.3)} {{}_{2^{n-1}}}),
        - []!{+(1,-.5)} \circ ([]!{+(0,-.3)} {{}_{2^{n-1}}})
)}}
$\\
$E_6$:&
${\xygraph{
    \circ ([]!{+(0,-.3)}  {{}_{27}}) - [r]
    \circ ([]!{+(0,-.3)}  {{}_{351}}) - [r]
    \circ ([]!{+(.3,-.3)} {\hspace{-2em}{}_{2925}}) (
        - [u] \circ ([]!{+(.3,0)}  {{}_{78}}),
        - [r] \circ ([]!{+(0,-.3)} {{}_{351}})
        - [r] \circ ([]!{+(0,-.3)} {{}_{27}})
)}}
$\\
$E_7$:&
${\xygraph{
    \circ ([]!{+(0,-.3)}  {{}_{133}}) - [r]
    \circ ([]!{+(0,-.3)}  {{}_{8645}}) - [r]
    \circ ([]!{+(.3,-.3)} {\hspace{-2em}{}_{365750}}) (
        - [u] \circ ([]!{+(.3,0)}  {{}_{912}}),
        - [r] \circ ([]!{+(0,-.3)} {{}_{27664}})
        - [r] \circ ([]!{+(0,-.3)} {{}_{1539}})
        - [r] \circ ([]!{+(0,-.3)} {{}_{56}}))
}}
$\\
$E_8$:&
${\xygraph{
    \circ ([]!{+(0,-.3)}  {\hspace{-0.5em}{}_{3875}}) - [r]
    \circ ([]!{+(0,-.3)}  {\hspace{-2em}{}_{6696000}}) - [r]
    \circ ([]!{+(.3,-.3)} {\hspace{-3em}{}_{6899079264}}) (
        - [u] \circ ([]!{+(.3,0)}  {\hspace{1em}{}_{147250}}),
        - [r] \circ ([]!{+(0,-.3)} {\hspace{1em}{}_{146325270}})
        - [r] \circ ([]!{+(0,-.3)} {\hspace{2em}{}_{2450240}})
        - [r] \circ ([]!{+(0,-.3)} {\hspace{1em}{}_{30380}})
        - [r] \circ ([]!{+(0,-.3)} {\hspace{0em}{}_{248}})
)}}
$\\
$F_4$:&
${\xygraph{!~:{@{=}|@{}}
    \circ   ([]!{+(0,-.3)} {{}_{52}}) - [r]
    \circ   ([]!{+(0,-.3)} {{}_{1274}}) : [r]
    \bullet ([]!{+(0,-.3)} {{}_{273}}) - [r]
    \bullet ([]!{+(0,-.3)} {{}_{26}})}}
$\\
$G_2$:&
${\xygraph{!~:{@3{-}|@{}}
    \circ   ([]!{+(0,-.3)} {{}_{14}}) : [r]
    \bullet ([]!{+(0,-.3)} {{}_{7}})}}
$\\
\end{longtable}
\end{center}
It can be calculated by using Weyl dimension formula discussed in
Sec.~\ref{Sec:Weyl-dimensional-formula}. 
The Dynkin indices of the fundamental representations of the Lie
algebras are given in Ref.~\cite{Dynkin:1957ek}.
\begin{center}
\begin{longtable}{cl}
\caption{Dynkin indices of fundamental representations}
\label{Table:Representaitons-fundamental-indices}\\
\hline\hline
Algebra $\mathfrak{g}$&Dynkin diagrams with Dynkin indices\\\hline\hline
\endfirsthead
\multicolumn{2}{c}{Table~\ref{Table:Representaitons-fundamental-indices} (continued)}\\\hline\hline
Algebra $\mathfrak{g}$&Dynkin diagrams with Dynkin indices\\\hline\hline
\endhead
\hline
\endfoot
$A_n$:&
${\xygraph{
    \circ ([]!{+(0,-.3)} {{}_{\frac{1}{2}}})  - [r]
    \circ ([]!{+(0,-.3)} {{}_{\frac{n-1}{2}}})  - [r]
    \cdots                                      - [r]
    \circ ([]!{+(0,-.3)} {{}_{\frac{{}_{n-1}C_{k-1}}{2}}})  - [r] 
    \cdots                        - [r]
    \circ ([]!{+(0,-.3)} {{}_{\frac{n-1}{2}}}) - [r]
    \circ ([]!{+(0,-.3)} {{}_{\frac{1}{2}}})}}
$\\
$B_n$:&
${\xygraph{!~:{@{=}|@{}}
    \circ   ([]!{+(0,-.3)} {{}_{1}}) - [r]
    \circ   ([]!{+(0,-.3)} {{}_{2n-1}})  - [r] 
    \cdots                             - [r]
    \circ   ([]!{+(0,-.3)} {{}_{{}_{2n-1}C_{k-1}}})  - [r] 
    \cdots                             - [r]
    \circ   ([]!{+(0,-.3)} {{}_{\frac{(2n-1)!}{n!(n-1)!}}}) : [r]
    \bullet ([]!{+(0,-.3)} {{}_{2^{n-3}}})}}
$\\
$C_n$:& 
${\xygraph{!~:{@{=}|@{}}
    \bullet ([]!{+(0,-.3)} {{}_{\frac{1}{2}}}) - [r]
    \bullet ([]!{+(0,-.3)} {{}_{n-1}}) - [r] 
    \cdots                      - [r]
    \bullet ([]!{+(0,-.3)} {{}_{\frac{{}_{2n-1}C_{k-1}-{}_{2n-1}C_{k-2}}{2}}}) - [r] 
    \cdots                      - [r]
    \bullet ([]!{+(0,-.3)} {\hspace{-0.5em}{}_{\frac{2(2n-1)!}{(n-2)!(n+2)!}}}) : [r]
    \circ   ([]!{+(0,-.3)} {\hspace{1.5em}{}_{\frac{(2n-1)!}{(n-1)!(n+1)!}}})}}
$\\
$D_n$:&
${\xygraph{
    \circ ([]!{+(0,-.3)} {{}_{1}})  - [r] 
    \circ ([]!{+(0,-.3)} {{}_{2n-2}})  - [r] 
    \cdots                        - [r]
    \circ ([]!{+(0,-.3)} {{}_{{}_{2n-2}C_{k-1}}})  - [r] 
    \cdots                        - [r]
    \circ ([]!{+(0,-.3)} {{}_{\hspace{-1em}\frac{(2n-2)!}{(n-3)!(n+1)!}}}) (
        - []!{+(1,.5)}  \circ ([]!{+(0,-.3)} {{}_{2^{n-4}}}),
        - []!{+(1,-.5)} \circ ([]!{+(0,-.3)} {{}_{2^{n-4}}})
)}}
$\\
$E_6$:&
${\xygraph{
    \circ ([]!{+(0,-.3)}  {{}_{3}}) - [r]
    \circ ([]!{+(0,-.3)}  {{}_{75}}) - [r]
    \circ ([]!{+(.3,-.3)} {\hspace{-2em}{}_{900}}) (
        - [u] \circ ([]!{+(.3,0)}  {{}_{12}}),
        - [r] \circ ([]!{+(0,-.3)} {{}_{75}})
        - [r] \circ ([]!{+(0,-.3)} {{}_{3}})
)}}
$\\
$E_7$:&
${\xygraph{
    \circ ([]!{+(0,-.3)}  {{}_{18}}) - [r]
    \circ ([]!{+(0,-.3)}  {{}_{2340}}) - [r]
    \circ ([]!{+(.3,-.3)} {\hspace{-2em}{}_{145800}}) (
        - [u] \circ ([]!{+(.3,0)}  {{}_{180}}),
        - [r] \circ ([]!{+(0,-.3)} {{}_{8580}})
        - [r] \circ ([]!{+(0,-.3)} {{}_{324}})
        - [r] \circ ([]!{+(0,-.3)} {{}_{6}}))
}}
$\\
$E_8$:&
${\xygraph{
    \circ ([]!{+(0,-.3)}  {\hspace{-0.5em}{}_{750}}) - [r]
    \circ ([]!{+(0,-.3)}  {\hspace{-2em}{}_{2646000}}) - [r]
    \circ ([]!{+(.3,-.3)} {\hspace{-3em}{}_{4172830200}}) (
        - [u] \circ ([]!{+(.3,0)}  {\hspace{1em}{}_{42750}}),
        - [r] \circ ([]!{+(0,-.3)} {\hspace{1em}{}_{70802550}})
        - [r] \circ ([]!{+(0,-.3)} {\hspace{2em}{}_{889200}})
        - [r] \circ ([]!{+(0,-.3)} {\hspace{1em}{}_{7350}})
        - [r] \circ ([]!{+(0,-.3)} {\hspace{0em}{}_{30}})
)}}
$\\
$F_4$:&
${\xygraph{!~:{@{=}|@{}}
    \circ   ([]!{+(0,-.3)} {{}_{9}}) - [r]
    \circ   ([]!{+(0,-.3)} {{}_{441}}) : [r]
    \bullet ([]!{+(0,-.3)} {{}_{63}}) - [r]
    \bullet ([]!{+(0,-.3)} {{}_{3}})}}
$\\
$G_2$:&
${\xygraph{!~:{@3{-}|@{}}
    \circ   ([]!{+(0,-.3)} {{}_{4}}) : [r]
    \bullet ([]!{+(0,-.3)} {{}_{1}})}}
$\\
\end{longtable}
\end{center}
It can be calculated by using the relations between the Dynkin indices
discussed in Sec.~\ref{Sec:Dynkin-index}. 

\subsection{$A_n=\mathfrak{su}_{n+1}$}

Let us summarize representations of the Lie algebra 
$A_n=\mathfrak{su}_{n+1}$. 
From Table~\ref{Table:Dynkin-diagrams}, 
the extended Dynkin diagram of the Lie algebra $A_n=\mathfrak{su}_{n+1}$
is 
\begin{align}
\xygraph{
    \circ ([]!{+(0,-.3)} {{}_1}) (
        - []!{+(2.5,+1.0)}  \circ ([]!{+(0,-.3)} {{}_x})
        - []!{+(2.5,-1.0)}  \circ ([]!{+(0,-.3)} {{}}))
        - [r]
    \circ ([]!{+(0,-.3)} {{}_2}) 
        - [r]
    \circ ([]!{+(0,-.3)} {{}_3}) 
        - [r] \cdots - [r]
    \circ ([]!{+(0,-.3)} {{}_{n-1}}) - [r]
    \circ ([]!{+(0,-.3)} {{}_n})}.
\end{align}
From Table~\ref{Table:Cartan-matrices}, the Cartan matrix of
$A_n=\mathfrak{su}_{n+1}$ is 
\begin{align}
A(A_n)=
\left(
\begin{array}{ccccccc}
2&-1&0&\cdots&0&0&0\\
-1&2&-1&\cdots&0&0&0\\
0&-1&2&\cdots&0&0&0\\
\vdots&\vdots&\vdots&\ddots&\vdots&\vdots&\vdots\\
0&0&0&\cdots&2&-1&0\\
0&0&0&\cdots&-1&2&-1\\
0&0&0&\cdots&0&-1&2\\
\end{array}
\right).
\end{align}
From Tables~\ref{Table:Complex-representations-1} and 
\ref{Table:Self-conjugate-representations},
types of representations of
$A_n=\mathfrak{su}_{n+1}$ are given in the following table.
\begin{center}
\begin{longtable}{clll}
\caption{Types of representations of $A_n$}
\label{Table:Types-representations-A}\\
\hline\hline
Algebra&Rank&Condition&C/R/PR\\\hline\hline
\endfirsthead
\multicolumn{4}{c}{Table~\ref{Table:Types-representations-A} (continued)}\\\hline\hline
Algebra&Rank&Condition&C/R/PR\\\hline\hline
\endhead
\hline
\endfoot
$\mathfrak{su}_{n+1}$ 
&$n=0,2,3\ (\bmod.\ 4)$
&$(a_1\cdots a_n)\not=(a_n\cdots a_1)$
&Complex\\\cline{3-4}
&
&$(a_1\cdots a_n)=(a_n\cdots a_1)$
&Real\\\cline{2-4}
&$n=1\ (\bmod.\ 4)$
&$(a_1\cdots a_n)\not=(a_n\cdots a_1)$
&Complex\\\cline{3-4}
&
&$(a_1\cdots a_n)=(a_n\cdots a_1)$
&Real for $a_{(n+1)/2}=0\ (\bmod.\ 2)$\\\cline{4-4}
&
&
&PR\ \ \ for $a_{(n+1)/2}=1\ (\bmod.\ 2)$\\
\end{longtable}
\end{center}
where Condition stands for a condition for Dynkin labels, 
and C/R/PR represent complex, real, and pseudo-real
representations.
By using Dynkin diagram, 
from Tables~\ref{Table:Complex-representations-2},
the complex representations of $A_n=\mathfrak{su}_{n+1}$ satisfy
\begin{align}
{\xygraph{
    \circ ([]!{+(0,-.3)} {a_1})  - [r]
    \circ ([]!{+(0,-.3)} {a_2})  - [r]
    \cdots                        - [r]
    \circ ([]!{+(0,-.3)} {a_{n-1}}) - [r]
    \circ ([]!{+(0,-.3)} {a_n})}}
\not=
{\xygraph{
    \circ ([]!{+(0,-.3)} {a_{n}})  - [r]
    \circ ([]!{+(0,-.3)} {a_{n-1}})  - [r]
    \cdots                        - [r]
    \circ ([]!{+(0,-.3)} {a_2}) - [r]
    \circ ([]!{+(0,-.3)} {a_1})}}.
\end{align}
From Eq.~(\ref{Eq:Weyl-formula-A}), the Weyl dimension formula of
$A_n=\mathfrak{su}_{n+1}$ is  
\begin{align}
d(R)=&
\prod_{k=1}^n\prod_{i=1}^k
\left(\frac{\sum_{j=i}^{k}a_j}{k-i+1}+1\right).
\end{align}
For the Lie algebra $A_{n}=\mathfrak{su}_{n+1}$, from
Eqs.~(\ref{Eq:2-Casimir}) and (\ref{Eq:Dynkin-index}), 
we get the Casimir invariant and the Dynkin index of a representation
$(a_1a_2a_3\cdots a_n)$:
\begin{align}
C_2(R)=&\frac{1}{2}
\sum_{i,j=1}^{n}(a_i+2g_i)G(A_n)_{ij}a_j,\\
T(R)=&\frac{d(R)}{d(G)}C_2(R)
=\frac{d(R)}{2 d(G)}
\sum_{i,j=1}^{n}(a_i+2g_i)G(A_n)_{ij}a_j,
\end{align}
where $G(A_n)$ is the metric matrix in
Table~\ref{Table:Cartan-matrices-inverse}
\begin{align}
G(A_n)&=\frac{1}{n+1}
\left(
\begin{array}{cccccc}
1\cdot n    &1\cdot (n-1)&1\cdot (n-2)&\cdots&1\cdot 2    &1\cdot 1\\
1\cdot (n-1)&2\cdot (n-1)&2\cdot (n-2)&\cdots&2\cdot 2    &2\cdot 1\\
1\cdot (n-2)&2\cdot (n-2)&3\cdot (n-2)&\cdots&3\cdot 2    &3\cdot 1\\
\vdots      &\vdots      &\vdots      &\ddots&\vdots      &\vdots  \\
1\cdot 2    &2\cdot 2    &3\cdot 2    &\cdots&(n-1)\cdot 2&(n-1)\cdot 1\\
1\cdot 1    &2\cdot 1    &3\cdot 1    &\cdots&(n-1)\cdot 1&n\cdot 1\\
\end{array}
\right).
\end{align}
From Eqs.~(\ref{Eq:anomaly-coefficients}) and 
(\ref{Eq:anomaly-coefficients-a})-(\ref{Eq:anomaly-coefficients-R3}),
the anomaly coefficient of a representation $R$ of
$A_{n-1}=\mathfrak{su}_{n}$ is  
\begin{align}
\frac{A(R)}{d(R)}:=
\sum_{i,j,k=1}^{n-1}a_{ijk}(a_i+1)(a_j+1)(a_k+1),
\end{align}
where $a_{ijk}$ is a completely symmetric under exchanging between $i$, 
$j$ and $k$, and for $i\leq j\leq k$,
\begin{align}
a_{ijk}=N_n i(n-2j)(n-k),
\end{align}
where $N_n$ is a normalization factor depending on its rank $n$ but
independent of $i,j,k$ because we take it as satisfying 
$A(10\cdots 0)/d(10\cdots 0)=+1$.  
The anomaly coefficient $A(R)$ satisfies the following properties:
\begin{align}
&A(R_i)=-A(\overline{R_i}),\\
&A(R_1\oplus R_2)=A(R_1)+A(R_2),\\
&A(R_1\otimes R_2)=d(R_2)A(R_1)+d(R_1)A(R_2),
\end{align}
where $R_1$ and $R_2$ are representations of $\mathfrak{su}_n$.
From Table~\ref{Table:Conjugacy-classes}, the conjugacy classes of the Lie
algebra $A_n=\mathfrak{su}_{n+1}$ is
\begin{align}
C_c(R):=a_1+2a_2+3a_3+\cdots\ \ \  (\bmod.\ n+1).
\end{align}
The Lie algebra $A_n=\mathfrak{su}_{n+1}$ has $(n+1)$ conjugacy classes.

The above features of representations of 
$A_{n-1}=\mathfrak{su}_n$ $(n\geq 2)$ 
are summarized in the following table: 
\begin{center}
\begin{longtable}{ccccccc}
\caption{Representations of $A_{n-1}=\mathfrak{su}_{n}$}
\label{Table:Representations-summary-A}\\
\hline\hline
$\mathfrak{su}_n$ irrep.&$d(R)$&$C_2(R)$&$T(R)$&$A(R)$&$C_c(R)$&C/R/PR\\\hline\hline
\endfirsthead
\multicolumn{7}{c}{Table~\ref{Table:Representations-summary-A} (continued)}\\\hline\hline
$\mathfrak{su}_n$ irrep.&$d(R)$&$C_2(R)$&$T(R)$&$A(R)$&$C_c(R)$&C/R/PR\\\hline\hline
\endhead
\hline
\endfoot
$(0000\cdots 0000)$&$1$   &$0$               &$0$&$0$&$0$&R\\ 
$(1000\cdots 0000)$&$n$   &$\frac{n^2-1}{2n}$&$\frac{1}{2}$&$+1$&$1$&C/PR$_{n=2}$\\ 
$(0000\cdots 0001)$&$n$   &$\frac{n^2-1}{2n}$&$\frac{1}{2}$&$-1$&$n-1$&C/PR$_{n=2}$\\ 
$(0100\cdots 0000)$&$\frac{n(n-1)}{2}$&$\frac{(n+1)(n-2)}{n}$&$\frac{n-2}{2}$&$n-4$&$2$&C/R$_{n=4}$\\ 
$(0000\cdots 0010)$&$\frac{n(n-1)}{2}$&$\frac{(n+1)(n-2)}{n}$&$\frac{n-2}{2}$&$-n+4$&$n-2$&C/R$_{n=4}$\\ 
$(2000\cdots 0000)$&$\frac{n(n+1)}{2}$&$\frac{(n-1)(n+2)}{n}$&$\frac{n+2}{2}$&$n+4$&$2$&C/R$_{n=2}$\\ 
$(0000\cdots 0002)$&$\frac{n(n+1)}{2}$&$\frac{(n-1)(n+2)}{n}$&$\frac{n+2}{2}$&$-n-4$&$n-2$&C/R$_{n=2}$\\ 
$(1000\cdots 0001)$&$n^2-1$&$n$&$n$&$0$&$0$&R\\ 
$(0010\cdots 0000)$&$\frac{n(n-1)(n-2)}{6}$&$\frac{(n-2)(n-3)(n-4)}{12}$&$\frac{(n-3)(n-2)}{4}$&$\frac{(n-3)(n-6)}{2}$&$3$&C/PR$_{n=6}$\\
$(0000\cdots 0100)$&$\frac{n(n-1)(n-2)}{6}$&$\frac{(n-2)(n-3)(n-4)}{12}$&$\frac{(n-3)(n-2)}{4}$&$-\frac{(n-3)(n-6)}{2}$&$n-3$&C/PR$_{n=6}$\\
$(1100\cdots 0000)$&$\frac{n(n-1)(n+1)}{3}$&$\frac{3(n^2-3)}{2n}$&$\frac{n^2-3}{2}$&$n^2-9$&$3$&C/R$_{n=3}$\\
$(0000\cdots 0011)$&$\frac{n(n-1)(n+1)}{3}$&$\frac{3(n^2-3)}{2n}$&$\frac{n^2-3}{2}$&$-n^2+9$&$n-3$&C/R$_{n=3}$\\
$(3000\cdots 0000)$&$\frac{n(n+1)(n+2)}{6}$&$\frac{(n+2)(n+3)(n+4)}{12}$&$\frac{(n+2)(n+3)}{4}$&$\frac{(n+3)(n+6)}{2}$&$3$&C/PR$_{n=2}$\\
$(0000\cdots 0003)$&$\frac{n(n+1)(n+2)}{6}$&$\frac{(n+2)(n+3)(n+4)}{12}$&$\frac{(n+2)(n+3)}{4}$&$-\frac{(n+3)(n+6)}{2}$&$n-3$&C/PR$_{n=2}$\\
$(0200\cdots 0000)$&$\frac{n^2(n+1)(n-1)}{12}$&$\frac{n(n^2-16)}{3}$&$\frac{n(n-2)(n+2)}{6}$&$\frac{n(n-4)(n+4)}{3}$&$4$&C/R$_{n=4}$\\
$(0000\cdots 0020)$&$\frac{n^2(n+1)(n-1)}{12}$&$\frac{n(n^2-16)}{3}$&$\frac{n(n-2)(n+2)}{6}$&$-\frac{n(n-4)(n+4)}{3}$&$n-4$&C/R$_{n=4}$\\
\end{longtable}
\end{center}
where $\mathfrak{su}_{n}$ irrep., $d(R)$, $C_2(R)$, $T(R)$, $A(R)$,
C/R/PR stand for the Dynkin label of the irreducible representations of
$A_{n-1}=\mathfrak{su}_n$, their 
dimension, their quadratic Casimir invariant, their Dynkin index, 
their triangle anomaly number, and complex, real or pseudo-real
representations, respectively, where
$T(R)d(G)=C_2(R)d(R)$ and $d(G)$ is the dimension of the adjoint
representation. The anomaly number of a representation is the same
magnitude and its opposite sign of that of its conjugate representation:
$A(\overline{R})=-A(R)$. For $\mathfrak{su}_n$ representations, 
$\overline{(a_1a_2\cdots a_{n-1}a_n)}=(a_n a_{n-1}\cdots a_2a_1)$. 
It is the conjugate representation 
$(a_1a_2\cdots a_{n-1}a_n)$.
For $\mathfrak{su}_{2}$, $A(R)=0$ for any representation because
any representation of $\mathfrak{su}_2$ $R$ is a self-conjugate (real
or pseudo-real) representation satisfying $R=\overline{R}$.
See Appendix~\ref{Sec:Representations} for further information.

From Table~\ref{Table:Maximal-subalgebra},
maximal $R$- and $S$-subalgebras of $A_n=\mathfrak{su}_{n+1}$ 
with their rank up to 16 are listed in the following table.
\begin{center}
\begin{longtable}{crcp{10cm}c}
\caption{Maximal subalgebras of $A_n$}
\label{Table:Maximal-subalgebra-A}\\
\hline\hline
Rank&\multicolumn{2}{l}{Algebra $\mathfrak{g}$}&Maximal subalgebras $\mathfrak{h}$&Type\\\hline\hline
\endfirsthead
\multicolumn{4}{c}{Table~\ref{Table:Maximal-subalgebra-A} (continued)}\\
\hline
Rank&\multicolumn{2}{l}{Algebra $\mathfrak{g}$}&Maximal subalgebras $\mathfrak{h}$&Type\\\hline\hline
\endhead
\hline
\endfoot
$1$ &$\mathfrak{su}_2$&$\supset$&$\mathfrak{u}_1$&$(R)$\\
$2$ &$\mathfrak{su}_3$&$\supset$&$\mathfrak{su}_2\oplus\mathfrak{u}_1$&$(R)$\\
    &       &$\supset$&$\mathfrak{su}_2$&$(S)$\\
$3$ &$\mathfrak{su}_4$&$\supset$
&$\mathfrak{su}_3\oplus\mathfrak{u}_1;\
\mathfrak{su}_2\oplus\mathfrak{su}_2\oplus\mathfrak{u}_1$
&$(R)$\\
    &       &$\supset$&$\mathfrak{usp}_4;\
\mathfrak{su}_2\oplus\mathfrak{su}_2$&$(S)$\\
$4$ &$\mathfrak{su}_5$&$\supset$
&$\mathfrak{su}_4\oplus\mathfrak{u}_1;\ 
\mathfrak{su}_3\oplus\mathfrak{su}_2\oplus\mathfrak{u}_1$
&$(R)$\\
    &       &$\supset$&$\mathfrak{usp}_4$&$(S)$\\
$5$ &$\mathfrak{su}_6$&$\supset$
&$\mathfrak{su}_5\oplus\mathfrak{u}_1;\
\mathfrak{su}_4\oplus\mathfrak{su}_2\oplus\mathfrak{u}_1;\
\mathfrak{su}_3\oplus\mathfrak{su}_3\oplus\mathfrak{u}_1$
&$(R)$\\
    &       &$\supset$
&$\mathfrak{su}_3;\ \mathfrak{su}_4;\ \mathfrak{usp}_6;\
\mathfrak{su}_3\oplus\mathfrak{su}_2$&$(S)$\\
$6$ &$\mathfrak{su}_7$&$\supset$
&$\mathfrak{su}_6\oplus\mathfrak{u}_1;\
\mathfrak{su}_5\oplus\mathfrak{su}_2\oplus\mathfrak{u}_1;\
\mathfrak{su}_4\oplus\mathfrak{su}_3\oplus\mathfrak{u}_1$
&$(R)$\\
    &       &$\supset$&$\mathfrak{so}_7$&$(S)$\\
$7$ &$\mathfrak{su}_8$&$\supset$
&$\mathfrak{su}_7\oplus\mathfrak{u}_1;\
\mathfrak{su}_6\oplus\mathfrak{su}_2\oplus\mathfrak{u}_1;\
\mathfrak{su}_5\oplus\mathfrak{su}_3\oplus\mathfrak{u}_1;\
\mathfrak{su}_4\oplus\mathfrak{su}_4\oplus\mathfrak{u}_1$
&$(R)$\\
    &       &$\supset$
&$\mathfrak{so}_8;\ \mathfrak{usp}_8;\
\mathfrak{su}_4\oplus\mathfrak{su}_2$&$(S)$\\
$8$ &$\mathfrak{su}_9$&$\supset$
&$\mathfrak{su}_8\oplus\mathfrak{u}_1;\
\mathfrak{su}_7\oplus\mathfrak{su}_2\oplus\mathfrak{u}_1;\
\mathfrak{su}_6\oplus\mathfrak{su}_3\oplus\mathfrak{u}_1;\
\mathfrak{su}_5\oplus\mathfrak{su}_4\oplus\mathfrak{u}_1$&$(R)$\\
    &       &$\supset$&$\mathfrak{so}_9;\
\mathfrak{su}_3\oplus\mathfrak{su}_3$&$(S)$\\
$9$ &$\mathfrak{su}_{10}$&$\supset$
&$\mathfrak{su}_9\oplus\mathfrak{u}_1;\
\mathfrak{su}_8\oplus\mathfrak{su}_2\oplus\mathfrak{u}_1;\
\mathfrak{su}_7\oplus\mathfrak{su}_3\oplus\mathfrak{u}_1;\
\mathfrak{su}_6\oplus\mathfrak{su}_4\oplus\mathfrak{u}_1;\
\mathfrak{su}_5\oplus\mathfrak{su}_5\oplus\mathfrak{u}_1$
&$(R)$\\
    &       &$\supset$&
$\mathfrak{su}_3;\ \mathfrak{su}_4;\ \mathfrak{su}_5;\ 
\mathfrak{so}_{10};\ \mathfrak{usp}_{10};\
\mathfrak{su}_5\oplus\mathfrak{su}_2
$&$(S)$\\
$10$&$\mathfrak{su}_{11}$&$\supset$
&$\mathfrak{su}_{10}\oplus\mathfrak{u}_1;\
\mathfrak{su}_9\oplus\mathfrak{su}_2\oplus\mathfrak{u}_1;\
\mathfrak{su}_8\oplus\mathfrak{su}_3\oplus\mathfrak{u}_1;\
\mathfrak{su}_7\oplus\mathfrak{su}_4\oplus\mathfrak{u}_1;\
\mathfrak{su}_6\oplus\mathfrak{su}_5\oplus\mathfrak{u}_1$&$(R)$\\
    &       &$\supset$&$\mathfrak{so}_{11}$&$(S)$\\
$11$&$\mathfrak{su}_{12}$&$\supset$
&$\mathfrak{su}_{11}\oplus\mathfrak{u}_1;\
\mathfrak{su}_{10}\oplus\mathfrak{su}_2\oplus\mathfrak{u}_1;\
\mathfrak{su}_9\oplus\mathfrak{su}_3\oplus\mathfrak{u}_1;\
\mathfrak{su}_8\oplus\mathfrak{su}_4\oplus\mathfrak{u}_1;\
\mathfrak{su}_7\oplus\mathfrak{su}_5\oplus\mathfrak{u}_1;\
\mathfrak{su}_6\oplus\mathfrak{su}_6\oplus\mathfrak{u}_1
$&$(R)$\\
    &       &$\supset$
&$\mathfrak{so}_{12};\
\mathfrak{usp}_{12};\
\mathfrak{su}_6\oplus\mathfrak{su}_2;\
\mathfrak{su}_4\oplus\mathfrak{su}_3
$&$(S)$\\
$12$&$\mathfrak{su}_{13}$&$\supset$
&$\mathfrak{su}_{12}\oplus\mathfrak{u}_1;\
\mathfrak{su}_{11}\oplus\mathfrak{su}_2\oplus\mathfrak{u}_1;\
\mathfrak{su}_{10}\oplus\mathfrak{su}_3\oplus\mathfrak{u}_1;\
\mathfrak{su}_9\oplus\mathfrak{su}_4\oplus\mathfrak{u}_1;\
\mathfrak{su}_8\oplus\mathfrak{su}_5\oplus\mathfrak{u}_1;\
\mathfrak{su}_7\oplus\mathfrak{su}_6\oplus\mathfrak{u}_1
$&$(R)$\\
    &       &$\supset$
&$\mathfrak{so}_{13}$&$(S)$\\
$13$&$\mathfrak{su}_{14}$&$\supset$
&$\mathfrak{su}_{13}\oplus\mathfrak{u}_1;\
\mathfrak{su}_{12}\oplus\mathfrak{su}_2\oplus\mathfrak{u}_1;\
\mathfrak{su}_{11}\oplus\mathfrak{su}_3\oplus\mathfrak{u}_1;\
\mathfrak{su}_{10}\oplus\mathfrak{su}_4\oplus\mathfrak{u}_1;\
\mathfrak{su}_9\oplus\mathfrak{su}_5\oplus\mathfrak{u}_1;\
\mathfrak{su}_8\oplus\mathfrak{su}_6\oplus\mathfrak{u}_1;\
\mathfrak{su}_7\oplus\mathfrak{su}_7\oplus\mathfrak{u}_1
$&$(R)$\\
    &       &$\supset$
&$\mathfrak{so}_{14};\
\mathfrak{usp}_{14};\
\mathfrak{su}_7\oplus\mathfrak{su}_2
$&$(S)$\\
$14$&$\mathfrak{su}_{15}$&$\supset$
&$\mathfrak{su}_{14}\oplus\mathfrak{u}_1;\
\mathfrak{su}_{13}\oplus\mathfrak{su}_2\oplus\mathfrak{u}_1;\
\mathfrak{su}_{12}\oplus\mathfrak{su}_3\oplus\mathfrak{u}_1;\
\mathfrak{su}_{11}\oplus\mathfrak{su}_4\oplus\mathfrak{u}_1;\
\mathfrak{su}_{10}\oplus\mathfrak{su}_5\oplus\mathfrak{u}_1;\
\mathfrak{su}_9\oplus\mathfrak{su}_6\oplus\mathfrak{u}_1;\
\mathfrak{su}_8\oplus\mathfrak{su}_7\oplus\mathfrak{u}_1
$&$(R)$\\
    &       &$\supset$
&$\mathfrak{so}_{15};\
\mathfrak{su}_5\oplus\mathfrak{su}_3;\
\mathfrak{su}_3;\
\mathfrak{su}_5;\
\mathfrak{su}_6
$&$(S)$\\
$15$&$\mathfrak{su}_{16}$&$\supset$
&$\mathfrak{su}_{15}\oplus\mathfrak{u}_1;\
\mathfrak{su}_{14}\oplus\mathfrak{su}_2\oplus\mathfrak{u}_1;\
\mathfrak{su}_{13}\oplus\mathfrak{su}_3\oplus\mathfrak{u}_1;\
\mathfrak{su}_{12}\oplus\mathfrak{su}_4\oplus\mathfrak{u}_1;\
\mathfrak{su}_{11}\oplus\mathfrak{su}_5\oplus\mathfrak{u}_1;\
\mathfrak{su}_{10}\oplus\mathfrak{su}_6\oplus\mathfrak{u}_1;\
\mathfrak{su}_9\oplus\mathfrak{su}_7\oplus\mathfrak{u}_1;\
\mathfrak{su}_8\oplus\mathfrak{su}_8\oplus\mathfrak{u}_1
$&$(R)$\\
    &       &$\supset$
&$\mathfrak{so}_{16};\
\mathfrak{usp}_{16};\
\mathfrak{so}_{10};\
\mathfrak{su}_8\oplus\mathfrak{su}_2;\
\mathfrak{su}_4\oplus\mathfrak{su}_4
$&$(S)$\\
$16$&$\mathfrak{su}_{17}$&$\supset$
&$\mathfrak{su}_{16}\oplus\mathfrak{u}_1;\
\mathfrak{su}_{15}\oplus\mathfrak{su}_2\oplus\mathfrak{u}_1;\
\mathfrak{su}_{14}\oplus\mathfrak{su}_3\oplus\mathfrak{u}_1;\
\mathfrak{su}_{13}\oplus\mathfrak{su}_4\oplus\mathfrak{u}_1;\
\mathfrak{su}_{12}\oplus\mathfrak{su}_5\oplus\mathfrak{u}_1;\
\mathfrak{su}_{11}\oplus\mathfrak{su}_6\oplus\mathfrak{u}_1;\
\mathfrak{su}_{10}\oplus\mathfrak{su}_7\oplus\mathfrak{u}_1;\
\mathfrak{su}_9\oplus\mathfrak{su}_8\oplus\mathfrak{u}_1
$&$(R)$\\
    &       &$\supset$
&$\mathfrak{so}_{17}
$&$(S)$\\
\end{longtable}
\end{center}
where Type $(R)$ and $(S)$ stand for $R$- and $S$-subalgebras,
respectively. 

We summarize generic projection matrices of the Lie algebra
$A_n=\mathfrak{su}_{n+1}$ and any maximal regular subalgebras, 
which are calculated by using a method given in
Sec.~\ref{Sec:Representations-subalgebras}. 
The Dynkin diagram of 
$A_{n}\supset A_{n-1}\oplus \mathfrak{u}_1$ $(n\geq 2)$
is given by
\begin{align}
&\underset{A_n=\mathfrak{su}_{n+1}}
{\xygraph{
    \circ ([]!{+(0,-.3)}  {{}_1})       - [r]
    \circ ([]!{+(0,-.3)}  {{}_2})       - [r]
    \cdots                              - [r] 
    \circ ([]!{+(0,-.3)}  {{}_{n-2}})   - [r] 
    \circ ([]!{+(0,-.3)}  {{}_{n-1}})   - [r] 
    \circ ([]!{+(0,-.3)}  {{}_{n}})
}}\nonumber\\
\underset{\rm Removing}{\Longrightarrow}
& \underset{A_{n-1}\oplus\mathfrak{u}_1
=\mathfrak{su}_{n}\oplus\mathfrak{u}_1}
{\xygraph{
    \circ ([]!{+(0,-.3)}  {{}_1})       - [r]
    \circ ([]!{+(0,-.3)}  {{}_2})       - [r]
    \cdots                              - [r] 
    \circ ([]!{+(0,-.3)}  {{}_{n-2}})   - [r] 
    \circ ([]!{+(0,-.3)}  {{}_{n-1}})   - [r] 
    {\circ\hspace{-0.645em}\times} ([]!{+(0,-.3)}  {{}_{n}})   
}}.
\end{align}
Its corresponding projection matrix is
\begin{align}
P_{\mathfrak{su}_{n+1}\supset
\mathfrak{su}_{n}\oplus\mathfrak{u}_1}
&=
\left(
\begin{array}{cc}
\multicolumn{1}{c|}{I_{n-1}}&\multicolumn{1}{c}{O}\\\hline
\multicolumn{1}{c}{1\ 2\cdots n-2\ n-1}&n\\
\end{array}
\right),
\end{align}
where 
$P_{\mathfrak{su}_{n+1}\supset\mathfrak{su}_{n}\oplus\mathfrak{u}_1}$ 
is an $n\times n$ matrix, 
$I_{n-1}$ is an $(n-1)\times (n-1)$ matrix, where its matrix
elements $(I_{n-1})_{ii}=1 (i=1,2,\cdots,n-1)$ and the other matrix
elements $(I_{n-1})_{ij}=0 (i,j=1,2,\cdots,n-1;i\not=j)$; i.e., 
$I_{n-1}=\mbox{diag}(1,1,\cdots,1)$.
In the matrix, $O$ is an $(n-1)\times 1$ matrix whose matrix elements
are zero. In the following discussion, any $O$ in projection matrices is
an appropriate size matrix whose matrix elements are zero.
The Dynkin diagram of 
$A_{n}\supset A_{k}\oplus A_{n-k-1}\oplus \mathfrak{u}_1$
$(n\geq 2; 1\leq k \leq n-2)$
is given by
\begin{align}
&\underset{A_n=\mathfrak{su}_{n+1}}
{\xygraph{
    \circ ([]!{+(0,-.3)}  {{}_1})     - [r]
    \cdots                            - [r] 
    \circ ([]!{+(0,-.3)}  {{}_{k-1}}) - [r] 
    \circ ([]!{+(0,-.3)}  {{}_{k}})   - [r] 
    \circ ([]!{+(0,-.3)}  {{}_{k+1}}) - [r] 
    \cdots                            - [r] 
    \circ ([]!{+(0,-.3)}  {{}_{n}})
}}\nonumber\\
\underset{\rm Removing}{\Longrightarrow}&
\underset{A_{k}\oplus A_{n-k-1}\oplus\mathfrak{u}_1
=\mathfrak{su}_{k+1}\oplus\mathfrak{su}_{n-k}\oplus\mathfrak{u}_1}
{\xygraph{
    \circ ([]!{+(0,-.3)}  {{}_1})     - [r]
    \cdots                            - [r] 
    \circ ([]!{+(0,-.3)}  {{}_{k-1}}) - [r] 
    {\circ\hspace{-0.645em}\times} ([]!{+(0,-.3)}  {{}_{k}})   - [r] 
    \circ ([]!{+(0,-.3)}  {{}_{k+1}}) - [r] 
    \cdots                            - [r] 
    \circ ([]!{+(0,-.3)}  {{}_{n}})
}}.
\end{align}
Its corresponding projection matrix is
\begin{align}
&P_{\mathfrak{su}_{n+1}\supset
\mathfrak{su}_{k+1}\oplus\mathfrak{su}_{n-k}\oplus\mathfrak{u}_1}
\nonumber\\
&=
\left(
\begin{array}{cccccc}
\multicolumn{3}{c|}{I_{k}}&\multicolumn{3}{c}{O}\\\hline
\multicolumn{4}{c|}{O}        &\multicolumn{2}{c}{I_{n-k-1}}\\\hline
\multicolumn{3}{c|}{(n-k)\ 2(n-k)\cdots k(n-k)}
&(k+1)(n-k)
&\multicolumn{2}{|c}{(n-k-1)(k+1)\cdots 2(k+1)\ (k+1)}\\
\end{array}
\right),
\end{align}
where 
$P_{\mathfrak{su}_{n+1}\supset
\mathfrak{su}_{k+1}\oplus\mathfrak{su}_{n-k}\oplus\mathfrak{u}_1}$ 
is an $n\times n$ matrix.

We summarize generic projection matrices of the Lie algebra
$A_n=\mathfrak{su}_{n+1}$ and several maximal special subalgebras
given in Tables~\ref{Table:Maximal-S-sub-classical-1}
and \ref{Table:Maximal-S-sub-classical-2}, which are also calculated by
using a method given in Sec.~\ref{Sec:Representations-subalgebras}.  
(The following projection matrices cannot cover all maximal special
subalgebras, so we need to calculate exceptions one by one.)
The Dynkin diagram of $A_{2n}\supset B_n$ $(n\geq 3)$ is given by
\begin{align}
&\underset{A_{2n}=\mathfrak{su}_{2n+1}}
{\xygraph{
    \circ ([]!{+(0,-.3)}  {{}_1})       - [r]
    \circ ([]!{+(0,-.3)}  {{}_2})       - [r]
    \cdots                              - [r] 
    \circ ([]!{+(0,-.3)}  {{}_{n-1}}) - [r] 
    \circ ([]!{+(0,-.3)}  {{}_{n}})   - [r] 
    \circ ([]!{+(0,-.3)}  {{}_{n+1}}) - [r] 
    \circ ([]!{+(0,-.3)}  {{}_{n+2}}) - [r] 
    \cdots                        - [r] 
    \circ ([]!{+(0,-.3)}  {{}_{2n-1}})   - [r] 
    \circ ([]!{+(0,-.3)}  {{}_{2n}})
}}\nonumber\\
\underset{\rm Folding}{\Longrightarrow}
&\underset{B_n=\mathfrak{so}_{2n+1}}
{\xygraph{!~:{@{=}|@{}}
    \circ ([]!{+(0,-.3)} {{}_{1,2n}}) - [r]
    \circ ([]!{+(0,-.3)} {{}_{2,2n-1}}) - [r]
    \cdots                        - [r] 
    \circ   ([]!{+(0,-.3)} {{}_{n+\frac{1}{2}\pm\frac{5}{2}}}) - [r]
    \circ   ([]!{+(0,-.3)} {{}_{n+\frac{1}{2}\pm\frac{3}{2}}}) : [r]
    \bullet ([]!{+(0,-.3)} {{}_{n+\frac{1}{2}\pm\frac{1}{2}}})}
}.
\end{align}
Its corresponding projection matrix is
\begin{align}
P_{\mathfrak{su}_{2n+1}\supset\mathfrak{so}_{2n+1}}
&=
\left(
\begin{array}{cccccc}
\multicolumn{2}{c|}{I_{n-1}}
&\multicolumn{2}{c}{O}
&\multicolumn{2}{|c}{E_{n-1}}\\\hline
\multicolumn{2}{c|}{O}&
2&2&\multicolumn{2}{|c}{O}\\
\end{array}
\right),
\end{align}
where 
$P_{\mathfrak{su}_{2n+1}\supset\mathfrak{so}_{2n+1}}$
is an $n\times 2n$ matrix;
$E_{n-1}$ is an $(n-1)\times (n-1)$ matrix, where its matrix
elements $(E_{n-1})_{i,n-i}=1 (i=1,2,\cdots,n-1)$ and the other matrix
elements $(E_{n-1})_{i,n-j}=0 (i,j=1,2,\cdots,n-1;i\not=j)$.
The Dynkin diagram of 
$A_{2n-1}\supset C_n$ $(n\geq 2)$
is given by
\begin{align}
&\underset{A_{2n-1}=\mathfrak{su}_{2n}}
{\xygraph{
    \circ ([]!{+(0,-.3)}  {{}_1})       - [r]
    \circ ([]!{+(0,-.3)}  {{}_2})       - [r]
    \cdots                              - [r] 
    \circ ([]!{+(0,-.3)}  {{}_{n-1}})   - [r] 
    \circ ([]!{+(0,-.3)}  {{}_{n}})     - [r] 
    \circ ([]!{+(0,-.3)}  {{}_{n+1}})   - [r] 
    \cdots                              - [r] 
    \circ ([]!{+(0,-.3)}  {{}_{2n-2}})  - [r] 
    \circ ([]!{+(0,-.3)}  {{}_{2n-1}})
}}\nonumber\\
\underset{\rm Folding}{\Longrightarrow}
&\underset{C_n=\mathfrak{usp}_{2n}}
{\xygraph{!~:{@{=}|@{}}
    \bullet ([]!{+(0,-.3)} {{}_{1,2n-1}})     - [r]
    \bullet ([]!{+(0,-.3)} {{}_{2,2n-2}})     - [r] 
    \cdots                             - [r] 
    \bullet ([]!{+(0,-.3)} {{}_{n\pm 2}}) - [r] 
    \bullet ([]!{+(0,-.3)} {{}_{n\pm 1}}) : [r]
    \circ   ([]!{+(0,-.3)} {{}_{n}})}
}.
\end{align}
Its corresponding projection matrix is
\begin{align}
P_{\mathfrak{su}_{2n}\supset\mathfrak{usp}_{2n}}
&=
\left(
\begin{array}{ccccc}
\multicolumn{2}{c|}{I_{n-1}}
&\multicolumn{1}{c}{O}
&\multicolumn{2}{|c}{E_{n-1}}\\\hline
\multicolumn{2}{c|}{O}&
1&\multicolumn{2}{|c}{O}\\
\end{array}
\right),
\end{align}
where 
$P_{\mathfrak{su}_{2n}\supset\mathfrak{usp}_{2n}}$
is an $n\times (2n-1)$ matrix.
The Dynkin diagram of 
$A_{2n-1}\supset D_n$ $(n\geq 4)$
is given by
\begin{align}
&\underset{A_{2n-1}=\mathfrak{su}_{2n}}
{\xygraph{
    \circ ([]!{+(0,-.3)}  {{}_1})      - [r]
    \circ ([]!{+(0,-.3)}  {{}_2})      - [r]
    \cdots                             - [r] 
    \circ ([]!{+(0,-.3)}  {{}_{n-1}})  - [r] 
    \circ ([]!{+(0,-.3)}  {{}_{n}})    - [r] 
    \circ ([]!{+(0,-.3)}  {{}_{n+1}})  - [r] 
    \cdots                             - [r] 
    \circ ([]!{+(0,-.3)}  {{}_{2n-2}}) - [r] 
    \circ ([]!{+(0,-.3)}  {{}_{2n-1}})
}}\nonumber\\
\underset{\rm Folding}{\Longrightarrow}
&\underset{D_n=\mathfrak{so}_{2n}}
{\xygraph{
    \circ ([]!{+(0,-.3)} {{}_{1,2n-1}})   - [r] 
    \circ ([]!{+(0,-.3)} {{}_{2,2n-2}}) - [r] 
    \cdots                              - [r] 
    \circ ([]!{+(0,-.3)} {{}_{n\pm 3}}) - [r] 
    \circ ([]!{+(0,-.3)} {{}_{n\pm 2}}) (
        - []!{+(1,.5)}  \circ ([]!{+(0,-.3)} {{}_{n\pm 1}}),
        - []!{+(1,-.5)} \circ ([]!{+(0,-.3)} {{}_{n,n\pm 1}})
)}}.
\end{align}
Its corresponding projection matrix is
\begin{align}
P_{\mathfrak{su}_{2n}\supset\mathfrak{so}_{2n}}
&=
\left(
\begin{array}{ccccc}
\multicolumn{2}{c|}{I_{n-1}}
&\multicolumn{1}{c}{O}
&\multicolumn{2}{|c}{E_{n-1}}\\\hline
\multicolumn{1}{c|}{O}&1&\multicolumn{1}{c}{2}
&1&\multicolumn{1}{|c}{O}\\
\end{array}
\right),
\end{align}
where 
$P_{\mathfrak{su}_{2n}\supset\mathfrak{so}_{2n}}$
is an $n\times (2n-1)$ matrix.
The Dynkin diagram of 
$A_{mn-1}\supset A_{m-1}\oplus A_{n-1}$ $(m,n\geq 2)$
is given by
\begin{align}
&\underset{A_{mn-1}=\mathfrak{su}_{mn}}
{\xygraph{
    \circ ([]!{+(0,-.3)}  {{}_1})      - [r]
    \circ ([]!{+(0,-.3)}  {{}_2})      - [r]
    \cdots                             - [r] 
    \circ ([]!{+(0,-.3)}  {{}_{n-1}})  - [r] 
    \circ ([]!{+(0,-.3)}  {{}_{n}})    - [r] 
    \circ ([]!{+(0,-.3)}  {{}_{n+1}})  - [r] 
    \cdots                             - [r] 
    \circ ([]!{+(0,-.3)}  {{}_{mn-2}}) - [r] 
    \circ ([]!{+(0,-.3)}  {{}_{mn-1}})
}}\nonumber\\
\underset{\rm Folding}{\Longrightarrow}
&\underset{A_{m-1}\oplus A_{n-1}=\mathfrak{su}_{m}\oplus\mathfrak{su}_{n}}
{\xygraph{
    \circ ([]!{+(0,-.3)}  {{}_{1}})      - [r]
    \circ ([]!{+(0,-.3)}  {{}_{2}})    - [r]
    \cdots                             - [r] 
    \circ ([]!{+(0,-.3)}  {{}_{m-1}})      [r] 
    \circ ([]!{+(0,-.3)}  {{}_{1}})    - [r] 
    \circ ([]!{+(0,-.3)}  {{}_{2}})    - [r] 
    \cdots                             - [r] 
    \circ ([]!{+(0,-.3)}  {{}_{n-1}})
}}.
\end{align}
Its corresponding projection matrix is
\begin{align}
&P_{\mathfrak{su}_{mn}\supset\mathfrak{su}_{m}\oplus\mathfrak{su}_{n}}
\nonumber\\
&=
\left(
\begin{array}{ccccccccccccc}
 \multicolumn{1}{c}{1\ 2\cdots n-1}
&\multicolumn{1}{c}{n}
&\multicolumn{1}{c}{n-1\cdots 2\ 1}
&\multicolumn{6}{|c}{O}\\
\hline
 \multicolumn{2}{c|}{O}
&\multicolumn{1}{c}{1\ 2\cdots n-1}
&\multicolumn{1}{c}{n}
&\multicolumn{1}{c}{\cdots}
&\multicolumn{4}{c}{O}\\
\hline
 \multicolumn{4}{c}{\vdots}
&\multicolumn{1}{c}{\ddots}
&\multicolumn{4}{c}{\vdots}\\
\hline
 \multicolumn{4}{c}{O}
&\multicolumn{1}{c}{\cdots}
&\multicolumn{1}{c}{n}
&\multicolumn{1}{c|}{n-1 \cdots 2\ 1}
&\multicolumn{2}{c}{O}\\
\hline
 \multicolumn{6}{c|}{O}
&\multicolumn{1}{c}{1\ 2\cdots n-1}
&\multicolumn{1}{c}{n}
&\multicolumn{1}{c}{n-1 \cdots 2\ 1}\\
\hline
 \multicolumn{1}{c}{I_{n-1}}
&\multicolumn{1}{|c|}{O}
&\multicolumn{1}{c}{I_{n-1}}
&\multicolumn{1}{|c|}{O}
&\multicolumn{1}{c}{\cdots}
&\multicolumn{1}{|c|}{O}
&\multicolumn{1}{c}{I_{n-1}}
&\multicolumn{1}{|c|}{O}
&\multicolumn{1}{c}{I_{n-1}}\\
\end{array}
\right),
\end{align}
where 
$P_{\mathfrak{su}_{mn}\supset\mathfrak{su}_{m}\oplus\mathfrak{su}_{n}}$
is an $(mn-1)\times (m+n-2)$ matrix.

\subsection{$B_n=\mathfrak{so}_{2n+1}$}

Let us summarize some properties of representations of the Lie algebra
$B_n=\mathfrak{so}_{2n+1}$. 
From Table~\ref{Table:Dynkin-diagrams}, 
the extended Dynkin diagram of the Lie algebra
$B_n=\mathfrak{so}_{2n+1}$ is 
\begin{align}
\xygraph{!~:{@{=}|@{}}
    \circ ([]!{+(0,-.3)} {{}_2}) (
        - []!{+(-1,.5)}  \circ ([]!{+(0,-.3)} {{}_1}),
        - []!{+(-1,-.5)} \circ ([]!{+(0,-.3)} {{}_x}))
        - [r]
    \circ ([]!{+(0,-.3)} {{}_3}) 
        - [r] \cdots - [r]
    \circ   ([]!{+(0,-.3)} {{}_{n-1}}) : [r]
    \bullet ([]!{+(0,-.3)} {{}_n})}.
\end{align}
From Table~\ref{Table:Cartan-matrices}, 
the Cartan matrix of the Lie algebra $B_n=\mathfrak{so}_{2n+1}$ is
\begin{align}
A(B_n)=
\left(
\begin{array}{ccccccc}
2&-1&0&\cdots&0&0&0\\
-1&2&-1&\cdots&0&0&0\\
0&-1&2&\cdots&0&0&0\\
\vdots&\vdots&\vdots&\ddots&\vdots&\vdots&\vdots\\
0&0&0&\cdots&2&-1&0\\
0&0&0&\cdots&-1&2&-2\\
0&0&0&\cdots&0&-1&2\\
\end{array}
\right).
\end{align}
From Tables~\ref{Table:Complex-representations-1} and 
\ref{Table:Self-conjugate-representations},
the
types of representations of $B_n=\mathfrak{so}_{2n+1}$
are in the following table.
\begin{center}
\begin{longtable}{clll}
\caption{Types of representations of $B_n$}
\label{Table:Types-representations-B}\\
\hline\hline
Algebra&Rank&Condition&C/R/PR\\\hline\hline
\endfirsthead
\multicolumn{4}{c}{Table~\ref{Table:Types-representations-B} (continued)}\\\hline\hline
Algebra&Rank&Condition&C/R/PR\\\hline\hline
\endhead
\hline
\endfoot
$\mathfrak{so}_{2n+1}$ 
&$n=0,3\ (\bmod.\ 4)$
&None
&Real
\\\cline{2-4}
&$n=1,2\ (\bmod.\ 4)$
&$a_n=0\ (\bmod.\ 2)$
&Real\\
&
&$a_n=1\ (\bmod.\ 2)$ 
&PR\\
\end{longtable}
\end{center}
where $a_n$ is the short root,
Condition stands for a condition for Dynkin labels, 
and C/R/PR represent complex, real, and pseudo-real 
representations.
From Eq.~(\ref{Eq:Weyl-formula-B}), the Weyl dimension formula of the 
Lie algebra $B_n=\mathfrak{so}_{2n+1}$ is  
\begin{align}
d(R)=&
\left[
\prod_{k=1}^n\prod_{i=1}^k
\left(\frac{\sum_{j=i}^{k}a_j}{k-i+1}+1\right)
\right]
\prod_{k=0}^{n-2}
\prod_{i=k}^{n-2}
\left(
\frac{\left(\sum_{j=k+1}^{i}a_{n-1-j}\right)
+\left(\sum_{j=0}^{k}2a_{n-1-j}\right)+a_n}{k+i+3}+1\right).
\end{align}
For the Lie algebra $B_n=\mathfrak{so}_{2n+1}$, from
Eqs.~(\ref{Eq:2-Casimir}) and (\ref{Eq:Dynkin-index}), 
we get the Casimir invariant and the Dynkin index of a representation
$(a_1a_2a_3\cdots a_n)$:
\begin{align}
C_2(R)=&\frac{1}{2}
\sum_{i,j=1}^{n}(a_i+2g_i)G(B_n)_{ij}a_j,\\
T(R)=&\frac{d(R)}{d(G)}C_2(R)=
\frac{d(R)}{2 d(G)}
\sum_{i,j=1}^{n}(a_i+2g_i)G(B_n)_{ij}a_j,
\end{align}
where $G(B_n)$ is the metric matrix in 
Table~\ref{Table:Cartan-matrices-inverse}
\begin{align}
G(B_n)&=\frac{1}{2}
\left(
\begin{array}{cccccc}
2&2&2&\cdots&2&1\\
2&4&4&\cdots&4&2\\
2&4&6&\cdots&6&3\\
\vdots&\vdots&\vdots&\ddots&\vdots&\vdots  \\
2&4&6&\cdots&2(n-1)&n-1\\
1&2&3&\cdots&n-1&n/2\\
\end{array}
\right).
\end{align}
From Table~\ref{Table:Conjugacy-classes}, the conjugacy classes of the Lie
algebra $B_n=\mathfrak{so}_{2n+1}$ is
\begin{align}
C_c(R):=a_n\ \ \  (\bmod.\ 2).
\end{align}
The Lie algebra $B_n=\mathfrak{so}_{2n+1}$ has two conjugacy classes.
 
Some features of the Lie algebra 
$B_n=\mathfrak{so}_{2n+1}$ are summarized 
in the following table.
\begin{center}
\begin{longtable}{cccccc}
\caption{Representations of $B_n=\mathfrak{so}_{2n+1}$}
\label{Table:Representations-summary-B}\\
\hline\hline
$\mathfrak{so}_{2n+1}$ irrep.&$d(R)$&$C_2(R)$&$T(R)$&$C_c(R)$&C/R/PR\\\hline\hline
\endfirsthead
\multicolumn{6}{c}{Table~\ref{Table:Representations-summary-B} (continued)}\\\hline\hline
$\mathfrak{so}_{2n+1}$ irrep.&$d(R)$&$C_2(R)$&$T(R)$&$C_c(R)$&C/R/PR\\\hline\hline
\endhead
\hline
\endfoot
$(0000\cdots 00)$&$1$                   &$0$               &$0$&$0$&R\\ 
$(1000\cdots 00)$&$2n+1$                &$n$               &$1$&$0$&R\\ 
$(0000\cdots 01)$&$2^n$                 &$\frac{n(n+1)}{8}$&$2^{n-3}$&$1$&R$_{n=2k}$/PR$_{n=2k+1}$\\ 
$(0100\cdots 00)$&$n(2n+1)$             &$2n-1$            &$2n-1$&$0$&R\\ 
$(2000\cdots 00)$&$n(2n+3)$             &$2n+1$           &$2n+3$&$0$&R\\ 
$(0010\cdots 00)$&$\frac{n(2n-1)(2n+1)}{3}$&$3(n-1)$       &$(n-1)(2n-1)$&$0$&R\\ 
$(3000\cdots 00)$&$\frac{n(2n+1)(2n+5)}{3}$&$3(n+1)$       &$(n+1)(2n+5)$&$0$&R\\ 
$(1000\cdots 01)$&$n2^{n+1}$&$\frac{(n+4)(2n+1)}{8}$&$(n+4)2^{n-2}$&$1$&R$_{n=2k}$/PR$_{n=2k+1}$\\
\end{longtable}
\end{center}
where $\mathfrak{so}_{2n+1}$ irrep., $d(R)$, $C_2(R)$, $T(R)$, $C_c(R)$, and C/P/PR stand for 
the Dynkin label of the irreducible representation of
$B_n=\mathfrak{so}_{2n+1}$, 
its dimension, its quadratic Casimir invariant, its Dynkin indices, 
its conjugacy class, and 
complex, real, and pseudo-real representations,
respectively. 
Representations such as $(0000\cdots 01)$ and $(1000\cdots 01)$ are real
representations for $n=2k$ and pseudo-real representations for $n=2k+1$.
The representation $(010\cdots 0)$ is the adjoint representation of
$\mathfrak{so}_{2n+1}$. 
See Appendix~\ref{Sec:Representations} for further information.

From Table~\ref{Table:Maximal-subalgebra},
maximal $R$- and $S$-subalgebras of $B_n=\mathfrak{so}_{2n+1}$
with their rank up to 16 are listed in the following table.
\begin{center}
\begin{longtable}{crcp{10cm}c}
\caption{Maximal subalgebras of $B_n$}
\label{Table:Maximal-subalgebra-B}\\
\hline\hline
Rank&\multicolumn{2}{l}{Algebra $\mathfrak{g}$}&Maximal subalgebras $\mathfrak{h}$&Type\\\hline\hline
\endfirsthead
\multicolumn{4}{c}{Table~\ref{Table:Maximal-subalgebra-B} (continued)}\\
\hline
Rank&\multicolumn{2}{l}{Algebra $\mathfrak{g}$}&Maximal subalgebras $\mathfrak{h}$&Type\\\hline\hline
\endhead
\hline
\endfoot
$3$ &$\mathfrak{so}_7$&$\supset$
&$\mathfrak{su}_4;\
\mathfrak{su}_2\oplus\mathfrak{su}_2\oplus\mathfrak{su}_2;\
\mathfrak{usp}_4\oplus\mathfrak{u}_1$&$(R)$\\
    &       &$\supset$&$G_2$&$(S)$\\
$4$ &$\mathfrak{so}_9$&$\supset$
&$\mathfrak{so}_8;\ 
\mathfrak{su}_2\oplus\mathfrak{su}_2\oplus\mathfrak{usp}_4;\
\mathfrak{su}_4\oplus\mathfrak{su}_2;\ \mathfrak{so}_7\oplus\mathfrak{u}_1$
&$(R)$\\
    &       &$\supset$&$\mathfrak{su}_2;\ 
\mathfrak{su}_2\oplus\mathfrak{su}_2$&$(S)$\\
$5$ &$\mathfrak{so}_{11}$&$\supset$
&$\mathfrak{so}_{10};\ \mathfrak{so}_8\oplus\mathfrak{su}_2;\
\mathfrak{su}_4\oplus\mathfrak{usp}_4;\
\mathfrak{su}_2\oplus\mathfrak{su}_2\oplus\mathfrak{so}_7;\
\mathfrak{so}_9\oplus\mathfrak{u}_1$&$(R)$\\
    &       &$\supset$&$\mathfrak{su}_2$&$(S)$\\
$6$ &$\mathfrak{so}_{13}$&$\supset$
&$\mathfrak{so}_{12};\ \mathfrak{so}_{10}\oplus\mathfrak{su}_2;\
\mathfrak{so}_8\oplus\mathfrak{usp}_4;\
\mathfrak{su}_4\oplus\mathfrak{so}_7;\
\mathfrak{su}_2\oplus\mathfrak{su}_2\oplus\mathfrak{so}_9;\
\mathfrak{so}_{11}\oplus\mathfrak{u}_1$
&$(R)$\\
    &       &$\supset$&$\mathfrak{su}_2$&$(S)$\\
$7$ &$\mathfrak{so}_{15}$&$\supset$
&$\mathfrak{so}_{14};\ \mathfrak{so}_{12}\oplus\mathfrak{su}_2;\
\mathfrak{so}_{10}\oplus\mathfrak{usp}_4;\
\mathfrak{so}_{8}\oplus\mathfrak{so}_7;\
\mathfrak{su}_4\oplus\mathfrak{so}_9;\
\mathfrak{su}_2\oplus\mathfrak{su}_2\oplus\mathfrak{so}_{11};\
\mathfrak{so}_{13}\oplus\mathfrak{u}_1$&$(R)$\\
    &       &$\supset$&$\mathfrak{su}_2;\ \mathfrak{su}_4;\
\mathfrak{su}_2\oplus\mathfrak{usp}_4$&$(S)$\\
$8$ &$\mathfrak{so}_{17}$&$\supset$
&$\mathfrak{so}_{16};\ \mathfrak{so}_{14}\oplus\mathfrak{su}_2;\
\mathfrak{so}_{12}\oplus\mathfrak{usp}_4;\
\mathfrak{so}_{10}\oplus\mathfrak{so}_7;\
\mathfrak{so}_8\oplus\mathfrak{so}_9;\
\mathfrak{su}_4\oplus\mathfrak{so}_{11};\
\mathfrak{su}_2\oplus\mathfrak{su}_2\oplus\mathfrak{so}_{13};\
\mathfrak{so}_{15}\oplus\mathfrak{u}_1$&$(R)$\\
    &       &$\supset$&$\mathfrak{su}_2$&$(S)$\\
$9$ &$\mathfrak{so}_{19}$&$\supset$
&$\mathfrak{so}_{18};\ \mathfrak{so}_{16}\oplus\mathfrak{su}_2;\
\mathfrak{so}_{14}\oplus\mathfrak{usp}_4;\
\mathfrak{so}_{12}\oplus\mathfrak{so}_7;\
\mathfrak{so}_{10}\oplus\mathfrak{so}_9;\
\mathfrak{so}_8\oplus\mathfrak{so}_{11};\
\mathfrak{su}_4\oplus\mathfrak{so}_{13};\
\mathfrak{su}_2\oplus\mathfrak{su}_2\oplus\mathfrak{so}_{15};\
\mathfrak{so}_{17}\oplus\mathfrak{u}_1$
&$(R)$\\
    &       &$\supset$&$\mathfrak{su}_2$&$(S)$\\
$10$&$\mathfrak{so}_{21}$&$\supset$
&$\mathfrak{so}_{20};\ \mathfrak{so}_{18}\oplus\mathfrak{su}_2;\
\mathfrak{so}_{16}\oplus\mathfrak{usp}_{4};\
\mathfrak{so}_{14}\oplus\mathfrak{so}_7;\
\mathfrak{so}_{12}\oplus\mathfrak{so}_9;\
\mathfrak{so}_{10}\oplus\mathfrak{so}_{11};\
\mathfrak{so}_8\oplus\mathfrak{so}_{13};\
\mathfrak{su}_4\oplus\mathfrak{so}_{15};\
\mathfrak{su}_2\oplus\mathfrak{su}_2\oplus\mathfrak{so}_{17};\
\mathfrak{so}_{19}\oplus\mathfrak{u}_1$
&$(R)$\\
    &       &$\supset$&$\mathfrak{su}_2;\
\mathfrak{su}_2\oplus\mathfrak{so}_7;\
\mathfrak{so}_7;\mathfrak{usp}_6$&$(S)$\\
$11$&$\mathfrak{so}_{23}$&$\supset$
&$\mathfrak{so}_{22};\ 
\mathfrak{so}_{20}\oplus\mathfrak{su}_2;\
\mathfrak{so}_{18}\oplus\mathfrak{usp}_{4};\
\mathfrak{so}_{16}\oplus\mathfrak{so}_7;\
\mathfrak{so}_{14}\oplus\mathfrak{so}_9;\
\mathfrak{so}_{12}\oplus\mathfrak{so}_{11};\
\mathfrak{so}_{10}\oplus\mathfrak{so}_{13};\
\mathfrak{so}_{8}\oplus\mathfrak{so}_{15};\
\mathfrak{su}_4\oplus\mathfrak{so}_{17};\
\mathfrak{su}_2\oplus\mathfrak{su}_2\oplus\mathfrak{so}_{19};\
\mathfrak{so}_{21}\oplus\mathfrak{u}_1$
&$(R)$\\
    &       &$\supset$&$\mathfrak{su}_2$&$(S)$\\
$12$&$\mathfrak{so}_{25}$&$\supset$
&$\mathfrak{so}_{24};\
\mathfrak{so}_{22}\oplus\mathfrak{su}_2;\
\mathfrak{so}_{20}\oplus\mathfrak{usp}_{4};\
\mathfrak{so}_{18}\oplus\mathfrak{so}_7;\
\mathfrak{so}_{16}\oplus\mathfrak{so}_9;\
\mathfrak{so}_{14}\oplus\mathfrak{so}_{11};\
\mathfrak{so}_{12}\oplus\mathfrak{so}_{13};\
\mathfrak{so}_{10}\oplus\mathfrak{so}_{15};\
\mathfrak{so}_{8}\oplus\mathfrak{so}_{17};\
\mathfrak{su}_4\oplus\mathfrak{so}_{19};\
\mathfrak{su}_2\oplus\mathfrak{su}_2\oplus\mathfrak{so}_{21};\
\mathfrak{so}_{23}\oplus\mathfrak{u}_1$
&$(R)$\\
    &       &$\supset$
&$\mathfrak{su}_2;\
\mathfrak{usp}_4\oplus\mathfrak{usp}_4
$&$(S)$\\
$13$&$\mathfrak{so}_{27}$&$\supset$
&$\mathfrak{so}_{26};\
\mathfrak{so}_{24}\oplus\mathfrak{su}_2;\
\mathfrak{so}_{22}\oplus\mathfrak{usp}_{4};\
\mathfrak{so}_{20}\oplus\mathfrak{so}_7;\
\mathfrak{so}_{18}\oplus\mathfrak{so}_9;\
\mathfrak{so}_{16}\oplus\mathfrak{so}_{11};\
\mathfrak{so}_{14}\oplus\mathfrak{so}_{13};\
\mathfrak{so}_{12}\oplus\mathfrak{so}_{15};\
\mathfrak{so}_{10}\oplus\mathfrak{so}_{17};\
\mathfrak{so}_{8}\oplus\mathfrak{so}_{19};\
\mathfrak{su}_4\oplus\mathfrak{so}_{21};\
\mathfrak{su}_2\oplus\mathfrak{su}_2\oplus\mathfrak{so}_{23};\
\mathfrak{so}_{25}\oplus\mathfrak{u}_1$
&$(R)$\\
    &       &$\supset$
&$\mathfrak{su}_2;\
\mathfrak{su}_3;\
\mathfrak{so}_7;\
\mathfrak{su}_2\oplus\mathfrak{so}_9;\
\mathfrak{usp}_8
$&$(S)$\\
$14$&$\mathfrak{so}_{29}$&$\supset$
&$\mathfrak{so}_{28};\
\mathfrak{so}_{26}\oplus\mathfrak{su}_2;\
\mathfrak{so}_{24}\oplus\mathfrak{usp}_{4};\
\mathfrak{so}_{22}\oplus\mathfrak{so}_7;\
\mathfrak{so}_{20}\oplus\mathfrak{so}_9;\
\mathfrak{so}_{18}\oplus\mathfrak{so}_{11};\
\mathfrak{so}_{16}\oplus\mathfrak{so}_{13};\
\mathfrak{so}_{14}\oplus\mathfrak{so}_{15};\
\mathfrak{so}_{12}\oplus\mathfrak{so}_{17};\
\mathfrak{so}_{10}\oplus\mathfrak{so}_{19};\
\mathfrak{so}_{8}\oplus\mathfrak{so}_{21};\
\mathfrak{su}_4\oplus\mathfrak{so}_{23};\
\mathfrak{su}_2\oplus\mathfrak{su}_2\oplus\mathfrak{so}_{25};\
\mathfrak{so}_{27}\oplus\mathfrak{u}_1$
&$(R)$\\
    &       &$\supset$
&$\mathfrak{su}_2$&$(S)$\\
$15$&$\mathfrak{so}_{31}$&$\supset$
&$\mathfrak{so}_{30};\
\mathfrak{so}_{28}\oplus\mathfrak{su}_2;\
\mathfrak{so}_{26}\oplus\mathfrak{usp}_{4};\
\mathfrak{so}_{24}\oplus\mathfrak{so}_7;\
\mathfrak{so}_{22}\oplus\mathfrak{so}_9;\
\mathfrak{so}_{20}\oplus\mathfrak{so}_{11};\
\mathfrak{so}_{18}\oplus\mathfrak{so}_{13};\
\mathfrak{so}_{16}\oplus\mathfrak{so}_{15};\
\mathfrak{so}_{14}\oplus\mathfrak{so}_{17};\
\mathfrak{so}_{12}\oplus\mathfrak{so}_{19};\
\mathfrak{so}_{10}\oplus\mathfrak{so}_{21};\
\mathfrak{so}_{8}\oplus\mathfrak{so}_{23};\
\mathfrak{su}_4\oplus\mathfrak{so}_{25};\
\mathfrak{su}_2\oplus\mathfrak{su}_2\oplus\mathfrak{so}_{27};\
\mathfrak{so}_{29}\oplus\mathfrak{u}_1$
&$(R)$\\
    &       &$\supset$
&$\mathfrak{su}_2$&$(S)$\\
$16$&$\mathfrak{so}_{33}$&$\supset$
&$\mathfrak{so}_{32};\
\mathfrak{so}_{30}\oplus\mathfrak{su}_2;\
\mathfrak{so}_{28}\oplus\mathfrak{usp}_{4};\
\mathfrak{so}_{26}\oplus\mathfrak{so}_7;\
\mathfrak{so}_{24}\oplus\mathfrak{so}_9;\
\mathfrak{so}_{22}\oplus\mathfrak{so}_{11};\
\mathfrak{so}_{20}\oplus\mathfrak{so}_{13};\
\mathfrak{so}_{18}\oplus\mathfrak{so}_{15};\
\mathfrak{so}_{16}\oplus\mathfrak{so}_{17};\
\mathfrak{so}_{14}\oplus\mathfrak{so}_{19};\
\mathfrak{so}_{12}\oplus\mathfrak{so}_{21};\
\mathfrak{so}_{10}\oplus\mathfrak{so}_{23};\
\mathfrak{so}_{8}\oplus\mathfrak{so}_{25};\
\mathfrak{su}_4\oplus\mathfrak{so}_{27};\
\mathfrak{su}_2\oplus\mathfrak{su}_2\oplus\mathfrak{so}_{29};\
\mathfrak{so}_{31}\oplus\mathfrak{u}_1$
&$(R)$\\
    &       &$\supset$
&$\mathfrak{su}_2;\
\mathfrak{su}_2\oplus\mathfrak{so}_{11}
$&$(S)$\\
\end{longtable}
\end{center}
where Type $(R)$ and $(S)$ stand for $R$- and $S$-subalgebras,
respectively. 

We summarize generic projection matrices of the Lie algebra
$B_n=\mathfrak{so}_{2n+1}$ and any maximal regular subalgebras, which
are calculated by using a method given in
Sec.~\ref{Sec:Representations-subalgebras}. 
The Dynkin diagram of 
$B_n\supset B_{n-1}\oplus\mathfrak{u}_1$ $(n\geq 3)$
is given by
\begin{align}
&\underset{B_n=\mathfrak{so}_{2n+1}}
{\xygraph{!~:{@{=}|@{}}
    \circ ([]!{+(0,-.3)} {{}_2}) (
        - []!{+(-1,.5)}  \circ ([]!{+(0,-.3)} {{}_1}),
        - []!{+(-1,-.5)} \circ ([]!{+(0,-.3)} {{}_x}))
        - [r]
    \circ   ([]!{+(0,-.3)} {{}_3})     - [r]
    \cdots                             - [r] 
    \circ   ([]!{+(0,-.3)} {{}_{n-2}})   - [r]
    \circ   ([]!{+(0,-.3)} {{}_{n-1}}) : [r]
    \bullet ([]!{+(0,-.3)} {{}_{n}})}
}\nonumber\\
\underset{\rm Removing}{\Longrightarrow}&
\underset{B_{n-1}=\mathfrak{so}_{2n-1}\oplus\mathfrak{u}_1}
{\xygraph{!~:{@{=}|@{}}
    \circ ([]!{+(0,-.3)} {{}_2}) (
        - []!{+(-1,.5)}  {\circ\hspace{-0.645em}\times} ([]!{+(0,-.3)} {{}_1}),
        - []!{+(-1,-.5)} {\circ\hspace{-0.645em}\times} ([]!{+(0,-.3)} {{}_x}))
        - [r]
    \circ   ([]!{+(0,-.3)} {{}_3})     - [r]
    \cdots                             - [r] 
    \circ   ([]!{+(0,-.3)} {{}_{n-2}})   - [r]
    \circ   ([]!{+(0,-.3)} {{}_{n-1}}) : [r]
    \bullet ([]!{+(0,-.3)} {{}_{n}})}
}.
\end{align}
Its corresponding projection matrix is
\begin{align}
P_{\mathfrak{so}_{2n+1}\supset\mathfrak{so}_{2n-1}\oplus\mathfrak{u}_1}
&=
\left(
\begin{array}{c|c}
O&I_{n-1}\\\hline
\multicolumn{2}{c}{2\ \cdots\ 2\ 1}\\
\end{array}
\right),
\end{align}
where 
$P_{\mathfrak{so}_{2n+1}\supset\mathfrak{so}_{2n-1}\oplus\mathfrak{u}_1}$
is an $n\times n$ matrix;
$I_{n-1}$ is an $(n-1)\times (n-1)$ matrix, where its matrix
elements $(I_{n-1})_{ii}=1 (i=1,2,\cdots,n-1)$ and the other matrix
elements $(I_{n-1})_{ij}=0 (i,j=1,2,\cdots,n-1;i\not=j)$.
The Dynkin diagram of 
$B_n\supset D_{n}$ $(n\geq 4)$
is given by
\begin{align}
&\underset{B_n=\mathfrak{so}_{2n+1}}
{\xygraph{!~:{@{=}|@{}}
    \circ ([]!{+(0,-.3)} {{}_2}) (
        - []!{+(-1,.5)}  \circ ([]!{+(0,-.3)} {{}_1}),
        - []!{+(-1,-.5)} \circ ([]!{+(0,-.3)} {{}_x}))
        - [r]
    \circ   ([]!{+(0,-.3)} {{}_3})     - [r]
    \cdots                             - [r] 
    \circ   ([]!{+(0,-.3)} {{}_{n-2}})   - [r]
    \circ   ([]!{+(0,-.3)} {{}_{n-1}}) : [r]
    \bullet ([]!{+(0,-.3)} {{}_{n}})}
}\nonumber\\
\underset{\rm Removing}{\Longrightarrow}&
\underset{D_{n}=\mathfrak{so}_{2n}}
{\xygraph{!~:{@{=}|@{}}
    \circ ([]!{+(0,-.3)} {{}_2}) (
        - []!{+(-1,.5)}  \circ ([]!{+(0,-.3)} {{}_1}),
        - []!{+(-1,-.5)} \circ ([]!{+(0,-.3)} {{}_x}))
        - [r]
    \circ   ([]!{+(0,-.3)} {{}_3})     - [r]
    \cdots                             - [r] 
    \circ   ([]!{+(0,-.3)} {{}_{n-2}})   - [r]
    \circ   ([]!{+(0,-.3)} {{}_{n-1}}) : [r]
    {\bullet\hspace{-0.645em}\times} ([]!{+(0,-.3)} {{}_{n}})}
}.
\end{align}
Its corresponding projection matrix is
\begin{align}
P_{\mathfrak{so}_{2n+1}\supset\mathfrak{so}_{2n}}
&=
\left(
\begin{array}{c|c}
E_{n-1}&O\\\hline
\multicolumn{1}{c}{-1\ -2\cdots\ -2}&-1\\
\end{array}
\right),
\end{align}
where 
$P_{\mathfrak{so}_{2n+1}\supset\mathfrak{so}_{2n}}$
is an $n\times n$ matrix;
$E_{n-1}$ is an $(n-1)\times (n-1)$ matrix, where its matrix
elements $(E_{n-1})_{i,n-i}=1 (i=1,2,\cdots,n-1)$ and the other matrix
elements $(E_{n-1})_{i,n-j}=0 (i,j=1,2,\cdots,n-1;i\not=j)$.
The Dynkin diagram of 
$B_n\supset D_{n-1}\oplus A_1$ $(n\geq 5)$
is given by
\begin{align}
&\underset{B_n=\mathfrak{so}_{2n+1}}
{\xygraph{!~:{@{=}|@{}}
    \circ ([]!{+(0,-.3)} {{}_2}) (
        - []!{+(-1,.5)}  \circ ([]!{+(0,-.3)} {{}_1}),
        - []!{+(-1,-.5)} \circ ([]!{+(0,-.3)} {{}_x}))
        - [r]
    \circ   ([]!{+(0,-.3)} {{}_3})     - [r]
    \cdots                             - [r] 
    \circ   ([]!{+(0,-.3)} {{}_{n-2}})   - [r]
    \circ   ([]!{+(0,-.3)} {{}_{n-1}}) : [r]
    \bullet ([]!{+(0,-.3)} {{}_{n}})}
}\nonumber\\
\underset{\rm Removing}{\Longrightarrow}&
\underset{D_{n-1}\oplus A_1=\mathfrak{so}_{2n-2}\oplus\mathfrak{su}_2}
{\xygraph{!~:{@{=}|@{}}
    \circ ([]!{+(0,-.3)} {{}_2}) (
        - []!{+(-1,.5)}  \circ ([]!{+(0,-.3)} {{}_1}),
        - []!{+(-1,-.5)} \circ ([]!{+(0,-.3)} {{}_x}))
        - [r]
    \circ   ([]!{+(0,-.3)} {{}_3})     - [r]
    \cdots                             - [r] 
    \circ   ([]!{+(0,-.3)} {{}_{n-2}})   - [r]
    {\circ\hspace{-0.645em}\times}    ([]!{+(0,-.3)} {{}_{n-1}}) : [r]
    \bullet ([]!{+(0,-.3)} {{}_{n}})}
}.
\end{align}
Its corresponding projection matrix is
\begin{align}
P_{\mathfrak{so}_{2n+1}\supset\mathfrak{so}_{2n-2}\oplus\mathfrak{su}_2}
&=
\left(
\begin{array}{ccc}
\multicolumn{1}{c|}{E_{n-2}}&\multicolumn{2}{c}{O}\\\hline
-1\ -2\ \cdots\ -2&-2&-1\\\hline
\multicolumn{2}{c|}{O}& 1\\
\end{array}
\right),
\end{align}
where 
$P_{\mathfrak{so}_{2n+1}\supset\mathfrak{so}_{2n-2}\oplus\mathfrak{su}_2}$
is an $n\times n$ matrix.
The Dynkin diagram of 
$B_n\supset A_1\oplus A_1\oplus B_{n-2}$ $(n\geq 4)$
is given by
\begin{align}
&\underset{B_n=\mathfrak{so}_{2n+1}}
{\xygraph{!~:{@{=}|@{}}
    \circ ([]!{+(0,-.3)} {{}_2}) (
        - []!{+(-1,.5)}  \circ ([]!{+(0,-.3)} {{}_1}),
        - []!{+(-1,-.5)} \circ ([]!{+(0,-.3)} {{}_x}))
        - [r]
    \circ   ([]!{+(0,-.3)} {{}_3})     - [r]
    \cdots                             - [r] 
    \circ   ([]!{+(0,-.3)} {{}_{n-2}})   - [r]
    \circ   ([]!{+(0,-.3)} {{}_{n-1}}) : [r]
    \bullet ([]!{+(0,-.3)} {{}_{n}})}
}\nonumber\\
\underset{\rm Removing}{\Longrightarrow}&
\underset{A_1\oplus A_1\oplus B_{n-2}=
\mathfrak{su}_2\oplus\mathfrak{su}_2\oplus\mathfrak{so}_{2n-3}}
{\xygraph{!~:{@{=}|@{}}
    {\circ\hspace{-0.645em}\times}  ([]!{+(0,-.3)} {{}_2}) (
        - []!{+(-1,.5)}  \circ ([]!{+(0,-.3)} {{}_1}),
        - []!{+(-1,-.5)} \circ ([]!{+(0,-.3)} {{}_x}))
        - [r]
    \circ   ([]!{+(0,-.3)} {{}_3})     - [r]
    \cdots                             - [r] 
    \circ   ([]!{+(0,-.3)} {{}_{n-2}})   - [r]
    \circ   ([]!{+(0,-.3)} {{}_{n-1}}) : [r]
    \bullet ([]!{+(0,-.3)} {{}_{n}})}
}.
\end{align}
Its corresponding projection matrix is
\begin{align}
P_{\mathfrak{so}_{2n+1}\supset
\mathfrak{su}_2\oplus\mathfrak{su}_2\oplus\mathfrak{so}_{2n-4}}
&=
\left(
\begin{array}{ccc}
1&\multicolumn{2}{|c}{O}\\\hline
-1&-2&-2\cdots\ -2\ -1\\\hline
\multicolumn{2}{c|}{O}&I_{n-2}\\
\end{array}
\right),
\end{align}
where 
$P_{\mathfrak{so}_{2n+1}\supset
\mathfrak{su}_2\oplus\mathfrak{su}_2\oplus\mathfrak{so}_{2n-4}}$
is an $n\times n$ matrix.
The Dynkin diagram of 
$B_n\supset A_3\oplus B_{n-3}$ $(n\geq 6)$
is given by
\begin{align}
&\underset{B_n=\mathfrak{so}_{2n+1}}
{\xygraph{!~:{@{=}|@{}}
    \circ ([]!{+(0,-.3)} {{}_2}) (
        - []!{+(-1,.5)}  \circ ([]!{+(0,-.3)} {{}_1}),
        - []!{+(-1,-.5)} \circ ([]!{+(0,-.3)} {{}_x}))
        - [r]
    \circ   ([]!{+(0,-.3)} {{}_3})     - [r]
    \cdots                             - [r] 
    \circ   ([]!{+(0,-.3)} {{}_{n-2}})   - [r]
    \circ   ([]!{+(0,-.3)} {{}_{n-1}}) : [r]
    \bullet ([]!{+(0,-.3)} {{}_{n}})}
}\nonumber\\
\underset{\rm Removing}{\Longrightarrow}&
\underset{A_3\oplus B_{n-3}=
\mathfrak{su}_{4}\oplus\mathfrak{so}_{2n-5}}
{\xygraph{!~:{@{=}|@{}}
    \circ ([]!{+(0,-.3)} {{}_2}) (
        - []!{+(-1,.5)}  \circ ([]!{+(0,-.3)} {{}_1}),
        - []!{+(-1,-.5)} \circ ([]!{+(0,-.3)} {{}_x}))
        - [r]
    {\circ\hspace{-0.645em}\times} ([]!{+(0,-.3)} {{}_3})     - [r]
    \cdots                             - [r] 
    \circ   ([]!{+(0,-.3)} {{}_{n-2}})   - [r]
    \circ   ([]!{+(0,-.3)} {{}_{n-1}}) : [r]
    \bullet ([]!{+(0,-.3)} {{}_{n}})}
}.
\end{align}
Its corresponding projection matrix is
\begin{align}
P_{\mathfrak{so}_{2n+1}\supset
\mathfrak{su}_4\oplus\mathfrak{so}_{2n-5}}
&=
\left(
\begin{array}{cccc}
\multicolumn{2}{c}{I_2}&\multicolumn{2}{|c}{O}\\\hline
-1&-2&-2&\cdots\ -2\ -1\\\hline
\multicolumn{3}{c|}{O}&I_{n-3}\\
\end{array}
\right),
\end{align}
where 
$P_{\mathfrak{so}_{2n+1}\supset
\mathfrak{su}_4\oplus\mathfrak{so}_{2n-5}}$
is an $n\times n$ matrix.
The Dynkin diagram of 
$B_n\supset D_k\oplus B_{n-k}$ $(4\leq k\leq n; n-k\geq 4)$
is given by
\begin{align}
&\underset{B_n=\mathfrak{so}_{2n+1}}
{\xygraph{!~:{@{=}|@{}}
    \circ ([]!{+(0,-.3)} {{}_2}) (
        - []!{+(-1,.5)}  \circ ([]!{+(0,-.3)} {{}_1}),
        - []!{+(-1,-.5)} \circ ([]!{+(0,-.3)} {{}_x}))
        - [r]
    \cdots                             - [r] 
    \circ   ([]!{+(0,-.3)} {{}_{k-1}}) - [r]
    \circ   ([]!{+(0,-.3)} {{}_{k}})   - [r]
    \circ   ([]!{+(0,-.3)} {{}_{k+1}}) - [r]
    \cdots                             - [r] 
    \circ   ([]!{+(0,-.3)} {{}_{n-2}})   - [r]
    \circ   ([]!{+(0,-.3)} {{}_{n-1}}) : [r]
    \bullet ([]!{+(0,-.3)} {{}_{n}})}
}\nonumber\\
\underset{\rm Removing}{\Longrightarrow}&
\underset{D_k\oplus B_{n-k}
=\mathfrak{so}_{2k}\oplus\mathfrak{so}_{2n-2k+1}}
{\xygraph{!~:{@{=}|@{}}
    \circ ([]!{+(0,-.3)} {{}_2}) (
        - []!{+(-1,.5)}  \circ ([]!{+(0,-.3)} {{}_1}),
        - []!{+(-1,-.5)} \circ ([]!{+(0,-.3)} {{}_x}))
        - [r]
    \cdots                             - [r] 
    \circ   ([]!{+(0,-.3)} {{}_{k-1}}) - [r]
    {\circ\hspace{-0.645em}\times} ([]!{+(0,-.3)} {{}_{k}})   - [r]
    \circ   ([]!{+(0,-.3)} {{}_{k+1}}) - [r]
    \cdots                             - [r] 
    \circ   ([]!{+(0,-.3)} {{}_{n-2}})   - [r]
    \circ   ([]!{+(0,-.3)} {{}_{n-1}}) : [r]
    \bullet ([]!{+(0,-.3)} {{}_{n}})}
}.
\end{align}
Its corresponding projection matrix is
\begin{align}
P_{\mathfrak{so}_{2n+1}\supset
\mathfrak{so}_{2k}\oplus\mathfrak{so}_{2n-2k+1}}
&=
\left(
\begin{array}{ccc}
\multicolumn{1}{c}{E_{k-1}}&\multicolumn{2}{|c}{O}\\\hline
-1\ -2\cdots\ -2&-2&-2\cdots\ -2\ -1\\\hline
\multicolumn{2}{c|}{O}&I_{n-k}\\
\end{array}
\right),
\end{align}
where 
$P_{\mathfrak{so}_{2n+1}\supset
\mathfrak{so}_{2k}\oplus\mathfrak{so}_{2n-2k+1}}$
is an $n\times n$ matrix.

We summarize generic projection matrices of the Lie algebra
$B_n=\mathfrak{so}_{2n+1}$ and several maximal special subalgebras
given in Tables~\ref{Table:Maximal-S-sub-classical-1}
and \ref{Table:Maximal-S-sub-classical-2}, 
which are also calculated by using a method given in
Sec.~\ref{Sec:Representations-subalgebras}. 
(The following projection matrices cannot cover all maximal special
subalgebras, so we need to calculate exceptions one by one.)
The Dynkin diagram of 
$B_n\supset A_1$ $(n\geq 4)$
is given by
\begin{align}
&\underset{B_n=\mathfrak{so}_{2n+1}}
{\xygraph{!~:{@{=}|@{}}
    \circ ([]!{+(0,-.3)} {{}_2}) (
        - []!{+(-1,.5)}  \circ ([]!{+(0,-.3)} {{}_1}),
        - []!{+(-1,-.5)} \circ ([]!{+(0,-.3)} {{}_x}))
        - [r]
    \circ   ([]!{+(0,-.3)} {{}_3})     - [r]
    \cdots                             - [r] 
    \circ   ([]!{+(0,-.3)} {{}_{n-2}})   - [r]
    \circ   ([]!{+(0,-.3)} {{}_{n-1}}) : [r]
    \bullet ([]!{+(0,-.3)} {{}_{n}})}
}
\ \ \underset{\rm Shrinking}{\Longrightarrow}\ \
\underset{A_{1}=\mathfrak{su}_2}
{\underset{{1,\cdots,n}}{\circ}}.
\end{align}
Its corresponding projection matrix is
\begin{align}
P_{\mathfrak{so}_{2n+1}\supset\mathfrak{su}_2}
&=
\left(
\begin{array}{ccccc|c}
2n&\cdots&k(2n-k+1)&\cdots&(n-1)(n+2)&(n+2)(n-1)/2+1\\
\end{array}
\right),
\end{align}
where 
$P_{\mathfrak{so}_{2n+1}\supset\mathfrak{su}_2}$
is a $1\times n$ matrix.
The Dynkin diagram of 
$B_{\left[\frac{mn}{2}\right]}\supset B_{\left[\frac{m}{2}\right]}\oplus B_{\left[\frac{n}{2}\right]}$ 
$(m,n\in \mathbb{Z}_{2k+1})$
is given by
\begin{align}
&\underset{B_{[mn/2]}=\mathfrak{so}_{mn}}
{\xygraph{!~:{@{=}|@{}}
    \circ   ([]!{+(0,-.3)} {{}_1})     - [r]
    \circ   ([]!{+(0,-.3)} {{}_2})     - [r]
    \circ   ([]!{+(0,-.3)} {{}_3})     - [r]
    \cdots                             - [r] 
    \circ   ([]!{+(0,-.3)} {{}_{\left[\frac{mn-2}{2}\right]}}) : [r]
    \bullet ([]!{+(0,-.3)} {{}_{\left[\frac{mn}{2}\right]}})}
}\nonumber\\
\underset{\rm Folding}{\Longrightarrow}
&\underset{B_{\left[\frac{m}{2}\right]}\oplus B_{\left[\frac{n}{2}\right]}=
\mathfrak{so}_{m}\oplus\mathfrak{so}_{n}}
{\xygraph{!~:{@{=}|@{}}
    \circ   ([]!{+(0,-.3)} {{}_1})     - [r]
    \circ   ([]!{+(0,-.3)} {{}_2})     - [r]
    \cdots                             - [r] 
    \circ   ([]!{+(0,-.3)} {{}_{\left[\frac{m-2}{2}\right]}}) : [r]
    \bullet ([]!{+(0,-.3)} {{}_{\left[\frac{m}{2}\right]}})    [r]
    \circ   ([]!{+(0,-.3)} {{}_1})     - [r]
    \circ   ([]!{+(0,-.3)} {{}_2})     - [r]
    \cdots                             - [r] 
    \circ   ([]!{+(0,-.3)} {{}_{\left[\frac{n-2}{2}\right]}}) : [r]
    \bullet ([]!{+(0,-.3)} {{}_{\left[\frac{n}{2}\right]}})}
}.
\end{align}
Its corresponding projection matrix is
\begin{align}
&P_{\mathfrak{so}_{mn}\supset\mathfrak{so}_{m}\oplus\mathfrak{so}_{n}}
\nonumber\\
&=
\left(
\begin{array}{cccccccccccccccccc}
 \multicolumn{3}{c}{1\ 2\cdots n-1}
&\multicolumn{1}{c}{n}
&\multicolumn{3}{c}{n-1\cdots 2\ 1}
&\multicolumn{6}{|c}{O}\\
\hline
 \multicolumn{4}{c|}{O}
&\multicolumn{3}{c}{1\ 2\cdots n-1}
&\multicolumn{1}{c}{n}
&\multicolumn{1}{c}{\cdots}
&\multicolumn{4}{c}{O}\\
\hline
 \multicolumn{8}{c}{\vdots}
&\multicolumn{1}{c}{\ddots}
&\multicolumn{4}{c}{\vdots}\\
\hline
 \multicolumn{8}{c}{O}
&\multicolumn{1}{c}{\cdots}
&\multicolumn{1}{c}{n}
&\multicolumn{3}{c|}{n-1 \cdots 2\ 1}
&\multicolumn{2}{c}{O}\\
\hline
 \multicolumn{8}{c}{O}
&\multicolumn{1}{c}{\cdots}
&\multicolumn{1}{c}{0}
&\multicolumn{3}{c}{2\ 4 \cdots 2n-2}
&\multicolumn{1}{|c|}{2n}
&\multicolumn{1}{c}{2n \cdots 2n}
&\multicolumn{1}{|c}{n}\\
\hline
 \multicolumn{3}{c}{I_{\left[\frac{n-2}{2}\right]}\ O\ E_{\left[\frac{n-2}{2}\right]}}
&\multicolumn{1}{|c|}{O}
&\multicolumn{3}{c}{I_{\left[\frac{n-2}{2}\right]}\ O\ E_{\left[\frac{n-2}{2}\right]}}
&\multicolumn{1}{|c|}{O}
&\multicolumn{1}{c}{\cdots}
&\multicolumn{1}{|c|}{O}
&\multicolumn{3}{c}{I_{\left[\frac{n-2}{2}\right]}\ O\ E_{\left[\frac{n-2}{2}\right]}}
&\multicolumn{1}{|c|}{O}
&\multicolumn{1}{c}{I_{\left[\frac{n-2}{2}\right]}}
&\multicolumn{1}{|c}{O}\\
\hline
 \multicolumn{3}{c}{O\ 22\ O}
&\multicolumn{1}{|c|}{O}
&\multicolumn{3}{c}{O\ 22\ O}
&\multicolumn{1}{|c|}{O}
&\multicolumn{1}{c}{\cdots}
&\multicolumn{1}{|c|}{O}
&\multicolumn{3}{c}{O\ 22\ O}
&\multicolumn{1}{|c|}{O}
&\multicolumn{1}{c}{O}
&\multicolumn{1}{|c}{1}\\
\end{array}
\right),
\end{align}
where 
$P_{\mathfrak{so}_{mn}\supset\mathfrak{so}_{m}\oplus\mathfrak{so}_{n}}$
is an $\left[\frac{mn}{2}\right]\times 
\left(\left[\frac{m}{2}\right]+\left[\frac{n}{2}\right]\right)$ matrix.

\subsection{$C_n=\mathfrak{usp}_{2n}$}

Let us summarize features of representations of the Lie algebra
$C_n=\mathfrak{usp}_{2n}$. 
From Table~\ref{Table:Dynkin-diagrams}, 
the extended Dynkin diagram of $C_n=\mathfrak{usp}_{2n}$ is 
\begin{align}
\xygraph{!~:{@{=}|@{}}
    \circ   ([]!{+(0,-.3)} {{}_x}) : [r]
    \bullet ([]!{+(0,-.3)} {{}_1}) - [r]
    \bullet ([]!{+(0,-.3)} {{}_2}) - [r] \cdots - [r]
    \bullet ([]!{+(0,-.3)} {{}_{n-1}}) : [r]
    \circ   ([]!{+(0,-.3)} {{}_n})}.
\end{align}
From Table~\ref{Table:Cartan-matrices}, 
the Cartan matrix of $C_n=\mathfrak{usp}_{2n}$ is
\begin{align}
A(C_n)=
\left(
\begin{array}{ccccccc}
2&-1&0&\cdots&0&0&0\\
-1&2&-1&\cdots&0&0&0\\
0&-1&2&\cdots&0&0&0\\
\vdots&\vdots&\vdots&\ddots&\vdots&\vdots&\vdots\\
0&0&0&\cdots&2&-1&0\\
0&0&0&\cdots&-1&2&-1\\
0&0&0&\cdots&0&-2&2\\
\end{array}
\right).
\end{align}
From Tables~\ref{Table:Complex-representations-1} and 
\ref{Table:Self-conjugate-representations},
types of representations of $C_n=\mathfrak{usp}_{2n}$
are given in the following table.
\begin{center}
\begin{longtable}{clll}
\caption{Types of representations of $C_n$}
\label{Table:Types-representations-C}\\
\hline\hline
Algebra&Rank&Condition&C/R/PR\\\hline\hline
\endfirsthead
\multicolumn{4}{c}{Table~\ref{Table:Types-representations-C} (continued)}\\\hline\hline
Algebra&Rank&Condition&C/R/PR\\\hline\hline
\endhead
\hline
\endfoot
$\mathfrak{usp}_{2n}$ 
&${}^\forall n$
&$a_1+a_3+\cdots+a_{2n-1}=0\ (\bmod.\ 2)$
&Real\\
&
&$a_1+a_3+\cdots+a_{2n-1}=1\ (\bmod.\ 2)$
&PR\\
\end{longtable}
\end{center}
where $a_n$ is the long root,
Condition stand for a condition for Dynkin labels, 
and C/R/PR represents complex, real, and  pseudo-real
representations.
From Eq.~(\ref{Eq:Weyl-formula-C}), the Weyl dimension formula of
$C_n=\mathfrak{usp}_{2n}$ is  
\begin{align}
d(R)=&
\left[
\prod_{k=1}^n\prod_{i=1}^k
\left(\frac{\sum_{j=i}^{k}a_j}{k-i+1}+1\right)
\right]
\prod_{k=0}^{n-2}
\prod_{i=k+1}^{n-1}
\left(
\frac{\left(\sum_{j=k+1}^{i}a_{n-j}\right)
+\left(\sum_{j=0}^{k}2a_{n-j}\right)}{k+i+2}+1\right).
\end{align}
For the Lie algebra $C_n=\mathfrak{usp}_{2n}$, from
Eqs.~(\ref{Eq:2-Casimir}) and (\ref{Eq:Dynkin-index}), we get the
Casimir invariant and the Dynkin index of a representation 
$(a_1a_2a_3\cdots a_n)$:
\begin{align}
C_2(R)=&\frac{1}{2}
\sum_{i,j=1}^{n}(a_i+2g_i)G(C_n)_{ij}a_j,\\
T(R)=&\frac{d(R)}{d(G)}C_2(R)
=\frac{d(R)}{2 d(G)}
\left[\sum_{i,j=1}^{n}(a_i+2g_i)G(C_n)_{ij}a_j\right],
\end{align}
where the metric matrix $G(C_n)$ in 
Table~\ref{Table:Cartan-matrices-inverse}
\begin{align}
G(C_n)=\frac{1}{2}
\left(
\begin{array}{cccccc}
1&1&1&\cdots&1&1\\
1&2&2&\cdots&2&2\\
1&2&3&\cdots&3&3\\
\vdots&\vdots&\vdots&\ddots&\vdots&\vdots\\
1&2&3&\cdots&n-1&n-1\\
1&2&3&\cdots&n-1&n\\
\end{array}
\right).
\end{align}
From Table~\ref{Table:Conjugacy-classes}, the conjugacy classes of the Lie
algebra $C_n=\mathfrak{usp}_{2n}$ is
\begin{align}
C_c(R):=a_1+a_3+a_5+\cdots\ \ \  (\bmod.\ 2).
\end{align}
The Lie algebra $C_n=\mathfrak{usp}_{2n}$ has two conjugacy classes.

The above features of representations of 
$C_n=\mathfrak{usp}_{2n}$ $(n\geq 2)$
are summarized in the following table: 
\begin{center}
\begin{longtable}{cccccc}
\caption{Representations of $C_n=\mathfrak{usp}_{2n}$}
\label{Table:Representations-summary-C}\\
\hline\hline
$\mathfrak{usp}_{2n}$ irrep.&$d(R)$&$C_2(R)$&$T(R)$&$C_c(R)$&C/R/PR\\\hline\hline
\endfirsthead
\multicolumn{6}{c}{Table~\ref{Table:Representations-summary-C} (continued)}\\\hline\hline
$\mathfrak{usp}_{2n}$ irrep.&$d(R)$&$C_2(R)$&$T(R)$&$C_c(R)$&C/R/PR\\\hline\hline
\endhead
\hline
\endfoot
$(0000\cdots 0000)$&$1$   &$0$               &$0$&$0$&R\\ 
$(1000\cdots 0000)$&$2n$         &$\frac{2n+1}{4}$&$\frac{1}{2}$&$0$&PR\\ 
$(0100\cdots 0000)$&$(n-1)(2n+1)$&$n$             &$n-1$&$0$&R\\ 
$(2000\cdots 0000)$&$n(2n+1)$    &$n+1$           &$n+1$&$0$&R\\ 
\end{longtable}
\end{center}
where $\mathfrak{usp}_{2n}$ irrep., $d(R)$, $C_2(R)$, $T(R)$, $C_c(R)$,
and C/R/PR stand for 
the Dynkin label of the irreducible representation of
$C_n=\mathfrak{usp}_{2n}$, its dimension, its quadratic Casimir
invariant, its Dynkin indices,  its conjugacy class, 
complex, real, and pseudo-real representations,
respectively. 
The representation $(20\cdots 0)$ is the adjoint representation of
$C_n=\mathfrak{usp}_{2n}$. 
See Appendix~\ref{Sec:Representations} for further information.

From Table~\ref{Table:Maximal-subalgebra},
maximal $R$- and $S$-subalgebras of $C_n=\mathfrak{usp}_{2n}$
with their rank up to 16 are listed in the following table.
\begin{center}
\begin{longtable}{crcp{10cm}c}
\caption{Maximal subalgebras of $C_n$}
\label{Table:Maximal-subalgebra-C}\\
\hline\hline
Rank&\multicolumn{2}{l}{Algebra $\mathfrak{g}$}&Maximal subalgebras $\mathfrak{h}$&Type\\\hline\hline
\endfirsthead
\multicolumn{4}{c}{Table~\ref{Table:Maximal-subalgebra-C} (continued)}\\
\hline
Rank&\multicolumn{2}{l}{Algebra $\mathfrak{g}$}&Maximal subalgebras $\mathfrak{h}$&Type\\\hline\hline
\endhead
\hline
\endfoot
$2$ &$\mathfrak{usp}_4$&$\supset$&$\mathfrak{su}_2\oplus\mathfrak{su}_2;\
     \mathfrak{su}_2\oplus\mathfrak{u}_1$&$(R)$\\
    &       &$\supset$&$\mathfrak{su}_2$&$(S)$\\
$3$ &$\mathfrak{usp}_6$&$\supset$
&$\mathfrak{su}_3\oplus\mathfrak{u}_1;\ \mathfrak{su}_2\oplus\mathfrak{usp}_4$
&$(R)$\\
    &       &$\supset$
&$\mathfrak{su}_2;\ \mathfrak{su}_2\oplus\mathfrak{su}_2$&$(S)$\\
$4$ &$\mathfrak{usp}_8$&$\supset$
&$\mathfrak{su}_4\oplus\mathfrak{u}_1;\
\mathfrak{su}_2\oplus\mathfrak{usp}_6;\ 
\mathfrak{usp}_4\oplus\mathfrak{usp}_4$&$(R)$\\
    &       &$\supset$&$\mathfrak{su}_2;\
\mathfrak{su}_2\oplus\mathfrak{su}_2\oplus\mathfrak{su}_2$&$(S)$\\
$5$ &$\mathfrak{usp}_{10}$&$\supset$
&$\mathfrak{su}_5\oplus\mathfrak{u}_1;\
\mathfrak{su}_2\oplus\mathfrak{usp}_8;\
\mathfrak{usp}_4\oplus\mathfrak{usp}_6$&$(R)$\\
    &       &$\supset$
&$\mathfrak{su}_2;\ \mathfrak{su}_2\oplus\mathfrak{usp}_4$&$(S)$\\
$6$ &$\mathfrak{usp}_{12}$&$\supset$
&$\mathfrak{su}_6\oplus\mathfrak{u}_1;\
\mathfrak{su}_2\oplus\mathfrak{usp}_{10};\
\mathfrak{usp}_4\oplus\mathfrak{usp}_8;\
\mathfrak{usp}_6\oplus\mathfrak{usp}_6$&$(R)$\\
    &       &$\supset$
&$\mathfrak{su}_2;\ \mathfrak{su}_2\oplus\mathfrak{su}_4;\
\mathfrak{su}_2\oplus\mathfrak{usp}_4$&$(S)$\\
$7$ &$\mathfrak{usp}_{14}$&$\supset$
&$\mathfrak{su}_7\oplus\mathfrak{u}_1;\
\mathfrak{su}_2\oplus\mathfrak{usp}_{12};\
\mathfrak{usp}_4\oplus\mathfrak{usp}_{10};\
\mathfrak{usp}_6\oplus\mathfrak{usp}_8$&$(R)$\\
    &       &$\supset$&$\mathfrak{su}_2;\
\mathfrak{su}_2\oplus\mathfrak{so}_7;\
\mathfrak{usp}_6$&$(S)$\\
$8$ &$\mathfrak{usp}_{16}$&$\supset$
&$\mathfrak{su}_8\oplus\mathfrak{u}_1;\
\mathfrak{su}_2\oplus\mathfrak{usp}_{14};\
\mathfrak{usp}_4\oplus\mathfrak{usp}_{12};\
\mathfrak{usp}_6\oplus\mathfrak{usp}_{10};\
\mathfrak{usp}_8\oplus\mathfrak{usp}_{8}$&$(R)$\\
    &       &$\supset$&$\mathfrak{su}_2;\ \mathfrak{usp}_4;\
\mathfrak{su}_2\oplus\mathfrak{so}_8$&$(S)$\\
$9$ &$\mathfrak{usp}_{18}$&$\supset$
&$\mathfrak{su}_9\oplus\mathfrak{u}_1;\
\mathfrak{su}_2\oplus\mathfrak{usp}_{16};\
\mathfrak{usp}_4\oplus\mathfrak{usp}_{14};\
\mathfrak{usp}_6\oplus\mathfrak{usp}_{12};\
\mathfrak{usp}_8\oplus\mathfrak{usp}_{10}$&$(R)$\\
    &       &$\supset$&$\mathfrak{su}_2;\
\mathfrak{su}_2\oplus\mathfrak{so}_9;\
\mathfrak{su}_2\oplus\mathfrak{usp}_6$&$(S)$\\
$10$&$\mathfrak{usp}_{20}$&$\supset$
&$\mathfrak{su}_{10}\oplus\mathfrak{u}_1;\
\mathfrak{su}_2\oplus\mathfrak{usp}_{18};\
\mathfrak{usp}_4\oplus\mathfrak{usp}_{16};\
\mathfrak{usp}_6\oplus\mathfrak{usp}_{14};\
\mathfrak{usp}_8\oplus\mathfrak{usp}_{12};\
\mathfrak{usp}_{10}\oplus\mathfrak{usp}_{10}$&$(R)$\\
    &       &$\supset$&
$\mathfrak{su}_2;\
\mathfrak{usp}_4\oplus\mathfrak{usp}_4;\
\mathfrak{su}_2\oplus\mathfrak{so}_{10};\ \mathfrak{su}_6$&$(S)$\\
$11$&$\mathfrak{usp}_{22}$&$\supset$
&$\mathfrak{su}_{11}\oplus\mathfrak{u}_1;\
\mathfrak{su}_2\oplus\mathfrak{usp}_{20};\
\mathfrak{usp}_4\oplus\mathfrak{usp}_{18};\
\mathfrak{usp}_6\oplus\mathfrak{usp}_{16};\
\mathfrak{usp}_8\oplus\mathfrak{usp}_{14};\
\mathfrak{usp}_{10}\oplus\mathfrak{usp}_{12}$&$(R)$\\
    &       &$\supset$&
$\mathfrak{su}_2;\
\mathfrak{su}_2\oplus\mathfrak{so}_{11}
 $&$(S)$\\
$12$&$\mathfrak{usp}_{24}$&$\supset$
&$\mathfrak{su}_{12}\oplus\mathfrak{u}_1;\
\mathfrak{su}_2\oplus\mathfrak{usp}_{22};\
\mathfrak{usp}_4\oplus\mathfrak{usp}_{20};\
\mathfrak{usp}_6\oplus\mathfrak{usp}_{18};\
\mathfrak{usp}_8\oplus\mathfrak{usp}_{16};\
\mathfrak{usp}_{10}\oplus\mathfrak{usp}_{14};\
\mathfrak{usp}_{12}\oplus\mathfrak{usp}_{12}
$&$(R)$\\
    &       &$\supset$&
$\mathfrak{su}_2;\
\mathfrak{su}_2\oplus\mathfrak{usp}_{8};\
\mathfrak{su}_4\oplus\mathfrak{usp}_{4};\
\mathfrak{su}_2\oplus\mathfrak{so}_{12}
$&$(S)$\\
$13$&$\mathfrak{usp}_{26}$&$\supset$
&$\mathfrak{su}_{13}\oplus\mathfrak{u}_1;\
\mathfrak{su}_2\oplus\mathfrak{usp}_{24};\
\mathfrak{usp}_4\oplus\mathfrak{usp}_{22};\
\mathfrak{usp}_6\oplus\mathfrak{usp}_{20};\
\mathfrak{usp}_8\oplus\mathfrak{usp}_{18};\
\mathfrak{usp}_{10}\oplus\mathfrak{usp}_{16};\
\mathfrak{usp}_{12}\oplus\mathfrak{usp}_{14}
$&$(R)$\\
    &       &$\supset$&
$\mathfrak{su}_2;\
\mathfrak{su}_2\oplus\mathfrak{so}_{13}$&$(S)$\\
$14$&$\mathfrak{usp}_{28}$&$\supset$
&$\mathfrak{su}_{14}\oplus\mathfrak{u}_1;\
\mathfrak{su}_2\oplus\mathfrak{usp}_{26};\
\mathfrak{usp}_4\oplus\mathfrak{usp}_{24};\
\mathfrak{usp}_6\oplus\mathfrak{usp}_{22};\
\mathfrak{usp}_8\oplus\mathfrak{usp}_{20};\
\mathfrak{usp}_{10}\oplus\mathfrak{usp}_{18};\
\mathfrak{usp}_{12}\oplus\mathfrak{usp}_{16};\
\mathfrak{usp}_{14}\oplus\mathfrak{usp}_{14}
$&$(R)$\\
    &       &$\supset$&
$\mathfrak{su}_2;\
\mathfrak{so}_{7}\oplus\mathfrak{usp}_{4};\
\mathfrak{su}_2\oplus\mathfrak{so}_{14}
$&$(S)$\\
$15$&$\mathfrak{usp}_{30}$&$\supset$
&$\mathfrak{su}_{15}\oplus\mathfrak{u}_1;\
\mathfrak{su}_2\oplus\mathfrak{usp}_{28};\
\mathfrak{usp}_4\oplus\mathfrak{usp}_{26};\
\mathfrak{usp}_6\oplus\mathfrak{usp}_{24};\
\mathfrak{usp}_8\oplus\mathfrak{usp}_{22};\
\mathfrak{usp}_{10}\oplus\mathfrak{usp}_{20};\
\mathfrak{usp}_{12}\oplus\mathfrak{usp}_{18};\
\mathfrak{usp}_{14}\oplus\mathfrak{usp}_{16}
$&$(R)$\\
    &       &$\supset$&
$\mathfrak{su}_2;\
\mathfrak{su}_2\oplus\mathfrak{usp}_{10};\
\mathfrak{usp}_4\oplus\mathfrak{usp}_{6};\
\mathfrak{su}_2\oplus\mathfrak{so}_{15}
$&$(S)$\\
$16$&$\mathfrak{usp}_{32}$&$\supset$
&$\mathfrak{su}_{16}\oplus\mathfrak{u}_1;\
\mathfrak{su}_2\oplus\mathfrak{usp}_{30};\
\mathfrak{usp}_4\oplus\mathfrak{usp}_{28};\
\mathfrak{usp}_6\oplus\mathfrak{usp}_{26};\
\mathfrak{usp}_8\oplus\mathfrak{usp}_{24};\
\mathfrak{usp}_{10}\oplus\mathfrak{usp}_{22};\
\mathfrak{usp}_{12}\oplus\mathfrak{usp}_{20};\
\mathfrak{usp}_{14}\oplus\mathfrak{usp}_{18};\
\mathfrak{usp}_{16}\oplus\mathfrak{usp}_{16}
$&$(R)$\\
    &       &$\supset$&
$\mathfrak{su}_2;\
\mathfrak{so}_{8}\oplus\mathfrak{usp}_{4};\
\mathfrak{so}_{12};\
\mathfrak{su}_2\oplus\mathfrak{so}_{16}
$&$(S)$\\
\end{longtable}
\end{center}
where Type $(R)$ and $(S)$ stand for $R$- and $S$-subalgebras,
respectively. 

We summarize generic projection matrices of the Lie algebra
$C_n=\mathfrak{usp}_{2n}$ and any maximal regular subalgebras, which are
calculated by using a method given in
Sec.~\ref{Sec:Representations-subalgebras}. 
The Dynkin diagram of 
$C_n\supset A_{n-1}\oplus\mathfrak{u}_1$ $(n\geq 2)$
is given by
\begin{align}
&\underset{C_n=\mathfrak{usp}_{2n}}
{\xygraph{!~:{@{=}|@{}}
    \circ   ([]!{+(0,-.3)} {{}_x})     : [r]
    \bullet ([]!{+(0,-.3)} {{}_1})     - [r]
    \bullet ([]!{+(0,-.3)} {{}_2})     - [r]
    \cdots                             - [r] 
    \bullet ([]!{+(0,-.3)} {{}_{n-2}}) - [r]
    \bullet ([]!{+(0,-.3)} {{}_{n-1}}) : [r]
    \circ   ([]!{+(0,-.3)} {{}_{n}})}
}\nonumber\\
\underset{\rm Removing}{\Longrightarrow}
&\underset{A_{n-1}\oplus\mathfrak{u}_1=\mathfrak{su}_{n}\oplus\mathfrak{u}_1}
{\xygraph{!~:{@{=}|@{}}
    {\circ\hspace{-0.645em}\times}   ([]!{+(0,-.3)} {{}_x})     : [r]
    \bullet ([]!{+(0,-.3)} {{}_1})     - [r]
    \bullet ([]!{+(0,-.3)} {{}_2})     - [r]
    \cdots                             - [r] 
    \bullet ([]!{+(0,-.3)} {{}_{n-2}}) - [r]
    \bullet ([]!{+(0,-.3)} {{}_{n-1}}) : [r]
    {\circ\hspace{-0.645em}\times}   ([]!{+(0,-.3)} {{}_{n}})}
}.
\end{align}
Its corresponding projection matrix is
\begin{align}
P_{\mathfrak{usp}_{2n}\supset\mathfrak{su}_{n}\oplus\mathfrak{u}_1}
&=
\left(
\begin{array}{cc}
\multicolumn{1}{c|}{I_{n-1}}&0\\\hline
\multicolumn{1}{c}{1\ 2\ 3\ \cdots\ n-1}&n\\
\end{array}
\right),
\end{align}
where 
$P_{\mathfrak{usp}_{2n}\supset\mathfrak{su}_{n}\oplus\mathfrak{u}_1}$
is an $n\times n$ matrix;
$I_{n-1}$ is an $(n-1)\times (n-1)$ matrix, where its matrix
elements $(I_{n-1})_{ii}=1 (i=1,2,\cdots,n-1)$ and the other matrix
elements $(I_{n-1})_{ij}=0 (i,j=1,2,\cdots,n-1;i\not=j)$.
The Dynkin diagram of 
$C_n\supset A_1\oplus C_{n-1}$ $(n\geq 2)$
is given by
\begin{align}
&\underset{C_n=\mathfrak{usp}_{2n}}
{\xygraph{!~:{@{=}|@{}}
    \circ   ([]!{+(0,-.3)} {{}_x})     : [r]
    \bullet ([]!{+(0,-.3)} {{}_1})     - [r]
    \bullet ([]!{+(0,-.3)} {{}_2})     - [r]
    \cdots                             - [r] 
    \bullet ([]!{+(0,-.3)} {{}_{n-2}}) : [r]
    \bullet ([]!{+(0,-.3)} {{}_{n-1}}) : [r]
    \circ   ([]!{+(0,-.3)} {{}_{n}})}
}\nonumber\\
\underset{\rm Removing}{\Longrightarrow}&
\underset{A_1\oplus  C_{n-1}
=\mathfrak{su}_{2}\oplus\mathfrak{usp}_{2n-2}}
{\xygraph{!~:{@{=}|@{}}
    \circ   ([]!{+(0,-.3)} {{}_x})     : [r]
    {\bullet\hspace{-0.645em}\times} ([]!{+(0,-.3)} {{}_1})     - [r]
    \bullet ([]!{+(0,-.3)} {{}_2})     - [r]
    \cdots                             - [r] 
    \bullet ([]!{+(0,-.3)} {{}_{n-2}}) - [r]
    \bullet ([]!{+(0,-.3)} {{}_{n-1}}) : [r]
    \circ   ([]!{+(0,-.3)} {{}_{n}})}
}.
\end{align}
Its corresponding projection matrix is
\begin{align}
P_{\mathfrak{usp}_{2n}\supset\mathfrak{su}_{2}\oplus\mathfrak{usp}_{2n-2}}
&=
\left(
\begin{array}{cc}
1&\multicolumn{1}{c}{1\cdots 1}\\\hline
O&\multicolumn{1}{|c}{I_{n-1}}\\
\end{array}
\right),
\end{align}
where 
$P_{\mathfrak{usp}_{2n}\supset\mathfrak{su}_{2}\oplus\mathfrak{usp}_{2n-2}}$
is an $n\times n$ matrix.
The Dynkin diagram of 
$C_n\supset C_k\oplus C_{n-k}$ $(n\geq 3;2\leq k\leq n-2)$,
is given by
\begin{align}
&\underset{C_n=\mathfrak{usp}_{2n}}
{\xygraph{!~:{@{=}|@{}}
    \circ   ([]!{+(0,-.3)} {{}_x})     : [r]
    \bullet ([]!{+(0,-.3)} {{}_1})     - [r]
    \bullet ([]!{+(0,-.3)} {{}_2})     - [r]
    \cdots                             - [r] 
    \bullet ([]!{+(0,-.3)} {{}_{k-1}}) - [r]
    \bullet ([]!{+(0,-.3)} {{}_{k}})   - [r]
    \bullet ([]!{+(0,-.3)} {{}_{k+1}}) - [r]
    \cdots                             - [r] 
    \bullet ([]!{+(0,-.3)} {{}_{n-2}}) : [r]
    \bullet ([]!{+(0,-.3)} {{}_{n-1}}) : [r]
    \circ   ([]!{+(0,-.3)} {{}_{n}})}
}\nonumber\\
\underset{\rm Removing}{\Longrightarrow}&
\underset{C_{k}\oplus  C_{n-k}
=\mathfrak{usp}_{2k}\oplus\mathfrak{usp}_{2n-2k}}
{\xygraph{!~:{@{=}|@{}}
    \bullet ([]!{+(0,-.3)} {{}_1})     - [r]
    \bullet ([]!{+(0,-.3)} {{}_2})     - [r]
    \cdots                             - [r] 
    \bullet ([]!{+(0,-.3)} {{}_{k-1}}) : [r]
    \circ   ([]!{+(0,-.3)} {{}_{k,\cdots,n}})       [r]
    \bullet ([]!{+(0,-.3)} {{}_{k+1}}) - [r]
    \cdots                             - [r] 
    \bullet ([]!{+(0,-.3)} {{}_{n-2}}) : [r]
    \bullet ([]!{+(0,-.3)} {{}_{n-1}}) : [r]
    \circ   ([]!{+(0,-.3)} {{}_{n}})}
}.
\end{align}
Its corresponding projection matrix is
\begin{align}
P_{\mathfrak{usp}_{2n}\supset\mathfrak{usp}_{2k}\oplus\mathfrak{usp}_{2n-2k}}
&=
\left(
\begin{array}{ccc}
\multicolumn{1}{c|}{I_{k-1}}&\multicolumn{2}{c}{O}\\\hline
\multicolumn{1}{c|}{O}&1&\multicolumn{1}{c}{1\cdots 1}\\\hline
\multicolumn{2}{c|}{O}&\multicolumn{1}{c}{I_{n-k}}\\
\end{array}
\right),
\end{align}
where 
$P_{\mathfrak{usp}_{2n}\supset\mathfrak{usp}_{2k}\oplus\mathfrak{usp}_{2n-2k}}$
is an $n\times n$ matrix.

We summarize generic projection matrices of the Lie algebra
$C_n=\mathfrak{usp}_{2n}$ and several maximal special subalgebras
given in Tables~\ref{Table:Maximal-S-sub-classical-1}
and \ref{Table:Maximal-S-sub-classical-2}, which is also calculated by
using a method given in Sec.~\ref{Sec:Representations-subalgebras}. 
(The following projection matrices cannot cover all maximal special
subalgebras, so we need to calculate exceptions one by one.)
The Dynkin diagram of 
$C_n\supset A_1$ $(n\geq 2)$,
is given by
\begin{align}
&\underset{C_n=\mathfrak{usp}_{2n}}
{\xygraph{!~:{@{=}|@{}}
    \circ   ([]!{+(0,-.3)} {{}_x})     : [r]
    \bullet ([]!{+(0,-.3)} {{}_1})     - [r]
    \cdots                             - [r] 
    \bullet ([]!{+(0,-.3)} {{}_{n-1}}) : [r]
    \circ   ([]!{+(0,-.3)} {{}_{n}})}
}
\underset{\rm Shrinking}{\Longrightarrow}
\underset{A_1=\mathfrak{su}_{2}}
{\underset{1,\cdots,n}{\circ}}.
\end{align}
Its corresponding projection matrix is
\begin{align}
P_{\mathfrak{usp}_{2n}\supset\mathfrak{su}_{2}}
&=
\left(
\begin{array}{cccccc}
2n-1&\cdots&k(2n-k)&\cdots&(n-1)(n+1)&n^2\\
\end{array}
\right),
\end{align}
where
$P_{\mathfrak{usp}_{2n}\supset\mathfrak{su}_{2}}$
is a $1\times n$ matrix.
The Dynkin diagram of 
$C_{mn}\supset B_{\left[\frac{m}{2}\right]}\oplus C_{n}$
$(m\in\mathbb{Z}_{2k+1}; n\in\mathbb{Z})$
is given by
\begin{align}
&\underset{C_n=\mathfrak{usp}_{2n}}
{\xygraph{!~:{@{=}|@{}}
    \circ   ([]!{+(0,-.3)} {{}_x})     : [r]
    \bullet ([]!{+(0,-.3)} {{}_1})     - [r]
    \cdots                             - [r] 
    \bullet ([]!{+(0,-.3)} {{}_{n-1}}) : [r]
    \circ   ([]!{+(0,-.3)} {{}_{n}})}
}\nonumber\\
\underset{\rm Folding}{\Longrightarrow}
&\underset{B_{\left[\frac{m}{2}\right]}\oplus C_{n}=
\mathfrak{so}_{m}\oplus\mathfrak{usp}_{2n}}
{\xygraph{!~:{@{=}|@{}}
    \circ   ([]!{+(0,-.3)} {{}_1}) - [r]
    \circ   ([]!{+(0,-.3)} {{}_2}) - [r] \cdots - [r]
    \circ   ([]!{+(0,-.3)} {{}_{\left[\frac{m-2}{2}\right]}}) : [r]
    \bullet ([]!{+(0,-.3)} {{}_{\left[\frac{m}{2}\right]}})   [r]
    \bullet ([]!{+(0,-.3)} {{}_1}) - [r]
    \bullet ([]!{+(0,-.3)} {{}_2}) - [r] \cdots - [r]
    \bullet ([]!{+(0,-.3)} {{}_{n-1}}) : [r]
    \circ   ([]!{+(0,-.3)} {{}_n}) [r]
}}.
\end{align}
Its corresponding projection matrix is
\begin{align}
&P_{\mathfrak{usp}_{2mn}\supset\mathfrak{so}_{m}\oplus\mathfrak{usp}_{2n}}
\nonumber\\
&=
\left(
\begin{array}{cccccccccccccccccc}
 \multicolumn{3}{c}{1\ 2\cdots 2n-1}
&\multicolumn{1}{c}{2n}
&\multicolumn{3}{c}{2n-1\cdots 2\ 1}
&\multicolumn{6}{|c}{O}\\
\hline
 \multicolumn{4}{c|}{O}
&\multicolumn{3}{c}{1\ 2\cdots 2n-1}
&\multicolumn{1}{c}{2n}
&\multicolumn{1}{c}{\cdots}
&\multicolumn{4}{c}{O}\\
\hline
 \multicolumn{8}{c}{\vdots}
&\multicolumn{1}{c}{\ddots}
&\multicolumn{4}{c}{\vdots}\\
\hline
 \multicolumn{8}{c}{O}
&\multicolumn{1}{c}{\cdots}
&\multicolumn{1}{c}{2n}
&\multicolumn{3}{c|}{2n-1 \cdots 2\ 1}
&\multicolumn{1}{c}{O}\\
\hline
 \multicolumn{8}{c}{O}
&\multicolumn{1}{c}{\cdots}
&\multicolumn{1}{c}{0}
&\multicolumn{3}{c|}{2\ \cdots 4n-2}
&\multicolumn{1}{c|}{4n}
&\multicolumn{1}{c}{4n \cdots 4n}\\
\hline
 \multicolumn{3}{c}{I_{n-1}\ O\ E_{n-1}}
&\multicolumn{1}{|c|}{O}
&\multicolumn{3}{c}{I_{n-1}\ O\ E_{n-1}}
&\multicolumn{1}{|c|}{O}
&\multicolumn{1}{c}{\cdots}
&\multicolumn{1}{|c|}{O}
&\multicolumn{3}{c}{I_{n-1}\ O\ E_{n-1}}
&\multicolumn{1}{|c|}{O}
&\multicolumn{1}{c}{I_{n-1}\ O}\\
\hline
 \multicolumn{3}{c}{O\ 1\ O}
&\multicolumn{1}{|c|}{O}
&\multicolumn{3}{c}{O\ 1\ O}
&\multicolumn{1}{|c|}{O}
&\multicolumn{1}{c}{\cdots}
&\multicolumn{1}{|c|}{O}
&\multicolumn{3}{c}{O\ 1\ O}
&\multicolumn{1}{|c|}{O}
&\multicolumn{1}{c}{O\ 1}\\
\end{array}
\right),
\end{align}
where 
$P_{\mathfrak{usp}_{2mn}\supset\mathfrak{so}_{m}\oplus\mathfrak{usp}_{2n}}$
is an $n\times n$ matrix;
$E_{n-1}$ is an $(n-1)\times (n-1)$ matrix, where its matrix
elements $(E_{n-1})_{i,n-i}=1 (i=1,2,\cdots,n-1)$ and the other matrix
elements $(E_{n-1})_{i,n-j}=0 (i,j=1,2,\cdots,n-1;i\not=j)$.
The Dynkin diagram of 
$C_{mn}\supset D_{\frac{m}{2}}\oplus C_{n}$
$(m\in \mathbb{Z}_{2k}; n\in\mathbb{Z})$
is given by
\begin{align}
&\underset{C_{mn}=\mathfrak{usp}_{2mn}}
{\xygraph{!~:{@{=}|@{}}
    \circ   ([]!{+(0,-.3)} {{}_x})     : [r]
    \bullet ([]!{+(0,-.3)} {{}_1})     - [r]
    \cdots                             - [r] 
    \bullet ([]!{+(0,-.3)} {{}_{mn-1}}) : [r]
    \circ   ([]!{+(0,-.3)} {{}_{mn}})}
}\nonumber\\
\underset{\rm Folding}{\Longrightarrow}
&\underset{D_{\frac{m}{2}}\oplus C_{n}=
\mathfrak{so}_{m}\oplus\mathfrak{usp}_{2n}}
{\xygraph{!~:{@{=}|@{}}
    \circ   ([]!{+(0,-.3)} {{}_1}) - [r]
    \circ   ([]!{+(0,-.3)} {{}_2}) - [r] \cdots - [r]
    \circ ([]!{+(0,-.3)} {{}_{\frac{m-4}{2}}}) (
        - []!{+(1,.5)}  \circ ([]!{+(0,-.3)} {{}_{\frac{m-2}{2}}}),
        - []!{+(1,-.5)} \circ ([]!{+(0,-.3)} {{}_{\frac{m}{2}}}))  [r] [r]
    \bullet ([]!{+(0,-.3)} {{}_1}) - [r]
    \bullet ([]!{+(0,-.3)} {{}_2}) - [r] \cdots - [r]
    \bullet ([]!{+(0,-.3)} {{}_{n-1}}) : [r]
    \circ   ([]!{+(0,-.3)} {{}_n}) [r]
}}.
\end{align}
Its corresponding projection matrix is
{\small
\begin{align}
&P_{\mathfrak{usp}_{2mn}\supset\mathfrak{so}_{m}\oplus\mathfrak{usp}_{2n}}
\nonumber\\
&=
\left(
\begin{array}{cccccccccccccccccc}
 \multicolumn{3}{c}{1\ 2\cdots 2n-1}
&\multicolumn{1}{c}{2n}
&\multicolumn{3}{c}{2n-1\cdots 2\ 1}
&\multicolumn{6}{|c}{O}\\
\hline
 \multicolumn{4}{c|}{O}
&\multicolumn{3}{c}{1\ 2\cdots 2n-1}
&\multicolumn{1}{c}{2n}
&\multicolumn{1}{c}{\cdots}
&\multicolumn{4}{c}{O}\\
\hline
 \multicolumn{8}{c}{\vdots}
&\multicolumn{1}{c}{\ddots}
&\multicolumn{4}{c}{\vdots}\\
\hline
 \multicolumn{8}{c}{O}
&\multicolumn{1}{c}{\cdots}
&\multicolumn{1}{c}{2n}
&\multicolumn{3}{c|}{2n-1 \cdots 2\ 1}
&\multicolumn{3}{c}{O}\\
\hline
 \multicolumn{8}{c}{O}
&\multicolumn{1}{c}{\cdots}
&\multicolumn{1}{c}{0}
&\multicolumn{3}{c|}{1\ 2\ \cdots 2n-1}
&\multicolumn{1}{c|}{2n}
&\multicolumn{1}{c}{2n-1 \cdots 1}
&\multicolumn{1}{|c}{0}\\
\hline
 \multicolumn{8}{c}{O}
&\multicolumn{1}{c}{\cdots}
&\multicolumn{1}{c}{0}
&\multicolumn{3}{c|}{1\ 2\ \cdots 2n-1}
&\multicolumn{1}{c|}{2n}
&\multicolumn{1}{c}{2n+1 \cdots 4n-1}
&\multicolumn{1}{|c}{4n}\\
\hline
 \multicolumn{3}{c}{I_{n-1}\ O\ E_{n-1}}
&\multicolumn{1}{|c|}{O}
&\multicolumn{3}{c}{I_{n-1}\ O\ E_{n-1}}
&\multicolumn{1}{|c|}{O}
&\multicolumn{1}{c}{\cdots}
&\multicolumn{1}{|c|}{O}
&\multicolumn{3}{c}{I_{n-1}\ O\ E_{n-1}}
&\multicolumn{1}{|c|}{O}
&\multicolumn{1}{c}{I_{n-1}\ O\ E_{n-1}}
&\multicolumn{1}{|c}{O}\\
\hline
 \multicolumn{3}{c}{O\ 1\ O}
&\multicolumn{1}{|c|}{O}
&\multicolumn{3}{c}{O\ 1\ O}
&\multicolumn{1}{|c|}{O}
&\multicolumn{1}{c}{\cdots}
&\multicolumn{1}{|c|}{O}
&\multicolumn{3}{c}{O\ 1\ O}
&\multicolumn{1}{|c|}{O}
&\multicolumn{1}{c}{O\ 1\ O}
&\multicolumn{1}{|c}{O}\\
\end{array}
\right),
\end{align}}
where 
$P_{\mathfrak{usp}_{2mn}\supset\mathfrak{so}_{m}\oplus\mathfrak{usp}_{2n}}$
is an $(mn)\times (\frac{m}{2}+n)$ matrix.
The Dynkin diagram of 
$C_{mn}\supset C_{m}\oplus B_{\left[\frac{n}{2}\right]}$
$(m\in \mathbb{Z}; n\in \mathbb{Z}_{2k+1})$
is given by
\begin{align}
&\underset{C_{mn}=\mathfrak{usp}_{2mn}}
{\xygraph{!~:{@{=}|@{}}
    \circ   ([]!{+(0,-.3)} {{}_x})     : [r]
    \bullet ([]!{+(0,-.3)} {{}_1})     - [r]
    \cdots                             - [r] 
    \bullet ([]!{+(0,-.3)} {{}_{mn-1}}) : [r]
    \circ   ([]!{+(0,-.3)} {{}_{mn}})}
}\nonumber\\
\underset{\rm Folding}{\Longrightarrow}
&\underset{C_{m}\oplus B_{\left[\frac{n}{2}\right]}=
\mathfrak{usp}_{2m}\oplus\mathfrak{so}_{n}}
{\xygraph{!~:{@{=}|@{}}
    \bullet ([]!{+(0,-.3)} {{}_1}) - [r]
    \bullet ([]!{+(0,-.3)} {{}_2}) - [r] \cdots - [r]
    \bullet ([]!{+(0,-.3)} {{}_{m-1}}) : [r]
    \circ   ([]!{+(0,-.3)} {{}_m}) [r]
    \circ   ([]!{+(0,-.3)} {{}_1}) - [r]
    \circ   ([]!{+(0,-.3)} {{}_2}) - [r] \cdots - [r]
    \circ   ([]!{+(0,-.3)} {{}_{\left[\frac{n-2}{2}\right]}}) : [r]
    \bullet ([]!{+(0,-.3)} {{}_{\left[\frac{n}{2}\right]}})   
}}.
\end{align}
Its corresponding projection matrix is
{\small
\begin{align}
&P_{\mathfrak{usp}_{2mn}\supset\mathfrak{usp}_{2m}\oplus\mathfrak{so}_{n}}
\nonumber\\
&=
\left(
\begin{array}{cccccccccccccccccc}
 \multicolumn{3}{c}{1\ 2\cdots n-1}
&\multicolumn{1}{c}{n}
&\multicolumn{3}{c}{n-1\cdots 2\ 1}
&\multicolumn{6}{|c}{O}\\
\hline
 \multicolumn{4}{c|}{O}
&\multicolumn{3}{c}{1\ 2\cdots n-1}
&\multicolumn{1}{c}{n}
&\multicolumn{1}{c}{\cdots}
&\multicolumn{4}{c}{O}\\
\hline
 \multicolumn{8}{c}{\vdots}
&\multicolumn{1}{c}{\ddots}
&\multicolumn{4}{c}{\vdots}\\
\hline
 \multicolumn{8}{c}{O}
&\multicolumn{1}{c}{\cdots}
&\multicolumn{1}{c}{n}
&\multicolumn{3}{c|}{n-1 \cdots 2\ 1}
&\multicolumn{3}{c}{O}\\
\hline
 \multicolumn{8}{c}{O}
&\multicolumn{1}{c}{\cdots}
&\multicolumn{1}{c}{0}
&\multicolumn{3}{c|}{1\ 2\ \cdots n-1}
&\multicolumn{1}{c|}{n}
&\multicolumn{1}{c}{n-1 \cdots 2\ 1}
&\multicolumn{1}{|c}{0}\\
\hline
 \multicolumn{8}{c}{O}
&\multicolumn{1}{c}{\cdots}
&\multicolumn{1}{c}{0}
&\multicolumn{3}{c|}{O}
&\multicolumn{1}{c|}{0}
&\multicolumn{1}{c}{1\ 2\ \cdots n-1}
&\multicolumn{1}{|c}{n}\\
\hline
 \multicolumn{3}{c}{I_{\left[\frac{n-2}{2}\right]}\ O\ E_{\left[\frac{n-2}{2}\right]}}
&\multicolumn{1}{|c|}{O}
&\multicolumn{3}{c}{I_{\left[\frac{n-2}{2}\right]}\ O\ E_{\left[\frac{n-2}{2}\right]}}
&\multicolumn{1}{|c|}{O}
&\multicolumn{1}{c}{\cdots}
&\multicolumn{1}{|c|}{O}
&\multicolumn{3}{c}{I_{\left[\frac{n-2}{2}\right]}\ O\ E_{\left[\frac{n-2}{2}\right]}}
&\multicolumn{1}{|c|}{O}
&\multicolumn{1}{c}{I_{\left[\frac{n-2}{2}\right]}\ O\ E_{\left[\frac{n-2}{2}\right]}}
&\multicolumn{1}{|c}{O}\\
\hline
 \multicolumn{3}{c}{O\ 22\ O}
&\multicolumn{1}{|c|}{O}
&\multicolumn{3}{c}{O\ 22\ O}
&\multicolumn{1}{|c|}{O}
&\multicolumn{1}{c}{\cdots}
&\multicolumn{1}{|c|}{O}
&\multicolumn{3}{c}{O\ 22\ O}
&\multicolumn{1}{|c|}{O}
&\multicolumn{1}{c}{O\ 22\ O}
&\multicolumn{1}{|c}{O}\\
\end{array}
\right),
\end{align}}
where 
$P_{\mathfrak{usp}_{2mn}\supset\mathfrak{usp}_{2m}\oplus\mathfrak{so}_{n}}$
is an $(mn)\times \left(m+\left[\frac{n}{2}\right]\right)$ matrix.
The Dynkin diagram of 
$C_{mn}\supset C_{m}\oplus D_{\frac{n}{2}}$
$(m\in \mathbb{Z}; n\in \mathbb{Z}_{2k})$
is given by
\begin{align}
&\underset{C_{mn}=\mathfrak{usp}_{2mn}}
{\xygraph{!~:{@{=}|@{}}
    \circ   ([]!{+(0,-.3)} {{}_x})     : [r]
    \bullet ([]!{+(0,-.3)} {{}_1})     - [r]
    \cdots                             - [r] 
    \bullet ([]!{+(0,-.3)} {{}_{mn-1}}) : [r]
    \circ   ([]!{+(0,-.3)} {{}_{mn}})}
}\nonumber\\
\underset{\rm Folding}{\Longrightarrow}
&\underset{C_{m}\oplus D_{\frac{n}{2}}=
\mathfrak{usp}_{2m}\oplus\mathfrak{so}_{n}}
{\xygraph{!~:{@{=}|@{}}
    \bullet ([]!{+(0,-.3)} {{}_1}) - [r]
    \bullet ([]!{+(0,-.3)} {{}_2}) - [r] \cdots - [r]
    \bullet ([]!{+(0,-.3)} {{}_{m-1}}) : [r]
    \circ   ([]!{+(0,-.3)} {{}_m}) [r]
    \circ   ([]!{+(0,-.3)} {{}_1}) - [r]
    \circ   ([]!{+(0,-.3)} {{}_2}) - [r] \cdots - [r]
    \circ ([]!{+(0,-.3)} {{}_{\frac{n-4}{2}}}) (
        - []!{+(1,.5)}  \circ ([]!{+(0,-.3)} {{}_{\frac{n-2}{2}}}),
        - []!{+(1,-.5)} \circ ([]!{+(0,-.3)} {{}_{\frac{n}{2}}}))
}}.
\end{align}
Its corresponding projection matrix is
{\small
\begin{align}
&P_{\mathfrak{usp}_{2mn}\supset\mathfrak{usp}_{2m}\oplus\mathfrak{so}_{n}}
\nonumber\\
&=
\left(
\begin{array}{cccccccccccccccccc}
 \multicolumn{3}{c}{1\ 2\cdots n-1}
&\multicolumn{1}{c}{n}
&\multicolumn{3}{c}{n-1\cdots 2\ 1}
&\multicolumn{6}{|c}{O}\\
\hline
 \multicolumn{4}{c|}{O}
&\multicolumn{3}{c}{1\ 2\cdots n-1}
&\multicolumn{1}{c}{n}
&\multicolumn{1}{c}{\cdots}
&\multicolumn{4}{c}{O}\\
\hline
 \multicolumn{8}{c}{\vdots}
&\multicolumn{1}{c}{\ddots}
&\multicolumn{4}{c}{\vdots}\\
\hline
 \multicolumn{8}{c}{O}
&\multicolumn{1}{c}{\cdots}
&\multicolumn{1}{c}{n}
&\multicolumn{3}{c|}{n-1 \cdots 2\ 1}
&\multicolumn{3}{c}{O}\\
\hline
 \multicolumn{8}{c}{O}
&\multicolumn{1}{c}{\cdots}
&\multicolumn{1}{c}{0}
&\multicolumn{3}{c|}{1\ 2\ \cdots n-1}
&\multicolumn{1}{c|}{n}
&\multicolumn{1}{c}{n-1 \cdots 2\ 1}
&\multicolumn{1}{|c}{0}\\
\hline
 \multicolumn{8}{c}{O}
&\multicolumn{1}{c}{\cdots}
&\multicolumn{1}{c}{0}
&\multicolumn{3}{c|}{O}
&\multicolumn{1}{c|}{0}
&\multicolumn{1}{c}{1\ 2\ \cdots n-1}
&\multicolumn{1}{|c}{n}\\
\hline
 \multicolumn{3}{c}{I_{\frac{n-2}{2}}\ O\ E_{\frac{n-2}{2}}}
&\multicolumn{1}{|c|}{O}
&\multicolumn{3}{c}{I_{\frac{n-2}{2}}\ O\ E_{\frac{n-2}{2}}}
&\multicolumn{1}{|c|}{O}
&\multicolumn{1}{c}{\cdots}
&\multicolumn{1}{|c|}{O}
&\multicolumn{3}{c}{I_{\frac{n-2}{2}}\ O\ E_{\frac{n-2}{2}}}
&\multicolumn{1}{|c|}{O}
&\multicolumn{1}{c}{I_{\frac{n-2}{2}}\ O\ E_{\frac{n-2}{2}}}
&\multicolumn{1}{|c}{O}\\
\hline
 \multicolumn{3}{c}{O\ 121\ O}
&\multicolumn{1}{|c|}{O}
&\multicolumn{3}{c}{O\ 121\ O}
&\multicolumn{1}{|c|}{O}
&\multicolumn{1}{c}{\cdots}
&\multicolumn{1}{|c|}{O}
&\multicolumn{3}{c}{O\ 121\ O}
&\multicolumn{1}{|c|}{O}
&\multicolumn{1}{c}{O\ 121\ O}
&\multicolumn{1}{|c}{O}\\
\end{array}
\right),
\end{align}}
where 
$P_{\mathfrak{usp}_{2mn}\supset\mathfrak{usp}_{2m}\oplus\mathfrak{so}_{n}}$
is an $(mn)\times \left(m+\frac{n}{2}\right)$ matrix.

\subsection{$D_n=\mathfrak{so}_{2n}$}

Let us summarize features of representations of the 
Lie algebra $D_n=\mathfrak{so}_{2n}$.
From Table~\ref{Table:Dynkin-diagrams}, 
the extended Dynkin diagram of the Lie algebra $D_n=\mathfrak{so}_{2n}$
is 
\begin{align}
\xygraph{
    \circ ([]!{+(0,-.3)} {{}_2}) (
        - []!{+(-1,.5)}  \circ ([]!{+(0,-.3)} {{}_1}),
        - []!{+(-1,-.5)} \circ ([]!{+(0,-.3)} {{}_x}))
        - [r] 
    \circ ([]!{+(0,-.3)} {{}_3})
        - [r] \cdots - [r]
    \circ ([]!{+(0,-.3)} {{}_{n-2}}) (
        - []!{+(1,.5)}  \circ ([]!{+(0,-.3)} {{}_{n-1}}),
        - []!{+(1,-.5)} \circ ([]!{+(0,-.3)} {{}_n})
)}.
\end{align}
From Table~\ref{Table:Cartan-matrices}, 
the Cartan matrix of the Lie algebra $D_n=\mathfrak{so}_{2n}$ is 
\begin{align}
A(D_n)=
\left(
\begin{array}{ccccccc}
2&-1&0&\cdots&0&0&0\\
-1&2&-1&\cdots&0&0&0\\
0&-1&2&\cdots&0&0&0\\
\vdots&\vdots&\vdots&\ddots&\vdots&\vdots&\vdots\\
0&0&0&\cdots&2&-1&-1\\
0&0&0&\cdots&-1&2&0\\
0&0&0&\cdots&-1&0&2\\
\end{array}
\right).
\end{align}
The complex representations of the Lie algebra
$D_{2n+1}=\mathfrak{so}_{4n+2}$ satisfy the following 
condition
\begin{align}
{\xygraph{
    \circ ([]!{+(0,-.3)} {a_1})  - [r] 
    \circ ([]!{+(0,-.3)} {a_2})  - [r] 
    \cdots                        - [r]
    \circ ([]!{+(0,-.3)} {a_{2n-1}}) (
        - []!{+(1,.5)}  \circ ([]!{+(0,-.3)} {a_{2n}}),
        - []!{+(1,-.5)} \circ ([]!{+(0,-.3)} {a_{2n+1}})
)}}
\not=
{\xygraph{
    \circ ([]!{+(0,-.3)} {a_1})  - [r] 
    \circ ([]!{+(0,-.3)} {a_2})  - [r] 
    \cdots                        - [r]
    \circ ([]!{+(0,-.3)} {a_{2n-1}}) (
        - []!{+(1,.5)}  \circ ([]!{+(0,-.3)} {a_{2n+1}}),
        - []!{+(1,-.5)} \circ ([]!{+(0,-.3)} {a_{2n}})
)}}.
\end{align}
From Tables~\ref{Table:Complex-representations-1} and 
\ref{Table:Self-conjugate-representations},
types of representations of $D_n=\mathfrak{so}_{2n}$
are given in the following table.
\begin{center}
\begin{longtable}{clll}
\caption{Types of representations of $D_n$}
\label{Table:Types-representations-D}\\
\hline\hline
Algebra&Rank&Condition&C/R/PR\\\hline\hline
\endfirsthead
\multicolumn{4}{c}{Table~\ref{Table:Types-representations-D} (continued)}\\\hline\hline
Algebra&Rank&Condition&C/R/PR\\\hline\hline
\endhead
\hline
\endfoot
$\mathfrak{so}_{2n}$ 
&$n=1\ (\bmod.\ 2)$
&$a_{n-1}\not=a_n$
&Complex
\\\cline{3-4}
&
&$a_{n-1}=a_n$
&Real
\\\cline{2-4}
&$n=0\ (\bmod.\ 4)$
&None
&Real
\\\cline{2-4}
&$n=2\ (\bmod.\ 4)$
&$a_{n-1}+a_n=0\ (\bmod.\ 2)$
&Real
\\\cline{3-4}
&
&$a_{n-1}+a_n=1\ (\bmod.\ 2)$
&PR\\
\end{longtable}
\end{center}
where $\alpha_n$ and $\alpha_{n-1}$ are the spinor roots,
Condition stands for a condition for Dynkin labels, 
and C/R/PR represents complex, real, and pseudo-real
representations.
From Eq.~(\ref{Eq:Weyl-formula-D}), the Weyl dimension formula of the
Lie algebra $D_{n}=\mathfrak{so}_{2n}$ is  
\begin{align}
d(R)=&
\left[
\prod_{k=1}^{n-2}\prod_{i=1}^k
\left(\frac{\sum_{j=i}^{k}a_j}{k-i+1}+1\right)
\right]
\left[
\prod_{k=0}^{n-2}
\prod_{\ell=n-1}^{n}
\left(
\frac{\left(\sum_{j=1}^{k}a_{n-1-j}\right)+a_\ell}{k+1}+1\right)
\right]\nonumber\\
&\times
\left[
\prod_{k=0}^{n-3}
\prod_{i=k+1}^{n-2}
\left(
\frac{\left(\sum_{j=k+1}^{i}a_{n-1-j}\right)
 +\left(\sum_{j=1}^{k}2a_{n-1-j}\right)+a_{n-1}+a_n}{i+k+2}+1\right)
\right].
\end{align}
For the Lie algebra $D_n=\mathfrak{so}_{2n}$, from
Eqs.~(\ref{Eq:2-Casimir}) and (\ref{Eq:Dynkin-index}), 
we get the Casimir invariant and the Dynkin index of a representation
$(a_1a_2a_3\cdots a_n)$:
\begin{align}
C_2(R)=&\frac{1}{2}
\sum_{i,j=1}^{n}(a_i+2g_i)G(D_n)_{ij}a_j,\\
T(R)=&\frac{d(R)}{d(G)}C_2(R)
=\frac{d(R)}{2d(G)}
\sum_{i,j=1}^{n}(a_i+2g_i)G(D_n)_{ij}a_j,
\end{align}
where $G(D_4)$ is the metric matrix in 
Table~\ref{Table:Cartan-matrices-inverse}
\begin{align}
G(D_n)&=\frac{1}{2}
\left(
\begin{array}{ccccccc}
2&2&2&\cdots&2&1&1\\
2&4&4&\cdots&4&2&2\\
2&4&6&\cdots&6&3&3\\
\vdots&\vdots&\vdots&\ddots&\vdots&\vdots&\vdots  \\
2&4&6&\cdots&2(n-2)&n-2&n-2\\
1&2&3&\cdots&n-2&n/2&(n-2)/2\\
1&2&3&\cdots&n-2&(n-2)/2&n/2\\
\end{array}
\right).
\end{align}
From Table~\ref{Table:Conjugacy-classes}, 
the conjugacy classes of the Lie algebra 
$D_n=\mathfrak{so}_{2n}$ is 
\begin{align}
C_{c1}(R):=&a_{n-1}+a_n\ \ \  (\bmod.\ 2),\\
C_{c2}(R):=&
\left\{
\begin{array}{ll}
2a_1+2a_3+\cdots+2a_{n-2}+(n-2)a_{n-1}+n a_n\ (\bmod.\ 4)
&\mbox{for}\ n=1\ (\bmod.\ 2)\\
2a_1+2a_3+\cdots+2a_{n-3}
+(n-2)a_{n-1}+n a_n\ (\bmod.\ 4)
&\mbox{for}\ n=0\ (\bmod.\ 2)\\
\end{array}
\right..
\end{align}
The algebra $D_n=\mathfrak{so}_{2n}$ has four conjugacy classes. 

The above features of representations of 
$D_n=\mathfrak{so}_{2n}$ $(n\geq 4)$
are summarized in the following table.
\begin{center}
\begin{longtable}{cccccc}
\caption{Representations of $D_n=\mathfrak{so}_{2n}$}
\label{Table:Representations-summary-D}\\
\hline\hline
$\mathfrak{so}_{2n}$ irrep.&$d(R)$&$C_2(R)$&$T(R)$&$C_c(R)$&C/R/PR\\\hline\hline
\endfirsthead
\multicolumn{6}{c}{Table~\ref{Table:Representations-summary-D} (continued)}\\\hline\hline
$\mathfrak{so}_{2n}$ irrep.&$d(R)$&$C_2(R)$&$T(R)$&$C_c(R)$&C/R/PR\\\hline\hline
\endhead
\hline
\endfoot
$(0000\cdots 0000)$&$1$          &$0$                &$0$&$(0,0)$&R\\ 
$(1000\cdots 0000)$&$2n$         &$\frac{2n-1}{2}$   &$1$&$(0,2)$&R\\ 
$(0100\cdots 0000)$&$n(2n-1)$    &$2(n-1)$           &$2(n-1)$&$(0,0)$&R\\ 
$(0010\cdots 0000)$&$\frac{2n(2n-1)(n-1)}{3}$&$\frac{3(2n-3)}{2}$&$(n-1)(2n-3)$&$(0,2)$&R/PR\\ 
$(0000\cdots 0001)$&$2^{n-1}$    &$\frac{n(2n-1)}{8}$&$2^{n-4}$&$(1,n)$&C/R/PR\\ 
$(0000\cdots 0010)$&$2^{n-1}$    &$\frac{n(2n-1)}{8}$&$2^{n-4}$&$(1,n-2)$&C/R/PR\\ 
$(2000\cdots 0000)$&$(n+1)(2n-1)$&$2n$               &$2(n+1)$&$(0,0)$&R\\ 
\end{longtable}
\end{center}
where $\mathfrak{so}_{2n}$ irrep., $d(R)$, $C_2(R)$, $T(R)$, $C_c(R)$,
and C/R/PR stand for  
the Dynkin label of the irreducible representation of
$D_n=\mathfrak{so}_{2n}$, 
its dimension, its quadratic Casimir invariant, its Dynkin indices, 
its conjugacy classes, 
and complex, real, and pseudo-real representations,
respectively. 
$C_c(R):=(C_{c1},C_{c2})$.
The representation $(010\cdots 0)$ is the adjoint representation of
$D_n=\mathfrak{so}_{2n}$. 
See Appendix~\ref{Sec:Representations} for further information.

From Table~\ref{Table:Maximal-subalgebra},
maximal $R$- and $S$-subalgebras of $D_n=\mathfrak{so}_{2n}$
with their rank up to 16 are listed in the following table.
\begin{center}
\begin{longtable}{crcp{10cm}c}
\caption{Maximal subalgebras of $D_n$}
\label{Table:Maximal-subalgebra-D}\\
\hline\hline
Rank&\multicolumn{2}{l}{Algebra $\mathfrak{g}$}&Maximal subalgebras $\mathfrak{h}$&Type\\\hline\hline
\endfirsthead
\multicolumn{4}{c}{Table~\ref{Table:Maximal-subalgebra-D} (continued)}\\
\hline
Rank&\multicolumn{2}{l}{Algebra $\mathfrak{g}$}&Maximal subalgebras $\mathfrak{h}$&Type\\\hline\hline
\endhead
\hline
\endfoot
$4$ &$\mathfrak{so}_8$&$\supset$
&$\mathfrak{su}_2\oplus\mathfrak{su}_2\oplus\mathfrak{su}_2
\oplus\mathfrak{su}_2;\ \mathfrak{su}_4\oplus\mathfrak{u}_1$&$(R)$\\
    &       &$\supset$
&$\mathfrak{su}_3;\ \mathfrak{so}_7;\ 
\mathfrak{su}_2\oplus\mathfrak{usp}_4$&$(S)$\\

$5$ &$\mathfrak{so}_{10}$&$\supset$
&$\mathfrak{su}_5\oplus\mathfrak{u}_1;\
\mathfrak{su}_2\oplus\mathfrak{su}_2\oplus\mathfrak{su}_4;\
\mathfrak{so}_8\oplus\mathfrak{u}_1$&$(R)$\\
    &       &$\supset$
&$\mathfrak{usp}_4;\ \mathfrak{so}_9;\ \mathfrak{su}_2\oplus\mathfrak{so}_7;\
\mathfrak{usp}_4\oplus\mathfrak{usp}_4$&$(S)$\\
$6$ &$\mathfrak{so}_{12}$&$\supset$
&$\mathfrak{su}_6\oplus\mathfrak{u}_1;\
\mathfrak{su}_2\oplus\mathfrak{su}_2\oplus\mathfrak{so}_8;\
\mathfrak{su}_4\oplus\mathfrak{su}_4;\
\mathfrak{so}_{10}\oplus\mathfrak{u}_1$&$(R)$\\
    &       &$\supset$
&$\mathfrak{su}_2\oplus\mathfrak{usp}_6;\
\mathfrak{so}_{11};\
\mathfrak{su}_2\oplus\mathfrak{so}_9;\
\mathfrak{usp}_4\oplus\mathfrak{so}_7$&$(S)$\\
$7$ &$\mathfrak{so}_{14}$&$\supset$
&$\mathfrak{su}_7\oplus\mathfrak{u}_1;\
\mathfrak{su}_2\oplus\mathfrak{su}_2\oplus\mathfrak{so}_{10};\
\mathfrak{su}_4\oplus\mathfrak{so}_8;\
\mathfrak{so}_{12}\oplus\mathfrak{u}_1$&$(R)$\\
    &       &$\supset$&
$\mathfrak{usp}_4;\ \mathfrak{usp}_6;\ G_2;\
\mathfrak{so}_{13};\ \mathfrak{su}_2\oplus\mathfrak{so}_{11};\
\mathfrak{usp}_4\oplus\mathfrak{so}_9;\
\mathfrak{so}_7\oplus\mathfrak{so}_7$&$(S)$\\
$8$ &$\mathfrak{so}_{16}$&$\supset$
&$\mathfrak{su}_8\oplus\mathfrak{u}_1;\
\mathfrak{su}_2\oplus\mathfrak{su}_2\oplus\mathfrak{so}_{12};\
\mathfrak{su}_4\oplus\mathfrak{so}_{10};\
\mathfrak{so}_8\oplus\mathfrak{so}_{8};\
\mathfrak{so}_{14}\oplus\mathfrak{u}_1$&$(R)$\\
    &       &$\supset$&
$\mathfrak{so}_9;\
\mathfrak{su}_2\oplus\mathfrak{usp}_8;\
\mathfrak{usp}_4\oplus\mathfrak{usp}_4;\
\mathfrak{so}_{15};\
\mathfrak{su}_2\oplus\mathfrak{so}_{13};\
\mathfrak{usp}_4\oplus\mathfrak{so}_{11};\
\mathfrak{so}_7\oplus\mathfrak{so}_9$&$(S)$\\
$9$ &$\mathfrak{so}_{18}$&$\supset$
&$\mathfrak{su}_9\oplus\mathfrak{u}_1;\
\mathfrak{su}_2\oplus\mathfrak{su}_2\oplus\mathfrak{so}_{14};\
\mathfrak{su}_4\oplus\mathfrak{so}_{12};\
\mathfrak{so}_8\oplus\mathfrak{so}_{10};\
\mathfrak{so}_{16}\oplus\mathfrak{u}_1$&$(R)$\\
    &       &$\supset$&
$\mathfrak{su}_2\oplus\mathfrak{su}_4;\ \mathfrak{so}_{17};\
\mathfrak{su}_2\oplus\mathfrak{so}_{15};\
\mathfrak{usp}_4\oplus\mathfrak{so}_{13};\
\mathfrak{so}_7\oplus\mathfrak{so}_{11};\
\mathfrak{so}_9\oplus\mathfrak{so}_9$&$(S)$\\
$10$&$\mathfrak{so}_{20}$&$\supset$&
$\mathfrak{su}_{10}\oplus\mathfrak{u}_1;\
\mathfrak{su}_2\oplus\mathfrak{su}_2\oplus\mathfrak{so}_{16};\
\mathfrak{su}_4\oplus\mathfrak{so}_{14};\
\mathfrak{so}_{8}\oplus\mathfrak{so}_{12};\
\mathfrak{so}_{10}\oplus\mathfrak{so}_{10};\
\mathfrak{so}_{18}\oplus\mathfrak{u}_1$&$(R)$\\
    &       &$\supset$&
$\mathfrak{su}_2\oplus\mathfrak{usp}_{10};\
\mathfrak{so}_{19};\
\mathfrak{su}_2\oplus\mathfrak{so}_{17};\
\mathfrak{usp}_4\oplus\mathfrak{so}_{15};\
\mathfrak{so}_7\oplus\mathfrak{so}_{13};\
\mathfrak{so}_9\oplus\mathfrak{so}_{11};\
\mathfrak{su}_4
$&$(S)$\\
$11$&$\mathfrak{so}_{22}$&$\supset$&
$\mathfrak{su}_{11}\oplus\mathfrak{u}_1;\
\mathfrak{su}_2\oplus\mathfrak{su}_2\oplus\mathfrak{so}_{18};\
\mathfrak{su}_4\oplus\mathfrak{so}_{16};\
\mathfrak{so}_{8}\oplus\mathfrak{so}_{14};\
\mathfrak{so}_{10}\oplus\mathfrak{so}_{12};\
\mathfrak{so}_{20}\oplus\mathfrak{u}_1$&$(R)$\\
    &       &$\supset$&
$\mathfrak{so}_{21};\
\mathfrak{su}_2\oplus\mathfrak{so}_{19};\
\mathfrak{usp}_4\oplus\mathfrak{so}_{17};\
\mathfrak{so}_7\oplus\mathfrak{so}_{15};\
\mathfrak{so}_9\oplus\mathfrak{so}_{13};\
\mathfrak{so}_{11}\oplus\mathfrak{so}_{11}$&$(S)$\\
$12$&$\mathfrak{so}_{24}$&$\supset$&
$\mathfrak{su}_{12}\oplus\mathfrak{u}_1;\
\mathfrak{su}_2\oplus\mathfrak{su}_2\oplus\mathfrak{so}_{20};\
\mathfrak{su}_4\oplus\mathfrak{so}_{18};\
\mathfrak{so}_{8}\oplus\mathfrak{so}_{16};\
\mathfrak{so}_{10}\oplus\mathfrak{so}_{14};\
\mathfrak{so}_{12}\oplus\mathfrak{so}_{12};\
\mathfrak{so}_{22}\oplus\mathfrak{u}_1
$&$(R)$\\
    &       &$\supset$&
$\mathfrak{so}_{23};\
\mathfrak{su}_2\oplus\mathfrak{so}_{21};\
\mathfrak{usp}_4\oplus\mathfrak{so}_{19};\
\mathfrak{so}_7\oplus\mathfrak{so}_{17};\
\mathfrak{so}_9\oplus\mathfrak{so}_{15};\
\mathfrak{so}_{11}\oplus\mathfrak{so}_{13};\
\mathfrak{usp}_6\oplus\mathfrak{usp}_4;\
\mathfrak{su}_2\oplus\mathfrak{so}_8;\
\mathfrak{su}_5;\
\mathfrak{su}_2\oplus\mathfrak{usp}_{12}
$&$(S)$\\
$13$&$\mathfrak{so}_{26}$&$\supset$&
$\mathfrak{su}_{13}\oplus\mathfrak{u}_1;\
\mathfrak{su}_2\oplus\mathfrak{su}_2\oplus\mathfrak{so}_{22};\
\mathfrak{su}_4\oplus\mathfrak{so}_{20};\
\mathfrak{so}_{8}\oplus\mathfrak{so}_{18};\
\mathfrak{so}_{10}\oplus\mathfrak{so}_{16};\
\mathfrak{so}_{12}\oplus\mathfrak{so}_{14};\
\mathfrak{so}_{24}\oplus\mathfrak{u}_1
$&$(R)$\\
    &       &$\supset$&
$\mathfrak{so}_{25};\
\mathfrak{su}_2\oplus\mathfrak{so}_{23};\
\mathfrak{usp}_4\oplus\mathfrak{so}_{21};\
\mathfrak{so}_7\oplus\mathfrak{so}_{19};\
\mathfrak{so}_9\oplus\mathfrak{so}_{17};\
\mathfrak{so}_{11}\oplus\mathfrak{so}_{15};\
\mathfrak{so}_{13}\oplus\mathfrak{so}_{13};\
F_4
$&$(S)$\\
$14$&$\mathfrak{so}_{28}$&$\supset$&
$\mathfrak{su}_{14}\oplus\mathfrak{u}_1;\
\mathfrak{su}_2\oplus\mathfrak{su}_2\oplus\mathfrak{so}_{24};\
\mathfrak{su}_4\oplus\mathfrak{so}_{22};\
\mathfrak{so}_{8}\oplus\mathfrak{so}_{20};\
\mathfrak{so}_{10}\oplus\mathfrak{so}_{18};\
\mathfrak{so}_{12}\oplus\mathfrak{so}_{16};\
\mathfrak{so}_{14}\oplus\mathfrak{so}_{14};\
\mathfrak{so}_{26}\oplus\mathfrak{u}_1
$&$(R)$\\
    &       &$\supset$&
$\mathfrak{so}_{27};\
\mathfrak{su}_2\oplus\mathfrak{so}_{25};\
\mathfrak{usp}_4\oplus\mathfrak{so}_{23};\
\mathfrak{so}_7\oplus\mathfrak{so}_{21};\
\mathfrak{so}_9\oplus\mathfrak{so}_{19};\
\mathfrak{so}_{11}\oplus\mathfrak{so}_{17};\
\mathfrak{so}_{13}\oplus\mathfrak{so}_{15};\
\mathfrak{su}_2\oplus\mathfrak{usp}_{14}
$&$(S)$\\
$15$&$\mathfrak{so}_{30}$&$\supset$&
$\mathfrak{su}_{15}\oplus\mathfrak{u}_1;\
\mathfrak{su}_2\oplus\mathfrak{su}_2\oplus\mathfrak{so}_{26};\
\mathfrak{su}_4\oplus\mathfrak{so}_{24};\
\mathfrak{so}_{8}\oplus\mathfrak{so}_{22};\
\mathfrak{so}_{10}\oplus\mathfrak{so}_{20};\
\mathfrak{so}_{12}\oplus\mathfrak{so}_{18};\
\mathfrak{so}_{14}\oplus\mathfrak{so}_{16};\
\mathfrak{so}_{28}\oplus\mathfrak{u}_1
$&$(R)$\\
    &       &$\supset$&
$\mathfrak{so}_{29};\
\mathfrak{su}_2\oplus\mathfrak{so}_{27};\
\mathfrak{usp}_4\oplus\mathfrak{so}_{25};\
\mathfrak{so}_7\oplus\mathfrak{so}_{23};\
\mathfrak{so}_9\oplus\mathfrak{so}_{21};\
\mathfrak{so}_{11}\oplus\mathfrak{so}_{19};\
\mathfrak{so}_{13}\oplus\mathfrak{so}_{17};\
\mathfrak{so}_{15}\oplus\mathfrak{so}_{15};\
\mathfrak{su}_2\oplus\mathfrak{so}_{10};\
\mathfrak{usp}_{4}\oplus\mathfrak{su}_{4}
$&$(S)$\\
$16$&$\mathfrak{so}_{32}$&$\supset$&
$\mathfrak{su}_{16}\oplus\mathfrak{u}_1;\
\mathfrak{su}_2\oplus\mathfrak{su}_2\oplus\mathfrak{so}_{28};\
\mathfrak{su}_4\oplus\mathfrak{so}_{26};\
\mathfrak{so}_{8}\oplus\mathfrak{so}_{24};\
\mathfrak{so}_{10}\oplus\mathfrak{so}_{22};\
\mathfrak{so}_{12}\oplus\mathfrak{so}_{20};\
\mathfrak{so}_{14}\oplus\mathfrak{so}_{18};\
\mathfrak{so}_{16}\oplus\mathfrak{so}_{16};\
\mathfrak{so}_{30}\oplus\mathfrak{u}_1
$&$(R)$\\
    &       &$\supset$&
$\mathfrak{so}_{31};\
\mathfrak{su}_2\oplus\mathfrak{so}_{29};\
\mathfrak{usp}_4\oplus\mathfrak{so}_{27};\
\mathfrak{so}_7\oplus\mathfrak{so}_{25};\
\mathfrak{so}_9\oplus\mathfrak{so}_{23};\
\mathfrak{so}_{11}\oplus\mathfrak{so}_{21};\
\mathfrak{so}_{13}\oplus\mathfrak{so}_{19};\
\mathfrak{so}_{15}\oplus\mathfrak{so}_{17};\
\mathfrak{usp}_{4}\oplus\mathfrak{usp}_{8};\
\mathfrak{su}_2\oplus\mathfrak{usp}_{16}
$&$(S)$\\
\end{longtable}
\end{center}
where Type $(R)$ and $(S)$ stand for $R$- and $S$-subalgebras,
respectively. 

We summarize generic projection matrices of the Lie algebra
$D_n=\mathfrak{so}_{2n}$ and any maximal regular subalgebras, which are
calculated by using a method given in
Sec.~\ref{Sec:Representations-subalgebras}. 
The Dynkin diagram of 
$D_{n}\supset A_{n-1}\oplus\mathfrak{u}_1$ $(n\geq 4)$
is given by
\begin{align}
&\underset{D_{n}=\mathfrak{so}_{2n}}
{\xygraph{
    \circ ([]!{+(0,-.3)} {{}_2}) (
        - []!{+(-1,.5)}  \circ ([]!{+(0,-.3)} {{}_1}),
        - []!{+(-1,-.5)} \circ ([]!{+(0,-.3)} {{}_x}))
        - [r] 
    \circ ([]!{+(0,-.3)} {{}_3})      - [r]
    \cdots                            - [r]
    \circ ([]!{+(0,-.3)} {{}_{n-3}})  - [r]
    \circ ([]!{+(0,-.3)} {{}_{n-2}}) (
        - []!{+(1,.5)}  \circ ([]!{+(0,-.3)} {{}_{n-1}}),
        - []!{+(1,-.5)} \circ ([]!{+(0,-.3)} {{}_{n}})
)}}\nonumber\\
\underset{\rm Removing}{\Longrightarrow}
&\underset{A_{n-1}\oplus\mathfrak{u}_1=
\mathfrak{su}_{n}\oplus\mathfrak{u}_1}
{\xygraph{
    \circ ([]!{+(0,-.3)} {{}_2}) (
        - []!{+(-1,.5)}  \circ ([]!{+(0,-.3)} {{}_1}),
        - []!{+(-1,-.5)} {\circ\hspace{-0.645em}\times} ([]!{+(0,-.3)} {{}_x}))
        - [r] 
    \circ ([]!{+(0,-.3)} {{}_3})      - [r]
    \cdots                            - [r]
    \circ ([]!{+(0,-.3)} {{}_{n-3}})  - [r]
    \circ ([]!{+(0,-.3)} {{}_{n-2}}) (
        - []!{+(1,.5)}  \circ ([]!{+(0,-.3)} {{}_{n-1}}),
        - []!{+(1,-.5)} {\circ\hspace{-0.645em}\times} ([]!{+(0,-.3)} {{}_{n}})
)}}.
\end{align}
Its corresponding projection matrix is
\begin{align}
P_{\mathfrak{so}_{2n}\supset\mathfrak{su}_{n}\oplus\mathfrak{u}_1}
&=
\left(
\begin{array}{cc}
\multicolumn{1}{c|}{I_{n-1}}&0\\\hline
\multicolumn{1}{c}{2\ 4\ 6\ \cdots\ 2n-4\ n-2}&n\\
\end{array}
\right),
\end{align}
where 
$P_{\mathfrak{so}_{2n}\supset\mathfrak{su}_{n}\oplus\mathfrak{u}_1}$
is an $n\times n$ matrix;
$I_{n-1}$ is an $(n-1)\times (n-1)$ matrix, where its matrix
elements $(I_{n-1})_{ii}=1 (i=1,2,\cdots,n-1)$ and the other matrix
elements $(I_{n-1})_{ij}=0 (i,j=1,2,\cdots,n-1;i\not=j)$.
The Dynkin diagram of 
$D_{n}\supset D_{n-1}\oplus\mathfrak{u}_1$ $(n\geq 4)$
is given by
\begin{align}
&\underset{D_{n}=\mathfrak{so}_{2n}}
{\xygraph{
    \circ ([]!{+(0,-.3)} {{}_2}) (
        - []!{+(-1,.5)}  \circ ([]!{+(0,-.3)} {{}_1}),
        - []!{+(-1,-.5)} \circ ([]!{+(0,-.3)} {{}_x}))
        - [r] 
    \circ ([]!{+(0,-.3)} {{}_{3}})    - [r]
    \cdots                            - [r]
    \circ ([]!{+(0,-.3)} {{}_{n-3}})  - [r]
    \circ ([]!{+(0,-.3)} {{}_{n-2}}) (
        - []!{+(1,.5)}  \circ ([]!{+(0,-.3)} {{}_{n-1}}),
        - []!{+(1,-.5)} \circ ([]!{+(0,-.3)} {{}_{n}})
)}}\nonumber\\
\underset{\rm Removing}{\Longrightarrow}
&\underset{D_{n-1}\oplus\mathfrak{u}_1=
\mathfrak{so}_{2n-2}\oplus\mathfrak{u}_1}
{\xygraph{
    \circ ([]!{+(0,-.3)} {{}_2}) (
        - []!{+(-1,.5)}  {\circ\hspace{-0.645em}\times} ([]!{+(0,-.3)} {{}_1}),
        - []!{+(-1,-.5)} {\circ\hspace{-0.645em}\times} ([]!{+(0,-.3)} {{}_x}))
        - [r] 
    \circ ([]!{+(0,-.3)} {{}_{3}})    - [r]
    \cdots                            - [r]
    \circ ([]!{+(0,-.3)} {{}_{n-3}})  - [r]
    \circ ([]!{+(0,-.3)} {{}_{n-2}}) (
        - []!{+(1,.5)}  \circ ([]!{+(0,-.3)} {{}_{n-1}}),
        - []!{+(1,-.5)} \circ ([]!{+(0,-.3)} {{}_{n}})
)}}.
\end{align}
Its corresponding projection matrix is
\begin{align}
P_{\mathfrak{so}_{2n}\supset\mathfrak{so}_{2n-2}\oplus\mathfrak{u}_1}
&=
\left(
\begin{array}{cc}
\multicolumn{1}{c|}{O}&\multicolumn{1}{c}{I_{n-1}}\\\hline
\multicolumn{1}{c}{2}&{2\ \cdots\ 2\ 1\ 1}\\
\end{array}
\right),
\end{align}
where 
$P_{\mathfrak{so}_{2n}\supset\mathfrak{so}_{2n-2}\oplus\mathfrak{u}_1}$
is an $n\times n$ matrix.
The Dynkin diagram of 
$D_{n}\supset A_1\oplus A_1\oplus D_{n-2}$ $(n\geq 4)$
is given by
\begin{align}
&\underset{D_{n}=\mathfrak{so}_{2n}}
{\xygraph{
    \circ ([]!{+(0,-.3)} {{}_2}) (
        - []!{+(-1,.5)}  \circ ([]!{+(0,-.3)} {{}_1}),
        - []!{+(-1,-.5)} \circ ([]!{+(0,-.3)} {{}_x}))
        - [r] 
    \circ ([]!{+(0,-.3)} {{}_3})      - [r]
    \cdots                            - [r]
    \circ ([]!{+(0,-.3)} {{}_{n-3}})  - [r]
    \circ ([]!{+(0,-.3)} {{}_{n-2}}) (
        - []!{+(1,.5)}  \circ ([]!{+(0,-.3)} {{}_{n-1}}),
        - []!{+(1,-.5)} \circ ([]!{+(0,-.3)} {{}_{n}})
)}}\nonumber\\
\underset{\rm Removing}{\Longrightarrow}&
\underset{A_1\oplus A_1\oplus D_{n-2}
=\mathfrak{su}_{2}\oplus\mathfrak{su}_2\oplus\mathfrak{so}_{2n-4}}
{\xygraph{
    {\circ\hspace{-0.645em}\times} ([]!{+(0,-.3)} {{}_2}) (
        - []!{+(-1,.5)}  \circ ([]!{+(0,-.3)} {{}_1}),
        - []!{+(-1,-.5)} \circ ([]!{+(0,-.3)} {{}_x}))
        - [r] 
    \circ ([]!{+(0,-.3)} {{}_3})      - [r]
    \cdots                            - [r]
    \circ ([]!{+(0,-.3)} {{}_{n-3}})  - [r]
    \circ ([]!{+(0,-.3)} {{}_{n-2}}) (
        - []!{+(1,.5)}  \circ ([]!{+(0,-.3)} {{}_{n-1}}),
        - []!{+(1,-.5)} \circ ([]!{+(0,-.3)} {{}_{n}})
)}}.
\end{align}
Its corresponding projection matrix is
\begin{align}
P_{\mathfrak{so}_{2n}\supset
\mathfrak{su}_{2}\oplus\mathfrak{su}_2\oplus\mathfrak{so}_{2n-4}}
&=
\left(
\begin{array}{ccc}
\multicolumn{1}{c|}{1}&\multicolumn{2}{c}{O}\\\hline
\multicolumn{1}{c}{-1}&-2&\multicolumn{1}{c}{-2\ \cdots -2\ -1\ -1}\\\hline
\multicolumn{2}{c|}{O}&I_{n-2}\\
\end{array}
\right),
\end{align}
where 
$P_{\mathfrak{so}_{2n}\supset
\mathfrak{su}_{2}\oplus\mathfrak{su}_2\oplus\mathfrak{so}_{2n-4}}$
is an $n\times n$ matrix.
The Dynkin diagram of 
$D_{n}\supset A_3\oplus D_{n-3}$ $(n\geq 4)$
is given by
\begin{align}
&\underset{D_{n}=\mathfrak{so}_{2n}}
{\xygraph{
    \circ ([]!{+(0,-.3)} {{}_2}) (
        - []!{+(-1,.5)}  \circ ([]!{+(0,-.3)} {{}_1}),
        - []!{+(-1,-.5)} \circ ([]!{+(0,-.3)} {{}_x}))
        - [r] 
    \circ ([]!{+(0,-.3)} {{}_3})      - [r]
    \cdots                            - [r]
    \circ ([]!{+(0,-.3)} {{}_{n-3}})  - [r]
    \circ ([]!{+(0,-.3)} {{}_{n-2}}) (
        - []!{+(1,.5)}  \circ ([]!{+(0,-.3)} {{}_{n-1}}),
        - []!{+(1,-.5)} \circ ([]!{+(0,-.3)} {{}_{n}})
)}}\nonumber\\
\underset{\rm Removing}{\Longrightarrow}&
\underset{A_3\oplus D_{n-3}=\mathfrak{su}_{4}\oplus\mathfrak{so}_{2n-6}}
{\xygraph{
    \circ ([]!{+(0,-.3)} {{}_2}) (
        - []!{+(-1,.5)}  \circ ([]!{+(0,-.3)} {{}_1}),
        - []!{+(-1,-.5)} \circ ([]!{+(0,-.3)} {{}_x}))
        - [r] 
    {\circ\hspace{-0.645em}\times} ([]!{+(0,-.3)} {{}_3}) - [r]
    \cdots                            - [r]
    \circ ([]!{+(0,-.3)} {{}_{n-3}})  - [r]
    \circ ([]!{+(0,-.3)} {{}_{n-2}}) (
        - []!{+(1,.5)}  \circ ([]!{+(0,-.3)} {{}_{n-1}}),
        - []!{+(1,-.5)} \circ ([]!{+(0,-.3)} {{}_{n}})
)}}.
\end{align}
Its corresponding projection matrix is
\begin{align}
P_{\mathfrak{so}_{2n}\supset
\mathfrak{su}_{4}\oplus\mathfrak{so}_{2n-6}}
&=
\left(
\begin{array}{ccc}
\multicolumn{1}{c|}{I_2}&\multicolumn{2}{c}{O}\\\hline
\multicolumn{1}{c}{-1\ -2}&-2&\multicolumn{1}{c}{-2\ \cdots -2\ -1\ -1}\\\hline
\multicolumn{2}{c|}{O}&I_{n-3}\\
\end{array}
\right),
\end{align}
where 
$P_{\mathfrak{so}_{2n}\supset
\mathfrak{su}_{4}\oplus\mathfrak{so}_{2n-6}}$
is an $n\times n$ matrix.
The Dynkin diagram of 
$D_{n}\supset D_{k}\oplus D_{n-k}$ $(n\geq 4)$
is given by
\begin{align}
&\underset{D_{n}=\mathfrak{so}_{2n}}
{\xygraph{
    \circ ([]!{+(0,-.3)} {{}_2}) (
        - []!{+(-1,.5)}  \circ ([]!{+(0,-.3)} {{}_1}),
        - []!{+(-1,-.5)} \circ ([]!{+(0,-.3)} {{}_x}))
        - [r] 
    \cdots                           - [r]
    \circ ([]!{+(0,-.3)} {{}_{k-1}}) - [r]
    \circ ([]!{+(0,-.3)} {{}_k})     - [r]
    \circ ([]!{+(0,-.3)} {{}_{k+1}}) - [r]
    \cdots                       - [r]
    \circ ([]!{+(0,-.3)} {{}_{n-2}}) (
        - []!{+(1,.5)}  \circ ([]!{+(0,-.3)} {{}_{n-1}}),
        - []!{+(1,-.5)} \circ ([]!{+(0,-.3)} {{}_{n}})
)}}\nonumber\\
\underset{\rm Removing}{\Longrightarrow}&
\underset{D_{k}\oplus D_{n-k}
=\mathfrak{so}_{2k}\oplus\mathfrak{so}_{2n-2k}}
{\xygraph{
    \circ ([]!{+(0,-.3)} {{}_2}) (
        - []!{+(-1,.5)}  \circ ([]!{+(0,-.3)} {{}_1}),
        - []!{+(-1,-.5)} \circ ([]!{+(0,-.3)} {{}_x}))
        - [r] 
    \cdots                           - [r]
    \circ ([]!{+(0,-.3)} {{}_{k-1}}) - [r]
    {\circ\hspace{-0.645em}\times} ([]!{+(0,-.3)} {{}_k})     - [r]
    \circ ([]!{+(0,-.3)} {{}_{k+1}}) - [r]
    \cdots                       - [r]
    \circ ([]!{+(0,-.3)} {{}_{n-2}}) (
        - []!{+(1,.5)}  \circ ([]!{+(0,-.3)} {{}_{n-1}}),
        - []!{+(1,-.5)} \circ ([]!{+(0,-.3)} {{}_{n}})
)}}.
\end{align}
Its corresponding projection matrix is
\begin{align}
P_{\mathfrak{so}_{2n}\supset
\mathfrak{so}_{2k}\oplus\mathfrak{so}_{2n-2k}}
&=
\left(
\begin{array}{ccc}
\multicolumn{1}{c|}{E_{k-1}}&\multicolumn{2}{c}{O}\\\hline
\multicolumn{1}{c}{-1\ -2\cdots\ -2}&-2&\multicolumn{1}{c}{-2\ \cdots -2\ -1\ -1}\\\hline
\multicolumn{2}{c|}{O}&I_{n-k}\\
\end{array}
\right),
\end{align}
where 
$P_{\mathfrak{so}_{2n}\supset
\mathfrak{so}_{2k}\oplus\mathfrak{so}_{2n-2k}}$
is an $n\times n$ matrix;
$E_{k-1}$ is an $(k-1)\times (k-1)$ matrix, where its matrix
elements $(E_{k-1})_{i,n-i}=1 (i=1,2,\cdots,k-1)=1$ and the other matrix
elements $(E_{k-1})_{i,n-j}=0 (i,j=1,2,\cdots,k-1;i\not=j)$.

We summarize generic projection matrices of the Lie algebra
$D_n=\mathfrak{so}_{2n}$ and several maximal special subalgebras
given in Tables~\ref{Table:Maximal-S-sub-classical-1}
and \ref{Table:Maximal-S-sub-classical-2}, which are also calculated by
using a method given in Sec.~\ref{Sec:Representations-subalgebras}. 
(The following projection matrices cannot cover all maximal special
subalgebras, so we need to calculate exceptions one by one.)
The Dynkin diagram of 
$D_{n}\supset B_{n-1}$ $(n\geq 4)$
is given by
\begin{align}
&\underset{D_{n}=\mathfrak{so}_{2n}}
{\xygraph{
    \circ ([]!{+(0,-.3)} {{}_2}) (
        - []!{+(-1,.5)}  \circ ([]!{+(0,-.3)} {{}_1}),
        - []!{+(-1,-.5)} \circ ([]!{+(0,-.3)} {{}_x}))
        - [r] 
    \circ   ([]!{+(0,-.3)} {{}_{3}})   - [r]
    \cdots                           - [r]
    \circ ([]!{+(0,-.3)} {{}_{n-2}}) (
        - []!{+(1,.5)}  \circ ([]!{+(0,-.3)} {{}_{n-1}}),
        - []!{+(1,-.5)} \circ ([]!{+(0,-.3)} {{}_{n}})
)}}\nonumber\\
\underset{\rm Removing+Folding}{\Longrightarrow}&
\underset{B_{n-1}=\mathfrak{so}_{2n-1}}
{\xygraph{!~:{@{=}|@{}}
    \circ ([]!{+(0,-.3)} {{}_2}) (
        - []!{+(-1,.5)}  \circ ([]!{+(0,-.3)} {{}_1}),
        - []!{+(-1,-.5)} {\circ\hspace{-0.645em}\times} ([]!{+(0,-.3)} {{}_x}))
        - [r] 
    \circ   ([]!{+(0,-.3)} {{}_{3}})   - [r]
    \cdots                             - [r]
    \circ   ([]!{+(0,-.3)} {{}_{n-2}}) : [r]
    \bullet ([]!{+(0,-.3)} {{}_{n,n-1}})
}}.
\end{align}
Its corresponding projection matrix is
\begin{align}
P_{\mathfrak{so}_{2n}\supset\mathfrak{so}_{2n-1}}
&=
\left(
\begin{array}{ccc}
\multicolumn{1}{c|}{I_{n-2}}&\multicolumn{2}{c}{O}\\\hline
\multicolumn{1}{c|}{O}&1&1\\
\end{array}
\right),
\end{align}
where
$P_{\mathfrak{so}_{2n}\supset\mathfrak{so}_{2n-1}}$
is an $(n-1)\times n$ matrix.
The Dynkin diagram of 
$D_{n}\supset A_1\oplus B_{n-2}$ $(n\geq 4)$
is given by
\begin{align}
&\underset{D_{n}=\mathfrak{so}_{2n}}
{\xygraph{
    \circ ([]!{+(0,-.3)} {{}_2}) (
        - []!{+(-1,.5)}  \circ ([]!{+(0,-.3)} {{}_1}),
        - []!{+(-1,-.5)} \circ ([]!{+(0,-.3)} {{}_x}))
        - [r] 
    \circ   ([]!{+(0,-.3)} {{}_{3}})   - [r]
    \cdots                           - [r]
    \circ ([]!{+(0,-.3)} {{}_{n-2}}) (
        - []!{+(1,.5)}  \circ ([]!{+(0,-.3)} {{}_{n-1}}),
        - []!{+(1,-.5)} \circ ([]!{+(0,-.3)} {{}_{n}})
)}}\nonumber\\
\underset{\rm Folding+Removing}{\Longrightarrow}
&\underset{A_{1}\oplus B_{n-2}=\mathfrak{su}_{2}\oplus\mathfrak{so}_{2n-3}}
{\xygraph{!~:{@{=}|@{}}
    \bullet ([]!{+(0,-.3)} {{}_{1,\cdots,n}})     [r]
    \circ   ([]!{+(0,-.3)} {{}_{k}})   - [r]
    \cdots                             - [r]
    \circ   ([]!{+(0,-.3)} {{}_{n-2}}) : [r]
    \bullet ([]!{+(0,-.3)} {{}_{n,n-1}})
}}.
\end{align}
Its corresponding projection matrix is
\begin{align}
P_{\mathfrak{so}_{2n}\supset\mathfrak{su}_2\oplus\mathfrak{so}_{2k-3}}
&=
\left(
\begin{array}{cccc}
2&\multicolumn{1}{c}{2\ \cdots 2}&1&1\\\hline
O&\multicolumn{1}{|c|}{I_{n-3}}&\multicolumn{2}{c}{O}\\\hline
\multicolumn{2}{c|}{O}&1&1\\
\end{array}
\right),
\end{align}
where 
$P_{\mathfrak{so}_{2n}\supset\mathfrak{su}_2\oplus\mathfrak{so}_{2k-3}}$
is an $(n-1)\times n$ matrix.
The Dynkin diagram of 
$D_{n}\supset C_2\oplus B_{n-3}$ $(n\geq 4)$
is given by
\begin{align}
&\underset{D_{n}=\mathfrak{so}_{2n}}
{\xygraph{
    \circ ([]!{+(0,-.3)} {{}_2}) (
        - []!{+(-1,.5)}  \circ ([]!{+(0,-.3)} {{}_1}),
        - []!{+(-1,-.5)} \circ ([]!{+(0,-.3)} {{}_x}))
        - [r] 
    \circ ([]!{+(0,-.3)} {{}_3}) - [r]
    \cdots                       - [r]
    \circ ([]!{+(0,-.3)} {{}_{n-2}}) (
        - []!{+(1,.5)}  \circ ([]!{+(0,-.3)} {{}_{n-1}}),
        - []!{+(1,-.5)} \circ ([]!{+(0,-.3)} {{}_{n}})
)}}\nonumber\\
\underset{\rm Folding+Removing}{\Longrightarrow}&
\underset{A_{1}\oplus B_{n-2}=\mathfrak{su}_{2}\oplus\mathfrak{so}_{2n-5}}
{\xygraph{!~:{@{=}|@{}}
    \circ   ([]!{+(0,-.3)} {{}_{1,2,\cdots,n}}) : [r]
    \bullet ([]!{+(0,-.3)} {{}_{1}})              [r]
    \circ   ([]!{+(0,-.3)} {{}_{3}})   - [r]
    \cdots                             - [r]
    \circ   ([]!{+(0,-.3)} {{}_{n-2}}) : [r]
    \bullet ([]!{+(0,-.3)} {{}_{n,n-1}})
}}.
\end{align}
Its corresponding projection matrix is
\begin{align}
P_{\mathfrak{so}_{2n}\supset\mathfrak{usp}_{4}\oplus\mathfrak{so}_{2n-5}}
&=
\left(
\begin{array}{ccccc}
0&2\ \ 2 &\multicolumn{1}{c}{2\ \cdots 2}&1&1\\\hline
1&\multicolumn{4}{|c}{O}\\\hline
\multicolumn{2}{c}{O}&\multicolumn{2}{|c|}{I_{n-4}}
&\multicolumn{1}{c}{O}\\\hline
\multicolumn{3}{c|}{O}&-1&1\\
\end{array}
\right),
\end{align}
where 
$P_{\mathfrak{so}_{2n}\supset\mathfrak{usp}_{4}\oplus\mathfrak{so}_{2n-5}}$
is an $(n-1)\times n$ matrix.
The Dynkin diagram of 
$D_{n}\supset B_k\oplus B_{n-k-1}$ $(n\geq 4)$
is given by
\begin{align}
&\underset{D_{n}=\mathfrak{so}_{2n}}
{\xygraph{
    \circ ([]!{+(0,-.3)} {{}_2}) (
        - []!{+(-1,.5)}  \circ ([]!{+(0,-.3)} {{}_1}),
        - []!{+(-1,-.5)} \circ ([]!{+(0,-.3)} {{}_x}))
        - [r] 
    \cdots                           - [r]
    \circ ([]!{+(0,-.3)} {{}_{k-1}}) - [r]
    \circ ([]!{+(0,-.3)} {{}_k})     - [r]
    \circ ([]!{+(0,-.3)} {{}_{k+1}}) - [r]
    \cdots                       - [r]
    \circ ([]!{+(0,-.3)} {{}_{n-2}}) (
        - []!{+(1,.5)}  \circ ([]!{+(0,-.3)} {{}_{n-1}}),
        - []!{+(1,-.5)} \circ ([]!{+(0,-.3)} {{}_{n}})
)}}\nonumber\\
\underset{\rm Folding+Removing}{\Longrightarrow}&
\underset{B_{k}\oplus B_{n-k-1}=
\mathfrak{so}_{2k+1}\oplus\mathfrak{so}_{2n-2k-1}}
{\xygraph{!~:{@{=}|@{}}
    \circ   ([]!{+(0,-.3)} {{}_{1}})           - [r]
    \circ   ([]!{+(0,-.3)} {{}_{2}})           - [r]
    \cdots                             - [r]
    \circ   ([]!{+(0,-.3)} {{}_{k-1}})         : [r]
    \bullet ([]!{+(0,-.3)} {{}_{k,\cdots,n}})    [r]
    \circ   ([]!{+(0,-.3)} {{}_{k+1}}) - [r]
    \cdots                             - [r]
    \circ   ([]!{+(0,-.3)} {{}_{n-2}}) : [r]
    \bullet ([]!{+(0,-.3)} {{}_{n,n-1}})
}}.
\end{align}
Its corresponding projection matrix is
\begin{align}
P_{\mathfrak{so}_{2n}\supset
\mathfrak{so}_{2k+1}\oplus\mathfrak{so}_{2n-2k-1}}
&=
\left(
\begin{array}{ccccc}
I_{k-1}&\multicolumn{4}{|c}{O}\\\hline
O&\multicolumn{1}{c!}{2\ \ 2}&\multicolumn{1}{c}{2\ \cdots 2}&1&1\\\hline
\multicolumn{2}{c}{O}&\multicolumn{2}{|c|}{I_{n-k-2}}
&\multicolumn{1}{c}{O}\\\hline
\multicolumn{3}{c|}{O}&-1&1\\
\end{array}
\right),
\end{align}
where 
$P_{\mathfrak{so}_{2n}\supset
\mathfrak{so}_{2k+1}\oplus\mathfrak{so}_{2n-2k-1}}$
is an $(n-1)\times n$ matrix.
The Dynkin diagram of 
$D_{\frac{mn}{2}}\supset D_{\frac{m}{2}}\oplus D_{\frac{n}{2}}$ 
$(m,n\in\mathbb{Z}_{2k})$
is given by
\begin{align}
&\underset{D_{\frac{mn}{2}}=\mathfrak{so}_{mn}}
{\xygraph{
    \circ ([]!{+(0,-.3)} {{}_2}) (
        - []!{+(-1,.5)}  \circ ([]!{+(0,-.3)} {{}_1}),
        - []!{+(-1,-.5)} \circ ([]!{+(0,-.3)} {{}_x}))
        - [r] 
    \cdots                           - [r]
    \circ ([]!{+(0,-.3)} {{}_{k-1}}) - [r]
    \circ ([]!{+(0,-.3)} {{}_k})     - [r]
    \circ ([]!{+(0,-.3)} {{}_{k+1}}) - [r]
    \cdots                       - [r]
    \circ ([]!{+(0,-.3)} {{}_{\frac{mn-4}{2}}}) (
        - []!{+(1,.5)}  \circ ([]!{+(0,-.3)} {{}_{\frac{mn-2}{2}}}),
        - []!{+(1,-.5)} \circ ([]!{+(0,-.3)} {{}_{\frac{mn}{2}}})
)}}\nonumber\\
\underset{\rm Folding+Removing}{\Longrightarrow}
&\underset{D_{\frac{m}{2}}\oplus D_{\frac{n}{2}}=
\mathfrak{so}_{m}\oplus\mathfrak{so}_{n}}
{\xygraph{!~:{@{=}|@{}}
    \circ   ([]!{+(0,-.3)} {{}_1})     - [r]
    \circ   ([]!{+(0,-.3)} {{}_2})     - [r]
    \cdots                             - [r] 
    \circ ([]!{+(0,-.3)} {{}_{\frac{m-4}{2}}}) (
        - []!{+(1,.5)}  \circ ([]!{+(0,-.3)} {{}_{\frac{m-2}{2}}}),
        - []!{+(1,-.5)} \circ ([]!{+(0,-.3)} {{}_{\frac{m}{2}}}))  [r] [r]
    \circ   ([]!{+(0,-.3)} {{}_1})     - [r]
    \circ   ([]!{+(0,-.3)} {{}_2})     - [r]
    \cdots                             - [r] 
    \circ ([]!{+(0,-.3)} {{}_{\frac{n-4}{2}}}) (
        - []!{+(1,.5)}  \circ ([]!{+(0,-.3)} {{}_{\frac{n-2}{2}}}),
        - []!{+(1,-.5)} \circ ([]!{+(0,-.3)} {{}_{\frac{n}{2}}})
)}}.
\end{align}
Its corresponding projection matrix is
{\small
\begin{align}
&P_{\mathfrak{so}_{mn}\supset\mathfrak{so}_{m}\oplus\mathfrak{so}_{n}}
\nonumber\\
&=
\left(
\begin{array}{cccccccccccccccccc}
 \multicolumn{3}{c}{1\ 2\cdots n-1}
&\multicolumn{1}{c}{n}
&\multicolumn{3}{c}{n-1\cdots 2\ 1}
&\multicolumn{6}{|c}{O}\\
\hline
 \multicolumn{4}{c|}{O}
&\multicolumn{3}{c}{1\ 2\cdots n-1}
&\multicolumn{1}{c}{n}
&\multicolumn{1}{c}{\cdots}
&\multicolumn{4}{c}{O}\\
\hline
 \multicolumn{8}{c}{\vdots}
&\multicolumn{1}{c}{\ddots}
&\multicolumn{4}{c}{\vdots}\\
\hline
 \multicolumn{8}{c}{O}
&\multicolumn{1}{c}{\cdots}
&\multicolumn{1}{c}{n}
&\multicolumn{3}{c|}{n-1 \cdots 2\ 1}
&\multicolumn{2}{c}{O}\\
\hline
 \multicolumn{8}{c}{O}
&\multicolumn{1}{c}{\cdots}
&\multicolumn{1}{c}{0}
&\multicolumn{3}{c|}{1\ 2 \cdots n-1}
&\multicolumn{1}{c|}{n}
&\multicolumn{1}{c}{n-1 \cdots 2\ 1}
&\multicolumn{1}{c}{0}\\
 \multicolumn{8}{c}{O}
&\multicolumn{1}{c}{\cdots}
&\multicolumn{1}{c}{0}
&\multicolumn{3}{c|}{1\ 2 \cdots n-1}
&\multicolumn{1}{c|}{n}
&\multicolumn{1}{c}{n+1 \cdots 2n-2\ n-1}
&\multicolumn{1}{c}{n}\\
\hline
 \multicolumn{3}{c}{I_{\frac{n-4}{2}}\ O\ E_{\frac{n-4}{2}}}
&\multicolumn{1}{|c|}{O}
&\multicolumn{3}{c}{I_{\frac{n-4}{2}}\ O\ E_{\frac{n-4}{2}}}
&\multicolumn{1}{|c|}{O}
&\multicolumn{1}{c}{\cdots}
&\multicolumn{1}{|c|}{O}
&\multicolumn{3}{c}{I_{\frac{n-4}{2}}\ O\ E_{\frac{n-4}{2}}}
&\multicolumn{1}{|c|}{O}
&\multicolumn{1}{c}{I_{\frac{n-4}{2}}\ O\ E_{\frac{n-4}{2}}}
&\multicolumn{1}{|c}{O}\\
\hline
 \multicolumn{3}{c}{O\ 101\ O}
&\multicolumn{1}{|c|}{O}
&\multicolumn{3}{c}{O\ 101\ O}
&\multicolumn{1}{|c|}{O}
&\multicolumn{1}{c}{\cdots}
&\multicolumn{1}{|c|}{O}
&\multicolumn{3}{c}{O\ 101\ O}
&\multicolumn{1}{|c|}{O}
&\multicolumn{1}{c}{O\ 101\ O}
&\multicolumn{1}{|c}{O}\\
\hline
 \multicolumn{3}{c}{O\ 121\ O}
&\multicolumn{1}{|c|}{O}
&\multicolumn{3}{c}{O\ 121\ O}
&\multicolumn{1}{|c|}{O}
&\multicolumn{1}{c}{\cdots}
&\multicolumn{1}{|c|}{O}
&\multicolumn{3}{c}{O\ 121\ O}
&\multicolumn{1}{|c|}{O}
&\multicolumn{1}{c}{O\ 121\ O}
&\multicolumn{1}{|c}{O}\\
\end{array}
\right),
\end{align}}
where 
$P_{\mathfrak{so}_{mn}\supset\mathfrak{so}_{m}\oplus\mathfrak{so}_{n}}$
is an $\frac{mn}{2}\times \frac{m+n}{2}$ matrix.
The Dynkin diagram of 
$D_{\frac{mn}{2}}\supset B_{\left[\frac{m}{2}\right]}\oplus 
D_{\frac{n}{2}}$ 
$(m\in \mathbb{Z}_{2k+1}; n\in \mathbb{Z}_{2k})$
is given by
\begin{align}
&\underset{D_{\frac{mn}{2}}=\mathfrak{so}_{mn}}
{\xygraph{
    \circ ([]!{+(0,-.3)} {{}_2}) (
        - []!{+(-1,.5)}  \circ ([]!{+(0,-.3)} {{}_1}),
        - []!{+(-1,-.5)} \circ ([]!{+(0,-.3)} {{}_x}))
        - [r] 
    \cdots                           - [r]
    \circ ([]!{+(0,-.3)} {{}_{k-1}}) - [r]
    \circ ([]!{+(0,-.3)} {{}_k})     - [r]
    \circ ([]!{+(0,-.3)} {{}_{k+1}}) - [r]
    \cdots                       - [r]
    \circ ([]!{+(0,-.3)} {{}_{\frac{mn-4}{2}}}) (
        - []!{+(1,.5)}  \circ ([]!{+(0,-.3)} {{}_{\frac{mn-2}{2}}}),
        - []!{+(1,-.5)} \circ ([]!{+(0,-.3)} {{}_{\frac{mn}{2}}})
)}}\nonumber\\
\underset{\rm Folding+Removing}{\Longrightarrow}
&\underset{B_{\left[\frac{m}{2}\right]}\oplus D_{\frac{n}{2}}=
\mathfrak{so}_{m}\oplus\mathfrak{so}_{n}}
{\xygraph{!~:{@{=}|@{}}
    \circ   ([]!{+(0,-.3)} {{}_1})     - [r]
    \circ   ([]!{+(0,-.3)} {{}_2})     - [r]
    \cdots                             - [r] 
    \circ   ([]!{+(0,-.3)} {{}_{\left[\frac{m-2}{2}\right]}}) : [r]
    \bullet ([]!{+(0,-.3)} {{}_{\left[\frac{m}{2}\right]}})    [r]
    \circ   ([]!{+(0,-.3)} {{}_1})     - [r]
    \circ   ([]!{+(0,-.3)} {{}_2})     - [r]
    \cdots                             - [r] 
    \circ ([]!{+(0,-.3)} {{}_{\frac{n-4}{2}}}) (
        - []!{+(1,.5)}  \circ ([]!{+(0,-.3)} {{}_{\frac{n-2}{2}}}),
        - []!{+(1,-.5)} \circ ([]!{+(0,-.3)} {{}_{\frac{n}{2}}})
)}}.
\end{align}
Its corresponding projection matrix is
\begin{align}
&P_{\mathfrak{so}_{mn}\supset\mathfrak{so}_{m}\oplus\mathfrak{so}_{n}}
\nonumber\\
&=
\left(
\begin{array}{cccccccccccccccccc}
 \multicolumn{3}{c}{1\ 2\cdots n-1}
&\multicolumn{1}{c}{n}
&\multicolumn{3}{c}{n-1\cdots 2\ 1}
&\multicolumn{6}{|c}{O}\\
\hline
 \multicolumn{4}{c|}{O}
&\multicolumn{3}{c}{1\ 2\cdots n-1}
&\multicolumn{1}{c}{n}
&\multicolumn{1}{c}{\cdots}
&\multicolumn{4}{c}{O}\\
\hline
 \multicolumn{8}{c}{\vdots}
&\multicolumn{1}{c}{\ddots}
&\multicolumn{4}{c}{\vdots}\\
\hline
 \multicolumn{8}{c}{O}
&\multicolumn{1}{c}{\cdots}
&\multicolumn{1}{c}{n}
&\multicolumn{3}{c|}{n-1 \cdots 2\ 1}
&\multicolumn{2}{c}{O}\\
\hline
 \multicolumn{8}{c}{O}
&\multicolumn{1}{c}{\cdots}
&\multicolumn{1}{c}{0}
&\multicolumn{3}{c|}{2\ 4 \cdots 2(n-1)}
&\multicolumn{1}{c|}{2n}
&\multicolumn{1}{c}{2n \cdots 2n}
&\multicolumn{1}{|c}{nn}\\
\hline
 \multicolumn{3}{c}{I_{\frac{n-4}{2}}\ O\ E_{\frac{n-4}{2}}}
&\multicolumn{1}{|c|}{O}
&\multicolumn{3}{c}{I_{\frac{n-4}{2}}\ O\ E_{\frac{n-4}{2}}}
&\multicolumn{1}{|c|}{O}
&\multicolumn{1}{c}{\cdots}
&\multicolumn{1}{|c|}{O}
&\multicolumn{3}{c}{I_{\frac{n-4}{2}}\ O\ E_{\frac{n-4}{2}}}
&\multicolumn{1}{|c|}{O}
&\multicolumn{1}{c}{I_{\frac{n-4}{2}}}
&\multicolumn{1}{|c}{O}\\
\hline
 \multicolumn{3}{c}{O\ 101\ O}
&\multicolumn{1}{|c|}{O}
&\multicolumn{3}{c}{O\ 101\ O}
&\multicolumn{1}{|c|}{O}
&\multicolumn{1}{c}{\cdots}
&\multicolumn{1}{|c|}{O}
&\multicolumn{3}{c}{O\ 101\ O}
&\multicolumn{1}{|c|}{O}
&\multicolumn{1}{c}{O}
&\multicolumn{1}{|c}{10}\\
\hline
 \multicolumn{3}{c}{O\ 121\ O}
&\multicolumn{1}{|c|}{O}
&\multicolumn{3}{c}{O\ 121\ O}
&\multicolumn{1}{|c|}{O}
&\multicolumn{1}{c}{\cdots}
&\multicolumn{1}{|c|}{O}
&\multicolumn{3}{c}{O\ 121\ O}
&\multicolumn{1}{|c|}{O}
&\multicolumn{1}{c}{O}
&\multicolumn{1}{|c}{01}\\
\end{array}
\right),
\end{align}
where 
$P_{\mathfrak{so}_{mn}\supset\mathfrak{so}_{m}\oplus\mathfrak{so}_{n}}$
is an 
$\frac{mn}{2}\times \left(\left[\frac{m}{2}\right]+\frac{n}{2}\right)$
matrix. 
The Dynkin diagram of 
$D_{\frac{mn}{2}}\supset D_{\frac{m}{2}}\oplus B_{\left[\frac{n}{2}\right]}$ 
$(m\in\mathbb{Z}_{2k}; n\in\mathbb{Z}_{2k+1})$
is given by
\begin{align}
&\underset{D_{\frac{mn}{2}}=\mathfrak{so}_{mn}}
{\xygraph{
    \circ ([]!{+(0,-.3)} {{}_2}) (
        - []!{+(-1,.5)}  \circ ([]!{+(0,-.3)} {{}_1}),
        - []!{+(-1,-.5)} \circ ([]!{+(0,-.3)} {{}_x}))
        - [r] 
    \cdots                           - [r]
    \circ ([]!{+(0,-.3)} {{}_{k-1}}) - [r]
    \circ ([]!{+(0,-.3)} {{}_k})     - [r]
    \circ ([]!{+(0,-.3)} {{}_{k+1}}) - [r]
    \cdots                       - [r]
    \circ ([]!{+(0,-.3)} {{}_{\frac{mn-4}{2}}}) (
        - []!{+(1,.5)}  \circ ([]!{+(0,-.3)} {{}_{\frac{mn-2}{2}}}),
        - []!{+(1,-.5)} \circ ([]!{+(0,-.3)} {{}_{\frac{mn}{2}}})
)}}\nonumber\\
\underset{\rm Folding+Removing}{\Longrightarrow}
&\underset{D_{\frac{m}{2}}\oplus B_{\left[\frac{n}{2}\right]}=
\mathfrak{so}_{m}\oplus\mathfrak{so}_{n}}
{\xygraph{!~:{@{=}|@{}}
    \circ   ([]!{+(0,-.3)} {{}_1})     - [r]
    \circ   ([]!{+(0,-.3)} {{}_2})     - [r]
    \cdots                             - [r] 
    \circ ([]!{+(0,-.3)} {{}_{\frac{m-4}{2}}}) (
        - []!{+(1,.5)}  \circ ([]!{+(0,-.3)} {{}_{\frac{m-2}{2}}}),
        - []!{+(1,-.5)} \circ ([]!{+(0,-.3)} {{}_{\frac{m}{2}}}))    [r] [r]
    \circ   ([]!{+(0,-.3)} {{}_1})     - [r]
    \circ   ([]!{+(0,-.3)} {{}_2})     - [r]
    \cdots                             - [r] 
    \circ   ([]!{+(0,-.3)} {{}_{\left[\frac{n-2}{2}\right]}}) : [r]
    \bullet ([]!{+(0,-.3)} {{}_{\left[\frac{n}{2}\right]}})
}}.
\end{align}
Its corresponding projection matrix is
{\small
\begin{align}
&P_{\mathfrak{so}_{mn}\supset\mathfrak{so}_{m}\oplus\mathfrak{so}_{n}}
\nonumber\\
&=
\left(
\begin{array}{cccccccccccccccccc}
 \multicolumn{3}{c}{1\ 2\cdots n-1}
&\multicolumn{1}{c}{n}
&\multicolumn{3}{c}{n-1\cdots 2\ 1}
&\multicolumn{6}{|c}{O}\\
\hline
 \multicolumn{4}{c|}{O}
&\multicolumn{3}{c}{1\ 2\cdots n-1}
&\multicolumn{1}{c}{n}
&\multicolumn{1}{c}{\cdots}
&\multicolumn{4}{c}{O}\\
\hline
 \multicolumn{8}{c}{\vdots}
&\multicolumn{1}{c}{\ddots}
&\multicolumn{4}{c}{\vdots}\\
\hline
 \multicolumn{8}{c}{O}
&\multicolumn{1}{c}{\cdots}
&\multicolumn{1}{c}{n}
&\multicolumn{3}{c|}{n-1 \cdots 2\ 1}
&\multicolumn{2}{c}{O}\\
\hline
 \multicolumn{8}{c}{O}
&\multicolumn{1}{c}{\cdots}
&\multicolumn{1}{c}{0}
&\multicolumn{3}{c|}{1\ 2 \cdots n-1}
&\multicolumn{1}{c|}{n}
&\multicolumn{1}{c}{n-1 \cdots 2\ 1}
&\multicolumn{1}{|c}{0}\\
 \multicolumn{8}{c}{O}
&\multicolumn{1}{c}{\cdots}
&\multicolumn{1}{c}{0}
&\multicolumn{3}{c|}{1\ 2 \cdots n-1}
&\multicolumn{1}{c|}{n}
&\multicolumn{1}{c}{n+1 \cdots 2n-2\ n-1}
&\multicolumn{1}{|c}{n}\\
\hline
 \multicolumn{3}{c}{I_{\left[\frac{n-2}{2}\right]}\ O\ E_{\left[\frac{n-4}{2}\right]}}
&\multicolumn{1}{|c|}{O}
&\multicolumn{3}{c}{I_{\left[\frac{n-2}{2}\right]}\ O\ E_{\left[\frac{n-4}{2}\right]}}
&\multicolumn{1}{|c|}{O}
&\multicolumn{1}{c}{\cdots}
&\multicolumn{1}{|c|}{O}
&\multicolumn{3}{c}{I_{\left[\frac{n-2}{2}\right]}\ O\ E_{\left[\frac{n-4}{2}\right]}}
&\multicolumn{1}{|c|}{O}
&\multicolumn{1}{c}{I_{\left[\frac{n-2}{2}\right]}\ O\ E_{\left[\frac{n-4}{2}\right]}}
&\multicolumn{1}{|c}{O}\\
\hline
 \multicolumn{3}{c}{O\ 22\ O}
&\multicolumn{1}{|c|}{O}
&\multicolumn{3}{c}{O\ 22\ O}
&\multicolumn{1}{|c|}{O}
&\multicolumn{1}{c}{\cdots}
&\multicolumn{1}{|c|}{O}
&\multicolumn{3}{c}{O\ 22\ O}
&\multicolumn{1}{|c|}{O}
&\multicolumn{1}{c}{O\ 22\ O}
&\multicolumn{1}{|c}{O}\\
\end{array}
\right),
\end{align}}
where 
$P_{\mathfrak{so}_{mn}\supset\mathfrak{so}_{m}\oplus\mathfrak{so}_{n}}$
is an $\frac{mn}{2}\times 
\left(\frac{m}{2}+\left[\frac{n}{2}\right]\right)$ matrix.
The Dynkin diagram of 
$D_{2mn}\supset C_{m}\oplus C_{n}$ 
$(m,n\geq 2; m,n\in \mathbb{Z}_{2k})$
is given by
\begin{align}
&\underset{D_{2mn}=\mathfrak{so}_{4mn}}
{\xygraph{
    \circ ([]!{+(0,-.3)} {{}_2}) (
        - []!{+(-1,.5)}  \circ ([]!{+(0,-.3)} {{}_1}),
        - []!{+(-1,-.5)} \circ ([]!{+(0,-.3)} {{}_x}))
        - [r] 
    \cdots                           - [r]
    \circ ([]!{+(0,-.3)} {{}_{k-1}}) - [r]
    \circ ([]!{+(0,-.3)} {{}_k})     - [r]
    \circ ([]!{+(0,-.3)} {{}_{k+1}}) - [r]
    \cdots                       - [r]
    \circ ([]!{+(0,-.3)} {{}_{2mn-2}}) (
        - []!{+(1,.5)}  \circ ([]!{+(0,-.3)} {{}_{2mn-1}}),
        - []!{+(1,-.5)} \circ ([]!{+(0,-.3)} {{}_{2mn}})
)}}\nonumber\\
\underset{\rm Folding+Removing}{\Longrightarrow}
&\underset{C_{m}\oplus C_{n}=
\mathfrak{usp}_{2m}\oplus\mathfrak{usp}_{2n}}
{\xygraph{!~:{@{=}|@{}}
    \bullet ([]!{+(0,-.3)} {{}_1}) - [r]
    \bullet ([]!{+(0,-.3)} {{}_2}) - [r] \cdots - [r]
    \bullet ([]!{+(0,-.3)} {{}_{m-1}}) : [r]
    \circ   ([]!{+(0,-.3)} {{}_m})   [r]
    \bullet ([]!{+(0,-.3)} {{}_1}) - [r]
    \bullet ([]!{+(0,-.3)} {{}_2}) - [r] \cdots - [r]
    \bullet ([]!{+(0,-.3)} {{}_{n-1}}) : [r]
    \circ   ([]!{+(0,-.3)} {{}_n}) [r]
}}.
\end{align}
Its corresponding projection matrix is
{\small
\begin{align}
&P_{\mathfrak{so}_{4mn}\supset\mathfrak{usp}_{2m}\oplus\mathfrak{usp}_{2n}}
\nonumber\\
&=
\left(
\begin{array}{cccccccccccccccccc}
 \multicolumn{3}{c}{1\ 2\cdots 2n-1}
&\multicolumn{1}{c}{2n}
&\multicolumn{3}{c}{2n-1\cdots 2\ 1}
&\multicolumn{6}{|c}{O}\\
\hline
 \multicolumn{4}{c|}{O}
&\multicolumn{3}{c}{1\ 2\cdots 2n-1}
&\multicolumn{1}{c}{2n}
&\multicolumn{1}{c}{\cdots}
&\multicolumn{4}{c}{O}\\
\hline
 \multicolumn{8}{c}{\vdots}
&\multicolumn{1}{c}{\ddots}
&\multicolumn{4}{c}{\vdots}\\
\hline
 \multicolumn{8}{c}{O}
&\multicolumn{1}{c}{\cdots}
&\multicolumn{1}{c}{2n}
&\multicolumn{3}{c|}{2n-1 \cdots 2\ 1}
&\multicolumn{2}{c}{O}\\
\hline
 \multicolumn{8}{c}{O}
&\multicolumn{1}{c}{\cdots}
&\multicolumn{1}{c}{0}
&\multicolumn{3}{c|}{1\ 2 \cdots 2n-1}
&\multicolumn{1}{c|}{2n}
&\multicolumn{1}{c}{2n-1\ 2n-2\cdots 2\ 1}
&\multicolumn{1}{|c}{0}\\
\hline
 \multicolumn{8}{c}{O}
&\multicolumn{1}{c}{\cdots}
&\multicolumn{1}{c}{0}
&\multicolumn{3}{c|}{O}
&\multicolumn{1}{c|}{0}
&\multicolumn{1}{c}{1\ 2\cdots 2n-2\ n-1}
&\multicolumn{1}{|c}{n}\\
\hline
 \multicolumn{3}{c}{I_{n-1}\ O\ E_{n-1}}
&\multicolumn{1}{|c|}{O}
&\multicolumn{3}{c}{I_{n-1}\ O\ E_{n-1}}
&\multicolumn{1}{|c|}{O}
&\multicolumn{1}{c}{\cdots}
&\multicolumn{1}{|c|}{O}
&\multicolumn{3}{c}{I_{n-1}\ O\ E_{n-1}}
&\multicolumn{1}{|c|}{O}
&\multicolumn{1}{c}{I_{n-1}\ O\ E_{n-1}}
&\multicolumn{1}{|c}{O}\\
\hline
 \multicolumn{3}{c}{O\ 1\ O}
&\multicolumn{1}{|c|}{O}
&\multicolumn{3}{c}{O\ 1\ O}
&\multicolumn{1}{|c|}{O}
&\multicolumn{1}{c}{\cdots}
&\multicolumn{1}{|c|}{O}
&\multicolumn{3}{c}{O\ 1\ O}
&\multicolumn{1}{|c|}{O}
&\multicolumn{1}{c}{O\ 1\ O}
&\multicolumn{1}{|c}{O}\\
\end{array}
\right),
\end{align}}
where 
$P_{\mathfrak{so}_{4mn}\supset\mathfrak{usp}_{2m}\oplus\mathfrak{usp}_{2n}}$
is a $2mn\times (m+n)$ matrix.

\subsection{$E_6$}

Let us summarize features of representations of the Lie algebra $E_6$.
From Table~\ref{Table:Dynkin-diagrams}, 
the extended Dynkin diagram of the Lie algebra $E_6$ is 
\begin{align}
\xygraph{
    \circ ([]!{+(0,-.3)}  {{}_1}) - [r]
    \circ ([]!{+(0,-.3)}  {{}_2}) - [r]
    \circ ([]!{+(.3,-.3)} {\hspace{-2em}{}_3}) (
        - [u] \circ ([]!{+(.3,0)}  {{}_6}) (
        - [u] \circ ([]!{+(.3,0)}  {{}_x})
),
        - [r] \circ ([]!{+(0,-.3)} {{}_4})
        - [r] \circ ([]!{+(0,-.3)} {{}_5})
)}.
\end{align}
From Table~\ref{Table:Cartan-matrices}, 
the Cartan matrix of the Lie algebra $E_6$ is
\begin{align}
&A(E_6)=
\left(
\begin{array}{cccccc}
 2&-1& 0& 0& 0& 0\\
-1& 2&-1& 0& 0& 0\\
 0&-1& 2&-1& 0&-1\\
 0& 0&-1& 2&-1& 0\\
 0& 0& 0&-1& 2& 0\\
 0& 0&-1& 0& 0& 2\\
\end{array}
\right).
\end{align}
The complex representations of the Lie algebra $E_6$ satisfy the
following condition
\begin{align}
{\xygraph{
    \circ ([]!{+(0,-.3)}  {a_1}) - [r]
    \circ ([]!{+(0,-.3)}  {a_2}) - [r]
    \circ ([]!{+(.3,-.3)} {\hspace{-2em}a_3}) (
        - [u] \circ ([]!{+(.3,0)}  {a_6}),
        - [r] \circ ([]!{+(0,-.3)} {a_4})
        - [r] \circ ([]!{+(0,-.3)} {a_5})
)}}
\not=
{\xygraph{
    \circ ([]!{+(0,-.3)}  {a_5}) - [r]
    \circ ([]!{+(0,-.3)}  {a_4}) - [r]
    \circ ([]!{+(.3,-.3)} {\hspace{-2em}a_3}) (
        - [u] \circ ([]!{+(.3,0)}  {a_6}),
        - [r] \circ ([]!{+(0,-.3)} {a_2})
        - [r] \circ ([]!{+(0,-.3)} {a_1})
)}}.
\end{align}
From Tables~\ref{Table:Complex-representations-1} and 
\ref{Table:Self-conjugate-representations},
types of representations of $E_6$
are given in the following table.
\begin{center}
\begin{longtable}{clll}
\caption{Types of representations of $E_6$}
\label{Table:Types-representations-E6}\\
\hline\hline
Algebra&Rank&Condition&C/R/PR\\\hline\hline
\endfirsthead
\multicolumn{4}{c}{Table~\ref{Table:Types-representations-E6} (continued)}\\\hline\hline
Algebra&Rank&Condition&C/R/PR\\\hline\hline
\endhead
\hline
\endfoot
$E_6$
&6
&$(a_1a_2a_3a_4a_5a_6)\not=(a_5a_4a_3a_2a_1a_6)$
&Complex\\\cline{3-4}
&
&$(a_1a_2a_3a_4a_5a_6)=(a_5a_4a_3a_2a_1a_6)$
&Real\\
\end{longtable}
\end{center}
where Condition stands for a condition for Dynkin labels, 
and C/R/PR represents real, pseudo-real, and complex
representations.
For the Lie algebra $E_6$, from Eqs.~(\ref{Eq:2-Casimir}) and
(\ref{Eq:Dynkin-index}), 
we get the Casimir invariant and the Dynkin index of a representation
$(a_1a_2a_3a_4a_5a_6)$:
\begin{align}
C_2(R)=&\frac{1}{2}
\sum_{i,j=1}^{6}(a_i+2g_i)G(E_6)_{ij}a_j,\\
T(R)=&\frac{d(R)}{d(G)}C_2(R)
=\frac{d(R)}{2d(G)}
\sum_{i,j=1}^{6}(a_i+2g_i)G(E_6)_{ij}a_j,
\end{align}
where $G(E_6)$ is the metric matrix in
Table~\ref{Table:Cartan-matrices-inverse}
\begin{align}
G(E_6)&=\frac{1}{3}
\left(
\begin{array}{cccccc}
4&5&6&4&2&3\\
5&10&12&8&4&6\\
6&12&18&12&6&9\\
4&8&12&10&5&6\\
2&4&6&5&4&3\\
3&6&9&6&3&6\\
\end{array}
\right).
\end{align}
From Table~\ref{Table:Conjugacy-classes}, the conjugacy classes of the
Lie algebra $E_{6}$ are 
\begin{align}
C_c(R):=a_1-a_2+a_4-a_5\ \ \  (\bmod.\ 3).
\end{align}
The algebra $E_6$ has three conjugacy classes. 

The above features of representations of $E_6$
are summarized in the following table:.
\begin{center}
\begin{longtable}{cccccc}
\caption{Representations of $E_6$}
\label{Table:Representations-summary-E6}\\
\hline\hline
$E_6$ irrep.& $d(R)$& $C_2(R)$ &$T(R)$&$C_c(R)$&C/R/PR\\\hline\hline
\endfirsthead
\multicolumn{6}{c}{Table~\ref{Table:Representations-summary-E6} (continued)}\\\hline\hline
$E_6$ irrep.& $d(R)$& $C_2(R)$ &$T(R)$&$C_c(R)$&C/R/PR\\\hline\hline
\endhead
\hline
\endfoot
(000000)&${\bf 1}$                 &$0$&$0$&0&R\\
(100000)&${\bf 27}$                &$\frac{26}{3}$&$3$&1&C\\
(000010)&${\bf \overline{27}}$     &$\frac{26}{3}$&$3$&2&C\\
(000001)&${\bf 78}$                &$12$&$12$&0&R\\
(000100)&${\bf 351}$               &$\frac{50}{3}$&$75$&1&C\\
(010000)&${\bf \overline{351}}$    &$\frac{50}{3}$&$75$&2&C\\
(000020)&${\bf 351'}$              &$\frac{56}{3}$&$84$&1&C\\
(200000)&${\bf \overline{351'}}$   &$\frac{56}{3}$&$84$&2&C\\
(100010)&${\bf 650}$               &$18$&$150$&0&R\\
(100001)&${\bf 1728}$              &$\frac{65}{3}$&$480$&1&C\\
(000011)&${\bf \overline{1728}}$   &$\frac{65}{3}$&$480$&2&C\\
(000002)&${\bf 2430}$              &$26$&$810$&0&R\\
(001000)&${\bf 2925}$              &$24$&$900$&0&R\\
\end{longtable}
\end{center}
where $E_6$ irrep., $d(R)$, $C_2(R)$, $T(R)$, $C_c(R)$, and C/R/PR
stand for  
the Dynkin label of the irreducible representation of $E_6$,
its dimension, its quadratic Casimir invariant, its Dynkin indices, 
its conjugacy classes, 
complex, real, and pseudo-real representations,
respectively. 
The representation $(000001)$ is the adjoint representation of
$E_6$. 
See Appendix~\ref{Sec:Representations} for further information.

From Table~\ref{Table:Maximal-subalgebra},
maximal $R$- and $S$-subalgebras of $E_6$ are listed in the following
table.
\begin{center}
\begin{longtable}{crcp{10cm}c}
\caption{Maximal subalgebras of $E_6$}
\label{Table:Maximal-subalgebra-E6}\\
\hline\hline
Rank&\multicolumn{2}{l}{Algebra $\mathfrak{g}$}&Maximal subalgebras $\mathfrak{h}$&Type\\\hline\hline
\endfirsthead
\multicolumn{4}{c}{Table~\ref{Table:Maximal-subalgebra-E6} (continued)}\\
\hline
Rank&\multicolumn{2}{l}{Algebra $\mathfrak{g}$}&Maximal subalgebras $\mathfrak{h}$&Type\\\hline\hline
\endhead
\hline
\endfoot
$6$&$E_6$  &$\supset$
&$\mathfrak{so}_{10}\oplus\mathfrak{u}_1;\
\mathfrak{su}_2\oplus\mathfrak{su}_6;\
\mathfrak{su}_3\oplus\mathfrak{su}_3\oplus\mathfrak{su}_3$
&$(R)$\\
    &       &$\supset$
&$F_4;\ 
\mathfrak{su}_3\oplus G_2;\
\mathfrak{usp}_8;\
G_2;\
\mathfrak{su}_3$
&$(S)$\\
\end{longtable}
\end{center}
where Type $(R)$ and $(S)$ stand for $R$- and $S$-subalgebras,
respectively.

\subsection{$E_7$}

Let us summarize some features of representations of the Lie algebra
$E_7$. 
From Table~\ref{Table:Dynkin-diagrams}, 
the extended Dynkin diagram of the Lie algebra $E_7$ is 
\begin{align}
\xygraph{
    \circ ([]!{+(0,-.3)}  {{}_x}) - [r]
    \circ ([]!{+(0,-.3)}  {{}_1}) - [r]
    \circ ([]!{+(0,-.3)}  {{}_2}) - [r]
    \circ ([]!{+(.3,-.3)} {\hspace{-2em}{}_3}) (
        - [u] \circ ([]!{+(.3,0)}  {{}_7}),
        - [r] \circ ([]!{+(0,-.3)} {{}_4})
        - [r] \circ ([]!{+(0,-.3)} {{}_5})
        - [r] \circ ([]!{+(0,-.3)} {{}_6})
)}.
\end{align}
From Table~\ref{Table:Cartan-matrices},
the Cartan matrix of the Lie algebra $E_7$ is
\begin{align}
&A(E_7)=
\left(
\begin{array}{ccccccc}
 2&-1& 0& 0& 0& 0& 0\\
-1& 2&-1& 0& 0& 0& 0\\
 0&-1& 2&-1& 0& 0&-1\\
 0& 0&-1& 2&-1& 0& 0\\
 0& 0& 0&-1& 2&-1& 0\\
 0& 0& 0& 0&-1& 2& 0\\
 0& 0&-1& 0& 0& 0& 2\\
\end{array}
\right).
\end{align}
From Tables~\ref{Table:Complex-representations-1} and 
\ref{Table:Self-conjugate-representations},
types of representations of $E_7$
are given in the following table.
\begin{center}
\begin{longtable}{ccll}
\caption{Types of representations of $E_7$}
\label{Table:Types-representations-E7}\\
\hline\hline
Algebra&Rank&Condition&C/R/PR\\\hline\hline
\endfirsthead
\multicolumn{4}{c}{Table~\ref{Table:Types-representations-E7} (continued)}\\\hline\hline
Algebra&Rank&Condition&C/R/PR\\\hline\hline
\endhead
\hline
\endfoot
$E_7$ 
&7
&$a_4+a_6+a_7=0\ (\bmod.\ 2)$ 
&Real\\
&
&$a_4+a_6+a_7=1\ (\bmod.\ 2)$
&PR\\
\end{longtable}
\end{center}
where Condition stands for a condition for Dynkin labels,
and C/R/PR represent complex, real, and pseudo-real
representations.
For the Lie algebra $E_7$, from Eqs.~(\ref{Eq:2-Casimir}) and
(\ref{Eq:Dynkin-index}), 
we get the Casimir invariant and the Dynkin index of a representation
$(a_1a_2a_3a_4a_5a_6a_7)$:
\begin{align}
C_2(R)=&\frac{1}{2}
\sum_{i,j=1}^{7}(a_i+2g_i)G(E_7)_{ij}a_j,\\
T(R)=&\frac{d(R)}{d(G)}C_2(R)
=\frac{d(R)}{2d(G)}
\sum_{i,j=1}^{7}(a_i+2g_i)G(E_7)_{ij}a_j,
\end{align}
where $G(E_7)$ is the metric matrix in
Table~\ref{Table:Cartan-matrices-inverse}
\begin{align}
G(E_7)&=\frac{1}{2}
\left(
\begin{array}{ccccccc}
4&6&8&6&4&2&4\\
6&12&16&12&8&4&8\\
8&16&24&18&12&6&12\\
6&12&18&15&10&5&9\\
4&8&12&10&8&4&6\\
2&4&6&5&4&3&3\\
4&8&12&9&6&3&7\\
\end{array}
\right).
\end{align}
From Table~\ref{Table:Conjugacy-classes}, the conjugacy classes of  
the Lie algebra $E_7$ are
\begin{align}
C_c(R):=a_4+a_6+a_7\ \ \  (\bmod.\ 2).
\end{align}
The algebra $E_7$ has two conjugacy classes. 

The above features of representations of $E_7$
are summarized in the following table.
\begin{center}
\begin{longtable}{cccccc}
\caption{Representations of $E_7$}
\label{Table:Representations-summary-E7}\\
\hline\hline
$E_7$ irrep.& $d(R)$& $C_2(R)$ &$T(R)$&$C_c(R)$&C/R/PR\\\hline\hline
\endfirsthead
\multicolumn{6}{c}{Table~\ref{Table:Representations-summary-E7} (continued)}\\\hline\hline
$E_7$ irrep.& $d(R)$& $C_2(R)$ &$T(R)$&$C_c(R)$&C/R/PR\\\hline\hline
\endhead
\hline
\endfoot
(0000000)&${\bf 1}$      &$0$&$0$&0&R\\
(0000010)&${\bf 56}$     &$\frac{57}{4}$&$6$&1&PR\\
(1000000)&${\bf 133}$    &$18$&$18$&0&R\\
(0000001)&${\bf 912}$    &$\frac{105}{4}$&$180$&1&PR\\
(0000020)&${\bf 1463}$   &$30$&$330$&0&R\\
(0000100)&${\bf 1539}$   &$28$&$324$&0&R\\
(1000010)&${\bf 6480}$   &$\frac{133}{4}$&$1620$&1&PR\\
(2000000)&${\bf 7371}$   &$38$&$2106$&0&R\\
(0100000)&${\bf 8645}$   &$36$&$2340$&0&R\\
(0000030)&${\bf 24320}$  &$\frac{189}{4}$&$8640$&1&PR\\
(0001000)&${\bf 27664}$  &$\frac{165}{4}$&$8580$&1&PR\\
(0000011)&${\bf 40755}$  &$42$&$12870$&0&R\\
\end{longtable}
\end{center}
where $E_7$ irrep., $d(R)$, $C_2(R)$, $T(R)$, $C_c(R)$, C/R/PR stand
for the Dynkin label of the irreducible representation of $E_7$,
its dimension, its quadratic Casimir invariant, its Dynkin indices, 
its conjugacy classes, and 
complex, real, and pseudo-real representations,
respectively. 
The representation $(1000000)$ is the adjoint representation of
$E_7$. 
See Appendix~\ref{Sec:Representations} for further information.

From Table~\ref{Table:Maximal-subalgebra},
maximal $R$- and $S$-subalgebras of $E_7$ are listed in the following
table.
\begin{center}
\begin{longtable}{crcp{10cm}c}
\caption{Maximal subalgebras of $E_7$}
\label{Table:Maximal-subalgebra-E7}\\
\hline\hline
Rank&\multicolumn{2}{l}{Algebra $\mathfrak{g}$}&Maximal subalgebras $\mathfrak{h}$&Type\\\hline\hline
\endfirsthead
\multicolumn{4}{c}{Table~\ref{Table:Maximal-subalgebra-E7} (continued)}\\
\hline
Rank&\multicolumn{2}{l}{Algebra $\mathfrak{g}$}&Maximal subalgebras $\mathfrak{h}$&Type\\\hline\hline
\endhead
\hline
\endfoot
$7$ &$E_7$  &$\supset$&$E_6\oplus\mathfrak{u}_1;\
\mathfrak{su}_8;\
\mathfrak{so}_{12}\oplus\mathfrak{su}_2;\
\mathfrak{su}_3\oplus\mathfrak{su}_6$
&$(R)$\\
    &       &$\supset$&$\mathfrak{su}_2\oplus F_4;\
G_2\oplus\mathfrak{usp}_6;\
\mathfrak{su}_2\oplus G_2;\
\mathfrak{su}_3;\
\mathfrak{su}_2\oplus\mathfrak{su}_2;\
\mathfrak{su}_2;\
\mathfrak{su}_2$&$(S)$\\
\end{longtable}
\end{center}
where Type $(R)$ and $(S)$ stand for $R$- and $S$-subalgebras,
respectively.

\subsection{$E_8$}

Let us summarize features of representations of the
Lie algebra $E_8$.
From Table~\ref{Table:Dynkin-diagrams}, 
the extended Dynkin diagram of $E_8$ is 
\begin{align}
\xygraph{
    \circ ([]!{+(0,-.3)}  {{}_1}) - [r]
    \circ ([]!{+(0,-.3)}  {{}_2}) - [r]
    \circ ([]!{+(.3,-.3)} {\hspace{-2em}{}_3}) (
        - [u] \circ ([]!{+(.3,0)}  {{}_8}),
        - [r] \circ ([]!{+(0,-.3)} {{}_4})
        - [r] \circ ([]!{+(0,-.3)} {{}_5})
        - [r] \circ ([]!{+(0,-.3)} {{}_6})
        - [r] \circ ([]!{+(0,-.3)} {{}_7})
        - [r] \circ ([]!{+(0,-.3)} {{}_x}) 
)}.
\end{align}
From Table~\ref{Table:Cartan-matrices}, 
the Cartan matrix of $E_8$ is
\begin{align}
&A(E_8)=
\left(
\begin{array}{cccccccc}
 2&-1& 0& 0& 0& 0& 0& 0\\
-1& 2&-1& 0& 0& 0& 0& 0\\
 0&-1& 2&-1& 0& 0& 0&-1\\
 0& 0&-1& 2&-1& 0& 0& 0\\
 0& 0& 0&-1& 2&-1& 0& 0\\
 0& 0& 0& 0&-1& 2&-1& 0\\
 0& 0& 0& 0& 0&-1& 2& 0\\
 0& 0&-1& 0& 0& 0& 0& 2\\
\end{array}
\right).
\end{align}
From Tables~\ref{Table:Complex-representations-1} and 
\ref{Table:Self-conjugate-representations},
types of representations of $E_8$
are given in the following table.
\begin{center}
\begin{longtable}{clll}
\caption{Types of representations of $E_8$}
\label{Table:Types-representations-E8}\\
\hline\hline
Algebra&Rank&Condition&C/R/PR\\\hline\hline
\endfirsthead
\multicolumn{4}{c}{Table~\ref{Table:Types-representations-E8} (continued)}\\\hline\hline
Algebra&Rank&Condition&C/R/PR\\\hline\hline
\endhead
\hline
\endfoot
$E_8$ 
&8
&None
&Real\\
\end{longtable}
\end{center}
where Condition stand for a condition for Dynkin labels,
and C/R/PR represents complex, real, and pseudo-real
representations.
For $E_8$, from Eqs.~(\ref{Eq:2-Casimir}) and (\ref{Eq:Dynkin-index}),
we get the Casimir invariant and the Dynkin index of a representation
$(a_1a_2a_3a_4a_5a_6a_7a_8)$:
\begin{align}
C_2(R)=&\frac{1}{2}
\sum_{i,j=1}^{8}(a_i+2g_i)G(E_8)_{ij}a_j,\\
T(R)=&\frac{d(R)}{d(G)}C_2(R)
=\frac{d(R)}{2d(G)}
\sum_{i,j=1}^{8}(a_i+2g_i)G(E_8)_{ij}a_j,
\end{align}
where $G(E_8)$ is the metric matrix in 
Table~\ref{Table:Cartan-matrices-inverse}
\begin{align}
G(E_8)&=
\left(
\begin{array}{cccccccc}
4&7&10&8&6&4&2&5\\
7&14&20&16&12&8&4&10\\
10&20&30&24&18&12&6&15\\
8&16&24&20&15&10&5&12\\
6&12&18&15&12&8&4&9\\
4&8&12&10&8&6&3&6\\
2&4&6&5&4&3&2&3\\
5&10&15&12&9&6&3&8\\
\end{array}
\right).
\end{align}
From Table~\ref{Table:Conjugacy-classes}, the algebra $E_8$ has only one
conjugacy class. All representations of $E_8$ belong to the same
conjugacy class. 

The above features of representations of $E_8$
are summarized in the following table.
\begin{center}
\begin{longtable}{ccccc}
\caption{Representations of $E_8$}
\label{Table:Representations-summary-E8}\\
\hline\hline
$E_8$ irrep.& $d(R)$& $C_2(R)$ &$T(R)$&C/R/PR\\\hline\hline
\endfirsthead
\multicolumn{5}{c}{Table~\ref{Table:Representations-summary-E8} (continued)}\\\hline\hline
$E_8$ irrep.& $d(R)$& $C_2(R)$ &$T(R)$&C/R/PR\\\hline\hline
\endhead
\hline
\endfoot
(00000000)&${\bf 1}$       &$0$&$0$&R\\
(00000010)&${\bf 248}$     &$30$&$30$&R\\
(10000000)&${\bf 3875}$    &$48$&$750$&R\\
(00000020)&${\bf 27000}$   &$62$&$6750$&R\\
(00000100)&${\bf 30380}$   &$60$&$7350$&R\\
(00000001)&${\bf 147250}$  &$72$&$42750$&R\\
(10000010)&${\bf 779247}$  &$80$&$251370$&R\\
(00000030)&${\bf 1753125}$ &$96$&$682500$&R\\
(00001000)&${\bf 2450240}$ &$90$&$889200$&R\\
(00000110)&${\bf 4096000}$ &$93$&$1536000$&R\\
(20000000)&${\bf 4881384}$  &$100$&$1968300$&R\\
(01000000)&${\bf 6696000}$  &$98$ &$2646000$&R\\
\end{longtable}
\end{center}
where $E_8$ irrep., $d(R)$, $C_2(R)$, $T(R)$, C/R/PR stand for 
the Dynkin label of the irreducible representation of $E_8$,
its dimension, its quadratic Casimir invariant, its Dynkin indices, 
and complex, real, and pseudo-real representations,
respectively. 
The representation $(00000010)$ is the adjoint representation of
$E_8$. 
See Appendix~\ref{Sec:Representations} for further information.

From Table~\ref{Table:Maximal-subalgebra},
maximal $R$- and $S$-subalgebras of $E_8$ are listed in the following
table.
\begin{center}
\begin{longtable}{crcp{10cm}c}
\caption{Maximal subalgebras of $E_8$}
\label{Table:Maximal-subalgebra-E8}\\
\hline\hline
Rank&\multicolumn{2}{l}{Algebra $\mathfrak{g}$}&Maximal subalgebras $\mathfrak{h}$&Type\\\hline\hline
\endfirsthead
\multicolumn{4}{c}{Table~\ref{Table:Maximal-subalgebra-E8} (continued)}\\
\hline
Rank&\multicolumn{2}{l}{Algebra $\mathfrak{g}$}&Maximal subalgebras $\mathfrak{h}$&Type\\\hline\hline
\endhead
\hline
\endfoot
$8$ &$E_8$  &$\supset$&$\mathfrak{so}_{16};\
\mathfrak{su}_5\oplus\mathfrak{su}_5;\
\mathfrak{su}_3\oplus E_6;\
\mathfrak{su}_2\oplus E_7;\
\mathfrak{su}_9$
&$(R)$\\
    &       &$\supset$&
$G_2\oplus F_4;\
\mathfrak{su}_2\oplus\mathfrak{su}_3;\
\mathfrak{usp}_4;\
\mathfrak{su}_2;\
\mathfrak{su}_2;\
\mathfrak{su}_2
$&$(S)$\\
\end{longtable}
\end{center}
where Type $(R)$ and $(S)$ stand for $R$- and $S$-subalgebras,
respectively.

\subsection{$F_4$}

Let us summarize features of representations of the
Lie algebra $F_4$.
From Table~\ref{Table:Dynkin-diagrams}, 
the extended Dynkin diagram of $F_4$ is 
\begin{align}
\xygraph{!~:{@{=}|@{}}
    \circ   ([]!{+(0,-.3)} {{}_x}) - [r]
    \circ   ([]!{+(0,-.3)} {{}_1}) - [r]
    \circ   ([]!{+(0,-.3)} {{}_2}) : [r]
    \bullet ([]!{+(0,-.3)} {{}_3}) - [r]
    \bullet ([]!{+(0,-.3)} {{}_4})}.
\end{align}
From Table~\ref{Table:Cartan-matrices}, 
the Cartan matrix of $F_4$ is
\begin{align}
&A(F_4)=
\left(
\begin{array}{cccc}
 2&-1& 0& 0\\
-1& 2&-2& 0\\
 0&-1& 2&-1\\
 0& 0&-1& 2\\
\end{array}
\right).
\end{align}
From Tables~\ref{Table:Complex-representations-1} and 
\ref{Table:Self-conjugate-representations},
types of representations of $F_4$ 
are given in the following table.
\begin{center}
\begin{longtable}{ccll}
\caption{Types of representations of $F_4$}
\label{Table:Types-representations-F}\\
\hline\hline
Algebra&Rank&Condition&C/R/PR\\\hline\hline
\endfirsthead
\multicolumn{4}{c}{Table~\ref{Table:Types-representations-F} (continued)}\\\hline\hline
Algebra&Rank&Condition&C/R/PR\\\hline\hline
\endhead
\hline
\endfoot
$F_4$ 
&4
&None
&Real\\
\end{longtable}
\end{center}
where Condition stand for a condition for Dynkin labels,
and C/R/PR represent complex, real, and pseudo-real
representations.
For $F_4$, from Eqs.~(\ref{Eq:2-Casimir}) and (\ref{Eq:Dynkin-index}),
we get the Casimir invariant and the Dynkin index of a representation
$(a_1a_2a_3a_4)$:
\begin{align}
C_2(R)=&\frac{1}{2}
\sum_{i,j=1}^{4}(a_i+2g_i)G(F_4)_{ij}a_j,\\
T(R)=&\frac{d(R)}{d(G)}C_2(R)
=\frac{d(R)}{2d(G)}
\sum_{i,j=1}^{4}(a_i+2g_i)G(F_4)_{ij}a_j,
\end{align}
where $G(F_4)$ is the metric matrix in 
Table~\ref{Table:Cartan-matrices-inverse}
\begin{align}
G(F_4)&=\frac{1}{2}
\left(
\begin{array}{cccc}
4&6&4&2\\
6&12&8&4\\
4&8&6&3\\
2&4&3&2\\
\end{array}
\right).
\end{align}
From Table~\ref{Table:Conjugacy-classes}, the algebra $F_4$ has only one
conjugacy class. All representations of $F_4$ belong to the same
conjugacy class. 

The above features of representations of $F_4$
are summarized in the following table.
\begin{center}
\begin{longtable}{ccccc}
\caption{Representations of $F_4$}
\label{Table:Representations-summary-F}\\
\hline\hline
$F_4$ irrep.& $d(R)$& $C_2(R)$ &$T(R)$&C/R/PR\\\hline\hline
\endfirsthead
\multicolumn{5}{c}{Table~\ref{Table:Representations-summary-F} (continued)}\\\hline\hline
$F_4$ irrep.& $d(R)$& $C_2(R)$ &$T(R)$&C/R/PR\\\hline\hline
\endhead
\hline
\endfoot
(0000)&${\bf 1}$    &$0$&$0$&R\\
(0001)&${\bf 26}$   &$6$&$3$&R\\
(1000)&${\bf 52}$   &$9$&$9$&R\\
(0010)&${\bf 273}$  &$12$&$63$&R\\
(0002)&${\bf 324}$  &$13$&$81$&R\\
(1001)&${\bf 1053}$ &$16$&$324$&R\\
(2000)&${\bf 1053'}$&$20$&$405$&R\\  
(0100)&${\bf 1274}$ &$18$&$441$&R\\
(0003)&${\bf 2652}$ &$21$&$1071$&R\\
(0011)&${\bf 4096}$ &$\frac{39}{2}$&$1536$&R\\
(1010)&${\bf 8424}$ &$23$&$3726$&R\\
(1002)&${\bf 10829}$ &$24$&$4998$&R\\
\end{longtable}
\end{center}
where $F_4$ irrep., $d(R)$, $C_2(R)$, $T(R)$, and C/R/PR stand for 
the Dynkin label of the irreducible representation of $F_4$,
its dimension, its quadratic Casimir invariant, its Dynkin indices, 
and complex, real, and pseudo-real representations,
respectively. 
The representation $(1000)$ is the adjoint representation of
$F_4$. 
See Appendix~\ref{Sec:Representations} for further information.

From Table~\ref{Table:Maximal-subalgebra},
maximal $R$- and $S$-subalgebras of $F_4$ are listed in the following
table.
\begin{center}
\begin{longtable}{crcp{10cm}c}
\caption{Maximal subalgebras of $F_4$}
\label{Table:Maximal-subalgebra-F}\\
\hline\hline
Rank&\multicolumn{2}{l}{Algebra $\mathfrak{g}$}&Maximal subalgebras $\mathfrak{h}$&Type\\\hline\hline
\endfirsthead
\multicolumn{4}{c}{Table~\ref{Table:Maximal-subalgebra-F} (continued)}\\
\hline
Rank&\multicolumn{2}{l}{Algebra $\mathfrak{g}$}&Maximal subalgebras $\mathfrak{h}$&Type\\\hline\hline
\endhead
\hline
\endfoot
$4$ &$F_4$  &$\supset$
&$\mathfrak{so}_9;\
\mathfrak{su}_3\oplus\mathfrak{su}_3;\
\mathfrak{su}_2\oplus\mathfrak{usp}_6$&$(R)$\\
    &       &$\supset$
&$\mathfrak{su}_2;\
\mathfrak{su}_2\oplus G_2$&$(S)$\\
\end{longtable}
\end{center}
where Type $(R)$ and $(S)$ stand for $R$- and $S$-subalgebras,
respectively.

\subsection{$G_2$}

Let us summarize features of representations of the
Lie algebra $G_2$.
From Table~\ref{Table:Dynkin-diagrams}, 
the extended Dynkin diagram of $G_2$ is 
\begin{align}
\xygraph{!~:{@3{-}|@{}}
    \circ   ([]!{+(0,-.3)} {{}_x}) - [r]
    \circ   ([]!{+(0,-.3)} {{}_1}) : [r]
    \bullet ([]!{+(0,-.3)} {{}_2})}.
\end{align}
From Table~\ref{Table:Cartan-matrices}, 
the Cartan matrix of $G_2$ is
\begin{align}
&A(G_2)=
\left(
\begin{array}{cc}
 2&-3\\
-1& 2\\
\end{array}
\right).
\end{align}
From Tables~\ref{Table:Complex-representations-1} and 
\ref{Table:Self-conjugate-representations},
types of representations of $G_2$
are given in the following table.
\begin{center}
\begin{longtable}{ccll}
\caption{Types of representations of $G_2$}
\label{Table:Types-representations-G}\\
\hline\hline
Algebra&Rank&Condition&C/R/PR\\\hline\hline
\endfirsthead
\multicolumn{4}{c}{Table~\ref{Table:Types-representations-G} (continued)}\\\hline\hline
Algebra&Rank&Condition&C/R/PR\\\hline\hline
\endhead
\hline
\endfoot
$G_2$ 
&2
&None
&Real\\
\end{longtable}
\end{center}
where Condition stands for a condition for Dynkin labels,
and C/R/PR represents complex, real, and pseudo-real
representations.
For $G_2$, from Eqs.~(\ref{Eq:2-Casimir}) and (\ref{Eq:Dynkin-index}),
we get the Casimir invariant and the Dynkin index of a representation
$(a_1a_2)$:
\begin{align}
C_2(R)=&\frac{1}{2}
\sum_{i,j=1}^{2}(a_i+2g_i)G(G_2)_{ij}a_j,\\
T(R)=&\frac{d(R)}{d(G)}C_2(R)
=\frac{d(R)}{2d(G)}
\sum_{i,j=1}^{2}(a_i+2g_i)G(G_2)_{ij}a_j,
\end{align}
where $G(G_2)$ is the metric matrix in 
Table~\ref{Table:Cartan-matrices-inverse}
\begin{align}
G(G_2)&=\frac{1}{3}
\left(
\begin{array}{cc}
6&3\\
3&2\\
\end{array}
\right).
\end{align}
From Table~\ref{Table:Conjugacy-classes}, the algebra $G_2$ has only one
conjugacy class. All representations of $G_2$ belong to the same
conjugacy class. 

The above features of representations of $G_2$
are summarized in the following table.
\begin{center}
\begin{longtable}{ccccc}
\caption{Representations of $G_2$}
\label{Table:Representations-summary-G}\\
\hline\hline
$G_2$ irrep.& $d(R)$& $C_2(R)$ &$T(R)$&C/R/PR\\\hline\hline
\endfirsthead
\multicolumn{5}{c}{Table~\ref{Table:Representations-summary-G} (continued)}\\\hline\hline
$G_2$ irrep.& $d(R)$& $C_2(R)$ &$T(R)$&C/R/PR\\\hline\hline
\endhead
\hline
\endfoot
(00)&${\bf 1}$    &$0$&$0$&R\\
(01)&${\bf 7}$    &$2$&$1$&R\\
(10)&${\bf 14}$   &$4$&$4$&R\\
(02)&${\bf 27}$   &$\frac{14}{3}$&$9$&R\\
(11)&${\bf 64}$   &$7$&$32$&R\\
(03)&${\bf 77}$   &$8$&$44$&R\\
(20)&${\bf 77'}$  &$10$&$55$&R\\
(04)&${\bf 182}$  &$12$&$156$&R\\
(12)&${\bf 189}$  &$\frac{32}{3}$&$144$&R\\
(30)&${\bf 273}$  &$18$&$351$&R\\
(21)&${\bf 286}$  &$14$&$286$&R\\
(05)&${\bf 378}$  &$\frac{50}{3}$&$450$&R\\
\end{longtable}
\end{center}
where $G_2$ irrep., $d(R)$, $C_2(R)$, $T(R)$, and C/R/PR stand for 
the Dynkin label of the irreducible representation of $G_2$,
its dimension, its quadratic Casimir invariant, its Dynkin indices, 
and complex, real, and pseudo-real representations,
respectively. 
The representation $(10)$ is the adjoint representation of $G_2$. 
See Appendix~\ref{Sec:Representations} for further information.

From Table~\ref{Table:Maximal-subalgebra},
maximal $R$- and $S$-subalgebras of $G_2$ are listed in the following
table.
\begin{center}
\begin{longtable}{crcp{10cm}c}
\caption{Maximal subalgebras of $G_2$}
\label{Table:Maximal-subalgebra-G}\\
\hline\hline
Rank&\multicolumn{2}{l}{Algebra $\mathfrak{g}$}&Maximal subalgebras $\mathfrak{h}$&Type\\\hline\hline
\endfirsthead
\multicolumn{4}{c}{Table~\ref{Table:Maximal-subalgebra-G} (continued)}\\
\hline
Rank&\multicolumn{2}{l}{Algebra $\mathfrak{g}$}&Maximal subalgebras $\mathfrak{h}$&Type\\\hline\hline
\endhead
\hline
\endfoot
$2$ &$G_2$  &$\supset$&$\mathfrak{su}_3;\
    \mathfrak{su}_2\oplus\mathfrak{su}_2$&$(R)$\\
    &       &$\supset$&$\mathfrak{su}_2$&$(S)$\\
\end{longtable}
\end{center}
where Type $(R)$ and $(S)$ stand for $R$- and $S$-subalgebras,
respectively.

\section{Application for model building}
\label{Sec:Model-Building}

Let us consider what kind of Lie groups we can use in grand unified
models in four- and higher-dimensional, especially, five-dimensional
spacetime. 

First, the Standard Model is a chiral gauge theory in the
four-dimensional Minkowski spacetime, where its gauge groups are
$G_{\rm SM}:=SU(3)_C\times SU(2)_L\times U(1)_Y$. It is known that vectorlike
theories with the finite degrees of freedom of internal spaces in the
viewpoint of the four-dimensional spacetime cannot become chiral
theories in the four-dimensional spacetime.
Thus, to construct unified model beyond the SM in four-dimensional
spacetime, at least one gauge group must contain complex
representations. One of the most famous unified gauge groups is 
$G_{\rm PS}:=SU(4)_C\times SU(2)_L\times SU(2)_R$, where $SU(4)$ has complex
representations. $G_{\rm PS}$ is known as the
Pati-Salam gauge group \cite{Pati:1974yy}, which is not a grand unified
gauge group. Here we define a grand unified group as a group that
contains at least the SM gauge groups $G_{\rm SM}$.

The candidates of grand unified groups are only $A_n=SU(n+1)$,
$D_{2n+1}=SO(4n+2)$, and $E_6$ because the other Lie groups have no
complex representations. 
Thus, in four-dimensional gauge theories with the finite degree of
freedoms, we can regard the Lie groups $A_n=SU(n+1)$,
$D_{2n+1}=SO(4n+2)$, and $E_6$ as candidates for a grand unified (GUT)
gauge group. 
Of course, the candidate must contain the SM gauge group 
$SU(3)_C\times SU(2)_L\times U(1)_Y$, so less than rank-4 groups are
excluded. 
The candidates for a grand unified group in four-dimensional theories 
are summarized in the following table:
\begin{center}
\begin{longtable}{ccc}
\caption{Candidates for 4D GUT gauge group}
\label{Table:GUT-gauge-group-4D}\\
\hline\hline
Class      &Rank          &Compact group\\\hline\hline
\endfirsthead
\multicolumn{3}{c}{Table~\ref{Table:GUT-gauge-group-4D} (continued)}\\\hline\hline
Class      &Rank          &Compact group\\\hline\hline
\endhead
\hline
\endfoot
$A_n$      &$n\geq 4$     &$SU(n+1)$\\
$D_{2n+1}$ &$n\geq 2$     &$SO(4n+2)$\\
$E_6$      &$6$           &$E_6$\\
\end{longtable}
\end{center}
(See grand unified theories based on e.g.,
$SU(5)$\cite{Georgi:1974sy}, 
$SU(6)$\cite{Inoue:1977qd},
$SU(n)$\cite{Fonseca:2015aoa},
$SO(10)$\cite{Fritzsch:1974nn,Fukuyama2005},
$SO(14)$\cite{Ida:1980ea},
$SO(18)$\cite{Fujimoto:1981bv},
$SO(n)$\cite{Fonseca:2015aoa}, and
$E_6$\cite{Gursey:1975ki,Maekawa:2002bk}
gauge groups.)

On the other hand, if we can use some chiral source, e.g., orbifolding
extra dimension or coset spaces,
then we can use additional Lie groups as candidates for a grand unified 
group. Let us only consider one necessary condition: the candidate 
contains the SM gauge group $SU(3)_C\times SU(2)_L\times U(1)_Y$.
The candidates for a grand unified group are listed in the following 
table.
\begin{center}
\begin{longtable}{ccc}
\caption{Candidates for GUT gauge group in general}
\label{Table:GUT-gauge-group}\\
\hline\hline
Class      &Rank          &Compact group\\\hline\hline
\endfirsthead
\multicolumn{3}{c}{Table~\ref{Table:GUT-gauge-group} (continued)}\\\hline\hline
Class      &Rank          &Compact group\\\hline\hline
\endhead
\hline
\endfoot
$A_n$     &$n\geq 4$&$SU(n+1)$\\
$B_n$     &$n\geq 4$&$SO(2n+1)$\\
$C_n$     &$n\geq 4$&$USp(2n)$\\
$D_n$     &$n\geq 5$&$SO(2n)$\\
$E_n$     &$n=6,7,8$&$E_{6,7,8}$\\
$F_4$     &$4$      &$F_4$\\
\end{longtable}
\end{center}
(See five-dimensional gauge theories based on e.g., 
$SU(5)$\cite{Kojima:2011ad},
$SU(6)$\cite{Burdman:2002se,Lim:2007jv},
$SU(n)$\cite{Haba:2004qf,Yamamoto:2013oja},
$SO(10)$\cite{Kim:2002im,Fukuyama:2008pw,Fukuyama:2012rw},
$SO(11)$\cite{Hosotani:2015hoa,Yamatsu:2015rge,Furui:2016owe},
$SO(2n)$\cite{Kawamura:2009gr},
$E_6$\cite{Kawamura:2013rj}
gauge groups, 
and non-linear $\sigma$ model $G_{\rm global}/H_{\rm local}$ based on
e.g., 
$E_{7(+7)}/SU(8)$\cite{Cremmer:1979up},
$E_7/SU(5)\times SU(3)\times U(1)$\cite{Kugo:1983ai}.
See also string inspired models e.g.,
\cite{Dienes:1996du,Dienes:1996yh,Hebecker:2001jb,Hebecker:2003jt}.)

By constraining an additional necessary condition that the adjoint
representation of a grand unified group must contain the SM Higgs field
as an extra-dimensional gauge field, i.e., gauge-Higgs grand
unification,
we reduce the number of the candidate groups as the following table.
\begin{center}
\begin{longtable}{ccc}
\caption{Candidates for gauge-Higgs GUT gauge group}
\label{Table:GUT-gauge-group-GH}\\
\hline\hline
Class      &Rank          &Compact group\\\hline\hline
\endfirsthead
\multicolumn{3}{c}{Table~\ref{Table:GUT-gauge-group-GH} (continued)}\\\hline\hline
Class      &Rank          &Compact group\\\hline\hline
\endhead
\hline
\endfoot
$A_n$     &$n\geq 5$&$SU(n+1)$\\
$B_n$     &$n\geq 5$&$SO(2n+1)$\\
$C_n$     &$n\geq 5$&$USp(2n)$\\
$D_n$     &$n\geq 6$&$SO(2n)$\\
$E_n$     &$n=6,7,8$&$E_{6,7,8}$\\
$F_4$     &$4$      &$F_4$\\
\end{longtable}
\end{center}
(See five-dimensional gauge-Higgs grand unification based on e.g., 
$SU(6)$\cite{Burdman:2002se,Lim:2007jv},
$SO(11)$\cite{Hosotani:2015hoa,Yamatsu:2015rge,Furui:2016owe}.
gauge groups.)

In the rest of this section, we give only the projection matrices of
rank 4 and 5 GUT gauge groups to $SU(3)\times SU(2)\times U(1)(\times
U(1))$ and their corresponding branching rules as examples.
We can check 
Tables~\ref{Table:GUT-gauge-group-4D}, \ref{Table:GUT-gauge-group} and 
\ref{Table:GUT-gauge-group-GH}.

\subsection{Projection matrices of GUT gauge groups}

We give projection matrices of rank 4 and 5 GUT gauge groups to
$SU(3)\times SU(2)\times U(1)$ and 
$SU(3)\times SU(2)\times U(1)\times U(1)$, respectively.

\subsubsection{Rank 4}

The rank 4 groups $SU(5)$, $SO(9)$, $USp(8)$, and $F_4$ except $SO(8)$
contain $SU(3)\times SU(2)\times U(1)$.

The Dynkin diagram of 
$A_4=SU(5)\supset SU(3)\times SU(2)\times U(1)$ 
is given by
\begin{align}
\underset{SU(5)}
{\xygraph{
    \circ ([]!{+(0,-.3)}  {{}_1}) - [r]
    \circ ([]!{+(0,-.3)}  {{}_2}) - [r]
    \circ ([]!{+(0,-.3)}  {{}_3}) - [r] 
    \circ ([]!{+(0,-.3)}  {{}_4})
}}
&\Rightarrow
\underset{SU(3)\times SU(2)\times U(1)}
{\xygraph{
    \circ ([]!{+(0,-.3)}  {{}_1}) - [r]
    \circ ([]!{+(0,-.3)}  {{}_2}) - [r]
    {\circ\hspace{-0.645em}\times} ([]!{+(0,-.3)}  {{}_3}) - [r] 
    \circ ([]!{+(0,-.3)}  {{}_4})
}}.
\end{align}
Its corresponding projection matrix is
\begin{align}
P_{SU(5)\supset SU(3)\times SU(2)\times U(1)}&=
\left(
\begin{array}{cccc}
1&0&0&0\\
0&1&0&0\\
0&0&0&1\\
2&4&6&3\\
\end{array}
\right).
\label{Projection-matrix-SU5_GSM}
\end{align}

The Dynkin diagram of 
$B_4=SO(9)\supset SU(3)\times SU(2)\times U(1)$ 
is given by
\begin{align}
\underset{SO(9)}
{\xygraph{!~:{@{=}|@{}}
    \circ ([]!{+(0,-.3)} {{}_2}) (
        - []!{+(-1,.5)}  \circ ([]!{+(0,-.3)} {{}_1}),
        - []!{+(-1,-.5)} \circ ([]!{+(0,-.3)} {{}_x}))
        - [r]
    \circ   ([]!{+(0,-.3)} {{}_3}) : [r]
    \bullet ([]!{+(0,-.3)} {{}_4})}
}
&\Rightarrow
\underset{SU(3)\times SU(2)\times U(1)}
{\xygraph{!~:{@{=}|@{}}
    \circ ([]!{+(0,-.3)} {{}_2}) (
        - []!{+(-1,.5)}  \circ ([]!{+(0,-.3)} {{}_1}),
        - []!{+(-1,-.5)} {\circ\hspace{-0.645em}\times} ([]!{+(0,-.3)} {{}_x}))
        - [r]
    {\circ\hspace{-0.645em}\times}   ([]!{+(0,-.3)} {{}_3}) : [r]
    \bullet ([]!{+(0,-.3)} {{}_4})}
}.
\end{align}
Its corresponding projection matrix is
\begin{align}
P_{SO(9)\supset SU(3)\times SU(2)\times U(1)}&=
\left(
\begin{array}{cccc}
1&0&0&0\\
0&1&0&0\\
0&0&0&1\\
2&4&6&3\\
\end{array}
\right).
\label{Projection-matrix-SO9_GSM}
\end{align}

The Dynkin diagram of 
$C_4=USp(8)\supset SU(3)\times SU(2)\times U(1)$ 
is given by
\begin{align}
\underset{USp(8)}
{\xygraph{!~:{@{=}|@{}}
    \circ   ([]!{+(0,-.3)} {{}_x}) : [r]
    \bullet ([]!{+(0,-.3)} {{}_1}) - [r]
    \bullet ([]!{+(0,-.3)} {{}_2}) - [r] 
    \bullet ([]!{+(0,-.3)} {{}_3}) : [r]
    \circ   ([]!{+(0,-.3)} {{}_4})}
}
&\Rightarrow
\underset{SU(3)\times SU(2)\times U(1)}
{\xygraph{!~:{@{=}|@{}}
    \circ   ([]!{+(0,-.3)} {{}_x}) : [r]
    {\bullet\hspace{-0.645em}\times} ([]!{+(0,-.3)} {{}_1}) - [r]
    \bullet ([]!{+(0,-.3)} {{}_2}) - [r] 
    \bullet ([]!{+(0,-.3)} {{}_3}) : [r]
    {\circ\hspace{-0.645em}\times}  ([]!{+(0,-.3)} {{}_4})}
}.
\end{align}
Its corresponding projection matrix is
\begin{align}
P_{USp(8)\supset SU(3)\times SU(2)\times U(1)}&=
\left(
\begin{array}{cccc}
1&0&0&0\\
0&1&0&0\\
0&0&0&1\\
1&2&3&3\\
\end{array}
\right).
\label{Projection-matrix-USp8_GSM}
\end{align}

The Dynkin diagrams of 
$F_4\supset SU(3)\times SU(2)\times U(1)$ 
are given by
\begin{align}
\underset{F_4}
{\xygraph{!~:{@{=}|@{}}
    \circ   ([]!{+(0,-.3)} {{}_x}) - [r]
    \circ   ([]!{+(0,-.3)} {{}_1}) - [r]
    \circ   ([]!{+(0,-.3)} {{}_2}) : [r]
    \bullet ([]!{+(0,-.3)} {{}_3}) - [r]
    \bullet ([]!{+(0,-.3)} {{}_4})}
}
&\Rightarrow
\left\{
\begin{array}{c}
\underset{(1) SU(3)\times SU(2)\times U(1)}
{\xygraph{!~:{@{=}|@{}}
    {\circ\hspace{-0.645em}\times}   ([]!{+(0,-.3)} {{}_x}) - [r]
    \circ   ([]!{+(0,-.3)} {{}_1}) - [r]
    \circ   ([]!{+(0,-.3)} {{}_2}) : [r]
    {\bullet\hspace{-0.645em}\times} ([]!{+(0,-.3)} {{}_3}) - [r]
    \bullet ([]!{+(0,-.3)} {{}_4})}
}\\
\underset{(2) SU(3)\times SU(2)\times U(1)}
{\xygraph{!~:{@{=}|@{}}
    {\circ\hspace{-0.645em}\times}   ([]!{+(0,-.3)} {{}_x}) - [r]
    \circ   ([]!{+(0,-.3)} {{}_1}) - [r]
    {\circ\hspace{-0.645em}\times}   ([]!{+(0,-.3)} {{}_2}) : [r]
    \bullet ([]!{+(0,-.3)} {{}_3}) - [r]
    \bullet ([]!{+(0,-.3)} {{}_4})}
}\\
\end{array}
\right\}.
\end{align}
Their corresponding projection matrices are
\begin{align}
P_{F_4\supset SU(3)\times SU(2)\times U(1)}^{(1)}&=
\left(
\begin{array}{cccc}
1&0&0&0\\
0&1&0&0\\
0&0&0&1\\
4&8&6&3\\
\end{array}
\right),
\label{Projection-matrix-F4_GSM-1}\\
P_{F_4\supset SU(3)\times SU(2)\times U(1)}^{(2)}&=
\left(
\begin{array}{cccc}
0&0&1&0\\
0&0&0&1\\
1&0&0&0\\
3&6&4&2\\
\end{array}
\right),
\label{Projection-matrix-F4_GSM-2}
\end{align}
where there are two distinct projection matrices for
$F_4\supset SU(3)\times SU(2)\times U(1)$, 
and their superscript $(a)$ ($(a=1,2$) stands for the above corresponding
Dynkin diagram.

\subsubsection{Rank 5}

The rank 5 groups $SU(6)$, $SO(11)$, $USp(10)$, and $SO(10)$
contain $SU(3)\times SU(2)\times U(1)$.

By using projection matrices given in
Appendix~\ref{Sec:Projection-matrices} or
calculating the method discussed in 
Sec.~\ref{Sec:Representations-subalgebras},
we find the projection matrices 
$A_5=SU(6),B_5=SO(11),C_5=USp(10),D_5=SO(10)\supset 
SU(3)\times SU(2)\times U(1)\times U(1)$:
\begin{align}
P_{SU(6)\supset SU(3)\times SU(2)\times U(1) \times U(1)}&=
\left(
\begin{array}{ccccc}
1&0&0&0&0\\
0&1&0&0&0\\
0&0&0&1&0\\
2&4&6&3&0\\
1&2&3&4&5\\
\end{array}
\right),
\label{Projection-matrix-SU6_GSM}\\
P_{SO(11)\supset SU(3)\times SU(2)\times U(1) \times U(1)}&=
\left(
\begin{array}{ccccc}
0&0&0&1&0\\
0&0&1&0&0\\
1&0&0&0&0\\
3&6&4&2&0\\
2&4&6&8&5\\
\end{array}
\right),
\label{Projection-matrix-SO11_GSM}\\
P_{USp(10)\supset SU(3)\times SU(2)\times U(1) \times U(1)}^{(1)}&=
\left(
\begin{array}{ccccc}
1&0&0&0&0\\
0&1&0&0&0\\
0&0&0&1&0\\
2&4&6&3&0\\
1&2&3&4&5\\
\end{array}
\right),
\label{Projection-matrix-USp10_GSM-1}\\
P_{USp(10)\supset SU(3)\times SU(2)\times U(1) \times U(1)}^{(2)}&=
\left(
\begin{array}{ccccc}
0&1&0&0&0\\
0&0&1&0&0\\
0&0&0&0&1\\
1&1&1&1&1\\
0&1&2&3&3\\
\end{array}
\right),
\label{Projection-matrix-USp10_GSM-2}\\
P_{SO(10)\supset SU(3)\times SU(2)\times U(1) \times U(1)}&=
\left(
\begin{array}{ccccc}
0&0&0&1&0\\
0&0&1&0&0\\
1&0&0&0&0\\
3&6&4&2&0\\
2&4&6&3&5\\
\end{array}
\right).
\label{Projection-matrix-SO10_GSM}
\end{align}

\subsection{Branching rules of GUT gauge groups}

We give some examples for branching rules between rank 4 and 5 GUT gauge
groups and $SU(3)\times SU(2)\times U(1)(\times U(1))$.

\subsubsection{Rank 4}

The rank 4 groups $SU(5)$, $SO(9)$, $USp(8)$, and $F_4$ except $SO(8)$
contain $SU(3)\times SU(2)\times U(1)$.

We can decompose $SU(5)$ representations into
$SU(3)\times SU(2)\times U(1)$ representations
by using the projection matrix given in 
Eq.~(\ref{Projection-matrix-SU5_GSM}).
\begin{center}
\begin{longtable}{rcp{13cm}}
\caption{Branching rules of 
{$SU(5)\supset SU(3)\times SU(2)\times U(1)$}
}
\label{Table:Branching-rules-SU5_GSM}
\\
\hline\hline
\multicolumn{3}{p{14cm}}
{$SU(5)\supset SU(3)\times SU(2)\times U(1)$}
\\\hline\hline
\endfirsthead
\multicolumn{3}{c}
{Table~\ref{Table:Branching-rules-SU5_GSM}
(continued)}\\\hline\hline
\multicolumn{3}{p{14cm}}
{$SU(5)\supset SU(3)\times SU(2)\times U(1)$}
\\\hline\hline
\endhead
\hline
\endfoot
${\bf 1}$&=&
$({\bf 1,1})(0)$\\
${\bf 5}$&=&
$({\bf 3,1})(2)
\oplus({\bf 1,2})(-3)$\\
${\bf \overline{5}}$&=&
$({\bf \overline{3},1})(-2)
\oplus({\bf 1,2})(3)$\\
${\bf 10}$&=&
$({\bf 3,2})(-1)
\oplus({\bf \overline{3},1})(4)
\oplus({\bf 1,1})(-6)$\\
${\bf \overline{10}}$&=&
$({\bf 3,1})(-4)
\oplus({\bf \overline{3},2})(1)
\oplus({\bf 1,1})(6)$\\
${\bf 15}$&=&
$({\bf \overline{6},1})(4)
\oplus({\bf 3,2})(-1)
\oplus({\bf 1,3})(-6)$\\
${\bf \overline{15}}$&=&
$({\bf 6,1})(-4)
\oplus({\bf \overline{3},2})(1)
\oplus({\bf 1,3})(6)$\\
${\bf 24}$&=&
$({\bf 8,1})(0)
\oplus({\bf 3,2})(5)
\oplus({\bf \overline{3},2})(-5)
\oplus({\bf 1,3})(0)
\oplus({\bf 1,1})(0)$\\
${\bf 35}$&=&
$({\bf \overline{10},1})(-6)
\oplus({\bf 6,2})(-1)
\oplus({\bf \overline{3},3})(4)
\oplus({\bf 1,4})(9)$\\
${\bf \overline{35}}$&=&
$({\bf 10,1})(6)
\oplus({\bf \overline{6},2})(1)
\oplus({\bf 3,3})(-4)
\oplus({\bf 1,4})(-9)$\\
${\bf 40}$&=&
$({\bf 8,1})(-6)
\oplus({\bf 6,2})(-1)
\oplus({\bf 3,2})(-1)
\oplus({\bf \overline{3},3})(4)
\oplus({\bf \overline{3},1})(4)
\oplus({\bf 1,2})(9)$\\
${\bf \overline{40}}$&=&
$({\bf \overline{6},2})(1)
\oplus({\bf 8,1})(6)
\oplus({\bf 3,3})(-4)
\oplus({\bf 3,1})(-4)
\oplus({\bf \overline{3},2})(1)
\oplus({\bf 1,2})(-9)$\\
${\bf 45}$&=&
$({\bf 8,2})(-3)
\oplus({\bf 6,1})(2)
\oplus({\bf 3,3})(2)
\oplus({\bf 3,1})(2)
\oplus({\bf \overline{3},2})(7)
\oplus({\bf \overline{3},1})(-8)
\oplus({\bf 1,2})(-3)$\\
${\bf \overline{45}}$&=&
$({\bf \overline{6},1})(-2)
\oplus({\bf 8,2})(3)
\oplus({\bf 3,2})(-7)
\oplus({\bf 3,1})(8)
\oplus({\bf \overline{3},3})(-2)
\oplus({\bf \overline{3},1})(-2)
\oplus({\bf 1,2})(3)$\\
\end{longtable}
\end{center}
The $SU(5)$ group contains complex representations, and it can be used
in 4D GUT.
The representations of the SM fermion fields can be
embedded into $SU(5)$ ${\bf 10}$ and ${\bf \overline{5}}$
representations. 
The $SU(5)$ adjoint representation ${\bf 24}$ does not contain the 
$SU(3)\times SU(2)$ representation $({\bf 1,2})$ with a non-zero $U(1)$
charge, and it cannot be used in gauge-Higgs grand unification.
For more information, see Table~\ref{Table:Branching-rules-A4_A2-A1-u1}
in Appendix~\ref{Sec:Branching-rules}.

We can decompose $SO(9)$ representations into
$SU(3)\times SU(2)\times U(1)$ representations
by using the projection matrix given in 
Eq.~(\ref{Projection-matrix-SO9_GSM}).
\begin{center}
\begin{longtable}{rcp{13cm}}
\caption{Branching rules of 
{$SO(9)\supset SU(3)\times SU(2)\times U(1)$}
}
\label{Table:Branching-rules-SO9_GSM}
\\
\hline\hline
\multicolumn{3}{p{14cm}}
{$SO(9)\supset SU(3)\times SU(2)\times U(1)$}
\\\hline\hline
\endfirsthead
\multicolumn{3}{c}
{Table~\ref{Table:Branching-rules-SO9_GSM}
(continued)}\\\hline\hline
\multicolumn{3}{p{14cm}}
{$SO(9)\supset SU(3)\times SU(2)\times U(1)$}
\\\hline\hline
\endhead
\hline
\endfoot
${\bf 1}$&=&
$({\bf 1,1})(0)$\\
${\bf 9}$&=&
$({\bf 3,1})(-2)
\oplus({\bf \overline{3},1})(2)
\oplus({\bf 1,3})(0)$\\
${\bf 16}$&=&
$({\bf 3,2})(1)
\oplus({\bf \overline{3},2})(-1)
\oplus({\bf 1,2})(3)
\oplus({\bf 1,2})(-3)$\\
${\bf 36}$&=&
$({\bf 8,1})(0)
\oplus({\bf 3,3})(-2)
\oplus({\bf 3,1})(4)
\oplus({\bf \overline{3},3})(2)
\oplus({\bf \overline{3},1})(-4)
\oplus({\bf 1,3})(0)
\oplus({\bf 1,1})(0)$\\
${\bf 44}$&=&
$({\bf \overline{6},1})(-4)
\oplus({\bf 8,1})(0)
\oplus({\bf 6,1})(4)
\oplus({\bf 3,3})(-2)
\oplus({\bf \overline{3},3})(2)
\oplus({\bf 1,5})(0)
\oplus({\bf 1,1})(0)$\\
${\bf 84}$&=&
$({\bf \overline{6},1})(2)
\oplus({\bf 8,3})(0)
\oplus({\bf 6,1})(-2)
\oplus({\bf 3,3})(4)
\oplus({\bf 3,3})(-2)
\oplus({\bf 3,1})(-2)
\oplus({\bf \overline{3},3})(2)
\oplus({\bf \overline{3},3})(-4)
\oplus({\bf \overline{3},1})(2)
\oplus({\bf 1,3})(0)
\oplus({\bf 1,1})(6)
\oplus({\bf 1,1})(0)
\oplus({\bf 1,1})(-6)$\\
\end{longtable}
\end{center}
The $SO(9)$ group contains only real representations, and it cannot be
used in 4D GUT.
The representations of the SM fermion fields can be
embedded into e.g., $SO(9)$ ${\bf 9}$, ${\bf 16}$, ${\bf 36}$, and 
${\bf 84}$ representations.
One may achieve the SM matter content by using e.g., proper orbifold
boundary conditions.
The $SO(9)$ adjoint representation ${\bf 36}$ does not contain the 
$SU(3)\times SU(2)$ representation $({\bf 1,2})$ with a non-zero $U(1)$
charge, and it cannot be used in gauge-Higgs grand unification.

We can decompose $USp(8)$ representations into
$SU(3)\times SU(2)\times U(1)$ representations
by using the projection matrix given in 
Eq.~(\ref{Projection-matrix-USp8_GSM}).
\begin{center}
\begin{longtable}{rcp{13cm}}
\caption{Branching rules of 
{$USp(8)\supset SU(3)\times SU(2)\times U(1)$}
}
\label{Table:Branching-rules-USp8_GSM}
\\
\hline\hline
\multicolumn{3}{p{14cm}}
{$USp(8)\supset SU(3)\times SU(2)\times U(1)$}
\\\hline\hline
\endfirsthead
\multicolumn{3}{c}
{Table~\ref{Table:Branching-rules-USp8_GSM}
(continued)}\\\hline\hline
\multicolumn{3}{p{14cm}}
{$USp(8)\supset SU(3)\times SU(2)\times U(1)$}
\\\hline\hline
\endhead
\hline
\endfoot
${\bf 1}$&=&
$({\bf 1,1})(0)$\\
${\bf 8}$&=&
$({\bf 3,1})(1)
\oplus({\bf \overline{3},1})(-1)
\oplus({\bf 1,2})(0)$\\
${\bf 27}$&=&
$({\bf 8,1})(0)
\oplus({\bf 3,2})(1)
\oplus({\bf 3,1})(-2)
\oplus({\bf \overline{3},2})(-1)
\oplus({\bf \overline{3},1})(2)
\oplus({\bf 1,1})(0)$\\
${\bf 36}$&=&
$({\bf \overline{6},1})(2)
\oplus({\bf 8,1})(0)
\oplus({\bf 6,1})(-2)
\oplus({\bf 3,2})(1)
\oplus({\bf \overline{3},2})(-1)
\oplus({\bf 1,3})(0)
\oplus({\bf 1,1})(0)$\\
${\bf 42}$&=&
$({\bf \overline{6},2})(-1)
\oplus({\bf 8,1})(0)
\oplus({\bf 6,2})(1)
\oplus({\bf 3,1})(-2)
\oplus({\bf \overline{3},1})(2)
\oplus({\bf 1,2})(3)
\oplus({\bf 1,2})(-3)$\\
${\bf 48}$&=&
$({\bf \overline{6},1})(-1)
\oplus({\bf 8,2})(0)
\oplus({\bf 6,1})(1)
\oplus({\bf 3,2})(-2)
\oplus({\bf 3,1})(1)
\oplus({\bf \overline{3},2})(2)
\oplus({\bf \overline{3},1})(-1)
\oplus({\bf 1,1})(3)
\oplus({\bf 1,1})(-3)$\\
${\bf 120}$&=&
$({\bf 10,1})(3)
\oplus({\bf 15,1})(1)
\oplus({\bf \overline{15},1})(-1)
\oplus({\bf \overline{6},2})(2)
\oplus({\bf \overline{10},1})(-3)
\oplus({\bf 8,2})(0)
\oplus({\bf 6,2})(-2)
\oplus({\bf 3,3})(1)
\oplus({\bf 3,1})(1)
\oplus({\bf \overline{3},3})(-1)
\oplus({\bf \overline{3},1})(-1)
\oplus({\bf 1,4})(0)
\oplus({\bf 1,2})(0)$\\
\end{longtable}
\end{center}
The $USp(8)$ group contains only real and pseudo-real representations,
and it cannot be used in 4D GUT.
The representations of the SM fermion fields can be
embedded into e.g., $USp(8)$ ${\bf 27}$, ${\bf 36}$, ${\bf 315}$, and 
${\bf 825}$ representations. 
(Some of them are not listed in the above table.)
One may achieve the SM matter content by using e.g., proper orbifold
boundary conditions.
The $USp(8)$ adjoint representation ${\bf 36}$ does not contain the 
$SU(3)\times SU(2)$ representation $({\bf 1,2})$ with a non-zero $U(1)$
charge, and it cannot be used in gauge-Higgs grand unification.

We can decompose $F_4$ representations into
$SU(3)\times SU(2)\times U(1)$ representations
by using the projection matrix given in 
Eq.~(\ref{Projection-matrix-F4_GSM-1}).
\begin{center}
\begin{longtable}{rcp{13cm}}
\caption{Branching rules of 
{$F_4\supset SU(3)\times SU(2)\times U(1)$(1)}
}
\label{Table:Branching-rules-F4_GSM-1}
\\
\hline\hline
\multicolumn{3}{p{14cm}}
{$F_4\supset SU(3)\times SU(2)\times U(1)$(1)}
\\\hline\hline
\endfirsthead
\multicolumn{3}{c}
{Table~\ref{Table:Branching-rules-F4_GSM-1}
(continued)}\\\hline\hline
\multicolumn{3}{p{14cm}}
{$F_4\supset SU(3)\times SU(2)\times U(1)$(1)}
\\\hline\hline
\endhead
\hline
\endfoot
${\bf 1}$&=&
$({\bf 1,1})(0)$\\
${\bf 26}$&=&
$({\bf 3,2})(1)
\oplus({\bf 3,1})(-2)
\oplus({\bf \overline{3},2})(-1)
\oplus({\bf \overline{3},1})(2)
\oplus({\bf 1,3})(0)
\oplus({\bf 1,2})(3)
\oplus({\bf 1,2})(-3)
\oplus({\bf 1,1})(0)$\\
${\bf 52}$&=&
$({\bf 8,1})(0)
\oplus({\bf 3,3})(-2)
\oplus({\bf 3,2})(1)
\oplus({\bf 3,1})(4)
\oplus({\bf \overline{3},3})(2)
\oplus({\bf \overline{3},2})(-1)
\oplus({\bf \overline{3},1})(-4)
\oplus({\bf 1,3})(0)
\oplus({\bf 1,2})(3)
\oplus({\bf 1,2})(-3)
\oplus({\bf 1,1})(0)$\\
${\bf 273}$&=&
$({\bf \overline{6},2})(-1)
\oplus({\bf \overline{6},1})(2)
\oplus({\bf 8,3})(0)
\oplus({\bf 8,2})(3)
\oplus({\bf 8,2})(-3)
\oplus({\bf 8,1})(0)
\oplus({\bf 6,2})(1)
\oplus({\bf 6,1})(-2)
\oplus({\bf 3,4})(1)
\oplus({\bf 3,3})(4)
\oplus2({\bf 3,3})(-2)
\oplus3({\bf 3,2})(1)
\oplus({\bf 3,2})(-5)
\oplus({\bf 3,1})(4)
\oplus2({\bf 3,1})(-2)
\oplus({\bf \overline{3},4})(-1)
\oplus2({\bf \overline{3},3})(2)
\oplus({\bf \overline{3},3})(-4)
\oplus({\bf \overline{3},2})(5)
\oplus3({\bf \overline{3},2})(-1)
\oplus2({\bf \overline{3},1})(2)
\oplus({\bf \overline{3},1})(-4)
\oplus({\bf 1,4})(3)
\oplus({\bf 1,4})(-3)
\oplus3({\bf 1,3})(0)
\oplus2({\bf 1,2})(3)
\oplus2({\bf 1,2})(-3)
\oplus({\bf 1,1})(6)
\oplus2({\bf 1,1})(0)
\oplus({\bf 1,1})(-6)$\\
${\bf 324}$&=&
$({\bf \overline{6},3})(2)
\oplus({\bf \overline{6},2})(-1)
\oplus({\bf \overline{6},1})(-4)
\oplus({\bf 8,3})(0)
\oplus({\bf 8,2})(3)
\oplus({\bf 8,2})(-3)
\oplus2({\bf 8,1})(0)
\oplus({\bf 6,3})(-2)
\oplus({\bf 6,2})(1)
\oplus({\bf 6,1})(4)
\oplus({\bf 3,4})(1)
\oplus({\bf 3,3})(4)
\oplus2({\bf 3,3})(-2)
\oplus3({\bf 3,2})(1)
\oplus({\bf 3,2})(-5)
\oplus({\bf 3,1})(4)
\oplus2({\bf 3,1})(-2)
\oplus({\bf \overline{3},4})(-1)
\oplus2({\bf \overline{3},3})(2)
\oplus({\bf \overline{3},3})(-4)
\oplus({\bf \overline{3},2})(5)
\oplus3({\bf \overline{3},2})(-1)
\oplus2({\bf \overline{3},1})(2)
\oplus({\bf \overline{3},1})(-4)
\oplus({\bf 1,5})(0)
\oplus({\bf 1,4})(3)
\oplus({\bf 1,4})(-3)
\oplus({\bf 1,3})(6)
\oplus2({\bf 1,3})(0)
\oplus({\bf 1,3})(-6)
\oplus2({\bf 1,2})(3)
\oplus2({\bf 1,2})(-3)
\oplus3({\bf 1,1})(0)$\\
\end{longtable}
\end{center}
The $F_4$ group contains only real representations,
and it cannot be used in 4D GUT.
The representations of the SM fermion fields can be
embedded into e.g., $F_4$ ${\bf 26}$ and ${\bf 273}$ representations. 
One may achieve the SM matter content by using e.g., proper orbifold
boundary conditions.
The $F_4$ adjoint representation ${\bf 52}$ contains the 
$SU(3)\times SU(2)$ representation $({\bf 1,2})$ with a non-zero $U(1)$
charge, and it may be used in gauge-Higgs grand unification.

Also, we can decompose $F_4$ representations into
$SU(3)\times SU(2)\times U(1)$ representations
by using the projection matrix given in 
Eq.~(\ref{Projection-matrix-F4_GSM-2}).
\begin{center}
\begin{longtable}{rcp{13cm}}
\caption{Branching rules of 
{$F_4\supset SU(3)\times SU(2)\times U(1)$(2)}
}
\label{Table:Branching-rules-F4_GSM-2}
\\
\hline\hline
\multicolumn{3}{p{14cm}}
{$F_4\supset SU(3)\times SU(2)\times U(1)$(2)}
\\\hline\hline
\endfirsthead
\multicolumn{3}{c}
{Table~\ref{Table:Branching-rules-F4_GSM-2}
(continued)}\\\hline\hline
\multicolumn{3}{p{14cm}}
{$F_4\supset SU(3)\times SU(2)\times U(1)$(2)}
\\\hline\hline
\endhead
\hline
\endfoot
${\bf 1}$&=&
$({\bf 1,1})(0)$\\
${\bf 26}$&=&
$({\bf 8,1})(0)
\oplus({\bf 3,2})(1)
\oplus({\bf 3,1})(-2)
\oplus({\bf \overline{3},2})(-1)
\oplus({\bf \overline{3},1})(2)$\\
${\bf 52}$&=&
$({\bf \overline{6},2})(-1)
\oplus({\bf \overline{6},1})(2)
\oplus({\bf 8,1})(0)
\oplus({\bf 6,2})(1)
\oplus({\bf 6,1})(-2)
\oplus({\bf 1,3})(0)
\oplus({\bf 1,2})(3)
\oplus({\bf 1,2})(-3)
\oplus({\bf 1,1})(0)$\\
${\bf 273}$&=&
$({\bf 10,1})(0)
\oplus({\bf 15,2})(1)
\oplus({\bf 15,1})(-2)
\oplus({\bf \overline{15},2})(-1)
\oplus({\bf \overline{15},1})(2)
\oplus({\bf \overline{6},2})(-1)
\oplus({\bf \overline{6},1})(2)
\oplus({\bf \overline{10},1})(0)
\oplus({\bf 8,3})(0)
\oplus({\bf 8,2})(3)
\oplus({\bf 8,2})(-3)
\oplus2({\bf 8,1})(0)
\oplus({\bf 6,2})(1)
\oplus({\bf 6,1})(-2)
\oplus({\bf 3,3})(-2)
\oplus2({\bf 3,2})(1)
\oplus({\bf 3,1})(4)
\oplus({\bf 3,1})(-2)
\oplus({\bf \overline{3},3})(2)
\oplus2({\bf \overline{3},2})(-1)
\oplus({\bf \overline{3},1})(2)
\oplus({\bf \overline{3},1})(-4)
\oplus({\bf 1,1})(0)$\\
${\bf 324}$&=&
$({\bf 27,1})(0)
\oplus({\bf 15,2})(1)
\oplus({\bf 15,1})(-2)
\oplus({\bf \overline{15},2})(-1)
\oplus({\bf \overline{15},1})(2)
\oplus({\bf \overline{6},3})(2)
\oplus2({\bf \overline{6},2})(-1)
\oplus({\bf \overline{6},1})(2)
\oplus({\bf \overline{6},1})(-4)
\oplus({\bf 8,3})(0)
\oplus({\bf 8,2})(3)
\oplus({\bf 8,2})(-3)
\oplus2({\bf 8,1})(0)
\oplus({\bf 6,3})(-2)
\oplus2({\bf 6,2})(1)
\oplus({\bf 6,1})(4)
\oplus({\bf 6,1})(-2)
\oplus({\bf 3,2})(1)
\oplus({\bf 3,1})(-2)
\oplus({\bf \overline{3},2})(-1)
\oplus({\bf \overline{3},1})(2)
\oplus({\bf 1,3})(0)
\oplus({\bf 1,2})(3)
\oplus({\bf 1,2})(-3)
\oplus2({\bf 1,1})(0)$\\
\end{longtable}
\end{center}
The $F_4$ group contains only real representations,
and it cannot be used in 4D GUT.
The representations of the SM fermion fields can be
embedded into e.g., $F_4$ ${\bf 26}$, ${\bf 52}$, ${\bf 273}$, 
and ${\bf 1274}$ representations.
(some of them are not listed in the above table.)
One may achieve the SM matter content by using e.g., proper orbifold
boundary conditions.
The $F_4$ adjoint representation ${\bf 52}$ contains the 
$SU(3)\times SU(2)$ representation $({\bf 1,2})$ with a non-zero $U(1)$
charge, and it may be used in gauge-Higgs grand unification.

\subsubsection{Rank 5}

The rank 5 groups $SU(6)$, $SO(11)$, $USp(10)$, and $SO(10)$
contain $SU(3)\times SU(2)\times U(1)$.

We can decompose $SU(6)$ representations into
$SU(3)\times SU(2)\times U(1)\times U(1)$ representations
by using the projection matrix given in 
Eq.~(\ref{Projection-matrix-SU6_GSM}).
\begin{center}
\begin{longtable}{rcp{13cm}}
\caption{Branching rules of 
{$SU(6)\supset SU(3)\times SU(2)\times U(1)\times U(1)$}
}
\label{Table:Branching-rules-SU6_GSM}
\\
\hline\hline
\multicolumn{3}{p{14cm}}
{$SU(6)\supset SU(3)\times SU(2)\times U(1)\times U(1)$}
\\\hline\hline
\endfirsthead
\multicolumn{3}{c}
{Table~\ref{Table:Branching-rules-SU6_GSM}
(continued)}\\\hline\hline
\multicolumn{3}{p{14cm}}
{$SU(6)\supset SU(3)\times SU(2)\times U(1)\times U(1)$}
\\\hline\hline
\endhead
\hline
\endfoot
${\bf 1}$&=&
$({\bf 1,1})(0)(0)$\\
${\bf 6}$&=&
$({\bf 3,1})(2)(1)
\oplus({\bf 1,2})(-3)(1)
\oplus({\bf 1,1})(0)(-5)$\\
${\bf \overline{6}}$&=&
$({\bf \overline{3},1})(-2)(-1)
\oplus({\bf 1,2})(3)(-1)
\oplus({\bf 1,1})(0)(5)$\\
${\bf 15}$&=&
$({\bf 3,2})(-1)(2)
\oplus({\bf 3,1})(2)(-4)
\oplus({\bf \overline{3},1})(4)(2)
\oplus({\bf 1,2})(-3)(-4)
\oplus({\bf 1,1})(-6)(2)$\\
${\bf \overline{15}}$&=&
$({\bf 3,1})(-4)(-2)
\oplus({\bf \overline{3},2})(1)(-2)
\oplus({\bf \overline{3},1})(-2)(4)
\oplus({\bf 1,2})(3)(4)
\oplus({\bf 1,1})(6)(-2)$\\
${\bf 20}$&=&
$({\bf 3,2})(-1)(-3)
\oplus({\bf 3,1})(-4)(3)
\oplus({\bf \overline{3},2})(1)(3)
\oplus({\bf \overline{3},1})(4)(-3)
\oplus({\bf 1,1})(6)(3)
\oplus({\bf 1,1})(-6)(-3)$\\
${\bf 21}$&=&
$({\bf \overline{6},1})(4)(2)
\oplus({\bf 3,2})(-1)(2)
\oplus({\bf 3,1})(2)(-4)
\oplus({\bf 1,3})(-6)(2)
\oplus({\bf 1,2})(-3)(-4)
\oplus({\bf 1,1})(0)(-10)$\\
${\bf \overline{21}}$&=&
$({\bf 6,1})(-4)(-2)
\oplus({\bf \overline{3},2})(1)(-2)
\oplus({\bf \overline{3},1})(-2)(4)
\oplus({\bf 1,3})(6)(-2)
\oplus({\bf 1,2})(3)(4)
\oplus({\bf 1,1})(0)(10)$\\
${\bf 35}$&=&
$({\bf 8,1})(0)(0)
\oplus({\bf 3,2})(5)(0)
\oplus({\bf 3,1})(2)(6)
\oplus({\bf \overline{3},2})(-5)(0)
\oplus({\bf \overline{3},1})(-2)(-6)
\oplus({\bf 1,3})(0)(0)
\oplus({\bf 1,2})(-3)(6)
\oplus({\bf 1,2})(3)(-6)
\oplus2({\bf 1,1})(0)(0)$\\
${\bf 56}$&=&
$({\bf 10,1})(6)(3)
\oplus({\bf \overline{6},2})(1)(3)
\oplus({\bf \overline{6},1})(4)(-3)
\oplus({\bf 3,3})(-4)(3)
\oplus({\bf 3,2})(-1)(-3)
\oplus({\bf 3,1})(2)(-9)
\oplus({\bf 1,4})(-9)(3)
\oplus({\bf 1,3})(-6)(-3)
\oplus({\bf 1,2})(-3)(-9)
\oplus({\bf 1,1})(0)(-15)$\\
${\bf \overline{56}}$&=&
$({\bf \overline{10},1})(-6)(-3)
\oplus({\bf 6,2})(-1)(-3)
\oplus({\bf 6,1})(-4)(3)
\oplus({\bf \overline{3},3})(4)(-3)
\oplus({\bf \overline{3},2})(1)(3)
\oplus({\bf \overline{3},1})(-2)(9)
\oplus({\bf 1,4})(9)(-3)
\oplus({\bf 1,3})(6)(3)
\oplus({\bf 1,2})(3)(9)
\oplus({\bf 1,1})(0)(15)$\\
\end{longtable}
\end{center}
The $SU(6)$ group contains complex representations, and it can be used
in 4D GUT.
The representations of the SM fermion fields can be
embedded into e.g., $SU(6)$ ${\bf 15}$ and ${\bf \overline{6}}$ 
representations. 
The $SU(6)$ adjoint representation ${\bf 35}$ contains the 
$SU(3)\times SU(2)$ representation $({\bf 1,2})$ with a non-zero $U(1)$
charge, and it can be used in gauge-Higgs grand unification. 
(See e.g., Refs.~\cite{Burdman:2002se,Lim:2007jv}.) 

We can decompose $SO(11)$ representations into
$SU(3)\times SU(2)\times U(1)\times U(1)$ representations
by using the projection matrix given in 
Eq.~(\ref{Projection-matrix-SO11_GSM}).
\begin{center}
\begin{longtable}{rcp{13cm}}
\caption{Branching rules of 
{$SO(11)\supset SU(3)\times SU(2)\times U(1)\times U(1)$}
}
\label{Table:Branching-rules-SO11_GSM}
\\
\hline\hline
\multicolumn{3}{p{14cm}}
{$SO(11)\supset SU(3)\times SU(2)\times U(1)\times U(1)$}
\\\hline\hline
\endfirsthead
\multicolumn{3}{c}
{Table~\ref{Table:Branching-rules-SO11_GSM}
(continued)}\\\hline\hline
\multicolumn{3}{p{14cm}}
{$SO(11)\supset SU(3)\times SU(2)\times U(1)\times U(1)$}
\\\hline\hline
\endhead
\hline
\endfoot
${\bf 1}$&=&
$({\bf 1,1})(0)(0)$\\
${\bf 11}$&=&
$({\bf 3,1})(2)(-2)
\oplus({\bf \overline{3},1})(-2)(2)
\oplus({\bf 1,2})(3)(2)
\oplus({\bf 1,2})(-3)(-2)
\oplus({\bf 1,1})(0)(0)$\\
${\bf 32}$&=&
$({\bf 3,2})(-1)(1)
\oplus({\bf 3,1})(2)(3)
\oplus({\bf 3,1})(-4)(-1)
\oplus({\bf \overline{3},2})(1)(-1)
\oplus({\bf \overline{3},1})(4)(1)
\oplus({\bf \overline{3},1})(-2)(-3)
\oplus({\bf 1,2})(-3)(3)
\oplus({\bf 1,2})(3)(-3)
\oplus({\bf 1,1})(0)(5)
\oplus({\bf 1,1})(6)(-1)
\oplus({\bf 1,1})(-6)(1)
\oplus({\bf 1,1})(0)(-5)$\\
${\bf 55}$&=&
$({\bf 8,1})(0)(0)
\oplus({\bf 3,2})(5)(0)
\oplus({\bf 3,2})(-1)(-4)
\oplus({\bf 3,1})(-4)(4)
\oplus({\bf 3,1})(2)(-2)
\oplus({\bf \overline{3},2})(1)(4)
\oplus({\bf \overline{3},2})(-5)(0)
\oplus({\bf \overline{3},1})(-2)(2)
\oplus({\bf \overline{3},1})(4)(-4)
\oplus({\bf 1,3})(0)(0)
\oplus({\bf 1,2})(3)(2)
\oplus({\bf 1,2})(-3)(-2)
\oplus({\bf 1,1})(6)(4)
\oplus2({\bf 1,1})(0)(0)
\oplus({\bf 1,1})(-6)(-4)$\\
${\bf 65}$&=&
$({\bf \overline{6},1})(4)(-4)
\oplus({\bf 8,1})(0)(0)
\oplus({\bf 6,1})(-4)(4)
\oplus({\bf 3,2})(5)(0)
\oplus({\bf 3,2})(-1)(-4)
\oplus({\bf 3,1})(2)(-2)
\oplus({\bf \overline{3},2})(1)(4)
\oplus({\bf \overline{3},2})(-5)(0)
\oplus({\bf \overline{3},1})(-2)(2)
\oplus({\bf 1,3})(6)(4)
\oplus({\bf 1,3})(0)(0)
\oplus({\bf 1,3})(-6)(-4)
\oplus({\bf 1,2})(3)(2)
\oplus({\bf 1,2})(-3)(-2)
\oplus2({\bf 1,1})(0)(0)$\\
${\bf 165}$&=&
$({\bf \overline{6},1})(-2)(2)
\oplus({\bf 8,2})(3)(2)
\oplus({\bf 8,2})(-3)(-2)
\oplus({\bf 8,1})(0)(0)
\oplus({\bf 6,1})(2)(-2)
\oplus({\bf 3,3})(2)(-2)
\oplus({\bf 3,2})(-1)(6)
\oplus({\bf 3,2})(5)(0)
\oplus({\bf 3,2})(-7)(2)
\oplus({\bf 3,2})(-1)(-4)
\oplus({\bf 3,1})(8)(2)
\oplus({\bf 3,1})(-4)(4)
\oplus2({\bf 3,1})(2)(-2)
\oplus({\bf 3,1})(-4)(-6)
\oplus({\bf \overline{3},3})(-2)(2)
\oplus({\bf \overline{3},2})(1)(4)
\oplus({\bf \overline{3},2})(7)(-2)
\oplus({\bf \overline{3},2})(-5)(0)
\oplus({\bf \overline{3},2})(1)(-6)
\oplus({\bf \overline{3},1})(4)(6)
\oplus2({\bf \overline{3},1})(-2)(2)
\oplus({\bf \overline{3},1})(4)(-4)
\oplus({\bf \overline{3},1})(-8)(-2)
\oplus({\bf 1,3})(0)(0)
\oplus2({\bf 1,2})(3)(2)
\oplus2({\bf 1,2})(-3)(-2)
\oplus({\bf 1,1})(6)(4)
\oplus({\bf 1,1})(-6)(6)
\oplus2({\bf 1,1})(0)(0)
\oplus({\bf 1,1})(6)(-6)
\oplus({\bf 1,1})(-6)(-4)$\\
\end{longtable}
\end{center}
The $SO(11)$ group contains only real and pseudo-real representations,
and it cannot be used in 4D GUT.
The representations of the SM fermion fields can be
embedded into e.g., $SO(11)$ ${\bf 32}$ representation. 
The SM matter content can be achieved by using proper orbifold
boundary conditions.
The $SO(11)$ adjoint representation ${\bf 55}$ contains the 
$SU(3)\times SU(2)$ representation $({\bf 1,2})$ with a non-zero $U(1)$
charge, and it can be used in gauge-Higgs grand unification.
(See Ref.~\cite{Hosotani:2015hoa}.)

We can decompose $USp(10)$ representations into
$SU(3)\times SU(2)\times U(1)\times U(1)$ representations
by using the projection matrix given in 
Eq.~(\ref{Projection-matrix-USp10_GSM-1}).
\begin{center}
\begin{longtable}{rcp{13cm}}
\caption{Branching rules of 
{$USp(10)\supset SU(3)\times SU(2)\times U(1)\times U(1)$(1)}
}
\label{Table:Branching-rules-USp10_GSM-1}
\\
\hline\hline
\multicolumn{3}{p{14cm}}
{$USp(10)\supset SU(3)\times SU(2)\times U(1)\times U(1)$(1)}
\\\hline\hline
\endfirsthead
\multicolumn{3}{c}
{Table~\ref{Table:Branching-rules-USp10_GSM-1}
(continued)}\\\hline\hline
\multicolumn{3}{p{14cm}}
{$USp(10)\supset SU(3)\times SU(2)\times U(1)\times U(1)$(1)}
\\\hline\hline
\endhead
\hline
\endfoot
${\bf 1}$&=&
$({\bf 1,1})(0)(0)$\\
${\bf 10}$&=&
$({\bf 3,1})(2)(1)
\oplus({\bf \overline{3},1})(-2)(-1)
\oplus({\bf 1,2})(-3)(1)
\oplus({\bf 1,2})(3)(-1)$\\
${\bf 44}$&=&
$({\bf 8,1})(0)(0)
\oplus({\bf 3,2})(-1)(2)
\oplus({\bf 3,2})(5)(0)
\oplus({\bf 3,1})(-4)(-2)
\oplus({\bf \overline{3},2})(-5)(0)
\oplus({\bf \overline{3},2})(1)(-2)
\oplus({\bf \overline{3},1})(4)(2)
\oplus({\bf 1,3})(0)(0)
\oplus({\bf 1,1})(-6)(2)
\oplus({\bf 1,1})(0)(0)
\oplus({\bf 1,1})(6)(-2)$\\
${\bf 55}$&=&
$({\bf \overline{6},1})(4)(2)
\oplus({\bf 8,1})(0)(0)
\oplus({\bf 6,1})(-4)(-2)
\oplus({\bf 3,2})(-1)(2)
\oplus({\bf 3,2})(5)(0)
\oplus({\bf \overline{3},2})(-5)(0)
\oplus({\bf \overline{3},2})(1)(-2)
\oplus({\bf 1,3})(-6)(2)
\oplus({\bf 1,3})(0)(0)
\oplus({\bf 1,3})(6)(-2)
\oplus2({\bf 1,1})(0)(0)$\\
${\bf 110}$&=&
$({\bf \overline{6},1})(-2)(-1)
\oplus({\bf 8,2})(-3)(1)
\oplus({\bf 8,2})(3)(-1)
\oplus({\bf 6,1})(2)(1)
\oplus({\bf 3,3})(2)(1)
\oplus({\bf 3,2})(-7)(-1)
\oplus({\bf 3,2})(-1)(-3)
\oplus({\bf 3,1})(-4)(3)
\oplus({\bf 3,1})(2)(1)
\oplus({\bf 3,1})(8)(-1)
\oplus({\bf \overline{3},3})(-2)(-1)
\oplus({\bf \overline{3},2})(1)(3)
\oplus({\bf \overline{3},2})(7)(1)
\oplus({\bf \overline{3},1})(-8)(1)
\oplus({\bf \overline{3},1})(-2)(-1)
\oplus({\bf \overline{3},1})(4)(-3)
\oplus({\bf 1,2})(-3)(1)
\oplus({\bf 1,2})(3)(-1)
\oplus({\bf 1,1})(6)(3)
\oplus({\bf 1,1})(-6)(-3)$\\
${\bf 132}$&=&
$({\bf \overline{6},3})(-2)(-1)
\oplus({\bf \overline{6},1})(-8)(1)
\oplus({\bf \overline{6},1})(4)(-3)
\oplus({\bf 8,2})(-3)(1)
\oplus({\bf 8,2})(3)(-1)
\oplus({\bf 6,3})(2)(1)
\oplus({\bf 6,1})(-4)(3)
\oplus({\bf 6,1})(8)(-1)
\oplus({\bf 3,2})(-7)(-1)
\oplus({\bf 3,2})(-1)(-3)
\oplus({\bf 3,1})(2)(1)
\oplus({\bf \overline{3},2})(1)(3)
\oplus({\bf \overline{3},2})(7)(1)
\oplus({\bf \overline{3},1})(-2)(-1)
\oplus({\bf 1,3})(6)(3)
\oplus({\bf 1,3})(-6)(-3)
\oplus({\bf 1,1})(0)(5)
\oplus({\bf 1,1})(12)(1)
\oplus({\bf 1,1})(-12)(-1)
\oplus({\bf 1,1})(0)(-5)$\\
${\bf 165}$&=&
$({\bf \overline{6},2})(-5)(0)
\oplus({\bf \overline{6},2})(1)(-2)
\oplus({\bf 8,3})(0)(0)
\oplus({\bf 8,1})(-6)(2)
\oplus({\bf 8,1})(0)(0)
\oplus({\bf 8,1})(6)(-2)
\oplus({\bf 6,2})(-1)(2)
\oplus({\bf 6,2})(5)(0)
\oplus({\bf 3,3})(-4)(-2)
\oplus({\bf 3,2})(-1)(2)
\oplus({\bf 3,2})(5)(0)
\oplus({\bf 3,1})(-10)(0)
\oplus({\bf 3,1})(-4)(-2)
\oplus({\bf 3,1})(2)(-4)
\oplus({\bf \overline{3},3})(4)(2)
\oplus({\bf \overline{3},2})(-5)(0)
\oplus({\bf \overline{3},2})(1)(-2)
\oplus({\bf \overline{3},1})(-2)(4)
\oplus({\bf \overline{3},1})(4)(2)
\oplus({\bf \overline{3},1})(10)(0)
\oplus({\bf 1,2})(3)(4)
\oplus({\bf 1,2})(9)(2)
\oplus({\bf 1,2})(-9)(-2)
\oplus({\bf 1,2})(-3)(-4)
\oplus({\bf 1,1})(0)(0)$\\
\end{longtable}
\end{center}
The $USp(10)$ group contains only real and pseudo-real representations,
and it cannot be used in 4D GUT.
The representations of the SM fermion fields can be
embedded into e.g., $USp(10)$ ${\bf 10}$ and ${\bf 44}$ representations. 
One may achieve the SM matter content by using e.g., proper orbifold
boundary conditions.
The $USp(10)$ adjoint representation ${\bf 55}$  does not contain the 
$SU(3)\times SU(2)$ representation $({\bf 1,2})$ with a non-zero $U(1)$
charge, and it cannot be used in gauge-Higgs grand unification.

Also, we can decompose $USp(10)$ representations into
$SU(3)\times SU(2)\times U(1)\times U(1)$ representations
by using the projection matrix given in 
Eq.~(\ref{Projection-matrix-USp10_GSM-2}).
\begin{center}
\begin{longtable}{rcp{13cm}}
\caption{Branching rules of 
{$USp(10)\supset SU(3)\times SU(2)\times U(1)\times U(1)$(2)}
}
\label{Table:Branching-rules-USp10_GSM-2}
\\
\hline\hline
\multicolumn{3}{p{14cm}}
{$USp(10)\supset SU(3)\times SU(2)\times U(1)\times U(1)$(2)}
\\\hline\hline
\endfirsthead
\multicolumn{3}{c}
{Table~\ref{Table:Branching-rules-USp10_GSM-2}
(continued)}\\\hline\hline
\multicolumn{3}{p{14cm}}
{$USp(10)\supset SU(3)\times SU(2)\times U(1)\times U(1)$(2)}
\\\hline\hline
\endhead
\hline
\endfoot
${\bf 1}$&=&
$({\bf 1,1})(0)(0)$\\
${\bf 10}$&=&
$({\bf 3,1})(0)(1)
\oplus({\bf \overline{3},1})(0)(-1)
\oplus({\bf 1,2})(0)(0)
\oplus({\bf 1,1})(1)(0)
\oplus({\bf 1,1})(-1)(0)$\\
${\bf 44}$&=&
$({\bf 8,1})(0)(0)
\oplus({\bf 3,2})(0)(1)
\oplus({\bf 3,1})(1)(1)
\oplus({\bf 3,1})(0)(-2)
\oplus({\bf 3,1})(-1)(1)
\oplus({\bf \overline{3},2})(0)(-1)
\oplus({\bf \overline{3},1})(1)(-1)
\oplus({\bf \overline{3},1})(0)(2)
\oplus({\bf \overline{3},1})(-1)(-1)
\oplus({\bf 1,2})(1)(0)
\oplus({\bf 1,2})(-1)(0)
\oplus2({\bf 1,1})(0)(0)$\\
${\bf 55}$&=&
$({\bf \overline{6},1})(0)(2)
\oplus({\bf 8,1})(0)(0)
\oplus({\bf 6,1})(0)(-2)
\oplus({\bf 3,2})(0)(1)
\oplus({\bf 3,1})(1)(1)
\oplus({\bf 3,1})(-1)(1)
\oplus({\bf \overline{3},2})(0)(-1)
\oplus({\bf \overline{3},1})(1)(-1)
\oplus({\bf \overline{3},1})(-1)(-1)
\oplus({\bf 1,3})(0)(0)
\oplus({\bf 1,2})(1)(0)
\oplus({\bf 1,2})(-1)(0)
\oplus({\bf 1,1})(2)(0)
\oplus2({\bf 1,1})(0)(0)
\oplus({\bf 1,1})(-2)(0)$\\
${\bf 110}$&=&
$({\bf \overline{6},1})(0)(-1)
\oplus({\bf 8,2})(0)(0)
\oplus({\bf 8,1})(1)(0)
\oplus({\bf 8,1})(-1)(0)
\oplus({\bf 6,1})(0)(1)
\oplus({\bf 3,2})(1)(1)
\oplus({\bf 3,2})(0)(-2)
\oplus({\bf 3,2})(-1)(1)
\oplus({\bf 3,1})(1)(-2)
\oplus2({\bf 3,1})(0)(1)
\oplus({\bf 3,1})(-1)(-2)
\oplus({\bf \overline{3},2})(1)(-1)
\oplus({\bf \overline{3},2})(0)(2)
\oplus({\bf \overline{3},2})(-1)(-1)
\oplus({\bf \overline{3},1})(1)(2)
\oplus2({\bf \overline{3},1})(0)(-1)
\oplus({\bf \overline{3},1})(-1)(2)
\oplus({\bf 1,2})(0)(0)
\oplus({\bf 1,1})(1)(0)
\oplus({\bf 1,1})(0)(3)
\oplus({\bf 1,1})(0)(-3)
\oplus({\bf 1,1})(-1)(0)$\\
${\bf 132}$&=&
$({\bf \overline{6},2})(1)(-1)
\oplus({\bf \overline{6},2})(-1)(-1)
\oplus({\bf \overline{6},1})(0)(-1)
\oplus({\bf 8,2})(0)(0)
\oplus({\bf 8,1})(1)(0)
\oplus({\bf 8,1})(-1)(0)
\oplus({\bf 6,2})(1)(1)
\oplus({\bf 6,2})(-1)(1)
\oplus({\bf 6,1})(0)(1)
\oplus({\bf 3,2})(0)(-2)
\oplus({\bf 3,1})(1)(-2)
\oplus({\bf 3,1})(0)(1)
\oplus({\bf 3,1})(-1)(-2)
\oplus({\bf \overline{3},2})(0)(2)
\oplus({\bf \overline{3},1})(1)(2)
\oplus({\bf \overline{3},1})(0)(-1)
\oplus({\bf \overline{3},1})(-1)(2)
\oplus({\bf 1,2})(1)(3)
\oplus({\bf 1,2})(1)(-3)
\oplus({\bf 1,2})(-1)(3)
\oplus({\bf 1,2})(-1)(-3)
\oplus({\bf 1,1})(0)(3)
\oplus({\bf 1,1})(0)(-3)$\\
${\bf 165}$&=&
$({\bf \overline{6},2})(0)(-1)
\oplus({\bf \overline{6},1})(1)(-1)
\oplus({\bf \overline{6},1})(-1)(-1)
\oplus({\bf 8,2})(1)(0)
\oplus({\bf 8,2})(-1)(0)
\oplus2({\bf 8,1})(0)(0)
\oplus({\bf 6,2})(0)(1)
\oplus({\bf 6,1})(1)(1)
\oplus({\bf 6,1})(-1)(1)
\oplus({\bf 3,2})(1)(-2)
\oplus({\bf 3,2})(0)(1)
\oplus({\bf 3,2})(-1)(-2)
\oplus({\bf 3,1})(1)(1)
\oplus2({\bf 3,1})(0)(-2)
\oplus({\bf 3,1})(-1)(1)
\oplus({\bf \overline{3},2})(1)(2)
\oplus({\bf \overline{3},2})(0)(-1)
\oplus({\bf \overline{3},2})(-1)(2)
\oplus({\bf \overline{3},1})(1)(-1)
\oplus2({\bf \overline{3},1})(0)(2)
\oplus({\bf \overline{3},1})(-1)(-1)
\oplus({\bf 1,2})(0)(3)
\oplus({\bf 1,2})(0)(-3)
\oplus({\bf 1,1})(1)(3)
\oplus({\bf 1,1})(1)(-3)
\oplus({\bf 1,1})(0)(0)
\oplus({\bf 1,1})(-1)(3)
\oplus({\bf 1,1})(-1)(-3)$\\
\end{longtable}
\end{center}
The $USp(10)$ group contains only real and pseudo-real representations,
and it cannot be used in 4D GUT.
One may achieve the SM matter content by using e.g., proper orbifold
boundary conditions.
The $USp(10)$ adjoint representation ${\bf 55}$ contains the 
$SU(3)\times SU(2)$ representation $({\bf 1,2})$ with a non-zero $U(1)$
charge, and it may be used in gauge-Higgs grand unification. However, in this case it
is impossible to realize the SM matter content.

We can decompose $SO(10)$ representations into
$SU(3)\times SU(2)\times U(1)\times U(1)$ representations
by using the projection matrix given in 
Eq.~(\ref{Projection-matrix-SO10_GSM}).
\begin{center}
\begin{longtable}{rcp{13cm}}
\caption{Branching rules of 
{$SO(10)\supset SU(3)\times SU(2)\times U(1)\times U(1)$}
}
\label{Table:Branching-rules-SO10_GSM}
\\
\hline\hline
\multicolumn{3}{p{14cm}}
{$SO(10)\supset SU(3)\times SU(2)\times U(1)\times U(1)$}
\\\hline\hline
\endfirsthead
\multicolumn{3}{c}
{Table~\ref{Table:Branching-rules-SO10_GSM}
(continued)}\\\hline\hline
\multicolumn{3}{p{14cm}}
{$SO(10)\supset SU(3)\times SU(2)\times U(1)\times U(1)$}
\\\hline\hline
\endhead
\hline
\endfoot
${\bf 1}$&=&
$({\bf 1,1})(0)(0)$\\
${\bf 10}$&=&
$({\bf 3,1})(2)(-2)
\oplus({\bf \overline{3},1})(-2)(2)
\oplus({\bf 1,2})(3)(2)
\oplus({\bf 1,2})(-3)(-2)$\\
${\bf 16}$&=&
$({\bf 3,2})(-1)(1)
\oplus({\bf \overline{3},1})(-2)(-3)
\oplus({\bf \overline{3},1})(4)(1)
\oplus({\bf 1,2})(3)(-3)
\oplus({\bf 1,1})(-6)(1)
\oplus({\bf 1,1})(0)(5)$\\
${\bf \overline{16}}$&=&
$({\bf 3,1})(2)(3)
\oplus({\bf 3,1})(-4)(-1)
\oplus({\bf \overline{3},2})(1)(-1)
\oplus({\bf 1,2})(-3)(3)
\oplus({\bf 1,1})(6)(-1)
\oplus({\bf 1,1})(0)(-5)$\\
${\bf 45}$&=&
$({\bf 8,1})(0)(0)
\oplus({\bf 3,2})(5)(0)
\oplus({\bf 3,2})(-1)(-4)
\oplus({\bf 3,1})(-4)(4)
\oplus({\bf \overline{3},2})(1)(4)
\oplus({\bf \overline{3},2})(-5)(0)
\oplus({\bf \overline{3},1})(4)(-4)
\oplus({\bf 1,3})(0)(0)
\oplus({\bf 1,1})(6)(4)
\oplus2({\bf 1,1})(0)(0)
\oplus({\bf 1,1})(-6)(-4)$\\
${\bf 54}$&=&
$({\bf \overline{6},1})(4)(-4)
\oplus({\bf 8,1})(0)(0)
\oplus({\bf 6,1})(-4)(4)
\oplus({\bf 3,2})(5)(0)
\oplus({\bf 3,2})(-1)(-4)
\oplus({\bf \overline{3},2})(1)(4)
\oplus({\bf \overline{3},2})(-5)(0)
\oplus({\bf 1,3})(6)(4)
\oplus({\bf 1,3})(0)(0)
\oplus({\bf 1,3})(-6)(-4)
\oplus({\bf 1,1})(0)(0)$\\
${\bf 120}$&=&
$({\bf \overline{6},1})(-2)(2)
\oplus({\bf 8,2})(3)(2)
\oplus({\bf 8,2})(-3)(-2)
\oplus({\bf 6,1})(2)(-2)
\oplus({\bf 3,3})(2)(-2)
\oplus({\bf 3,2})(-1)(6)
\oplus({\bf 3,2})(-7)(2)
\oplus({\bf 3,1})(8)(2)
\oplus2({\bf 3,1})(2)(-2)
\oplus({\bf 3,1})(-4)(-6)
\oplus({\bf \overline{3},3})(-2)(2)
\oplus({\bf \overline{3},2})(7)(-2)
\oplus({\bf \overline{3},2})(1)(-6)
\oplus({\bf \overline{3},1})(4)(6)
\oplus2({\bf \overline{3},1})(-2)(2)
\oplus({\bf \overline{3},1})(-8)(-2)
\oplus2({\bf 1,2})(3)(2)
\oplus2({\bf 1,2})(-3)(-2)
\oplus({\bf 1,1})(-6)(6)
\oplus({\bf 1,1})(6)(-6)$\\
${\bf 126}$&=&
$({\bf \overline{6},1})(4)(6)
\oplus({\bf \overline{6},1})(-2)(2)
\oplus({\bf \overline{6},1})(-8)(-2)
\oplus({\bf 8,2})(3)(2)
\oplus({\bf 8,2})(-3)(-2)
\oplus({\bf 6,3})(2)(-2)
\oplus({\bf 3,2})(-1)(6)
\oplus({\bf 3,2})(-7)(2)
\oplus({\bf 3,1})(8)(2)
\oplus2({\bf 3,1})(2)(-2)
\oplus({\bf 3,1})(-4)(-6)
\oplus({\bf \overline{3},3})(-2)(2)
\oplus({\bf \overline{3},2})(7)(-2)
\oplus({\bf \overline{3},2})(1)(-6)
\oplus({\bf \overline{3},1})(-2)(2)
\oplus({\bf 1,3})(-6)(6)
\oplus({\bf 1,2})(3)(2)
\oplus({\bf 1,2})(-3)(-2)
\oplus({\bf 1,1})(12)(-2)
\oplus({\bf 1,1})(6)(-6)
\oplus({\bf 1,1})(0)(-10)$\\
${\bf \overline{126}}$&=&
$({\bf \overline{6},3})(-2)(2)
\oplus({\bf 8,2})(-3)(-2)
\oplus({\bf 8,2})(3)(2)
\oplus({\bf 6,1})(-4)(-6)
\oplus({\bf 6,1})(2)(-2)
\oplus({\bf 6,1})(8)(2)
\oplus({\bf 3,3})(2)(-2)
\oplus({\bf 3,2})(-7)(2)
\oplus({\bf 3,2})(-1)(6)
\oplus({\bf 3,1})(2)(-2)
\oplus({\bf \overline{3},2})(1)(-6)
\oplus({\bf \overline{3},2})(7)(-2)
\oplus({\bf \overline{3},1})(-8)(-2)
\oplus2({\bf \overline{3},1})(-2)(2)
\oplus({\bf \overline{3},1})(4)(6)
\oplus({\bf 1,3})(6)(-6)
\oplus({\bf 1,2})(-3)(-2)
\oplus({\bf 1,2})(3)(2)
\oplus({\bf 1,1})(-12)(2)
\oplus({\bf 1,1})(-6)(6)
\oplus({\bf 1,1})(0)(10)$\\
\end{longtable}
\end{center}
The $SO(10)$ group contains complex representations, and it can be used
in 4D GUT.
The representations of the SM fermion fields can be
embedded into e.g., $SO(10)$ ${\bf 16}$ representation. 
The $SO(10)$ adjoint representation ${\bf 45}$ does not contain the 
$SU(3)\times SU(2)$ representation $({\bf 1,2})$ with a non-zero $U(1)$
charge, and it cannot be used in gauge-Higgs grand unification.

\section{Summary and discussion}
\label{Sec:Summary-discussion}

We reviewed basics of finite-dimensional Lie algebras and their
subalgebras such as Dynkin diagrams, Cartan matrices, and notion of
regular, special, and maximal subalgebras. 
We saw several features of finite-dimensional representations of Lie
algebras such as conjugacy classes, types of representations, Weyl
dimension formulas, Dynkin indices, Casimir invariants, and anomaly
coefficients. 
We checked how to decompose irreducible representations of Lie algebras
into irreducible representations of their Lie subalgebras.
We discussed a method to calculate tensor products of two irreducible
representations of a Lie algebra mainly by using its Dynkin diagram. 
Several features of rank-$n$ classical Lie algebras and the five
exceptional Lie algebras are summarized. 
We showed which Lie algebras can be applied for grand unification in
general by using the above discussion. 

By using the above results, we showed tables of representations of
classical Lie algebras $A_n$, $B_n$, $C_n$, $D_n$ $(n=1,2,\cdots,20)$,
and the exceptional Lie algebras $E_6$, $E_7$, $E_8$,
$F_4$, and $G_2$, which include dimensions, conjugacy classes, Dynkin
indices, quadratic Casimir invariants, anomaly coefficients, and types
of representations.  
We showed tables of branching rules of classical Lie algebras $A_n$,
$B_n$, $C_n$, $D_n$ $(n=1,2,\cdots,20)$ and 
the exceptional Lie algebras $E_6$, $E_7$, $E_8$, $F_4$, and $G_2$ 
and their maximal regular and special subalgebras.
We summarized tables of tensor products of classical Lie algebras $A_n$,
$B_n$, $C_n$, $D_n$ $(n=1,2,\cdots,20)$ and
the exceptional Lie algebras $E_6$, $E_7$, $E_8$, $F_4$, and $G_2$.

We discussed regular and special subalgebras of simple Lie algebras.
At present, only regular subalgebras seem to be used for model building
in particle physics.
It may be interesting to use special subalgebras.
For example, the vector representation ${\bf 32}$ of 
$\mathfrak{so}_{32}$ can be corresponding to
spinor representations ${\bf 16}$ and ${\bf \overline{16}}$ of 
its non-maximal special subalgebra $\mathfrak{so}_{10}$.
\footnote{
After the first version of this paper was finished, the author proposed
new-type grand unified theories based on grand unified groups,
such as $SU(16)$ and $SO(32)$, broken into their special subgroups, as
well as their regular subgroups
\cite{Yamatsu:2017sgu,Yamatsu:2017ssg,Yamatsu:2018fsg,Yamatsu:2020usp}.
One may think that symmetry groups are broken to only
regular subgroups, not to special subgroups. 
However, the symmetry breaking of $SU(n)$ to its special subgroups such as
$SO(n)$ and $USp(2[n/2])$ are known to be realized by
a nonvanishing VEV of a fundamental
scalar field in rank-2 symmetric and anti-symmetric tensor
representations of $SU(n)$ 
\cite{Li:1973mq,Meljanac:1982rc,Abud:1984xn},
a nonvanishing VEV of a composite scalar field made by fermion
pair condensation in fundamental and rank-2 anti-symmetric tensor
representations of $SU(n)$ \cite{Kugo:2019isl} by examining the
dynamical symmetry breaking pattern in four-dimensional $SU(n)$
Nambu–-Jona-Lasinio-type
models\cite{Nambu:1961tp,Nambu:1961fr,Kugo:1994qr}, 
orbifold boundary conditions (BCs) by $\mathbb{Z}_2$ outer
automorphisms on $S^1/\mathbb{Z}_2$ orbifold space \cite{Hebecker:2001jb}.
Also, other symmetry groups such as $SO(n)$ and $E_6$ broken to their
special subgroups are discussed
in Ref.~\cite{Li:1973mq} for fundamental scalar fields;
in Refs.~\cite{Kugo:1994qr,Kugo:2019wge} for composite scalar fields;
in Ref.~\cite{Hebecker:2001jb} for $\mathbb{Z}_2$ orbifold space.
}

In this paper, we have discussed finite-dimensional Lie algebras and their
finite-dimensional representations to apply for model building.
The notion of grand unification is attractive to solve charge
quantization and particle zoo, but still there are a lot of issues in
particle physics. For example, the origin of quark's and lepton's
generations and their hierarchical mass structures 
seem to be impossible to solve by using finite-dimensional Lie algebras
and their finite-dimensional representations.
An attempt to solve this issue is discussed by using a
finite-dimensional Lie algebra and their infinite-dimensional
representations in 
Refs.~\cite{Inoue:1994qz,Inoue:2000ia,Inoue:2003qi,Yamatsu:2007,Yamatsu:2008,Yamatsu:2009,Yamatsu:2012,Yamatsu:2013}.
Also, finite discrete groups and their representations are used to
explain a neutrino mixing matrix, i.e., the Maki-Nakagawa-Sakata (MNS)
matrix \cite{Maki:1962mu}.
(For review, see e.g., Ref.~\cite{Ishimori:2010au}.)

\section*{Acknowledgments}

The author would like to thank Howard Haber, Yutaka Hosotani, Taichiro
Kugo, Kenji Nishiwaki, and Satoshi Yamaguchi for useful comments, 
Particle Physics Theory Group, Department of Physics at Osaka
University for hospitality, 
where main part of this work was carried out,
and Maskawa Institute for Science and Culture
at Kyoto Sangyo University, Kyoto University, and Kyushu university
where main correction was carried out.
This work was supported in part by the MEXT/JSPS KAKENHI Grant Number
23104009, JP18H05543 and JP19K23440.

\newpage

\bibliographystyle{utphys} 
\bibliography{../../arxiv/reference}

\appendix

\newpage
\input{representations}
\newpage
\input{positive-roots}

\newpage
\input{weight-diagrams}

\newpage
\input{subalgebras}
\newpage
\input{projection-matrices}
\newpage
\input{decomposition-reps}
\newpage
\input{tensor-products}

\end{document}

%% file: positive-roots.tex
\section{Positive roots}
\label{Sec:Positive-roots}

\subsection{$A_n=\mathfrak{su}_{n+1}$}

\begin{center}
\begin{longtable}{ccp{10cm}}
\caption{Positive roots of $A_n$}
\label{Table:Rositive-roots-A}\\
\hline\hline
Rank&Level/State&Coefficients of positive roots $(k_\alpha^1\cdots k_\alpha^n)$\\\hline\hline
\endfirsthead
\multicolumn{3}{c}{Table~\ref{Table:Positive-roots-A} (continued)}\\\hline\hline
Rank&Level/State&Coefficients of positive roots $(k_\alpha^1\cdots k_\alpha^n)$\\\hline\hline
\endhead
\hline
\endfoot
1
&0/1&(1)\\
\hline
2
&1/1&(11)\\
&0/2&(10), (01)\\
\hline
3
&2/1&(111)\\
&1/2&(110), (011)\\
&0/3&(100), (010), (001)\\
\hline
4
&3/1&(1111)\\
&2/2&(1110), (0111)\\
&1/3&(1100), (0110), (0011)\\
&0/4&(1000), (0100), (0010), (0001)\\
\hline
5
&4/1&(11111)\\
&3/2&(11110), (01111)\\
&2/3&(11100), (01110), (00111)\\
&1/4&(11000), (01100), (00110), (00011)\\
&0/5&(10000), (01000), (00100), (00010), (00001)\\
\hline
6
&5/1&(111111)\\
&4/2&(111110), (011111)\\
&3/3&(111100), (011110), (001111)\\
&2/4&(111000), (011100), (001110), (000111)\\
&4/5&(110000), (011000), (001100), (000110), (000011)\\
&5/6&(100000), (010000), (001000), (000100), (000010), (000001)\\
\hline
$\vdots$
&$\vdots$
&$\vdots$\\
\hline
$n$
&$n-1/1$&$(11\cdots 11)$\\
&$n-2/2$&$(11\cdots 10), (01\cdots 11)$\\
&$\vdots$&$\vdots$\\
&$1/n-1$&$(110\cdots 0), (0110\cdots 0), \cdots, (0\cdots 011)$\\
&$0/n$&$(10\cdots 0), (010\cdots 0), \cdots, (0\cdots 010), (0\cdots 01)$\\
\end{longtable}
where Dynkin diagram of $A_n$ with its coefficients of its positive roots 
\begin{align*}
\underset{A_n}{
\xygraph{
    \circ ([]!{+(0,-.3)} {k_\alpha^1})  - [r]
    \circ ([]!{+(0,-.3)} {k_\alpha^2})  - [r]
    \circ ([]!{+(0,-.3)} {k_\alpha^3})  - [r] 
    \cdots                        - [r]
    \circ ([]!{+(0,-.3)} {k_\alpha^{n-1}}) - [r]
    \circ ([]!{+(0,-.3)} {k_\alpha^n})}}.
\end{align*}
\end{center}

\newpage
\subsection{$B_n=\mathfrak{so}_{2n+1}$}

\begin{center}
\begin{longtable}{ccp{10cm}}
\caption{Positive roots of $B_n$}
\label{Table:Positive-roots-B}\\
\hline\hline
Rank&Level/State&Coefficients of positive roots $(k_\alpha^1\cdots k_\alpha^n)$\\\hline\hline
\endfirsthead
\multicolumn{3}{c}{Table~\ref{Table:Positive-roots-B} (continued)}\\\hline\hline
Rank&Level/State&Coefficients of positive roots $(k_\alpha^1\cdots k_\alpha^n)$\\\hline\hline
\endhead
\hline
\endfoot
2
&2/1&(12)\\
&1/1&(11)\\
&0/2&(10), (01)\\
\hline
3
&4/1&(122)\\
&3/1&(112)\\
&2/2&(111), (012)\\
&1/2&(110), (011)\\
&0/3&(100), (010), (001)\\
\hline
4
&6/1&(1222)\\
&5/1&(1122)\\
&4/2&(1112), (0122)\\
&3/2&(1111), (0112)\\
&2/3&(1110), (0111), (0012)\\
&1/3&(1100), (0110), (0011)\\
&0/4&(1000), (0100), (0010), (0001)\\
\hline
5
&8/1&(12222)\\
&7/1&(11222)\\
&6/2&(11122), (01222)\\
&5/2&(11112), (01122)\\
&4/3&(11111), (01112), (00122)\\
&3/3&(11110), (01111), (00112)\\
&2/4&(11100), (01110), (01110), (00012)\\
&1/4&(11000), (01100), (00110), (00011)\\
&0/5&(10000), (01000), (00100), (00010), (00001)\\
\hline
$\vdots$&$\vdots$&$\vdots$\\
\hline
$n$
&$2n-2/1$&$(1222\cdots 2)$\\
&$2n-3/1$&$(1122\cdots 2)$\\
&$2n-4/2$&$(1112\cdots 2)$, $(0122\cdots 2)$\\
$\vdots$&$\vdots$&$\vdots$\\
&$4/n-2$&$(0\cdots 0111110\cdots 0), (0\cdots 01112), (0\cdots 0122)$\\
&$3/n-2$&$(0\cdots 011110\cdots 0), (0\cdots 0112)$\\
&$2/n-1$&$(0\cdots 01110\cdots 0), (0\cdots 012)$\\
&$1/n-1$&$(0\cdots 0110\cdots 0)$\\
&$0/n$  &$(0\cdots 010\cdots 0)$\\
\end{longtable}
where Dynkin diagram $B_n$ with its coefficients of its positive roots 
\begin{align*}
\underset{B_n}{
\xygraph{!~:{@{=}|@{}}
    \circ   ([]!{+(0,-.3)} {k_\alpha^1})  - [r] 
    \circ   ([]!{+(0,-.3)} {k_\alpha^2})  - [r] 
    \circ   ([]!{+(0,-.3)} {k_\alpha^3})  - [r] 
    \cdots                             - [r]
    \circ   ([]!{+(0,-.3)} {k_\alpha^{n-1}}) : [r]
    \bullet ([]!{+(0,-.3)} {k_\alpha^n})}}.
\end{align*}
\end{center}

\newpage
\subsection{$C_n=\mathfrak{usp}_{2n}$}

\begin{center}
\begin{longtable}{ccp{10cm}}
\caption{Positive roots of $C_n$}
\label{Table:Positive-roots-C}\\
\hline\hline
Rank&Level/State&Coefficients of positive roots $(k_\alpha^1\cdots k_\alpha^n)$\\\hline\hline
\endfirsthead
\multicolumn{3}{c}{Table~\ref{Table:Positive-roots-C} (continued)}\\\hline\hline
Rank&Level/State&Coefficients of positive roots $(k_\alpha^1\cdots k_\alpha^n)$\\\hline\hline
\endhead
\hline
\endfoot
2
&2/1&(21)\\
&1/1&(11)\\
&0/2&(10), (01)\\
\hline
3
&4/1&(221)\\
&3/1&(121)\\
&2/2&(111), (021)\\
&1/2&(110), (011)\\
&0/3&(100), (010), (001)\\
\hline
4
&6/1&(2221)\\
&5/1&(1221)\\
&4/2&(1121), (0221)\\
&3/2&(1111), (0121)\\
&2/3&(1110), (0111), (0012)\\
&1/3&(1100), (0110), (0011)\\
&0/4&(1000), (0100), (0010), (0001)\\
\hline
5
&8/1&(22221)\\
&7/1&(12221)\\
&6/2&(11221), (02221)\\
&5/2&(11121), (01221)\\
&4/3&(11111), (01121), (00221)\\
&3/3&(11110), (01111), (00121)\\
&2/4&(11100), (01110), (00111), (00021)\\
&1/4&(11000), (01100), (00110), (00011)\\
&0/5&(10000), (01000), (00100), (00010), (00001)\\
\hline
$\vdots$&$\vdots$&$\vdots$\\
\hline
$n$
&$2n-2/1$&$(222\cdots 21)$\\
&$2n-3/1$&$(122\cdots 21)$\\
&$2n-4/2$&$(112\cdots 21), (012\cdots 21)$\\
$\vdots$&$\vdots$&$\vdots$\\
&$4/n-2$&$(0\cdots 0111110\cdots 0), (0\cdots 01121), (0\cdots 1221)$\\
&$3/n-2$&$(0\cdots 011110\cdots 0), (0\cdots 0121)$\\
&$2/n-1$&$(0\cdots 01110\cdots 0), (0\cdots 021)$\\
&$1/n-1$&$(0\cdots 0110\cdots 0)$\\
&$0/n$  &$(0\cdots 010\cdots 0)$\\
\end{longtable}
where Dynkin diagram of $C_n$ with its coefficients of its positive roots 
\begin{align*}
\underset{C_n}{
\xygraph{!~:{@{=}|@{}}
    \bullet ([]!{+(0,-.3)} {k_\alpha^1}) - [r]
    \bullet ([]!{+(0,-.3)} {k_\alpha^2}) - [r] \cdots - [r]
    \bullet ([]!{+(0,-.3)} {k_\alpha^{n-1}}) : [r]
    \circ   ([]!{+(0,-.3)} {k_\alpha^n})}}.
\end{align*}
\end{center}

\newpage
\subsection{$D_n=\mathfrak{so}_{2n}$}

\begin{center}
\begin{longtable}{ccp{10cm}}
\caption{Positive roots of $D_n$}
\label{Table:Positive-roots-D}\\
\hline\hline
Rank&Level/State&Coefficients of positive roots $(k_\alpha^1\cdots k_\alpha^n)$\\\hline\hline
\endfirsthead
\multicolumn{3}{c}{Table~\ref{Table:Positive-roots-D} (continued)}\\\hline\hline
Rank&Level/State&Coefficients of positive roots $(k_\alpha^1\cdots k_\alpha^n)$\\\hline\hline
\endhead
\hline
\endfoot
4
&4/1&(1211)\\
&3/1&(1111)\\
&2/3&(1110), (0111), (1101)\\
&1/3&(1100), (0110), (0011)\\
&0/4&(1000), (0100), (0010), (0001)\\
\hline
5
&6/1&(12211)\\
&5/1&(11211)\\
&4/2&(11111), (01211)\\
&3/3&(11110), (11101), (01111)\\
&2/4&(11100), (01110), (00111), (01101)\\
&1/4&(11000), (01100), (00110), (00101)\\
&0/5&(10000), (01000), (00100), (00010), (000001)\\
\hline
6
&8/1&(122211)\\
&7/1&(112211)\\
&6/2&(012211), (111211)\\
&5/2&(011211), (111111)\\
&4/4&(001211), (011111), (111110), (111101)\\
&3/4&(111100), (011110), (011101), (001111)\\
&2/5&(111000), (011100), (001110), (001101), (000111)\\
&1/5&(110000), (011000), (001100), (000110), (000101)\\
&0/6&(100000), (010000), (001000), (000100), (000010), (000001)\\
\hline
7
&10/1&(1222211)\\
&9/1&(1122211)\\
&8/2&(1112211), (0122211)\\
&7/2&(1111211), (0112211)\\
&6/3&(1111111), (0111211), (0012211)\\
&5/4&(1111110), (1111101), (0111111), (0011211)\\
&4/5&(1111100), (0111110), (0111101), (0011111), (0001211)\\
&3/5&(1111000), (0111100), (0011110), (0011101), (0001111)\\
&2/6&(1110000), (0111000), (0011100), (0001110), (0001101), (0000111)\\
&1/6&(1100000), (0110000), (0011000), (0001100), (0000110), (0000101)\\
&0/7&(1000000), (0100000), (0010000), (0001000), (0000100), (0000010),
	 (0000001)\\
\hline
$\vdots$&$\vdots$&$\vdots$\\
\end{longtable}
where Dynkin diagram of $D_n$ with its coefficients of its positive roots 
\begin{align*}
\underset{D_n}{
\xygraph{
    \circ ([]!{+(0,-.3)} {k_\alpha^1})  - [r] 
    \circ ([]!{+(0,-.3)} {k_\alpha^2})  - [r] 
    \circ ([]!{+(0,-.3)} {k_\alpha^3})  - [r] 
    \cdots                        - [r]
    \circ ([]!{+(0,-.3)} {k_\alpha^{n-2}}) (
        - []!{+(1,.5)}  \circ ([]!{+(0,-.3)} {k_\alpha^{n-1}}),
        - []!{+(1,-.5)} \circ ([]!{+(0,-.3)} {k_\alpha^n})
)}}
\end{align*}
\end{center}

\newpage
\subsection{$E_n$}

\begin{center}
\begin{longtable}{ccp{10cm}}
\caption{Positive roots of $E_n$}
\label{Table:Positive-roots-E}\\
\hline\hline
Rank&Level/State&Coefficients of positive roots $(k_\alpha^1\cdots k_\alpha^n)$\\\hline\hline
\endfirsthead
\multicolumn{3}{c}{Table~\ref{Table:Positive-roots-E} (continued)}\\\hline\hline
Rank&Level/State&Coefficients of positive roots $(k_\alpha^1\cdots k_\alpha^n)$\\\hline\hline
\endhead
\hline
\endfoot
6
&10/1&(123212)\\
&9/1 &(123211)\\
&8/1 &(122211)\\
&7/2 &(122111), (112211)\\
&6/3 &(122101), (112111), (012211)\\
&5/3 &(112101), (111111), (012111)\\
&4/4 &(111101), (012101), (111110), (011111)\\
&3/5 &(111100), (111001), (011101), (011110), (001111)\\
\end{longtable}
where Dynkin diagram of $E_n$ with its coefficients of its positive roots 
\begin{align*}
&\underset{E_n}{
\xygraph{
    \circ ([]!{+(0,-.3)}  {k_\alpha^1}) - [r]
    \circ ([]!{+(0,-.3)}  {k_\alpha^2}) - [r]
    \circ ([]!{+(.3,-.3)} {\hspace{-2em}k_\alpha^3}) (
        - [u] \circ ([]!{+(.3,0)}  {k_\alpha^{n}}),
        - [r] \circ ([]!{+(0,-.3)} {k_\alpha^4})
        - [r] \cdots                        
        - [r] \circ ([]!{+(0,-.3)} {k_\alpha^{n-1}})
)}}.
\end{align*}
\end{center}

\newpage
\subsection{$F_4$}

\begin{center}
\begin{longtable}{ccp{10cm}}
\caption{Positive roots of $F_4$}
\label{Table:Positive-roots-F}\\
\hline\hline
Rank&Level/State&Coefficients of positive roots $(k_\alpha^1\cdots k_\alpha^n)$\\\hline\hline
\endfirsthead
\multicolumn{3}{c}{Table~\ref{Table:Positive-roots-F} (continued)}\\\hline\hline
Rank&Level/State&Coefficients of positive roots $(k_\alpha^1\cdots k_\alpha^n)$\\\hline\hline
\endhead
\hline
\endfoot
4
&10/1&(2342)\\
&9/1 &(1342)\\
&8/1 &(1242)\\
&7/1 &(1232)\\
&6/2 &(1231), (1222)\\
&5/2 &(1221), (1122)\\
&4/3 &(1121), (1220), (0122)\\
&3/3 &(1111), (1120), (0121)\\
&2/3 &(1110), (0120), (0111)\\
&1/3 &(1100), (0110), (0011)\\
&0/4 &(1000), (0100), (0010), (0001)\\
\end{longtable}
where Dynkin diagram of $F_4$ with its coefficients of its positive roots 
\begin{align*}
\underset{F_4}{
\xygraph{!~:{@{=}|@{}}
    \circ   ([]!{+(0,-.3)} {k_\alpha^1}) - [r]
    \circ   ([]!{+(0,-.3)} {k_\alpha^2}) : [r]
    \bullet ([]!{+(0,-.3)} {k_\alpha^3}) - [r]
    \bullet ([]!{+(0,-.3)} {k_\alpha^4})}}.
\end{align*}
\end{center}

\newpage
\subsection{$G_2$}

\begin{center}
\begin{longtable}{ccp{10cm}}
\caption{Positive roots of $G_2$}
\label{Table:Positive-roots-G}\\
\hline\hline
Rank&Level/State&Coefficients of positive roots $(k_\alpha^1\cdots k_\alpha^n)$\\\hline\hline
\endfirsthead
\multicolumn{3}{c}{Table~\ref{Table:Positive-roots-G} (continued)}\\\hline\hline
Rank&Level/State&Coefficients of positive roots $(k_\alpha^1\cdots k_\alpha^n)$\\\hline\hline
\endhead
\hline
\endfoot
2
&4/1&(23)\\
&3/1&(13)\\
&2/1&(12)\\
&1/1&(11)\\
&0/2&(10), (01)\\
\end{longtable}
where Dynkin diagram of $G_2$ with its coefficients of its positive roots 
\begin{align*}
\underset{G_2}{
\xygraph{!~:{@3{-}|@{}}
    \circ   ([]!{+(0,-.3)} {k_\alpha^1}) : [r]
    \bullet ([]!{+(0,-.3)} {k_\alpha^2})}}.
\end{align*}
\end{center}

%% file: weight-diagrams.tex
\section{Weight diagrams}
\label{Sec:Weight-diagrams}

\subsection{$A_n=\mathfrak{su}_{n+1}$}

\subsubsection{Rank 1}

\begin{center}
\begin{longtable}{ccc}
\caption{Weight diagrams of $A_1$}
\label{Table:Weight-diagrams-A1}\\
\hline\hline
$A_1$ Irrep.&$d(R)$& Weight\\\hline\hline
\endfirsthead
\multicolumn{3}{c}{Table~\ref{Table:Weight-diagrams-A1} (continued)}\\\hline\hline
$A_1$ Irrep.&$d(R)$& Weight\\\hline\hline
\endhead
\hline
\endfoot
(0)&${\bf 1}$      & $(0)$\\
\hline
(1)&${\bf 2}$      & $(1)$\\
   &               & $(\bar{1})$\\
\hline
(2)&${\bf 3}$      & $(2)$\\
   &               & $(0)$\\
   &               & $(\bar{2})$\\
\hline
(3)&${\bf 4}$      & $(3)$\\
   &               & $(1)$\\
\end{longtable}
\end{center}

\newpage
\subsubsection{Rank 2}

\begin{center}
\begin{longtable}{ccc}
\caption{Weight diagrams of $A_2$}
\label{Table:Weight-diagrams-A2}\\
\hline\hline
$A_2$ Irrep.&$d(R)$& Weight\\\hline\hline
\endfirsthead
\multicolumn{3}{c}{Table~\ref{Table:Weight-diagrams-A2} (continued)}\\\hline\hline
$A_2$ Irrep.&$d(R)$& Weight\\\hline\hline
\endhead
\hline
\endfoot
(00)&${\bf 1}$
&$(00)$\\
\hline
(10)&${\bf 3}$
 &$(10)$\\
&&$(\bar{1}1)$\\
&&$(0\bar{1})$\\
\hline
(01)&$\overline{\bf 3}$
 &$(01)$\\
&&$(1\bar{1})$\\
&&$(\bar{1}0)$\\
\hline
(02)&${\bf 6}$
 &$(02)$\\
\end{longtable}
\end{center}

\newpage
\subsubsection{Rank 3}

\begin{center}
\begin{longtable}{ccc}
\caption{Weight diagrams of $A_3$}
\label{Table:Weight-diagrams-A3}\\
\hline\hline
$A_3$ Irrep.&$d(R)$& Weight\\\hline\hline
\endfirsthead
\multicolumn{3}{c}{Table~\ref{Table:Weight-diagrams-A3} (continued)}\\\hline\hline
$A_3$ Irrep.&$d(R)$& Weight\\\hline\hline
\endhead
\hline
\endfoot
(000)&${\bf 1}$
 &$(000)$\\
\hline
(100)&${\bf 4}$
&$(100)$\\
&&$(\bar{1}10)$\\
&&$(0\bar{1}1)$\\
&& $(00\bar{1})$\\
\hline
(001)&$\overline{\bf 4}$
 &$(001)$\\
&&$(01\bar{1})$\\
&&$(1\bar{1}0)$\\
\end{longtable}
\end{center}

\newpage
\subsubsection{Rank 4}

\begin{center}
\begin{longtable}{ccc}
\caption{Weight diagrams of $A_4$}
\label{Table:Weight-diagrams-A4}\\
\hline\hline
$A_4$ Irrep.&$d(R)$& Weight\\\hline\hline
\endfirsthead
\multicolumn{3}{c}{Table~\ref{Table:Weight-diagrams-A4} (continued)}\\\hline\hline
$A_4$ Irrep.&$d(R)$& Weight\\\hline\hline
\endhead
\hline
\endfoot
(0000)&${\bf 1}$
 &$(0000)$\\
\hline
(1000)&${\bf 5}$
 &$(1000)$\\
&&$(\bar{1}100)$\\
&&$(0\bar{1}10)$\\
&&$(00\bar{1}1)$\\
&&$(000\bar{1})$\\
\hline
(0001)&$\overline{\bf 5}$
 &$(0001)$\\
&&$(001\bar{1})$\\
\end{longtable}
\end{center}

\newpage
\subsubsection{Rank 5}

\begin{center}
\begin{longtable}{ccc}
\caption{Weight diagrams of $A_5$}
\label{Table:Weight-diagrams-A5}\\
\hline\hline
$A_5$ Irrep.&$d(R)$& Weight\\\hline\hline
\endfirsthead
\multicolumn{3}{c}{Table~\ref{Table:Weight-diagrams-A5} (continued)}\\\hline\hline
$A_5$ Irrep.&$d(R)$& Weight\\\hline\hline
\endhead
\hline
\endfoot
(00000)&${\bf 1}$
 &$(00000)$\\
\hline
(10000)&${\bf 6}$
 &$(10000)$\\
&&$(\bar{1}1000)$\\
&&$(0\bar{1}100)$\\
&&$(00\bar{1}10)$\\
&&$(000\bar{1}1)$\\
&&$(0000\bar{1})$\\
\hline
(00001)&$\overline{\bf 6}$
 &$(00001)$\\
\end{longtable}
\end{center}

\newpage
\subsubsection{Rank 6}

\begin{center}
\begin{longtable}{ccc}
\caption{Weight diagrams of $A_6$}
\label{Table:Weight-diagrams-A6}\\
\hline\hline
$A_6$ Irrep.&$d(R)$& Weight\\\hline\hline
\endfirsthead
\multicolumn{3}{c}{Table~\ref{Table:Weight-diagrams-A6} (continued)}\\\hline\hline
$A_6$ Irrep.&$d(R)$& Weight\\\hline\hline
\endhead
\hline
\endfoot
(000000)&${\bf 1}$
 &$(000000)$\\
\hline
(100000)&${\bf 7}$
 &$(100000)$\\
&&$(\bar{1}10000)$\\
&&$(0\bar{1}1000)$\\
&&$(00\bar{1}100)$\\
&&$(000\bar{1}10)$\\
&&$(0000\bar{1}1)$\\
&&$(00000\bar{1})$\\
\end{longtable}
\end{center}

\newpage
\subsubsection{Rank 7}

\begin{center}
\begin{longtable}{ccc}
\caption{Weight diagrams of $A_7$}
\label{Table:Weight-diagrams-A7}\\
\hline\hline
$A_7$ Irrep.&$d(R)$& Weight\\\hline\hline
\endfirsthead
\multicolumn{3}{c}{Table~\ref{Table:Weight-diagrams-A7} (continued)}\\\hline\hline
$A_7$ Irrep.&$d(R)$& Weight\\\hline\hline
\endhead
\hline
\endfoot
(0000000)&${\bf 1}$
 &$(0000000)$\\
\hline
(1000000)&${\bf 8}$
 &$(1000000)$\\
&&$(\bar{1}100000)$\\
&&$(0\bar{1}10000)$\\
&&$(00\bar{1}1000)$\\
&&$(000\bar{1}100)$\\
&&$(0000\bar{1}10)$\\
&&$(00000\bar{1}1)$\\
\end{longtable}
\end{center}

\newpage
\subsubsection{Rank 8}

\begin{center}
\begin{longtable}{ccc}
\caption{Weight diagrams of $A_8$}
\label{Table:Weight-diagrams-A8}\\
\hline\hline
$A_8$ Irrep.&$d(R)$& Weight\\\hline\hline
\endfirsthead
\multicolumn{3}{c}{Table~\ref{Table:Weight-diagrams-A8} (continued)}\\\hline\hline
$A_8$ Irrep.&$d(R)$& Weight\\\hline\hline
\endhead
\hline
\endfoot
(00000000)&${\bf 1}$
 &$(00000000)$\\
\hline
(10000000)&${\bf 9}$
 &$(10000000)$\\
&&$(\bar{1}1000000)$\\
&&$(0\bar{1}100000)$\\
&&$(00\bar{1}10000)$\\
&&$(000\bar{1}1000)$\\
&&$(0000\bar{1}100)$\\
&&$(00000\bar{1}10)$\\
\end{longtable}
\end{center}

\newpage
\subsubsection{Rank 9}

\begin{center}
\begin{longtable}{ccc}
\caption{Weight diagrams of $A_9$}
\label{Table:Weight-diagrams-A9}\\
\hline\hline
$A_9$ Irrep.&$d(R)$& Weight\\\hline\hline
\endfirsthead
\multicolumn{3}{c}{Table~\ref{Table:Weight-diagrams-A9} (continued)}\\\hline\hline
$A_9$ Irrep.&$d(R)$& Weight\\\hline\hline
\endhead
\hline
\endfoot
(000000000)&${\bf 1}$
 &$(000000000)$\\
\hline
   (100000000)&${\bf 10}$
 &$(100000000)$\\
&&$(\bar{1}10000000)$\\
&&$(0\bar{1}1000000)$\\
&&$(00\bar{1}100000)$\\
&&$(000\bar{1}10000)$\\
&&$(0000\bar{1}1000)$\\
&&$(00000\bar{1}100)$\\
\end{longtable}
\end{center}

\newpage
\subsubsection{Rank 10}

\begin{center}
\begin{longtable}{ccc}
\caption{Weight diagrams of $A_{10}$}
\label{Table:Weight-diagrams-A10}\\
\hline\hline
$A_{10}$ Irrep.&$d(R)$& Weight\\\hline\hline
\endfirsthead
\multicolumn{3}{c}{Table~\ref{Table:Weight-diagrams-A10} (continued)}\\\hline\hline
$A_{10}$ Irrep.&$d(R)$& Weight\\\hline\hline
\endhead
\hline
\endfoot
(0000000000)&${\bf 1}$
 &$(0000000000)$\\
\hline
(1000000000)&${\bf 11}$
 &$(1000000000)$\\
&&$(\bar{1}100000000)$\\
&&$(0\bar{1}10000000)$\\
&&$(00\bar{1}1000000)$\\
&&$(000\bar{1}100000)$\\
&&$(0000\bar{1}10000)$\\
&&$(00000\bar{1}1000)$\\
\end{longtable}
\end{center}

\newpage
\subsubsection{Rank 11}

\begin{center}
\begin{longtable}{ccc}
\caption{Weight diagrams of $A_{11}$}
\label{Table:Weight-diagrams-A11}\\
\hline\hline
$A_{11}$ Irrep.&$d(R)$& Weight\\\hline\hline
\endfirsthead
\multicolumn{3}{c}{Table~\ref{Table:Weight-diagrams-A11} (continued)}\\\hline\hline
$A_{11}$ Irrep.&$d(R)$& Weight\\\hline\hline
\endhead
\hline
\endfoot
(00000000000)&${\bf 1}$
 &$(00000000000)$\\
\hline
(10000000000)&${\bf 12}$
 &$(10000000000)$\\
&&$(\bar{1}1000000000)$\\
&&$(0\bar{1}100000000)$\\
&&$(00\bar{1}10000000)$\\
&&$(000\bar{1}1000000)$\\
&&$(0000\bar{1}100000)$\\
&&$(00000\bar{1}10000)$\\
&&$(000000\bar{1}1000)$\\
&&$(0000000\bar{1}100)$\\
&&$(00000000\bar{1}10)$\\
&&$(000000000\bar{1}1)$\\
&&$(0000000000\bar{1})$\\
\hline
(00000000001)&$\overline{\bf 12}$
 &$(00000000001)$\\
&&$(0000000001\bar{1})$\\
&&$(000000001\bar{1}0)$\\
&&$(00000001\bar{1}00)$\\
&&$(0000001\bar{1}000)$\\
&&$(000001\bar{1}0000)$\\
&&$(00001\bar{1}00000)$\\
&&$(0001\bar{1}000000)$\\
&&$(001\bar{1}0000000)$\\
&&$(01\bar{1}00000000)$\\
&&$(1\bar{1}000000000)$\\
&&$(\bar{1}0000000000)$\\
\end{longtable}
\end{center}

\newpage
\subsubsection{Rank 12}

\begin{center}
\begin{longtable}{ccc}
\caption{Weight diagrams of $A_{12}$}
\label{Table:Weight-diagrams-A12}\\
\hline\hline
$A_{12}$ Irrep.&$d(R)$& Weight\\\hline\hline
\endfirsthead
\multicolumn{3}{c}{Table~\ref{Table:Weight-diagrams-A12} (continued)}\\\hline\hline
$A_{12}$ Irrep.&$d(R)$& Weight\\\hline\hline
\endhead
\hline
\endfoot
(000000000000)&${\bf 1}$
 &$(000000000000)$\\
\hline
(100000000000)&${\bf 13}$
 &$(100000000000)$\\
&&$(\bar{1}10000000000)$\\
&&$(0\bar{1}1000000000)$\\
&&$(00\bar{1}100000000)$\\
&&$(000\bar{1}10000000)$\\
&&$(0000\bar{1}1000000)$\\
&&$(00000\bar{1}100000)$\\
&&$(000000\bar{1}10000)$\\
&&$(0000000\bar{1}1000)$\\
&&$(00000000\bar{1}100)$\\
&&$(000000000\bar{1}10)$\\
&&$(0000000000\bar{1}1)$\\
&&$(00000000000\bar{1})$\\
\hline
(000000000001)&$\overline{\bf 13}$
 &$(000000000001)$\\
&&$(00000000001\bar{1})$\\
&&$(0000000001\bar{1}0)$\\
&&$(000000001\bar{1}00)$\\
&&$(00000001\bar{1}000)$\\
&&$(0000001\bar{1}0000)$\\
&&$(000001\bar{1}00000)$\\
&&$(00001\bar{1}000000)$\\
&&$(0001\bar{1}0000000)$\\
&&$(001\bar{1}00000000)$\\
&&$(01\bar{1}000000000)$\\
&&$(1\bar{1}0000000000)$\\
&&$(\bar{1}00000000000)$\\
\end{longtable}
\end{center}

\newpage
\subsubsection{Rank 13}

\begin{center}
\begin{longtable}{ccc}
\caption{Weight diagrams of $A_{13}$}
\label{Table:Weight-diagrams-A13}\\
\hline\hline
$A_{13}$ Irrep.&$d(R)$& Weight\\\hline\hline
\endfirsthead
\multicolumn{3}{c}{Table~\ref{Table:Weight-diagrams-A13} (continued)}\\\hline\hline
$A_{13}$ Irrep.&$d(R)$& Weight\\\hline\hline
\endhead
\hline
\endfoot
(0000000000000)&${\bf 1}$
 &$(0000000000000)$\\
\hline
(1000000000000)&${\bf 14}$
 &$(1000000000000)$\\
&&$(\bar{1}100000000000)$\\
&&$(0\bar{1}10000000000)$\\
&&$(00\bar{1}1000000000)$\\
&&$(000\bar{1}100000000)$\\
&&$(0000\bar{1}10000000)$\\
&&$(00000\bar{1}1000000)$\\
&&$(000000\bar{1}100000)$\\
&&$(0000000\bar{1}10000)$\\
&&$(00000000\bar{1}1000)$\\
&&$(000000000\bar{1}100)$\\
&&$(0000000000\bar{1}10)$\\
&&$(00000000000\bar{1}1)$\\
&&$(000000000000\bar{1})$\\
\hline
(0000000000001)&$\overline{\bf 14}$
 &$(0000000000001)$\\
&&$(000000000001\bar{1})$\\
&&$(00000000001\bar{1}0)$\\
&&$(0000000001\bar{1}00)$\\
&&$(000000001\bar{1}000)$\\
&&$(00000001\bar{1}0000)$\\
&&$(0000001\bar{1}00000)$\\
&&$(000001\bar{1}000000)$\\
&&$(00001\bar{1}0000000)$\\
&&$(0001\bar{1}00000000)$\\
&&$(001\bar{1}000000000)$\\
&&$(01\bar{1}0000000000)$\\
&&$(1\bar{1}00000000000)$\\
&&$(\bar{1}000000000000)$\\
\end{longtable}
\end{center}

\newpage
\subsubsection{Rank 14}

\begin{center}
\begin{longtable}{ccc}
\caption{Weight diagrams of $A_{14}$}
\label{Table:Weight-diagrams-A14}\\
\hline\hline
$A_{14}$ Irrep.&$d(R)$& Weight\\\hline\hline
\endfirsthead
\multicolumn{3}{c}{Table~\ref{Table:Weight-diagrams-A14} (continued)}\\\hline\hline
$A_{14}$ Irrep.&$d(R)$& Weight\\\hline\hline
\endhead
\hline
\endfoot
(00000000000000)&${\bf 1}$
 &$(00000000000000)$\\
\hline
(10000000000000)&${\bf 15}$
 &$(10000000000000)$\\
&&$(\bar{1}1000000000000)$\\
&&$(0\bar{1}100000000000)$\\
&&$(00\bar{1}10000000000)$\\
&&$(000\bar{1}1000000000)$\\
&&$(0000\bar{1}100000000)$\\
&&$(00000\bar{1}10000000)$\\
&&$(000000\bar{1}1000000)$\\
&&$(0000000\bar{1}100000)$\\
&&$(00000000\bar{1}10000)$\\
&&$(000000000\bar{1}1000)$\\
&&$(0000000000\bar{1}100)$\\
&&$(00000000000\bar{1}10)$\\
&&$(000000000000\bar{1}1)$\\
&&$(0000000000000\bar{1})$\\
\hline
(00000000000001)&$\overline{\bf 15}$
 &$(00000000000001)$\\
&&$(0000000000001\bar{1})$\\
&&$(000000000001\bar{1}0)$\\
&&$(00000000001\bar{1}00)$\\
&&$(0000000001\bar{1}000)$\\
&&$(000000001\bar{1}0000)$\\
&&$(00000001\bar{1}00000)$\\
&&$(0000001\bar{1}000000)$\\
&&$(000001\bar{1}0000000)$\\
&&$(00001\bar{1}00000000)$\\
&&$(0001\bar{1}000000000)$\\
&&$(001\bar{1}0000000000)$\\
&&$(01\bar{1}00000000000)$\\
&&$(1\bar{1}000000000000)$\\
&&$(\bar{1}0000000000000)$\\
\end{longtable}
\end{center}

\newpage
\subsubsection{Rank 15}

\begin{center}
\begin{longtable}{ccc}
\caption{Weight diagrams of $A_{15}$}
\label{Table:Weight-diagrams-A15}\\
\hline\hline
$A_{15}$ Irrep.&$d(R)$& Weight\\\hline\hline
\endfirsthead
\multicolumn{3}{c}{Table~\ref{Table:Weight-diagrams-A15} (continued)}\\\hline\hline
$A_{15}$ Irrep.&$d(R)$& Weight\\\hline\hline
\endhead
\hline
\endfoot
(000000000000000)&${\bf 1}$
 &$(000000000000000)$\\
\hline
(100000000000000)&${\bf 16}$
 &$(100000000000000)$\\
&&$(\bar{1}10000000000000)$\\
&&$(0\bar{1}1000000000000)$\\
&&$(00\bar{1}100000000000)$\\
&&$(000\bar{1}10000000000)$\\
&&$(0000\bar{1}1000000000)$\\
&&$(00000\bar{1}100000000)$\\
&&$(000000\bar{1}10000000)$\\
&&$(0000000\bar{1}1000000)$\\
&&$(00000000\bar{1}100000)$\\
&&$(000000000\bar{1}10000)$\\
&&$(0000000000\bar{1}1000)$\\
&&$(00000000000\bar{1}100)$\\
&&$(000000000000\bar{1}10)$\\
&&$(0000000000000\bar{1}1)$\\
&&$(00000000000000\bar{1})$\\
\hline
(000000000000001)&$\overline{\bf 16}$
 &$(000000000000001)$\\
&&$(00000000000001\bar{1})$\\
&&$(0000000000001\bar{1}0)$\\
&&$(000000000001\bar{1}00)$\\
&&$(00000000001\bar{1}000)$\\
&&$(0000000001\bar{1}0000)$\\
&&$(000000001\bar{1}00000)$\\
&&$(00000001\bar{1}000000)$\\
&&$(0000001\bar{1}0000000)$\\
&&$(000001\bar{1}00000000)$\\
&&$(00001\bar{1}000000000)$\\
&&$(0001\bar{1}0000000000)$\\
&&$(001\bar{1}00000000000)$\\
&&$(01\bar{1}000000000000)$\\
&&$(1\bar{1}0000000000000)$\\
&&$(\bar{1}00000000000000)$\\
\end{longtable}
\end{center}

\newpage
\subsubsection{Rank 16}

\begin{center}
\begin{longtable}{ccc}
\caption{Weight diagrams of $A_{16}$}
\label{Table:Weight-diagrams-A16}\\
\hline\hline
$A_{16}$ Irrep.&$d(R)$& Weight\\\hline\hline
\endfirsthead
\multicolumn{3}{c}{Table~\ref{Table:Weight-diagrams-A16} (continued)}\\\hline\hline
$A_{16}$ Irrep.&$d(R)$& Weight\\\hline\hline
\endhead
\hline
\endfoot
(0000000000000000)&${\bf 1}$
 &$(0000000000000000)$\\
\hline
(1000000000000000)&${\bf 17}$
 &$(1000000000000000)$\\
&&$(\bar{1}100000000000000)$\\
&&$(0\bar{1}10000000000000)$\\
&&$(00\bar{1}1000000000000)$\\
&&$(000\bar{1}100000000000)$\\
&&$(0000\bar{1}10000000000)$\\
&&$(00000\bar{1}1000000000)$\\
&&$(000000\bar{1}100000000)$\\
&&$(0000000\bar{1}10000000)$\\
&&$(00000000\bar{1}1000000)$\\
&&$(000000000\bar{1}100000)$\\
&&$(0000000000\bar{1}10000)$\\
&&$(00000000000\bar{1}1000)$\\
&&$(000000000000\bar{1}100)$\\
&&$(0000000000000\bar{1}10)$\\
&&$(00000000000000\bar{1}1)$\\
&&$(000000000000000\bar{1})$\\
\hline
(0000000000000001)&$\overline{\bf 17}$
 &$(0000000000000001)$\\
&&$(000000000000001\bar{1})$\\
&&$(00000000000001\bar{1}0)$\\
&&$(0000000000001\bar{1}00)$\\
&&$(000000000001\bar{1}000)$\\
&&$(00000000001\bar{1}0000)$\\
&&$(0000000001\bar{1}00000)$\\
&&$(000000001\bar{1}000000)$\\
&&$(00000001\bar{1}0000000)$\\
&&$(0000001\bar{1}00000000)$\\
&&$(000001\bar{1}000000000)$\\
&&$(00001\bar{1}0000000000)$\\
&&$(0001\bar{1}00000000000)$\\
&&$(001\bar{1}000000000000)$\\
&&$(01\bar{1}0000000000000)$\\
&&$(1\bar{1}00000000000000)$\\
&&$(\bar{1}000000000000000)$\\
\end{longtable}
\end{center}

\newpage
\subsubsection{Rank 17}

\begin{center}
\begin{longtable}{ccc}
\caption{Weight diagrams of $A_{17}$}
\label{Table:Weight-diagrams-A17}\\
\hline\hline
$A_{17}$ Irrep.&$d(R)$& Weight\\\hline\hline
\endfirsthead
\multicolumn{3}{c}{Table~\ref{Table:Weight-diagrams-A17} (continued)}\\\hline\hline
$A_{17}$ Irrep.&$d(R)$& Weight\\\hline\hline
\endhead
\hline
\endfoot
(00000000000000000)&${\bf 1}$
 &$(00000000000000000)$\\
\hline
(10000000000000000)&${\bf 18}$
 &$(10000000000000000)$\\
&&$(\bar{1}1000000000000000)$\\
&&$(0\bar{1}100000000000000)$\\
&&$(00\bar{1}10000000000000)$\\
&&$(000\bar{1}1000000000000)$\\
&&$(0000\bar{1}100000000000)$\\
&&$(00000\bar{1}10000000000)$\\
&&$(000000\bar{1}1000000000)$\\
&&$(0000000\bar{1}100000000)$\\
&&$(00000000\bar{1}10000000)$\\
&&$(000000000\bar{1}1000000)$\\
&&$(0000000000\bar{1}100000)$\\
&&$(00000000000\bar{1}10000)$\\
&&$(000000000000\bar{1}1000)$\\
&&$(0000000000000\bar{1}100)$\\
&&$(00000000000000\bar{1}10)$\\
&&$(000000000000000\bar{1}1)$\\
&&$(0000000000000000\bar{1})$\\
\hline
(00000000000000001)&$\overline{\bf 18}$
 &$(00000000000000001)$\\
&&$(0000000000000001\bar{1})$\\
&&$(000000000000001\bar{1}0)$\\
&&$(00000000000001\bar{1}00)$\\
&&$(0000000000001\bar{1}000)$\\
&&$(000000000001\bar{1}0000)$\\
&&$(00000000001\bar{1}00000)$\\
&&$(0000000001\bar{1}000000)$\\
&&$(000000001\bar{1}0000000)$\\
&&$(00000001\bar{1}00000000)$\\
&&$(0000001\bar{1}000000000)$\\
&&$(000001\bar{1}0000000000)$\\
&&$(00001\bar{1}00000000000)$\\
&&$(0001\bar{1}000000000000)$\\
&&$(001\bar{1}0000000000000)$\\
&&$(01\bar{1}00000000000000)$\\
&&$(1\bar{1}000000000000000)$\\
&&$(\bar{1}0000000000000000)$\\
\end{longtable}
\end{center}

\newpage
\subsubsection{Rank 18}

\begin{center}
\begin{longtable}{ccc}
\caption{Weight diagrams of $A_{18}$}
\label{Table:Weight-diagrams-A18}\\
\hline\hline
$A_{18}$ Irrep.&$d(R)$& Weight\\\hline\hline
\endfirsthead
\multicolumn{3}{c}{Table~\ref{Table:Weight-diagrams-A18} (continued)}\\\hline\hline
$A_{18}$ Irrep.&$d(R)$& Weight\\\hline\hline
\endhead
\hline
\endfoot
(000000000000000000)&${\bf 1}$
 &$(000000000000000000)$\\
\hline
(100000000000000000)&${\bf 19}$
 &$(100000000000000000)$\\
&&$(\bar{1}10000000000000000)$\\
&&$(0\bar{1}1000000000000000)$\\
&&$(00\bar{1}100000000000000)$\\
&&$(000\bar{1}10000000000000)$\\
&&$(0000\bar{1}1000000000000)$\\
&&$(00000\bar{1}100000000000)$\\
&&$(000000\bar{1}10000000000)$\\
&&$(0000000\bar{1}1000000000)$\\
&&$(00000000\bar{1}100000000)$\\
&&$(000000000\bar{1}10000000)$\\
&&$(0000000000\bar{1}1000000)$\\
&&$(00000000000\bar{1}100000)$\\
&&$(000000000000\bar{1}10000)$\\
&&$(0000000000000\bar{1}1000)$\\
&&$(00000000000000\bar{1}100)$\\
&&$(000000000000000\bar{1}10)$\\
&&$(0000000000000000\bar{1}1)$\\
&&$(00000000000000000\bar{1})$\\
\hline
(000000000000000001)&$\overline{\bf 19}$
 &$(000000000000000001)$\\
&&$(00000000000000001\bar{1})$\\
&&$(0000000000000001\bar{1}0)$\\
&&$(000000000000001\bar{1}00)$\\
&&$(00000000000001\bar{1}000)$\\
&&$(0000000000001\bar{1}0000)$\\
&&$(000000000001\bar{1}00000)$\\
&&$(00000000001\bar{1}000000)$\\
&&$(0000000001\bar{1}0000000)$\\
&&$(000000001\bar{1}00000000)$\\
&&$(00000001\bar{1}000000000)$\\
&&$(0000001\bar{1}0000000000)$\\
&&$(000001\bar{1}00000000000)$\\
&&$(00001\bar{1}000000000000)$\\
&&$(0001\bar{1}0000000000000)$\\
&&$(001\bar{1}00000000000000)$\\
&&$(01\bar{1}000000000000000)$\\
&&$(1\bar{1}0000000000000000)$\\
&&$(\bar{1}00000000000000000)$\\
\end{longtable}
\end{center}

\newpage
\subsubsection{Rank 19}

\begin{center}
\begin{longtable}{ccc}
\caption{Weight diagrams of $A_{19}$}
\label{Table:Weight-diagrams-A19}\\
\hline\hline
$A_{19}$ Irrep.&$d(R)$& Weight\\\hline\hline
\endfirsthead
\multicolumn{3}{c}{Table~\ref{Table:Weight-diagrams-A19} (continued)}\\\hline\hline
$A_{19}$ Irrep.&$d(R)$& Weight\\\hline\hline
\endhead
\hline
\endfoot
(0000000000000000000)&${\bf 1}$
 &$(0000000000000000000)$\\
\hline
(1000000000000000000)&${\bf 20}$
 &$(1000000000000000000)$\\
&&$(\bar{1}100000000000000000)$\\
&&$(0\bar{1}10000000000000000)$\\
&&$(00\bar{1}1000000000000000)$\\
&&$(000\bar{1}100000000000000)$\\
&&$(0000\bar{1}10000000000000)$\\
&&$(00000\bar{1}1000000000000)$\\
&&$(000000\bar{1}100000000000)$\\
&&$(0000000\bar{1}10000000000)$\\
&&$(00000000\bar{1}1000000000)$\\
&&$(000000000\bar{1}100000000)$\\
&&$(0000000000\bar{1}10000000)$\\
&&$(00000000000\bar{1}1000000)$\\
&&$(000000000000\bar{1}100000)$\\
&&$(0000000000000\bar{1}10000)$\\
&&$(00000000000000\bar{1}1000)$\\
&&$(000000000000000\bar{1}100)$\\
&&$(0000000000000000\bar{1}10)$\\
&&$(00000000000000000\bar{1}1)$\\
&&$(000000000000000000\bar{1})$\\
\hline
(0000000000000000001)&$\overline{\bf 20}$
 &$(0000000000000000001)$\\
&&$(000000000000000001\bar{1})$\\
&&$(00000000000000001\bar{1}0)$\\
&&$(0000000000000001\bar{1}00)$\\
&&$(000000000000001\bar{1}000)$\\
&&$(00000000000001\bar{1}0000)$\\
&&$(0000000000001\bar{1}00000)$\\
&&$(000000000001\bar{1}000000)$\\
&&$(00000000001\bar{1}0000000)$\\
&&$(0000000001\bar{1}00000000)$\\
&&$(000000001\bar{1}000000000)$\\
&&$(00000001\bar{1}0000000000)$\\
&&$(0000001\bar{1}00000000000)$\\
&&$(000001\bar{1}000000000000)$\\
&&$(00001\bar{1}0000000000000)$\\
&&$(0001\bar{1}00000000000000)$\\
&&$(001\bar{1}000000000000000)$\\
&&$(01\bar{1}0000000000000000)$\\
&&$(1\bar{1}00000000000000000)$\\
&&$(\bar{1}000000000000000000)$\\
\end{longtable}
\end{center}

\newpage
\subsubsection{Rank 20}

\begin{center}
\begin{longtable}{ccc}
\caption{Weight diagrams of $A_{20}$}
\label{Table:Weight-diagrams-A20}\\
\hline\hline
$A_{20}$ Irrep.&$d(R)$& Weight\\\hline\hline
\endfirsthead
\multicolumn{3}{c}{Table~\ref{Table:Weight-diagrams-A20} (continued)}\\\hline\hline
$A_{20}$ Irrep.&$d(R)$& Weight\\\hline\hline
\endhead
\hline
\endfoot
(00000000000000000000)&${\bf 1}$
 &$(00000000000000000000)$\\
\hline
(10000000000000000000)&${\bf 21}$
 &$(10000000000000000000)$\\
&&$(\bar{1}1000000000000000000)$\\
&&$(0\bar{1}100000000000000000)$\\
&&$(00\bar{1}10000000000000000)$\\
&&$(000\bar{1}1000000000000000)$\\
&&$(0000\bar{1}100000000000000)$\\
&&$(00000\bar{1}10000000000000)$\\
&&$(000000\bar{1}1000000000000)$\\
&&$(0000000\bar{1}100000000000)$\\
&&$(00000000\bar{1}10000000000)$\\
&&$(000000000\bar{1}1000000000)$\\
&&$(0000000000\bar{1}100000000)$\\
&&$(00000000000\bar{1}10000000)$\\
&&$(000000000000\bar{1}1000000)$\\
&&$(0000000000000\bar{1}100000)$\\
&&$(00000000000000\bar{1}10000)$\\
&&$(000000000000000\bar{1}1000)$\\
&&$(0000000000000000\bar{1}100)$\\
&&$(00000000000000000\bar{1}10)$\\
&&$(000000000000000000\bar{1}1)$\\
&&$(0000000000000000000\bar{1})$\\
\hline
(00000000000000000001)&$\overline{\bf 21}$
 &$(00000000000000000001)$\\
&&$(0000000000000000001\bar{1})$\\
&&$(000000000000000001\bar{1}0)$\\
&&$(00000000000000001\bar{1}00)$\\
&&$(0000000000000001\bar{1}000)$\\
&&$(000000000000001\bar{1}0000)$\\
&&$(00000000000001\bar{1}00000)$\\
&&$(0000000000001\bar{1}000000)$\\
&&$(000000000001\bar{1}0000000)$\\
&&$(00000000001\bar{1}00000000)$\\
&&$(0000000001\bar{1}000000000)$\\
&&$(000000001\bar{1}0000000000)$\\
&&$(00000001\bar{1}00000000000)$\\
&&$(0000001\bar{1}000000000000)$\\
&&$(000001\bar{1}0000000000000)$\\
&&$(00001\bar{1}00000000000000)$\\
&&$(0001\bar{1}000000000000000)$\\
&&$(001\bar{1}0000000000000000)$\\
&&$(01\bar{1}00000000000000000)$\\
&&$(1\bar{1}000000000000000000)$\\
&&$(\bar{1}0000000000000000000)$\\
\end{longtable}
\end{center}

\newpage
\subsection{$B_n=\mathfrak{so}_{2n+1}$}

\subsubsection{Rank 2}

\begin{center}
\begin{longtable}{ccc}
\caption{Weight diagrams of $B_2$}
\label{Table:Weight-diagrams-B2}\\
\hline\hline
$B_2$ Irrep.&$d(R)$& Weight\\\hline\hline
\endfirsthead
\multicolumn{3}{c}{Table~\ref{Table:Weight-diagrams-B2} (continued)}\\\hline\hline
$B_2$ Irrep.&$d(R)$& Weight\\\hline\hline
\endhead
\hline
\endfoot
(00)&${\bf 1}$         & $(00)$\\
\hline
(01)&${\bf 4}$         & $(01)$\\
    &                  & $(1\bar{1})$\\
    &                  & $(\bar{1}1)$\\
    &                  & $(0\bar{1})$\\
\hline
(10)&${\bf 5}$         & $(10)$\\
    &                  & $(\bar{1}2)$\\
    &                  & $(00)$\\
\end{longtable}
\end{center}

\newpage
\subsubsection{Rank 3}

\begin{center}
\begin{longtable}{ccc}
\caption{Weight diagrams of $B_3$}
\label{Table:Weight-diagrams-B3}\\
\hline\hline
$B_3$ Irrep.&$d(R)$& Weight\\\hline\hline
\endfirsthead
\multicolumn{3}{c}{Table~\ref{Table:Weight-diagrams-B3} (continued)}\\\hline\hline
$B_3$ Irrep.&$d(R)$& Weight\\\hline\hline
\endhead
\hline
\endfoot
(000)&${\bf 1}$         & $(000)$\\
\hline
(100)&${\bf 7}$         & $(100)$\\
     &                  & $(\bar{1}10)$\\
     &                  & $(0\bar{1}2)$\\
     &                  & $(000)$\\
     &                  & $(01\bar{2})$\\
     &                  & $(1\bar{1}0)$\\
     &                  & $(\bar{1}00)$\\
\end{longtable}
\end{center}

\newpage
\subsubsection{Rank 4}

\begin{center}
\begin{longtable}{ccc}
\caption{Weight diagrams of $B_4$}
\label{Table:Weight-diagrams-B4}\\
\hline\hline
$B_4$ Irrep.&$d(R)$& Weight\\\hline\hline
\endfirsthead
\multicolumn{3}{c}{Table~\ref{Table:Weight-diagrams-B4} (continued)}\\\hline\hline
$B_4$ Irrep.&$d(R)$& Weight\\\hline\hline
\endhead
\hline
\endfoot
(0000)&${\bf 1}$         & $(0000)$\\
\hline
(1000)&${\bf 9}$         & $(1000)$\\
      &                  & $(\bar{1}100)$\\
      &                  & $(0\bar{1}10)$\\
      &                  & $(00\bar{1}2)$\\
      &                  & $(0000)$\\
      &                  & $(001\bar{2})$\\
      &                  & $(01\bar{1}0)$\\
\end{longtable}
\end{center}

\newpage
\subsubsection{Rank 5}

\begin{center}
\begin{longtable}{ccc}
\caption{Weight diagrams of $B_5$}
\label{Table:Weight-diagrams-B5}\\
\hline\hline
$B_5$ Irrep.&$d(R)$& Weight\\\hline\hline
\endfirsthead
\multicolumn{3}{c}{Table~\ref{Table:Weight-diagrams-B5} (continued)}\\\hline\hline
$B_5$ Irrep.&$d(R)$& Weight\\\hline\hline
\endhead
\hline
\endfoot
(00000)&${\bf 1}$         & $(00000)$\\
\hline
(10000)&${\bf 11}$        & $(10000)$\\
       &                  & $(\bar{1}1000)$\\
       &                  & $(0\bar{1}100)$\\
       &                  & $(00\bar{1}10)$\\
       &                  & $(000\bar{1}2)$\\
       &                  & $(00000)$\\
       &                  & $(0001\bar{2})$\\
\end{longtable}
\end{center}

\newpage
\subsubsection{Rank 6}

\begin{center}
\begin{longtable}{ccc}
\caption{Weight diagrams of $B_6$}
\label{Table:Weight-diagrams-B6}\\
\hline\hline
$B_6$ Irrep.&$d(R)$& Weight\\\hline\hline
\endfirsthead
\multicolumn{3}{c}{Table~\ref{Table:Weight-diagrams-B6} (continued)}\\\hline\hline
$B_6$ Irrep.&$d(R)$& Weight\\\hline\hline
\endhead
\hline
\endfoot
(000000)&${\bf 1}$
 &$(000000)$\\
\hline
(100000)&${\bf 13}$
 &$(100000)$\\
&&$(\bar{1}10000)$\\
&&$(0\bar{1}1000)$\\
&&$(00\bar{1}100)$\\
&&$(000\bar{1}10)$\\
&&$(0000\bar{1}2)$\\
&&$(000000)$\\
&&$(00001\bar{2})$\\
&&$(0001\bar{1}0)$\\
&&$(001\bar{1}00)$\\
&&$(01\bar{1}000)$\\
&&$(1\bar{1}0000)$\\
&&$(\bar{1}00000)$\\
\end{longtable}
\end{center}

\newpage
\subsubsection{Rank 7}

\begin{center}
\begin{longtable}{ccc}
\caption{Weight diagrams of $B_7$}
\label{Table:Weight-diagrams-B7}\\
\hline\hline
$B_7$ Irrep.&$d(R)$& Weight\\\hline\hline
\endfirsthead
\multicolumn{3}{c}{Table~\ref{Table:Weight-diagrams-B7} (continued)}\\\hline\hline
$B_7$ Irrep.&$d(R)$& Weight\\\hline\hline
\endhead
\hline
\endfoot
(0000000)&${\bf 1}$
 &$(0000000)$\\
\hline
(1000000)&${\bf 15}$
 &$(1000000)$\\
&&$(\bar{1}100000)$\\
&&$(0\bar{1}10000)$\\
&&$(00\bar{1}1000)$\\
&&$(000\bar{1}100)$\\
&&$(0000\bar{1}10)$\\
&&$(00000\bar{1}2)$\\
&&$(0000000)$\\
&&$(000001\bar{2})$\\
&&$(00001\bar{1}0)$\\
&&$(0001\bar{1}00)$\\
&&$(001\bar{1}000)$\\
&&$(01\bar{1}0000)$\\
&&$(1\bar{1}00000)$\\
&&$(\bar{1}000000)$\\
\end{longtable}
\end{center}

\newpage
\subsubsection{Rank 8}

\begin{center}
\begin{longtable}{ccc}
\caption{Weight diagrams of $B_8$}
\label{Table:Weight-diagrams-B8}\\
\hline\hline
$B_8$ Irrep.&$d(R)$& Weight\\\hline\hline
\endfirsthead
\multicolumn{3}{c}{Table~\ref{Table:Weight-diagrams-B8} (continued)}\\\hline\hline
$B_8$ Irrep.&$d(R)$& Weight\\\hline\hline
\endhead
\hline
\endfoot
(00000000)&${\bf 1}$
 &$(00000000)$\\
\hline
(10000000)&${\bf 17}$
 &$(10000000)$\\
&&$(\bar{1}1000000)$\\
&&$(0\bar{1}100000)$\\
&&$(00\bar{1}10000)$\\
&&$(000\bar{1}1000)$\\
&&$(0000\bar{1}100)$\\
&&$(00000\bar{1}10)$\\
&&$(000000\bar{1}2)$\\
&&$(00000000)$\\
&&$(0000001\bar{2})$\\
&&$(000001\bar{1}0)$\\
&&$(00001\bar{1}00)$\\
&&$(0001\bar{1}000)$\\
&&$(001\bar{1}0000)$\\
&&$(01\bar{1}00000)$\\
&&$(1\bar{1}000000)$\\
&&$(\bar{1}0000000)$\\
\end{longtable}
\end{center}

\newpage
\subsubsection{Rank 9}

\begin{center}
\begin{longtable}{ccc}
\caption{Weight diagrams of $B_9$}
\label{Table:Weight-diagrams-B9}\\
\hline\hline
$B_9$ Irrep.&$d(R)$& Weight\\\hline\hline
\endfirsthead
\multicolumn{3}{c}{Table~\ref{Table:Weight-diagrams-B9} (continued)}\\\hline\hline
$B_9$ Irrep.&$d(R)$& Weight\\\hline\hline
\endhead
\hline
\endfoot
(000000000)&${\bf 1}$
 &$(000000000)$\\
\hline
(100000000)&${\bf 19}$
 &$(100000000)$\\
&&$(\bar{1}10000000)$\\
&&$(0\bar{1}1000000)$\\
&&$(00\bar{1}100000)$\\
&&$(000\bar{1}10000)$\\
&&$(0000\bar{1}1000)$\\
&&$(00000\bar{1}100)$\\
&&$(000000\bar{1}10)$\\
&&$(0000000\bar{1}2)$\\
&&$(000000000)$\\
&&$(00000001\bar{2})$\\
&&$(0000001\bar{1}0)$\\
&&$(000001\bar{1}00)$\\
&&$(00001\bar{1}000)$\\
&&$(0001\bar{1}0000)$\\
&&$(001\bar{1}00000)$\\
&&$(01\bar{1}000000)$\\
&&$(1\bar{1}0000000)$\\
&&$(\bar{1}00000000)$\\
\end{longtable}
\end{center}

\newpage
\subsubsection{Rank 10}

\begin{center}
\begin{longtable}{ccc}
\caption{Weight diagrams of $B_{10}$}
\label{Table:Weight-diagrams-B10}\\
\hline\hline
$B_{10}$ Irrep.&$d(R)$& Weight\\\hline\hline
\endfirsthead
\multicolumn{3}{c}{Table~\ref{Table:Weight-diagrams-B10} (continued)}\\\hline\hline
$B_{10}$ Irrep.&$d(R)$& Weight\\\hline\hline
\endhead
\hline
\endfoot
(0000000000)&${\bf 1}$
 &$(0000000000)$\\
\hline
(1000000000)&${\bf 21}$
 &$(1000000000)$\\
&&$(\bar{1}100000000)$\\
&&$(0\bar{1}10000000)$\\
&&$(00\bar{1}1000000)$\\
&&$(000\bar{1}100000)$\\
&&$(0000\bar{1}10000)$\\
&&$(00000\bar{1}1000)$\\
&&$(000000\bar{1}100)$\\
&&$(0000000\bar{1}10)$\\
&&$(00000000\bar{1}2)$\\
&&$(0000000000)$\\
&&$(000000001\bar{2})$\\
&&$(00000001\bar{1}0)$\\
&&$(0000001\bar{1}00)$\\
&&$(000001\bar{1}000)$\\
&&$(00001\bar{1}0000)$\\
&&$(0001\bar{1}00000)$\\
&&$(001\bar{1}000000)$\\
&&$(01\bar{1}0000000)$\\
&&$(1\bar{1}00000000)$\\
&&$(\bar{1}000000000)$\\
\end{longtable}
\end{center}

\newpage
\subsubsection{Rank 11}

\begin{center}
\begin{longtable}{ccc}
\caption{Weight diagrams of $B_{11}$}
\label{Table:Weight-diagrams-B11}\\
\hline\hline
$B_{11}$ Irrep.&$d(R)$& Weight\\\hline\hline
\endfirsthead
\multicolumn{3}{c}{Table~\ref{Table:Weight-diagrams-B11} (continued)}\\\hline\hline
$B_{11}$ Irrep.&$d(R)$& Weight\\\hline\hline
\endhead
\hline
\endfoot
(00000000000)&${\bf 1}$
 &$(00000000000)$\\
\hline
(10000000000)&${\bf 23}$
 &$(10000000000)$\\
&&$(\bar{1}1000000000)$\\
&&$(0\bar{1}100000000)$\\
&&$(00\bar{1}10000000)$\\
&&$(000\bar{1}1000000)$\\
&&$(0000\bar{1}100000)$\\
&&$(00000\bar{1}10000)$\\
&&$(000000\bar{1}1000)$\\
&&$(0000000\bar{1}100)$\\
&&$(00000000\bar{1}10)$\\
&&$(000000000\bar{1}2)$\\
&&$(00000000000)$\\
&&$(0000000001\bar{2})$\\
&&$(000000001\bar{1}0)$\\
&&$(00000001\bar{1}00)$\\
&&$(0000001\bar{1}000)$\\
&&$(000001\bar{1}0000)$\\
&&$(00001\bar{1}00000)$\\
&&$(0001\bar{1}000000)$\\
&&$(001\bar{1}0000000)$\\
&&$(01\bar{1}00000000)$\\
&&$(1\bar{1}000000000)$\\
&&$(\bar{1}0000000000)$\\
\end{longtable}
\end{center}

\newpage
\subsubsection{Rank 12}

\begin{center}
\begin{longtable}{ccc}
\caption{Weight diagrams of $B_{12}$}
\label{Table:Weight-diagrams-B12}\\
\hline\hline
$B_{12}$ Irrep.&$d(R)$& Weight\\\hline\hline
\endfirsthead
\multicolumn{3}{c}{Table~\ref{Table:Weight-diagrams-B12} (continued)}\\\hline\hline
$B_{12}$ Irrep.&$d(R)$& Weight\\\hline\hline
\endhead
\hline
\endfoot
(000000000000)&${\bf 1}$
 &$(000000000000)$\\
\hline
(100000000000)&${\bf 25}$
 &$(100000000000)$\\
&&$(\bar{1}10000000000)$\\
&&$(0\bar{1}1000000000)$\\
&&$(00\bar{1}100000000)$\\
&&$(000\bar{1}10000000)$\\
&&$(0000\bar{1}1000000)$\\
&&$(00000\bar{1}100000)$\\
&&$(000000\bar{1}10000)$\\
&&$(0000000\bar{1}1000)$\\
&&$(00000000\bar{1}100)$\\
&&$(000000000\bar{1}10)$\\
&&$(0000000000\bar{1}2)$\\
&&$(000000000000)$\\
&&$(00000000001\bar{2})$\\
&&$(0000000001\bar{1}0)$\\
&&$(000000001\bar{1}00)$\\
&&$(00000001\bar{1}000)$\\
&&$(0000001\bar{1}0000)$\\
&&$(000001\bar{1}00000)$\\
&&$(00001\bar{1}000000)$\\
&&$(0001\bar{1}0000000)$\\
&&$(001\bar{1}00000000)$\\
&&$(01\bar{1}000000000)$\\
&&$(1\bar{1}0000000000)$\\
&&$(\bar{1}00000000000)$\\
\end{longtable}
\end{center}

\newpage
\subsubsection{Rank 13}

\begin{center}
\begin{longtable}{ccc}
\caption{Weight diagrams of $B_{13}$}
\label{Table:Weight-diagrams-B13}\\
\hline\hline
$B_{13}$ Irrep.&$d(R)$& Weight\\\hline\hline
\endfirsthead
\multicolumn{3}{c}{Table~\ref{Table:Weight-diagrams-B13} (continued)}\\\hline\hline
$B_{13}$ Irrep.&$d(R)$& Weight\\\hline\hline
\endhead
\hline
\endfoot
(0000000000000)&${\bf 1}$
 &$(0000000000000)$\\
\hline
(1000000000000)&${\bf 27}$
 &$(1000000000000)$\\
&&$(\bar{1}100000000000)$\\
&&$(0\bar{1}10000000000)$\\
&&$(00\bar{1}1000000000)$\\
&&$(000\bar{1}100000000)$\\
&&$(0000\bar{1}10000000)$\\
&&$(00000\bar{1}1000000)$\\
&&$(000000\bar{1}100000)$\\
&&$(0000000\bar{1}10000)$\\
&&$(00000000\bar{1}1000)$\\
&&$(000000000\bar{1}100)$\\
&&$(0000000000\bar{1}10)$\\
&&$(00000000000\bar{1}2)$\\
&&$(0000000000000)$\\
&&$(000000000001\bar{2})$\\
&&$(00000000001\bar{1}0)$\\
&&$(0000000001\bar{1}00)$\\
&&$(000000001\bar{1}000)$\\
&&$(00000001\bar{1}0000)$\\
&&$(0000001\bar{1}00000)$\\
&&$(000001\bar{1}000000)$\\
&&$(00001\bar{1}0000000)$\\
&&$(0001\bar{1}00000000)$\\
&&$(001\bar{1}000000000)$\\
&&$(01\bar{1}0000000000)$\\
&&$(1\bar{1}00000000000)$\\
&&$(\bar{1}000000000000)$\\
\end{longtable}
\end{center}

\newpage
\subsubsection{Rank 14}

\begin{center}
\begin{longtable}{ccc}
\caption{Weight diagrams of $B_{14}$}
\label{Table:Weight-diagrams-B14}\\
\hline\hline
$B_{14}$ Irrep.&$d(R)$& Weight\\\hline\hline
\endfirsthead
\multicolumn{3}{c}{Table~\ref{Table:Weight-diagrams-B14} (continued)}\\\hline\hline
$B_{14}$ Irrep.&$d(R)$& Weight\\\hline\hline
\endhead
\hline
\endfoot
(00000000000000)&${\bf 1}$
 &$(00000000000000)$\\
\hline
(10000000000000)&${\bf 29}$
 &$(10000000000000)$\\
&&$(\bar{1}1000000000000)$\\
&&$(0\bar{1}100000000000)$\\
&&$(00\bar{1}10000000000)$\\
&&$(000\bar{1}1000000000)$\\
&&$(0000\bar{1}100000000)$\\
&&$(00000\bar{1}10000000)$\\
&&$(000000\bar{1}1000000)$\\
&&$(0000000\bar{1}100000)$\\
&&$(00000000\bar{1}10000)$\\
&&$(000000000\bar{1}1000)$\\
&&$(0000000000\bar{1}100)$\\
&&$(00000000000\bar{1}10)$\\
&&$(000000000000\bar{1}2)$\\
&&$(00000000000000)$\\
&&$(0000000000001\bar{2})$\\
&&$(000000000001\bar{1}0)$\\
&&$(00000000001\bar{1}00)$\\
&&$(0000000001\bar{1}000)$\\
&&$(000000001\bar{1}0000)$\\
&&$(00000001\bar{1}00000)$\\
&&$(0000001\bar{1}000000)$\\
&&$(000001\bar{1}0000000)$\\
&&$(00001\bar{1}00000000)$\\
&&$(0001\bar{1}000000000)$\\
&&$(001\bar{1}0000000000)$\\
&&$(01\bar{1}00000000000)$\\
&&$(1\bar{1}000000000000)$\\
&&$(\bar{1}0000000000000)$\\
\end{longtable}
\end{center}

\newpage
\subsubsection{Rank 15}

\begin{center}
\begin{longtable}{ccc}
\caption{Weight diagrams of $B_{15}$}
\label{Table:Weight-diagrams-B15}\\
\hline\hline
$B_{15}$ Irrep.&$d(R)$& Weight\\\hline\hline
\endfirsthead
\multicolumn{3}{c}{Table~\ref{Table:Weight-diagrams-B15} (continued)}\\\hline\hline
$B_{15}$ Irrep.&$d(R)$& Weight\\\hline\hline
\endhead
\hline
\endfoot
(000000000000000)&${\bf 1}$
 &$(000000000000000)$\\
\hline
(100000000000000)&${\bf 31}$
 &$(100000000000000)$\\
&&$(\bar{1}10000000000000)$\\
&&$(0\bar{1}1000000000000)$\\
&&$(00\bar{1}100000000000)$\\
&&$(000\bar{1}10000000000)$\\
&&$(0000\bar{1}1000000000)$\\
&&$(00000\bar{1}100000000)$\\
&&$(000000\bar{1}10000000)$\\
&&$(0000000\bar{1}1000000)$\\
&&$(00000000\bar{1}100000)$\\
&&$(000000000\bar{1}10000)$\\
&&$(0000000000\bar{1}1000)$\\
&&$(00000000000\bar{1}100)$\\
&&$(000000000000\bar{1}10)$\\
&&$(0000000000000\bar{1}2)$\\
&&$(000000000000000)$\\
&&$(00000000000001\bar{2})$\\
&&$(0000000000001\bar{1}0)$\\
&&$(000000000001\bar{1}00)$\\
&&$(00000000001\bar{1}000)$\\
&&$(0000000001\bar{1}0000)$\\
&&$(000000001\bar{1}00000)$\\
&&$(00000001\bar{1}000000)$\\
&&$(0000001\bar{1}0000000)$\\
&&$(000001\bar{1}00000000)$\\
&&$(00001\bar{1}000000000)$\\
&&$(0001\bar{1}0000000000)$\\
&&$(001\bar{1}00000000000)$\\
&&$(01\bar{1}000000000000)$\\
&&$(1\bar{1}0000000000000)$\\
&&$(\bar{1}00000000000000)$\\
\end{longtable}
\end{center}

\newpage
\subsubsection{Rank 16}

\begin{center}
\begin{longtable}{ccc}
\caption{Weight diagrams of $B_{16}$}
\label{Table:Weight-diagrams-B16}\\
\hline\hline
$B_{16}$ Irrep.&$d(R)$& Weight\\\hline\hline
\endfirsthead
\multicolumn{3}{c}{Table~\ref{Table:Weight-diagrams-B16} (continued)}\\\hline\hline
$B_{16}$ Irrep.&$d(R)$& Weight\\\hline\hline
\endhead
\hline
\endfoot
(0000000000000000)&${\bf 1}$
 &$(0000000000000000)$\\
\hline
(1000000000000000)&${\bf 33}$
 &$(1000000000000000)$\\
&&$(\bar{1}100000000000000)$\\
&&$(0\bar{1}10000000000000)$\\
&&$(00\bar{1}1000000000000)$\\
&&$(000\bar{1}100000000000)$\\
&&$(0000\bar{1}10000000000)$\\
&&$(00000\bar{1}1000000000)$\\
&&$(000000\bar{1}100000000)$\\
&&$(0000000\bar{1}10000000)$\\
&&$(00000000\bar{1}1000000)$\\
&&$(000000000\bar{1}100000)$\\
&&$(0000000000\bar{1}10000)$\\
&&$(00000000000\bar{1}1000)$\\
&&$(000000000000\bar{1}100)$\\
&&$(0000000000000\bar{1}10)$\\
&&$(00000000000000\bar{1}2)$\\
&&$(0000000000000000)$\\
&&$(000000000000001\bar{2})$\\
&&$(00000000000001\bar{1}0)$\\
&&$(0000000000001\bar{1}00)$\\
&&$(000000000001\bar{1}000)$\\
&&$(00000000001\bar{1}0000)$\\
&&$(0000000001\bar{1}00000)$\\
&&$(000000001\bar{1}000000)$\\
&&$(00000001\bar{1}0000000)$\\
&&$(0000001\bar{1}00000000)$\\
&&$(000001\bar{1}000000000)$\\
&&$(00001\bar{1}0000000000)$\\
&&$(0001\bar{1}00000000000)$\\
&&$(001\bar{1}000000000000)$\\
&&$(01\bar{1}0000000000000)$\\
&&$(1\bar{1}00000000000000)$\\
&&$(\bar{1}000000000000000)$\\
\end{longtable}
\end{center}

\newpage
\subsubsection{Rank 17}

\begin{center}
\begin{longtable}{ccc}
\caption{Weight diagrams of $B_{17}$}
\label{Table:Weight-diagrams-B17}\\
\hline\hline
$B_{17}$ Irrep.&$d(R)$& Weight\\\hline\hline
\endfirsthead
\multicolumn{3}{c}{Table~\ref{Table:Weight-diagrams-B17} (continued)}\\\hline\hline
$B_{17}$ Irrep.&$d(R)$& Weight\\\hline\hline
\endhead
\hline
\endfoot
(00000000000000000)&${\bf 1}$
 &$(00000000000000000)$\\
\hline
(10000000000000000)&${\bf 35}$
 &$(10000000000000000)$\\
&&$(\bar{1}1000000000000000)$\\
&&$(0\bar{1}100000000000000)$\\
&&$(00\bar{1}10000000000000)$\\
&&$(000\bar{1}1000000000000)$\\
&&$(0000\bar{1}100000000000)$\\
&&$(00000\bar{1}10000000000)$\\
&&$(000000\bar{1}1000000000)$\\
&&$(0000000\bar{1}100000000)$\\
&&$(00000000\bar{1}10000000)$\\
&&$(000000000\bar{1}1000000)$\\
&&$(0000000000\bar{1}100000)$\\
&&$(00000000000\bar{1}10000)$\\
&&$(000000000000\bar{1}1000)$\\
&&$(0000000000000\bar{1}100)$\\
&&$(00000000000000\bar{1}10)$\\
&&$(000000000000000\bar{1}2)$\\
&&$(00000000000000000)$\\
&&$(0000000000000001\bar{2})$\\
&&$(000000000000001\bar{1}0)$\\
&&$(00000000000001\bar{1}00)$\\
&&$(0000000000001\bar{1}000)$\\
&&$(000000000001\bar{1}0000)$\\
&&$(00000000001\bar{1}00000)$\\
&&$(0000000001\bar{1}000000)$\\
&&$(000000001\bar{1}0000000)$\\
&&$(00000001\bar{1}00000000)$\\
&&$(0000001\bar{1}000000000)$\\
&&$(000001\bar{1}0000000000)$\\
&&$(00001\bar{1}00000000000)$\\
&&$(0001\bar{1}000000000000)$\\
&&$(001\bar{1}0000000000000)$\\
&&$(01\bar{1}00000000000000)$\\
&&$(1\bar{1}000000000000000)$\\
&&$(\bar{1}0000000000000000)$\\
\end{longtable}
\end{center}

\newpage
\subsubsection{Rank 18}

\begin{center}
\begin{longtable}{ccc}
\caption{Weight diagrams of $B_{18}$}
\label{Table:Weight-diagrams-B18}\\
\hline\hline
$B_{18}$ Irrep.&$d(R)$& Weight\\\hline\hline
\endfirsthead
\multicolumn{3}{c}{Table~\ref{Table:Weight-diagrams-B18} (continued)}\\\hline\hline
$B_{18}$ Irrep.&$d(R)$& Weight\\\hline\hline
\endhead
\hline
\endfoot
(000000000000000000)&${\bf 1}$
 &$(000000000000000000)$\\
\hline
(100000000000000000)&${\bf 37}$
 &$(100000000000000000)$\\
&&$(\bar{1}10000000000000000)$\\
&&$(0\bar{1}1000000000000000)$\\
&&$(00\bar{1}100000000000000)$\\
&&$(000\bar{1}10000000000000)$\\
&&$(0000\bar{1}1000000000000)$\\
&&$(00000\bar{1}100000000000)$\\
&&$(000000\bar{1}10000000000)$\\
&&$(0000000\bar{1}1000000000)$\\
&&$(00000000\bar{1}100000000)$\\
&&$(000000000\bar{1}10000000)$\\
&&$(0000000000\bar{1}1000000)$\\
&&$(00000000000\bar{1}100000)$\\
&&$(000000000000\bar{1}10000)$\\
&&$(0000000000000\bar{1}1000)$\\
&&$(00000000000000\bar{1}100)$\\
&&$(000000000000000\bar{1}10)$\\
&&$(0000000000000000\bar{1}2)$\\
&&$(000000000000000000)$\\
&&$(00000000000000001\bar{2})$\\
&&$(0000000000000001\bar{1}0)$\\
&&$(000000000000001\bar{1}00)$\\
&&$(00000000000001\bar{1}000)$\\
&&$(0000000000001\bar{1}0000)$\\
&&$(000000000001\bar{1}00000)$\\
&&$(00000000001\bar{1}000000)$\\
&&$(0000000001\bar{1}0000000)$\\
&&$(000000001\bar{1}00000000)$\\
&&$(00000001\bar{1}000000000)$\\
&&$(0000001\bar{1}0000000000)$\\
&&$(000001\bar{1}00000000000)$\\
&&$(00001\bar{1}000000000000)$\\
&&$(0001\bar{1}0000000000000)$\\
&&$(001\bar{1}00000000000000)$\\
&&$(01\bar{1}000000000000000)$\\
&&$(1\bar{1}0000000000000000)$\\
&&$(\bar{1}00000000000000000)$\\
\end{longtable}
\end{center}

\newpage
\subsubsection{Rank 19}

\begin{center}
\begin{longtable}{ccc}
\caption{Weight diagrams of $B_{19}$}
\label{Table:Weight-diagrams-B19}\\
\hline\hline
$B_{19}$ Irrep.&$d(R)$& Weight\\\hline\hline
\endfirsthead
\multicolumn{3}{c}{Table~\ref{Table:Weight-diagrams-B19} (continued)}\\\hline\hline
$B_{19}$ Irrep.&$d(R)$& Weight\\\hline\hline
\endhead
\hline
\endfoot
(0000000000000000000)&${\bf 1}$
 &$(0000000000000000000)$\\
\hline
(1000000000000000000)&${\bf 39}$
 &$(1000000000000000000)$\\
&&$(\bar{1}100000000000000000)$\\
&&$(0\bar{1}10000000000000000)$\\
&&$(00\bar{1}1000000000000000)$\\
&&$(000\bar{1}100000000000000)$\\
&&$(0000\bar{1}10000000000000)$\\
&&$(00000\bar{1}1000000000000)$\\
&&$(000000\bar{1}100000000000)$\\
&&$(0000000\bar{1}10000000000)$\\
&&$(00000000\bar{1}1000000000)$\\
&&$(000000000\bar{1}100000000)$\\
&&$(0000000000\bar{1}10000000)$\\
&&$(00000000000\bar{1}1000000)$\\
&&$(000000000000\bar{1}100000)$\\
&&$(0000000000000\bar{1}10000)$\\
&&$(00000000000000\bar{1}1000)$\\
&&$(000000000000000\bar{1}100)$\\
&&$(0000000000000000\bar{1}10)$\\
&&$(00000000000000000\bar{1}2)$\\
&&$(0000000000000000000)$\\
&&$(000000000000000001\bar{2})$\\
&&$(00000000000000001\bar{1}0)$\\
&&$(0000000000000001\bar{1}00)$\\
&&$(000000000000001\bar{1}000)$\\
&&$(00000000000001\bar{1}0000)$\\
&&$(0000000000001\bar{1}00000)$\\
&&$(000000000001\bar{1}000000)$\\
&&$(00000000001\bar{1}0000000)$\\
&&$(0000000001\bar{1}00000000)$\\
&&$(000000001\bar{1}000000000)$\\
&&$(00000001\bar{1}0000000000)$\\
&&$(0000001\bar{1}00000000000)$\\
&&$(000001\bar{1}000000000000)$\\
&&$(00001\bar{1}0000000000000)$\\
&&$(0001\bar{1}00000000000000)$\\
&&$(001\bar{1}000000000000000)$\\
&&$(01\bar{1}0000000000000000)$\\
&&$(1\bar{1}00000000000000000)$\\
&&$(\bar{1}000000000000000000)$\\
\end{longtable}
\end{center}

\newpage
\subsubsection{Rank 20}

\begin{center}
%
\end{center}

\newpage
\subsection{$C_n=\mathfrak{usp}_{2n}$}

\subsubsection{Rank 2}

\begin{center}
\begin{longtable}{ccc}
\caption{Weight diagrams of $C_2$}
\label{Table:Weight-diagrams-C2}\\
\hline\hline
$C_2$ Irrep.&$d(R)$& Weight\\\hline\hline
\endfirsthead
\multicolumn{3}{c}{Table~\ref{Table:Weight-diagrams-C2} (continued)}\\\hline\hline
$C_2$ Irrep.&$d(R)$& Weight\\\hline\hline
\endhead
\hline
\endfoot
(00)&${\bf 1}$         & $(00)$\\
\hline
(10)&${\bf 4}$         & $(10)$\\
    &                  & $(\bar{1}1)$\\
    &                  & $(1\bar{1})$\\
    &                  & $(\bar{1}0)$\\
\hline
(01)&${\bf 5}$         & $(01)$\\
    &                  & $(2\bar{1})$\\
    &                  & $(00)$\\
\end{longtable}
\end{center}

\newpage
\subsubsection{Rank 3}

\begin{center}
\begin{longtable}{ccc}
\caption{Weight diagrams of $C_3$}
\label{Table:Weight-diagrams-C3}\\
\hline\hline
$C_3$ Irrep.&$d(R)$& Weight\\\hline\hline
\endfirsthead
\multicolumn{3}{c}{Table~\ref{Table:Weight-diagrams-C3} (continued)}\\\hline\hline
$C_3$ Irrep.&$d(R)$& Weight\\\hline\hline
\endhead
\hline
\endfoot
(000)&${\bf 1}$         & $(000)$\\
\hline
(100)&${\bf 6}$         & $(100)$\\
     &                  & $(\bar{1}10)$\\
     &                  & $(0\bar{1}1)$\\
     &                  & $(01\bar{1})$\\
     &                  & $(1\bar{1}0)$\\
     &                  & $(\bar{1}00)$\\
\hline
(010)&${\bf 14}$
 &$(010)$\\
\end{longtable}
\end{center}

\newpage
\subsubsection{Rank 4}

\begin{center}
%
\end{center}

\newpage
\subsubsection{Rank 5}

\begin{center}
%
\end{center}

\newpage
\subsubsection{Rank 6}

\begin{center}
%
\end{center}

\newpage
\subsubsection{Rank 7}

\begin{center}
%
\end{center}

\newpage
\subsubsection{Rank 8}

\begin{center}
%
\end{center}

\newpage
\subsubsection{Rank 9}

\begin{center}
%
\end{center}

\newpage
\subsubsection{Rank 10}

\begin{center}
%
\end{center}

\newpage
\subsubsection{Rank 11}

\begin{center}
%
\end{center}

\newpage
\subsubsection{Rank 12}

\begin{center}
%
\end{center}

\newpage
\subsubsection{Rank 13}

\begin{center}
%
\end{center}

\newpage
\subsubsection{Rank 14}

\begin{center}
%
\end{center}

\newpage
\subsubsection{Rank 15}

\begin{center}
%
\end{center}

\newpage
\subsubsection{Rank 16}

\begin{center}
%
\end{center}

\newpage
\subsubsection{Rank 17}

\begin{center}
%
\end{center}

\newpage
\subsubsection{Rank 18}

\begin{center}
%
\end{center}

\newpage
\subsubsection{Rank 19}

\begin{center}
%
\end{center}

\newpage
\subsubsection{Rank 20}

\begin{center}
%
\end{center}

\newpage
\subsection{$D_n=\mathfrak{so}_{2n}$}

\subsubsection{Rank 3}

\begin{center}
%
\end{center}

\newpage
\subsubsection{Rank 4}

\begin{center}
%
\end{center}

\newpage
\subsubsection{Rank 5}

\begin{center}
%
\end{center}

\newpage
\subsubsection{Rank 6}

\begin{center}
%
\end{center}

\newpage
\subsubsection{Rank 7}

\begin{center}
%
\end{center}

\newpage
\subsubsection{Rank 8}

\begin{center}
%
\end{center}

\newpage
\subsubsection{Rank 9}

\begin{center}
%
\end{center}

\newpage
\subsubsection{Rank 10}

\begin{center}
%
\end{center}

\newpage
\subsubsection{Rank 11}

\begin{center}
%
\end{center}

\newpage
\subsubsection{Rank 12}

\begin{center}
%
\end{center}

\newpage
\subsubsection{Rank 13}

\begin{center}
%
\end{center}

\newpage
\subsubsection{Rank 14}

\begin{center}
%
\end{center}

\newpage
\subsubsection{Rank 15}

\begin{center}
%
\end{center}

\newpage
\subsubsection{Rank 16}

\begin{center}
%
\end{center}

\newpage
\subsubsection{Rank 17}

\begin{center}
%
\end{center}

\newpage
\subsubsection{Rank 18}

\begin{center}
%
\end{center}

\newpage
\subsubsection{Rank 19}

\begin{center}
%
\end{center}

\newpage
\subsubsection{Rank 20}

\begin{center}
%
\end{center}

\newpage
\subsection{$E_n$}

\subsubsection{Rank 6}

\begin{center}
%
\end{center}

\newpage
\subsubsection{Rank 7}

\begin{center}
%
\end{center}

\newpage
\subsubsection{Rank 8}

\begin{center}
%
\end{center}

\newpage
\subsection{$F_4$}

\begin{center}
%
\end{center}

\newpage
\subsection{$G_2$}

\begin{center}
%
\end{center}

%% file: subalgebras.tex
\section{Subalgebras}
\label{Sec:Subalgebras}

\begin{center}
%
\end{center}

\newpage

\subsection{$A_n$}

\subsubsection{Rank 1 (R)}

\begin{center}
%
\end{center}

\subsubsection{Rank 2 (R)}

\begin{center}
\begin{longtable}{llc}
\caption{Maximal subalgebras of $A_{2}$(R)}
\label{Table:Maximal-subalgebra-A2-R}\\
\hline\hline
 $\mathfrak{g}\supset\mathfrak{h}$&Branching rule&Type\\\hline\hline
\endfirsthead
\multicolumn{2}{c}{Table~\ref{Table:Maximal-subalgebra-A2-R} (continued)}\\
\hline
$\mathfrak{g}\supset\mathfrak{h}$&Branching rule&Type\\\hline\hline
\endhead
\hline
\endfoot
$\mathfrak{su}_3\supset\mathfrak{su}_2\oplus\mathfrak{u}_{1}$
&${\bf 3}=({\bf 2})(1)\oplus({\bf 1})(-2)$
&(R)\\
\end{longtable}
\end{center}

\subsubsection{Rank 2 (S)}

\begin{center}
%
\end{center}

\subsubsection{Rank 3 (R)}

\begin{center}
%
\end{center}

\subsubsection{Rank 3 (S)}

\begin{center}
%
\end{center}

\subsubsection{Rank 4 (R)}

\begin{center}
%
\end{center}

\subsubsection{Rank 4 (S)}

\begin{center}
%
\end{center}

\subsubsection{Rank 5 (R)}

\begin{center}
%
\end{center}

\subsubsection{Rank 5 (S)}

\begin{center}
%
\end{center}

\newpage
\subsubsection{Rank 6 (R)}

\begin{center}
%
\end{center}

\subsubsection{Rank 6 (S)}

\begin{center}
%
\end{center}

\subsubsection{Rank 7 (R)}

\begin{center}
%
\end{center}

\subsubsection{Rank 7 (S)}

\begin{center}
%
\end{center}

\newpage
\subsubsection{Rank 8 (R)}

\begin{center}
%
\end{center}

\subsubsection{Rank 8 (S)}

\begin{center}
%
\end{center}

\subsubsection{Rank 9 (R)}

\begin{center}
%
\end{center}

\subsubsection{Rank 9 (S)}

\begin{center}
%
\end{center}

\subsubsection{Rank 10 (R)}

\begin{center}
%
\end{center}

\subsubsection{Rank 10 (S)}

\begin{center}
%
\end{center}

\subsubsection{Rank 11 (R)}

\begin{center}
%
\end{center}

\subsubsection{Rank 11 (S)}

\begin{center}
%
\end{center}

\newpage
\subsubsection{Rank 12 (R)}

\begin{center}
%
\end{center}

\subsubsection{Rank 12 (S)}

\begin{center}
%
\end{center}

\subsubsection{Rank 13 (R)}

\begin{center}
%
\end{center}

\subsubsection{Rank 13 (S)}

\begin{center}
%
\end{center}

\subsubsection{Rank 14 (R)}

\begin{center}
%
\end{center}

\subsubsection{Rank 14 (S)}

\begin{center}
%
\end{center}

\newpage
\subsubsection{Rank 15 (R)}

\begin{center}
%
\end{center}

\subsubsection{Rank 15 (S)}

\begin{center}
%
\end{center}

\newpage

\newpage
\subsubsection{Rank 16 (R)}

\begin{center}
%
\end{center}

\subsubsection{Rank 16 (S)}

\begin{center}
%
\end{center}

\newpage
\subsubsection{Rank 17 (R)}

\begin{center}
%
\end{center}

\subsubsection{Rank 17 (S)}

\begin{center}
%
\end{center}

\newpage
\subsubsection{Rank 18 (R)}

\begin{center}
%
\end{center}

\subsubsection{Rank 18 (S)}

\begin{center}
%
\end{center}

\newpage
\subsubsection{Rank 19 (R)}

\begin{center}
%
\end{center}

\subsubsection{Rank 19 (S)}

\begin{center}
%
\end{center}

\newpage
\subsubsection{Rank 20 (R)}

\begin{center}
%
\end{center}

\subsubsection{Rank 20 (S)}

\begin{center}
%
\end{center}

\newpage

\subsection{$B_n$}

\subsubsection{Rank 3 (R)}

\begin{center}
%
\end{center}

\subsubsection{Rank 3 (S)}

\begin{center}
%
\end{center}

\subsubsection{Rank 4 (R)}

\begin{center}
%
\end{center}

\subsubsection{Rank 4 (S)}

\begin{center}
%
\end{center}

\newpage
\subsubsection{Rank 5 (R)}

\begin{center}
%
\end{center}

\subsubsection{Rank 5 (S)}

\begin{center}
%
\end{center}

\subsubsection{Rank 6 (R)}

\begin{center}
%
\end{center}

\subsubsection{Rank 6 (S)}

\begin{center}
%
\end{center}

\newpage
\subsubsection{Rank 7 (R)}

\begin{center}
%
\end{center}

\subsubsection{Rank 7 (S)}

\begin{center}
%
\end{center}

\subsubsection{Rank 8 (R)}

\begin{center}
%
\end{center}

\subsubsection{Rank 8 (S)}

\begin{center}
%
\end{center}

\subsubsection{Rank 9 (R)}

\begin{center}
%
\end{center}

\subsubsection{Rank 9 (S)}

\begin{center}
%
\end{center}

\newpage
\subsubsection{Rank 10 (R)}

\begin{center}
%
\end{center}

\subsubsection{Rank 10 (S)}

\begin{center}
%
\end{center}

\newpage
\subsubsection{Rank 11 (R)}

\begin{center}
%
\end{center}

\subsubsection{Rank 11 (S)}

\begin{center}
%
\end{center}

\newpage
\subsubsection{Rank 12 (R)}

\begin{center}
%
\end{center}

\subsubsection{Rank 12 (S)}

\begin{center}
%
\end{center}

\newpage
\subsubsection{Rank 13 (R)}

\begin{center}
%
\end{center}

\subsubsection{Rank 13 (S)}

\begin{center}
%
\end{center}

\newpage
\subsubsection{Rank 14 (R)}

\begin{center}
%
\end{center}

\subsubsection{Rank 14 (S)}

\begin{center}
%
\end{center}

\newpage
\subsubsection{Rank 15 (R)}

\begin{center}
%
\end{center}

\subsubsection{Rank 15 (S)}

\begin{center}
%
\end{center}


\newpage


\newpage
\subsubsection{Rank 16 (R)}

\begin{center}
%
\end{center}

\subsubsection{Rank 16 (S)}

\begin{center}
%
\end{center}

\newpage
\subsubsection{Rank 17 (R)}

\begin{center}
%
\end{center}

\subsubsection{Rank 17 (S)}

\begin{center}
%
\end{center}

\newpage
\subsubsection{Rank 18 (R)}

\begin{center}
%
\end{center}

\subsubsection{Rank 18 (S)}

\begin{center}
%
\end{center}

\newpage
\subsubsection{Rank 19 (R)}

\begin{center}
%
\end{center}

\subsubsection{Rank 19 (S)}

\begin{center}
%
\end{center}

\newpage
\subsubsection{Rank 20 (R)}

\begin{center}
%
\end{center}

\subsubsection{Rank 20 (S)}

\begin{center}
%
\end{center}

\newpage

\subsection{$C_n$}

\subsubsection{Rank 2 (R)}

\begin{center}
%
\end{center}

\subsubsection{Rank 2 (S)}

\begin{center}
%
\end{center}

\subsubsection{Rank 3 (R)}

\begin{center}
%
\end{center}

\subsubsection{Rank 3 (S)}

\begin{center}
%
\end{center}

\newpage
\subsubsection{Rank 4 (R)}

\begin{center}
%
\end{center}

\subsubsection{Rank 4 (S)}

\begin{center}
%
\end{center}

\subsubsection{Rank 5 (R)}

\begin{center}
%
\end{center}

\subsubsection{Rank 5 (S)}

\begin{center}
%
\end{center}

\newpage
\subsubsection{Rank 6 (R)}

\begin{center}
%
\end{center}

\subsubsection{Rank 6 (S)}

\begin{center}
%
\end{center}

\subsubsection{Rank 7 (R)}

\begin{center}
%
\end{center}

\subsubsection{Rank 7 (S)}

\begin{center}
%
\end{center}

\newpage
\subsubsection{Rank 8 (R)}

\begin{center}
%
\end{center}

\subsubsection{Rank 8 (S)}

\begin{center}
%
\end{center}

\subsubsection{Rank 9 (R)}

\begin{center}
%
\end{center}

\subsubsection{Rank 9 (S)}

\begin{center}
%
\end{center}

\newpage
\subsubsection{Rank 10 (R)}

\begin{center}
%
\end{center}

\subsubsection{Rank 10 (S)}

\begin{center}
%
\end{center}

\subsubsection{Rank 11 (R)}

\begin{center}
%
\end{center}

\subsubsection{Rank 11 (S)}

\begin{center}
%
\end{center}

\subsubsection{Rank 12 (R)}

\begin{center}
%
\end{center}

\subsubsection{Rank 12 (S)}

\begin{center}
%
\end{center}

\subsubsection{Rank 13 (R)}

\begin{center}
%
\end{center}

\subsubsection{Rank 13 (S)}

\begin{center}
%
\end{center}

\subsubsection{Rank 14 (R)}

\begin{center}
%
\end{center}

\subsubsection{Rank 14 (S)}

\begin{center}
%
\end{center}

\newpage
\subsubsection{Rank 15 (R)}

\begin{center}
%
\end{center}

\subsubsection{Rank 15 (S)}

\begin{center}
%
\end{center}

\newpage


\newpage
\subsubsection{Rank 16 (R)}

\begin{center}
%
\end{center}

\subsubsection{Rank 16 (S)}

\begin{center}
%
\end{center}

\newpage
\subsubsection{Rank 17 (R)}

\begin{center}
%
\end{center}

\subsubsection{Rank 17 (S)}

\begin{center}
%
\end{center}

\newpage
\subsubsection{Rank 18 (R)}

\begin{center}
%
\end{center}

\subsubsection{Rank 18 (S)}

\begin{center}
%
\end{center}

\newpage
\subsubsection{Rank 19 (R)}

\begin{center}
%
\end{center}

\subsubsection{Rank 19 (S)}

\begin{center}
%
\end{center}

\newpage
\subsubsection{Rank 20 (R)}

\begin{center}
%
\end{center}

\subsubsection{Rank 20 (S)}

\begin{center}
%
\end{center}

\newpage

\subsection{$D_n$}

\subsubsection{Rank 4 (R)}

\begin{center}
%
\end{center}

\subsubsection{Rank 4 (S)}

\begin{center}
%
\end{center}

\subsubsection{Rank 5 (R)}

\begin{center}
%
\end{center}

\subsubsection{Rank 5 (S)}

\begin{center}
%
\end{center}

\newpage
\subsubsection{Rank 6 (R)}

\begin{center}
%
\end{center}

\subsubsection{Rank 6 (S)}

\begin{center}
%
\end{center}

\subsubsection{Rank 7 (R)}

\begin{center}
%
\end{center}

\subsubsection{Rank 7 (S)}

\begin{center}
%
\end{center}

\subsubsection{Rank 8 (R)}

\begin{center}
%
\end{center}

\subsubsection{Rank 8 (S)}

\begin{center}
%
\end{center}

\newpage
\subsubsection{Rank 9 (R)}

\begin{center}
%
\end{center}

\subsubsection{Rank 9 (S)}

\begin{center}
%
\end{center}

\newpage
\subsubsection{Rank 10 (R)}

\begin{center}
%
\end{center}

\subsubsection{Rank 10 (S)}

\begin{center}
%
\end{center}

\newpage
\subsubsection{Rank 11 (R)}

\begin{center}
%
\end{center}

\subsubsection{Rank 11 (S)}

\begin{center}
%
\end{center}

\newpage
\subsubsection{Rank 12 (R)}

\begin{center}
%
\end{center}

\subsubsection{Rank 12 (S)}

\begin{center}
%
\end{center}

\newpage
\subsubsection{Rank 13 (R)}

\begin{center}
%
\end{center}

\subsubsection{Rank 13 (S)}

\begin{center}
%
\end{center}

\newpage
\subsubsection{Rank 14 (R)}

\begin{center}
%
\end{center}

\subsubsection{Rank 14 (S)}

\begin{center}
%
\end{center}

\newpage
\subsubsection{Rank 15 (R)}

\begin{center}
%
\end{center}

\subsubsection{Rank 15 (S)}

\begin{center}
%
\end{center}

\newpage
\subsubsection{Rank 16 (R)}

\begin{center}
%
\end{center}

\subsubsection{Rank 16 (S)}

\begin{center}
%
\end{center}

\newpage


\newpage
\subsubsection{Rank 17 (R)}

\begin{center}
%
\end{center}

\subsubsection{Rank 17 (S)}

\begin{center}
%
\end{center}

\newpage
\subsubsection{Rank 18 (R)}

\begin{center}
%
\end{center}

\subsubsection{Rank 18 (S)}

\begin{center}
%
\end{center}

\newpage
\subsubsection{Rank 19 (R)}

\begin{center}
%
\end{center}

\subsubsection{Rank 19 (S)}

\begin{center}
%
\end{center}

\newpage
\subsubsection{Rank 20 (R)}

\begin{center}
%
\end{center}

\subsubsection{Rank 20 (S)}

\begin{center}
%
\end{center}


\newpage

\subsection{$E_n$}

\subsubsection{Rank 6 (R)}

\begin{center}
%
\end{center}

\subsubsection{Rank 6 (S)}

\begin{center}
%
\end{center}

\newpage
\subsubsection{Rank 7 (R)}

\begin{center}
%
\end{center}

\subsubsection{Rank 7 (S)}

\begin{center}
%
\end{center}

\newpage
\subsubsection{Rank 8 (R)}

\begin{center}
%
\end{center}

\subsubsection{Rank 8 (S)}

{\small
\begin{center}
%
\end{center}}

\newpage
\subsection{$F_4$}

\subsubsection{Rank 4 (R)}

\begin{center}
%
\end{center}

\subsubsection{Rank 4 (S)}

\begin{center}
%
\end{center}

\newpage
\subsection{$G_2$}

\subsubsection{Rank 2 (R)}

\begin{center}
%
\end{center}

\subsubsection{Rank 2 (S)}

\begin{center}
%
\end{center}


%% file: projection-matrices.tex
\section{Projection matrices}
\label{Sec:Projection-matrices}

\subsection{$A_n$}

\subsubsection{Rank 1 (R)}

\begin{center}
%
\end{center}

\subsubsection{Rank 2 (R)}

\begin{center}
%
\end{center}

\subsubsection{Rank 2 (S)}

\begin{center}
%
\end{center}

\newpage
\subsubsection{Rank 3 (R)}

\begin{center}
%
\end{center}

\subsubsection{Rank 3 (S)}

\begin{center}
%
\end{center}

\newpage
\subsubsection{Rank 4 (R)}

\begin{center}
%
\end{center}

\subsubsection{Rank 4 (S)}

\begin{center}
%
\end{center}

\newpage
\subsubsection{Rank 5 (R)}

\begin{center}
%
\end{center}

\subsubsection{Rank 5 (S)}

\begin{center}
%
\end{center}

\newpage
\subsubsection{Rank 6 (R)}

\begin{center}
%
\end{center}

\subsubsection{Rank 6 (S)}

\begin{center}
%
\end{center}

\newpage
\subsubsection{Rank 7 (R)}

\begin{center}
%
\end{center}

\newpage
\subsubsection{Rank 7 (S)}

\begin{center}
%
\end{center}

\newpage
\subsubsection{Rank 8 (R)}

\begin{center}
%
\end{center}

\newpage
\subsubsection{Rank 8 (S)}

\begin{center}
%
\end{center}

\newpage
\subsubsection{Rank 9 (R)}

\begin{center}
%
\end{center}

\subsubsection{Rank 9 (S)}

\begin{center}
%
\end{center}

\newpage
\subsubsection{Rank 10 (R)}

\begin{center}
%
\end{center}

\newpage
\subsubsection{Rank 10 (S)}

\begin{center}
%
\end{center}

\newpage
\subsubsection{Rank 11 (R)}

\begin{center}
%
\end{center}

\newpage
\subsubsection{Rank 11 (S)}

\begin{center}
%
\end{center}

\newpage
\subsubsection{Rank 12 (R)}

\begin{center}
%
\end{center}

\newpage
\subsubsection{Rank 12 (S)}

\begin{center}
%
\end{center}

\newpage
\subsubsection{Rank 13 (R)}

{
\begin{center}
%
\end{center}

\newpage
\subsubsection{Rank 13 (S)}

{
\begin{center}
%
\end{center}

\newpage
\subsubsection{Rank 14 (R)}

\begin{center}
%
\end{center}

\newpage
\subsubsection{Rank 14 (S)}

\begin{center}
%
\end{center}

\newpage
\subsubsection{Rank 15 (R)}

\begin{center}
%
\end{center}

\newpage
\subsubsection{Rank 15 (S)}

\begin{center}
%
\end{center}
}

\newpage


\newpage
\subsubsection{Rank 16 (R)}

\begin{center}
%
\end{center}

\newpage
\subsubsection{Rank 16 (S)}

\begin{center}
%
\end{center}
}

\newpage
\subsubsection{Rank 17 (R)}

{\small 
\begin{center}
%
\end{center}
}

\newpage
\subsubsection{Rank 17 (S)}

{
\begin{center}
%
\end{center}
}

\newpage
\subsubsection{Rank 18 (R)}

{\footnotesize
\begin{center}
%
\end{center}
}

\newpage
\subsubsection{Rank 18 (S)}

{
\begin{center}
%
\end{center}
}

\newpage
\subsubsection{Rank 19 (R)}

{\footnotesize
\begin{center}
%
\end{center}
}

\newpage
\subsubsection{Rank 19 (S)}

{
\begin{center}
%
\end{center}
}

\newpage
\subsubsection{Rank 20 (R)}

{\fontsize{7pt}{10pt}\selectfont
\begin{center}
%
\end{center}
}

\newpage
\subsubsection{Rank 20 (S)}

{
\begin{center}
%
\end{center}
}

\newpage

\subsection{$B_n$}

\subsubsection{Rank 3 (R)}

\begin{center}
%
\end{center}

\newpage
\subsubsection{Rank 3 (S)}

\begin{center}
%
\end{center}

\newpage
\subsubsection{Rank 4 (R)}

\begin{center}
%
\end{center}

\newpage
\subsubsection{Rank 4 (S)}

\begin{center}
%
\end{center}

\newpage
\subsubsection{Rank 5 (R)}

\begin{center}
%
\end{center}

\newpage
\subsubsection{Rank 5 (S)}

\begin{center}
%
\end{center}

\newpage
\subsubsection{Rank 6 (R)}

\begin{center}
%
\end{center}

\newpage
\subsubsection{Rank 6 (S)}

\begin{center}
%
\end{center}

\newpage
\subsubsection{Rank 7 (R)}

\begin{center}
%
\end{center}

\newpage
\subsubsection{Rank 7 (S)}

\begin{center}
%
\end{center}

\newpage
\subsubsection{Rank 8 (R)}

\begin{center}
%
\end{center}

\newpage
\subsubsection{Rank 8 (S)}

\begin{center}
%
\end{center}

\newpage
\subsubsection{Rank 9 (R)}

\begin{center}
%
\end{center}

\newpage
\subsubsection{Rank 9 (S)}

\begin{center}
%
\end{center}

\newpage
\subsubsection{Rank 10 (R)}

\begin{center}
%
\end{center}

\newpage
\subsubsection{Rank 10 (S)}

\begin{center}
%
\end{center}

\newpage
\subsubsection{Rank 11 (R)}

\begin{center}
%
\end{center}

\newpage
\subsubsection{Rank 11 (S)}

\begin{center}
%
\end{center}

\newpage
\subsubsection{Rank 12 (R)}

\begin{center}
%
\end{center}

\newpage
\subsubsection{Rank 12 (S)}

\begin{center}
%
\end{center}

\newpage
\subsubsection{Rank 13 (R)}

\begin{center}
%
\end{center}

\newpage
\subsubsection{Rank 13 (S)}

\begin{center}
%
\end{center}

\newpage
\subsubsection{Rank 14 (R)}

\begin{center}
%
\end{center}

\newpage
\subsubsection{Rank 14 (S)}

\begin{center}
%
\end{center}

\newpage
\subsubsection{Rank 15 (R)}

{\footnotesize
\begin{center}
%
\end{center}
}

\newpage
\subsubsection{Rank 15 (S)}

{\footnotesize
\begin{center}
%
\end{center}
}

\newpage


\subsubsection{Rank 16 (R)}

{
\begin{center}
%
\end{center}
}

\newpage
\subsubsection{Rank 16 (S)}

{\fontsize{10pt}{12pt}\selectfont
\begin{center}
%
\end{center}
}

\newpage
\subsubsection{Rank 17 (R)}

{
\begin{center}
%
\end{center}
}

\newpage
\subsubsection{Rank 17 (S)}

{\fontsize{9pt}{14pt}\selectfont
\begin{center}
%
\end{center}
}

\newpage
\subsubsection{Rank 18 (R)}

{
\begin{center}
%
\end{center}
}

\newpage
\subsubsection{Rank 18 (S)}

{\fontsize{8pt}{14pt}\selectfont
\begin{center}
%
\end{center}
}

\newpage
\subsubsection{Rank 19 (R)}

{
\begin{center}
%
\end{center}
}

\newpage
\subsubsection{Rank 19 (S)}

{\fontsize{7.5pt}{12pt}\selectfont
\begin{center}
%
\end{center}
}

\newpage
\subsubsection{Rank 20 (R)}

{
\begin{center}
%
\end{center}
}

\newpage
\subsubsection{Rank 20 (S)}

{\fontsize{6pt}{12pt}\selectfont
\begin{center}
%
\end{center}
}

\newpage

\subsection{$C_n$}

\subsubsection{Rank 2 (R)}

\begin{center}
%
\end{center}

\subsubsection{Rank 2 (S)}

\begin{center}
%
\end{center}

\newpage
\subsubsection{Rank 3 (R)}

\begin{center}
%
\end{center}

\subsubsection{Rank 3 (S)}

\begin{center}
%
\end{center}

\newpage
\subsubsection{Rank 4 (R)}

\begin{center}
%
\end{center}

\subsubsection{Rank 4 (S)}

\begin{center}
%
\end{center}

\newpage
\subsubsection{Rank 5 (R)}

\begin{center}
%
\end{center}

\subsubsection{Rank 5 (S)}

\begin{center}
%
\end{center}

\newpage
\subsubsection{Rank 6 (R)}

\begin{center}
%
\end{center}

\subsubsection{Rank 6 (S)}

\begin{center}
%
\end{center}

\newpage
\subsubsection{Rank 7 (R)}

\begin{center}
%
\end{center}

\subsubsection{Rank 7 (S)}

\begin{center}
%
\end{center}

\newpage
\subsubsection{Rank 8 (R)}

\begin{center}
%
\end{center}

\newpage
\subsubsection{Rank 8 (S)}

\begin{center}
%
\end{center}

\newpage
\subsubsection{Rank 9 (R)}

\begin{center}
%
\end{center}

\newpage
\subsubsection{Rank 9 (S)}

\begin{center}
%
\end{center}

\newpage
\subsubsection{Rank 10 (R)}

\begin{center}
%
\end{center}

\newpage
\subsubsection{Rank 10 (S)}

\begin{center}
%
\end{center}

\newpage
\subsubsection{Rank 11 (R)}

\begin{center}
%
\end{center}

\newpage
\subsubsection{Rank 11 (S)}

\begin{center}
%
\end{center}

\newpage
\subsubsection{Rank 12 (R)}

\begin{center}
%
\end{center}

\newpage
\subsubsection{Rank 12 (S)}

\begin{center}
%
\end{center}

\newpage
\subsubsection{Rank 13 (R)}

\begin{center}
%
\end{center}

\newpage
\subsubsection{Rank 13 (S)}

\begin{center}
%
\end{center}

\newpage
\subsubsection{Rank 14 (R)}

\begin{center}
%
\end{center}

\newpage
\subsubsection{Rank 14 (S)}

\begin{center}
%
\end{center}

\newpage
\subsubsection{Rank 15 (R)}

\begin{center}
%
\end{center}

\newpage
\subsubsection{Rank 15 (S)}

\begin{center}
%
\end{center}

\newpage


\subsubsection{Rank 16 (R)}

\begin{center}
%
\end{center}

\newpage
\subsubsection{Rank 16 (S)}

{\small
\begin{center}
%
\end{center}
}

\newpage
\subsubsection{Rank 17 (R)}

\begin{center}
%
\end{center}

\newpage
\subsubsection{Rank 17 (S)}

{\fontsize{9pt}{12pt}\selectfont
\begin{center}
%
\end{center}
}

\newpage
\subsubsection{Rank 18 (R)}

\begin{center}
%
\end{center}

\newpage
\subsubsection{Rank 18 (S)}

{\fontsize{8pt}{12pt}\selectfont
\begin{center}
%
\end{center}
}

\newpage
\subsubsection{Rank 19 (R)}

\begin{center}
%
\end{center}

\newpage
\subsubsection{Rank 19 (S)}

{\fontsize{7pt}{12pt}\selectfont
\begin{center}
%
\end{center}
}

\newpage
\subsubsection{Rank 20 (R)}

{\fontsize{10pt}{12pt}\selectfont
\begin{center}
%
\end{center}
}

\newpage
\subsubsection{Rank 20 (S)}

{\fontsize{6.5pt}{12pt}\selectfont
\begin{center}
%
\end{center}
}

\newpage

\subsection{$D_n$}

\subsubsection{Rank 4 (R)}

\begin{center}
%
\end{center}

\subsubsection{Rank 4 (S)}

\begin{center}
%
\end{center}
Note that the basis for the projection matrices of
$D_4=\mathfrak{so}_{8}$
are selected as breaking the $\mathfrak{so}_{8}$
vector representation ${\bf 8_v}$ into the vector representations of its
orthogonal 
subgroups such as ${\bf 7_v}$ of $\mathfrak{so}_{7}$,
${\bf 6}$ of $\mathfrak{so}_{6}\simeq \mathfrak{su}_{4}$,
${\bf 5}$ of $\mathfrak{so}_{5}$, and
${\bf 3}$ of $\mathfrak{so}_{3}\simeq \mathfrak{su}_{2}$.

\newpage
\subsubsection{Rank 5 (R)}

\begin{center}
%
\end{center}

\subsubsection{Rank 5 (S)}

\begin{center}
%
\end{center} 

\newpage
\subsubsection{Rank 6 (R)}

\begin{center}
%
\end{center} 

\newpage
\subsubsection{Rank 6 (S)}

\begin{center}
%
\end{center} 

\newpage
\subsubsection{Rank 7 (R)}

\begin{center}
%
\end{center} 

\newpage
\subsubsection{Rank 7 (S)}

\begin{center}
%
\end{center} 

\newpage
\subsubsection{Rank 8 (R)}

\begin{center}
%
\end{center} 

\newpage
\subsubsection{Rank 8 (S)}

\begin{center}
%
\end{center} 

\newpage
\subsubsection{Rank 9 (R)}

\begin{center}
%
\end{center} 

\newpage
\subsubsection{Rank 9 (S)}

\begin{center}
%
\end{center} 

\newpage
\subsubsection{Rank 10 (R)}

\begin{center}
%
\end{center} 

\newpage
\subsubsection{Rank 10 (S)}

\begin{center}
%
\end{center} 

\newpage
\subsubsection{Rank 11 (R)}

\begin{center}
%
\end{center} 

\newpage
\subsubsection{Rank 11 (S)}

\begin{center}
%
\end{center}

\newpage
\subsubsection{Rank 12 (R)}

\begin{center}
%
\end{center}

\newpage
\subsubsection{Rank 12 (S)}

\begin{center}
%
\end{center}

\newpage
\subsubsection{Rank 13 (R)}

\begin{center}
%
\end{center}

\newpage
\subsubsection{Rank 13 (S)}

\begin{center}
%
\end{center}

\newpage
\subsubsection{Rank 14 (R)}

{
\begin{center}
%
\end{center}
}

Note that $\bar{n}$ stands for $-n$.

\newpage
\subsubsection{Rank 14 (S)}

\begin{center}
%
\end{center}

\newpage

\subsubsection{Rank 15 (R)}

{
\begin{center}
%
\end{center}
}

Note that $\bar{n}$ stands for $-n$.

\newpage
\subsubsection{Rank 15 (S)}

\begin{center}
%
\end{center}

\newpage

\subsubsection{Rank 16 (R)}

{
\begin{center}
%
\end{center}
}

Note that $\bar{n}$ stands for $-n$.

\newpage
\subsubsection{Rank 16 (S)}

{
\begin{center}
%
\end{center}
}

\newpage

\newpage

\subsubsection{Rank 17 (R)}

{
\begin{center}
%
\end{center}
}

Note that $\bar{n}$ stands for $-n$.

\newpage
\subsubsection{Rank 17 (S)}

{
\begin{center}
%
\end{center}
}

\newpage

\subsubsection{Rank 18 (R)}

{
\begin{center}
%
\end{center}
}

Note that $\bar{n}$ stands for $-n$.

\newpage
\subsubsection{Rank 18 (S)}

{
\begin{center}
%
\end{center}
}

\newpage
\subsubsection{Rank 19 (R)}

{\small
\begin{center}
%
\end{center}
}

Note that $\bar{n}$ stands for $-n$.

\newpage
\subsubsection{Rank 19 (S)}

{
\begin{center}
%
\end{center}
}

\newpage
\subsubsection{Rank 20 (R)}

{\footnotesize
\begin{center}
%
\end{center}
}

Note that $\bar{n}$ stands for $-n$.

\newpage
\subsubsection{Rank 20 (S)}

{\small
\begin{center}
%
\end{center}
}


\newpage

\subsection{$E_n$}

\subsubsection{Rank 6 (R)}

\begin{center}
%
\end{center}

\subsubsection{Rank 6 (S)}

\begin{center}
%
\end{center}

\newpage
\subsubsection{Rank 7 (R)}

\begin{center}
%
\end{center}

\newpage
\subsubsection{Rank 7 (S)}

\begin{center}
%
\end{center}

\newpage
\subsubsection{Rank 8 (R)}

\begin{center}
%
\end{center}

\newpage
\subsubsection{Rank 8 (S)}

\begin{center}
%
\end{center}

\newpage
\subsection{$F_4$}

\subsubsection{Rank 4 (R)}

\begin{center}
%
\end{center}

\subsubsection{Rank 4 (S)}

\begin{center}
%
\end{center}

\newpage
\subsection{$G_2$}

\subsubsection{Rank 2 (R)}

\begin{center}
%
\end{center}

\subsubsection{Rank 2 (S)}

\begin{center}
%
\end{center}


%% file: decomposition-reps.tex
\section{Branching rules}
\label{Sec:Branching-rules}

\input{branching-A}
\newpage
\input{branching-B}
\newpage
\input{branching-B-2}
\newpage
\input{branching-C}
\newpage
\input{branching-C-2}
\newpage
\input{branching-D}
\newpage
\input{branching-D-2}
\newpage
\input{branching-EFG}

%% file: branching-A.tex
\subsection{$A_n=\mathfrak{su}_{n+1}$}

\input{branching-A-2-3}
\newpage
\input{branching-A-4-5}
\newpage
\input{branching-A-6-8}
\newpage
\input{branching-A-9-11}
\newpage
\input{branching-A-12-15}
\newpage
\input{branching-A-16-20}

%% file: branching-B-2.tex
\subsubsection{Rank 16 (R)}

\subsubsubsection{$\mathfrak{so}_{32}$}
\begin{center}
%
\end{center}

\newpage
\subsubsection{Rank 16 (S)}

\subsubsubsection{$\mathfrak{su}_{2}$}
\begin{center}
%
\end{center}

\newpage
\subsubsection{Rank 17 (R)}

\subsubsubsection{$\mathfrak{so}_{34}$}
\begin{center}
%
\end{center}

\newpage
\subsubsection{Rank 17 (S)}

\subsubsubsection{$\mathfrak{su}_{2}$}
\begin{center}
%
\end{center}

\newpage
\subsubsection{Rank 18 (R)}

\subsubsubsection{$\mathfrak{so}_{36}$}
\begin{center}
%
\end{center}

\newpage
\subsubsection{Rank 18 (S)}

\subsubsubsection{$\mathfrak{su}_{2}$}
\begin{center}
%
\end{center}

\newpage
\subsubsection{Rank 19 (R)}

\subsubsubsection{$\mathfrak{so}_{38}$}
\begin{center}
%
\end{center}

\newpage
\subsubsection{Rank 19 (S)}

\subsubsubsection{$\mathfrak{su}_{2}$}
\begin{center}
%
\end{center}

\newpage
\subsubsection{Rank 20 (R)}

\subsubsubsection{$\mathfrak{so}_{40}$}
\begin{center}
%
\end{center}

\newpage
\subsubsection{Rank 20 (S)}

\subsubsubsection{$\mathfrak{su}_{2}$}
\begin{center}
%
\end{center}


%% file: branching-C-2.tex
\subsubsection{Rank 16 (R)}

\subsubsubsection{$\mathfrak{su}_{16}\oplus\mathfrak{u}_{1}$}
\begin{center}
%
\end{center}

\newpage
\subsubsection{Rank 16 (S)}

\subsubsubsection{$\mathfrak{su}_{2}$}
\begin{center}
%
\end{center}

\newpage
\subsubsection{Rank 17 (R)}

\subsubsubsection{$\mathfrak{su}_{17}\oplus\mathfrak{u}_{1}$}
\begin{center}
%
\end{center}

\newpage
\subsubsection{Rank 17 (S)}

\subsubsubsection{$\mathfrak{su}_{2}$}
\begin{center}
%
\end{center}

\newpage
\subsubsection{Rank 18 (R)}

\subsubsubsection{$\mathfrak{su}_{18}\oplus\mathfrak{u}_{1}$}
\begin{center}
%
\end{center}

\newpage
\subsubsection{Rank 18 (S)}

\subsubsubsection{$\mathfrak{su}_{2}$}
\begin{center}
%
\end{center}

\newpage
\subsubsection{Rank 19 (R)}

\subsubsubsection{$\mathfrak{su}_{19}\oplus\mathfrak{u}_{1}$}
\begin{center}
%
\end{center}

\newpage
\subsubsection{Rank 19 (S)}

\subsubsubsection{$\mathfrak{su}_{2}$}
\begin{center}
%
\end{center}

\newpage
\subsubsection{Rank 20 (R)}

\subsubsubsection{$\mathfrak{su}_{20}\oplus\mathfrak{u}_{1}$}
\begin{center}
%
\end{center}

\newpage
\subsubsection{Rank 20 (S)}

\subsubsubsection{$\mathfrak{su}_{2}$}
\begin{center}
%
\end{center}


%% file: branching-D-2.tex
\subsubsection{Rank 17 (R)}

\subsubsubsection{$\mathfrak{su}_{17}\oplus\mathfrak{u}_{1}$}
\begin{center}
%
\end{center}

\newpage
\subsubsection{Rank 17 (S)}

\subsubsubsection{$\mathfrak{so}_{33}$}
\begin{center}
%
\end{center}

\newpage
\subsubsection{Rank 18 (R)}

\subsubsubsection{$\mathfrak{su}_{18}\oplus\mathfrak{u}_{1}$}
\begin{center}
%
\end{center}

\newpage
\subsubsection{Rank 18 (S)}

\subsubsubsection{$\mathfrak{so}_{35}$}
\begin{center}
%
\end{center}

\newpage
\subsubsection{Rank 19 (R)}

\subsubsubsection{$\mathfrak{su}_{19}\oplus\mathfrak{u}_{1}$}
\begin{center}
%
\end{center}

\newpage
\subsubsection{Rank 19 (S)}

\subsubsubsection{$\mathfrak{so}_{37}$}
\begin{center}
%
\end{center}

\newpage
\subsubsection{Rank 20 (R)}

\subsubsubsection{$\mathfrak{su}_{20}\oplus\mathfrak{u}_{1}$}
\begin{center}
%
\end{center}

\newpage
\subsubsection{Rank 20 (S)}

\subsubsubsection{$\mathfrak{so}_{39}$}
\begin{center}
%
\end{center}


%% file: gut.bbl
\providecommand{\href}[2]{#2}\begingroup\raggedright\begin{thebibliography}{100}

\bibitem{Slansky:1981yr}
R.~Slansky, ``{Group Theory for Unified Model Building},''
\href{http://dx.doi.org/10.1016/0370-1573(81)90092-2}{{ Phys. Rept.} {\bfseries
  79} (1981) 1--128}.

\bibitem{Dynkin:1950um}
E.~Dynkin, ``{The Structure of Semi-Simple Lie Algebra},''
{ Amer. Math. Soc. Transl.} {\bfseries 1} (1950) 1.

\bibitem{Dynkin:1957um}
E.~Dynkin, ``{Semisimple Subalgebras of Semisimple Lie Algebras},''
\href{http://dx.doi.org/10.1090/trans2/006/02}{{ Amer. Math. Soc. Transl.}
  {\bfseries 6} (1957) 111}.

\bibitem{Dynkin:1957ek}
E.~Dynkin, ``{Maximal Subgroups of the Classical Groups},''
  \href{http://dx.doi.org/10.1090/trans2/006/03}{{ Amer. Math. Soc. Transl.}
  {\bfseries 6} (1957) 245}.

\bibitem{Dynkin:2000}
E.~B. Dynkin, { {Selected Papers of E.~B.~Dynkin with Commentary}}.
\newblock American Mathematical Society International Press, 2000.

\bibitem{Cahn:1985wk}
R.~Cahn, { {Semi-Simple Lie Algebras and Their Representations}}.
\newblock Benjamin-Cummings Publishing Company,
1985.
\newblock

\bibitem{Weyl:1946}
H.~Weyl, { The Classical Groups -Their Invariants and Representations-}.
\newblock Princeton University Press, 1946.

\bibitem{Gilmore:102082}
R.~Gilmore, { Lie Groups, Lie Algebras, and some of their Applications}.
\newblock John Wiley \& Sons, Inc., New York, 1974.

\bibitem{Fulton:1991}
W.~Fulton and J.~Harris, { Representation Theory}.
\newblock Springer, 1991.

\bibitem{Georgi:1982jb}
H.~Georgi, { {Lie Algebras in Particle Physics. From Isospin to Unified
  Theories}}.
\newblock Westview Press, USA,
1999.
\newblock

\bibitem{McKay:1981}
W.~G. McKay and J.~Patera, { Tables of Dimensions, Indices, and Branching Rules
  for Representations of Simple Lie Algebras}.
\newblock Marcel Dekker, Inc., New York, 1981.

\bibitem{Patera:1976yd}
J.~Patera, R.~T. Sharp, and P.~Winternitz, ``{Higher Indices of Group
  Representations},''
\href{http://dx.doi.org/10.1063/1.522836}{{ J. Math. Phys.} {\bfseries 17}
  (1976) 1972}.

\bibitem{Okubo:1981td}
S.~Okubo, ``{Modified Fourth Order Casimir Invariants and Indices for Simple
  Lie Algebras},''
\href{http://dx.doi.org/10.1063/1.525212}{{ J.Math.Phys.} {\bfseries 23} (1982)
  8}.

\bibitem{Mckay:1977}
W.~Mckay, J.~Patera, and D.~Sankoff, { {The Computation of Branching Rules for
  Representations of Semisimple Lie Algebras}}.
\newblock New York Academic Press, 1977.
\newblock in ``Computers in Nonassociative Rings and Algebras'' edited by
  R.~E.~Beck and B.~Kolman.

\bibitem{Hakova:2008ax}
L.~Hakova, M.~Larouche, and J.~Patera, ``{The Rings of $n$-Dimensional
  Polytopes},'' { J. Phys. A: Math. Theor.} (2008) 495202.

\bibitem{Larouche:2009ax}
M.~Larouche, M.~Nesterenko, and J.~Patera, ``{Branching Rules for the Weyl
  Group Orbits of the Lie Algebra $A_n$},'' { J. Phys. A: Math. Theor.}
  {\bfseries 42} (2009) 485203.

\bibitem{Larouche:2011bs}
M.~Larouche and J.~Patera, ``{Branching Rules for Weyl Group Orbits of Simple
  Lie Algebras $B_n$, $C_n$ and $D_n$},'' { J. Phys. A: Math. Theor.}
  {\bfseries 44} (2011) 115203.

\bibitem{Larouche:2011cy}
M.~Larouche, ``{Branching Rules for Weyl Group Orbits Involving the Lie Algebra
  $U(1)$},'' \href{http://dx.doi.org/10.1088/1742-6596/284/1/012043}{{ J.
  Phys.: Conf. Ser.} {\bfseries 284} (2011) 012043}.

\bibitem{Lemire:1979qd}
F.~Lemire and J.~Patera, ``{Congruence Number: A Generalization of SU(3)
  Triality},''
\href{http://dx.doi.org/10.1063/1.524711}{{ J.Math.Phys.} {\bfseries 21} (1980)
  2026}.

\bibitem{Banks1976}
J.~Banks and H.~Georgi, ``{Comment on Gauge Theories Without Anomalies},''
\href{http://dx.doi.org/10.1103/PhysRevD.14.1159}{{ Phys. Rev.} {\bfseries D14}
  (1976) 1159--1160}.

\bibitem{Okubo:1977sc}
S.~Okubo, ``{Gauge Groups Without Triangular Anomaly},''
\href{http://dx.doi.org/10.1103/PhysRevD.16.3528}{{ Phys.Rev.} {\bfseries D16}
  (1977) 3528}.

\bibitem{Patera:1981sc}
J.~Patera and R.~T. Sharp, ``{On the Triangle Anomaly Number of $SU(n)$
  Representaions},''
\href{http://dx.doi.org/10.1063/1.524815}{{ J. Math. Phys.} {\bfseries 22}
  (1981) 2352}.

\bibitem{Fonseca:2011sy}
R.~M. Fonseca, ``{Calculating the Renormalisation Group Equations of a SUSY
  Model with Susyno},'' \href{http://dx.doi.org/10.1016/j.cpc.2012.05.017}{{
  Comput.Phys.Commun.} {\bfseries 183} (2012) 2298--2306},
\href{http://arxiv.org/abs/1106.5016}{{\ttfamily arXiv:1106.5016 [hep-ph]}}.

\bibitem{Feger:2012bs}
R.~Feger and T.~W. Kephart, ``{LieART - A Mathematica Application for Lie
  Algebras and Representation Theory},''
  \href{http://dx.doi.org/10.1016/j.cpc.2014.12.023}{{ Comput.Phys.Commun.}
  {\bfseries 192} (2015) 166--195},
\href{http://arxiv.org/abs/1206.6379}{{\ttfamily arXiv:1206.6379 [math-ph]}}.

\bibitem{Feger:2019tvk}
R.~Feger, T.~W. Kephart, and R.~J. Saskowski, ``{LieART 2.0 -- A Mathematica
  Application for Lie Algebras and Representation Theory},'' { Comput. Phys.
  Commun.} {\bfseries 257} (2020) 107490,
\href{http://arxiv.org/abs/1912.10969}{{\ttfamily arXiv:1912.10969 [hep-th]}}.

\bibitem{Leeuwen:1992}
M.~A.~A. van Leeuwen, A.~M. Cohen, and B.~Lisser, { LiE, A Package for Lie
  Group Computations}.
\newblock Computer Algebra Nederland, Amsterdam, 1992.

\bibitem{Ball:1988xg}
R.~Ball, ``{Chiral Gauge Theory},''
\href{http://dx.doi.org/10.1016/0370-1573(89)90027-6}{{ Phys.Rept.} {\bfseries
  182} (1989) 1}.

\bibitem{Fujikawa:2004xa}
K.~Fujikawa and H.~Suzuki, ``Path Integrals and Quantum Anomalies,'' { Oxford
  University Press} (2004) .

\bibitem{Politzer:1974fr}
H.~D. Politzer, ``{Asymptotic Freedom: An Approach to Strong Interactions},''
\href{http://dx.doi.org/10.1016/0370-1573(74)90014-3}{{ Phys. Rept.} {\bfseries
  14} (1974) 129--180}.

\bibitem{Machacek:1983tz}
M.~E. Machacek and M.~T. Vaughn, ``{Two Loop Renormalization Group Equations in
  a General Quantum Field Theory. 1. Wave Function Renormalization},''
\href{http://dx.doi.org/10.1016/0550-3213(83)90610-7}{{ Nucl. Phys.} {\bfseries
  B222} (1983) 83}.

\bibitem{Machacek:1983fi}
M.~E. Machacek and M.~T. Vaughn, ``{Two Loop Renormalization Group Equations in
  a General Quantum Field Theory. 2. Yukawa Couplings},''
\href{http://dx.doi.org/10.1016/0550-3213(84)90533-9}{{ Nucl. Phys.} {\bfseries
  B236} (1984) 221}.

\bibitem{Machacek:1984zw}
M.~E. Machacek and M.~T. Vaughn, ``{Two Loop Renormalization Group Equations in
  a General Quantum Field Theory. 3. Scalar Quartic Couplings},''
\href{http://dx.doi.org/10.1016/0550-3213(85)90040-9}{{ Nucl. Phys.} {\bfseries
  B249} (1985) 70}.

\bibitem{Bremner1983}
M.~Bremner, R.~Moody, and J.~Patera, { Tables of Dominant Weight Multiplicities
  for Representations of Simple Lie Algebras}.
\newblock Marcel Dekker, 1983.

\bibitem{Klimyk:2006ax}
A.~Klimyk and J.~Patera, ``Orbit Functions,''
  \href{http://dx.doi.org/10.3842/SIGMA.2006.006}{{ SIGMA} {\bfseries 2} (2006)
  006}.

\bibitem{Georgi:1974sy}
H.~Georgi and S.~L. Glashow, ``{Unity of All Elementary Particle Forces},''
\href{http://dx.doi.org/10.1103/PhysRevLett.32.438}{{ Phys. Rev. Lett.}
  {\bfseries 32} (1974) 438--441}.

\bibitem{Fuchs:1992}
J.~Fuchs, { Affine Lie Algebras and Quantum Groups}.
\newblock Cambridge University Press, 1992.

\bibitem{Kac:1983}
V.~G. Kac, { Infinite Dimensional Lie Algebras}.
\newblock Cambridge University Press, 1983.

\bibitem{Eichten1982}
E.~Eichten, K.~Kang, and I.-G. Koh, ``{Anomaly Free Complex Representations in
  $SU(N)$},''
\href{http://dx.doi.org/10.1063/1.525299}{{ J. Math. Phys.} {\bfseries 23}
  (1982) 2529}.

\bibitem{Okubo:1982dt}
S.~Okubo and J.~Patera, ``{General Indices of Representations and Casimir
  Invariants},''
\href{http://dx.doi.org/10.1063/1.526143}{{ J.Math.Phys.} {\bfseries 25} (1984)
  219}.

\bibitem{Okubo:1983sv}
S.~Okubo and J.~Patera, ``{Symmetrization of Product Representations and
  General Indices and Simple Lie Algebras},''
\href{http://dx.doi.org/10.1063/1.525670}{{ J.Math.Phys.} {\bfseries 24} (1983)
  2722}.

\bibitem{Okubo:1985qk}
S.~Okubo and J.~Patera, ``{On Cancellation of Higher Order Anomalies},''
\href{http://dx.doi.org/10.1103/PhysRevD.31.2669}{{ Phys.Rev.} {\bfseries D31}
  (1985) 2669}.

\bibitem{Bilal:2008qx}
A.~Bilal, ``{Lectures on Anomalies},''
\href{http://arxiv.org/abs/0802.0634}{{\ttfamily arXiv:0802.0634 [hep-th]}}.

\bibitem{Witten:1982fp}
E.~Witten, ``{An SU(2) Anomaly},''
\href{http://dx.doi.org/10.1016/0370-2693(82)90728-6}{{ Phys.Lett.} {\bfseries
  B117} (1982) 324--328}.

\bibitem{Geng:1987fg}
C.-q. Geng, R.~Marshak, Z.-y. Zhao, and S.~Okubo, ``{Relation Between
  Triangular and Witten SU(2) Anomaly Cancellation for Gauge Groups},''
\href{http://dx.doi.org/10.1103/PhysRevD.36.1953}{{ Phys.Rev.} {\bfseries D36}
  (1987) 1953}.

\bibitem{Okubo:1987dh}
S.~Okubo, C.-q. Geng, R.~Marshak, and Z.-y. Zhao, ``{Conditions for Absence of
  Global SU(2) Anomaly for Four-dimensional Gauge Groups},''
\href{http://dx.doi.org/10.1103/PhysRevD.36.3268}{{ Phys.Rev.} {\bfseries D36}
  (1987) 3268}.

\bibitem{Okubo:1989vn}
S.~Okubo and Y.~Tosa, ``{Further Study of Global Gauge Anomalies of Simple
  Groups},''
\href{http://dx.doi.org/10.1103/PhysRevD.40.1925}{{ Phys.Rev.} {\bfseries D40}
  (1989) 1925}.

\bibitem{Borghini:2001sa}
N.~Borghini, Y.~Gouverneur, and M.~H. Tytgat, ``{Anomalies and Fermion Content
  of Grand Unified Theories in Extra Dimensions},''
  \href{http://dx.doi.org/10.1103/PhysRevD.65.025017}{{ Phys.Rev.} {\bfseries
  D65} (2002) 025017},
\href{http://arxiv.org/abs/hep-ph/0108094}{{\ttfamily arXiv:hep-ph/0108094
  [hep-ph]}}.

\bibitem{vonGersdorff:2006nt}
G.~von Gersdorff, ``{Anomalies on Six Dimensional Orbifolds},''
  \href{http://dx.doi.org/10.1088/1126-6708/2007/03/083}{{ JHEP} {\bfseries
  0703} (2007) 083},
\href{http://arxiv.org/abs/hep-th/0612212}{{\ttfamily arXiv:hep-th/0612212
  [hep-th]}}.

\bibitem{Fulton:1997}
W.~Fulton, { Young Tableaux}.
\newblock Cambridge University Press, 1997.

\bibitem{Black:1983}
G.~Black, R.~King, and B.~Wybourne, ``Kronecker Products for Compact Semisimple
  Lie Groups,'' \href{http://dx.doi.org/10.1088/0305-4470/16/8/006}{{ J. Phys.
  A: Math. Gen.} {\bfseries 16} (1983) 1555--1589}.

\bibitem{Koike:1987aa}
K.~Koike and I.~Terada, ``Young-Diagrammatic Methods for the Representation
  Theory of the Classical Groups of Type $B_n$, $C_n$, $D_n$,''
  \href{http://dx.doi.org/10.1016/0021-8693(87)90099-8}{{ J. of Algebra}
  {\bfseries 107} (1987) 466--511}.

\bibitem{Koike:1987bb}
K.~Koike, ``On New Multiplicity Formulas of Weights of Representations for the
  Classical Groups,'' \href{http://dx.doi.org/10.1016/0021-8693(87)90100-1}{{
  J. of Algebra} {\bfseries 107} (1987) 512--533}.

\bibitem{Pati:1974yy}
J.~C. Pati and A.~Salam, ``{Lepton Number as the Fourth Color},''
\href{http://dx.doi.org/10.1103/PhysRevD.10.275}{{ Phys. Rev.} {\bfseries D10}
  (1974) 275--289}.

\bibitem{Inoue:1977qd}
K.~Inoue, A.~Kakuto, and Y.~Nakano, ``{Unification of the Lepton-Quark World by
  the Gauge Group SU(6)},''
\href{http://dx.doi.org/10.1143/PTP.58.630}{{ Prog.Theor.Phys.} {\bfseries 58}
  (1977) 630}.

\bibitem{Fonseca:2015aoa}
R.~M. Fonseca, ``{On the Chirality of the SM and the Fermion Content of
  GUTs},'' \href{http://dx.doi.org/10.1016/j.nuclphysb.2015.06.012}{{ Nucl.
  Phys.} {\bfseries B897} (2015) 757--780},
\href{http://arxiv.org/abs/1504.03695}{{\ttfamily arXiv:1504.03695 [hep-ph]}}.

\bibitem{Fritzsch:1974nn}
H.~Fritzsch and P.~Minkowski, ``{Unified Interactions of Leptons and
  Hadrons},''
\href{http://dx.doi.org/10.1016/0003-4916(75)90211-0}{{ Ann. Phys.} {\bfseries
  93} (1975) 193--266}.

\bibitem{Fukuyama2005}
T.~Fukuyama, A.~Ilakovac, T.~Kikuchi, S.~Meljanac, and N.~Okada, ``{$SO(10)$
  Group Theory for the Unified Model Building},''
  \href{http://dx.doi.org/10.1063/1.1847709}{{ J. Math. Phys.} {\bfseries 46}
  (2005) 033505},
\href{http://arxiv.org/abs/hep-ph/0405300}{{\ttfamily arXiv:hep-ph/0405300}}.

\bibitem{Ida:1980ea}
M.~Ida, Y.~Kayama, and T.~Kitazoe, ``{Inclusion of Generations in SO(14)},''
\href{http://dx.doi.org/10.1143/PTP.64.1745}{{ Prog. Theor. Phys.} {\bfseries
  64} (1980) 1745}.

\bibitem{Fujimoto:1981bv}
Y.~Fujimoto, ``{SO(18) Unification},''
\href{http://dx.doi.org/10.1103/PhysRevD.26.3183}{{ Phys. Rev.} {\bfseries D26}
  (1982) 3183}.

\bibitem{Gursey:1975ki}
F.~Gursey, P.~Ramond, and P.~Sikivie, ``{A Universal Gauge Theory Model Based
  on $E_6$},''
\href{http://dx.doi.org/10.1016/0370-2693(76)90417-2}{{ Phys. Lett.} {\bfseries
  B60} (1976) 177}.

\bibitem{Maekawa:2002bk}
N.~Maekawa and T.~Yamashita, ``{$E_6$ Unification, Doublet-Triplet Splitting
  and Anomalous $U(1)_A$ Symmetry},''
  \href{http://dx.doi.org/10.1143/PTP.107.1201}{{ Prog. Theor. Phys.}
  {\bfseries 107} (2002) 1201--1233},
\href{http://arxiv.org/abs/hep-ph/0202050}{{\ttfamily arXiv:hep-ph/0202050}}.

\bibitem{Kojima:2011ad}
K.~Kojima, K.~Takenaga, and T.~Yamashita, ``{Grand Gauge-Higgs Unification},''
  \href{http://dx.doi.org/10.1103/PhysRevD.84.051701}{{ Phys. Rev.} {\bfseries
  D84} (2011) 051701},
\href{http://arxiv.org/abs/1103.1234}{{\ttfamily arXiv:1103.1234 [hep-ph]}}.

\bibitem{Burdman:2002se}
G.~Burdman and Y.~Nomura, ``{Unification of Higgs and Gauge Fields in
  Five-Dimensions},'' \href{http://dx.doi.org/10.1016/S0550-3213(03)00088-9}{{
  Nucl. Phys.} {\bfseries B656} (2003) 3--22},
\href{http://arxiv.org/abs/hep-ph/0210257}{{\ttfamily arXiv:hep-ph/0210257
  [hep-ph]}}.

\bibitem{Lim:2007jv}
C.~S. Lim and N.~Maru, ``{Towards a Realistic Grand Gauge-Higgs Unification},''
  \href{http://dx.doi.org/10.1016/j.physletb.2007.07.053}{{ Phys.Lett.}
  {\bfseries B653} (2007) 320--324},
\href{http://arxiv.org/abs/0706.1397}{{\ttfamily arXiv:0706.1397 [hep-ph]}}.

\bibitem{Haba:2004qf}
N.~Haba, Y.~Hosotani, Y.~Kawamura, and T.~Yamashita, ``{Dynamical Symmetry
  Breaking in Gauge Higgs Unification on Orbifold},''
  \href{http://dx.doi.org/10.1103/PhysRevD.70.015010}{{ Phys. Rev.} {\bfseries
  D70} (2004) 015010},
\href{http://arxiv.org/abs/hep-ph/0401183}{{\ttfamily arXiv:hep-ph/0401183
  [hep-ph]}}.

\bibitem{Yamamoto:2013oja}
K.~Yamamoto, ``{The Formulation of Gauge-Higgs Unification with Dynamical
  Boundary Conditions},''
  \href{http://dx.doi.org/10.1016/j.nuclphysb.2014.03.017}{{ Nucl. Phys.}
  {\bfseries B883} (2014) 45--58},
\href{http://arxiv.org/abs/1401.0466}{{\ttfamily arXiv:1401.0466 [hep-th]}}.

\bibitem{Kim:2002im}
H.~D. Kim and S.~Raby, ``{Unification in 5-D SO(10)},''
  \href{http://dx.doi.org/10.1088/1126-6708/2003/01/056}{{ JHEP} {\bfseries 01}
  (2003) 056},
\href{http://arxiv.org/abs/hep-ph/0212348}{{\ttfamily arXiv:hep-ph/0212348
  [hep-ph]}}.

\bibitem{Fukuyama:2008pw}
T.~Fukuyama and N.~Okada, ``{A Simple SO(10) GUT in Five Dimensions},''
  \href{http://dx.doi.org/10.1103/PhysRevD.78.015005}{{ Phys. Rev.} {\bfseries
  D78} (2008) 015005},
\href{http://arxiv.org/abs/0803.1758}{{\ttfamily arXiv:0803.1758 [hep-ph]}}.

\bibitem{Fukuyama:2012rw}
T.~Fukuyama, ``{SO(10) GUT in Four and Five Dimensions: A Review},''
  \href{http://dx.doi.org/10.1142/S0217751X13300081}{{ Int. J. Mod. Phys.}
  {\bfseries A28} (2013) 1330008},
\href{http://arxiv.org/abs/1212.3407}{{\ttfamily arXiv:1212.3407 [hep-ph]}}.

\bibitem{Hosotani:2015hoa}
Y.~Hosotani and N.~Yamatsu, ``{Gauge-Higgs Grand Unification},''
  \href{http://dx.doi.org/10.1093/ptep/ptv153}{{ Prog. Theor. Exp. Phys.}
  {\bfseries 2015} (2015) 111B01},
\href{http://arxiv.org/abs/1504.03817}{{\ttfamily arXiv:1504.03817 [hep-ph]}}.

\bibitem{Yamatsu:2015rge}
N.~Yamatsu, ``{Gauge Coupling Unification in Gauge-Higgs Grand Unification},''
  \href{http://dx.doi.org/10.1093/ptep/ptw023}{{ Prog. Theor. Exp. Phys.}
  {\bfseries 2016} (2016) 043B02},
\href{http://arxiv.org/abs/1512.05559}{{\ttfamily arXiv:1512.05559 [hep-ph]}}.

\bibitem{Furui:2016owe}
A.~Furui, Y.~Hosotani, and N.~Yamatsu, ``{Toward Realistic Gauge-Higgs Grand
  Unification},'' \href{http://dx.doi.org/10.1093/ptep/ptw116}{{ Prog. Theor.
  Exp. Phys.} {\bfseries 2016} (2016) 093B01},
\href{http://arxiv.org/abs/1606.07222}{{\ttfamily arXiv:1606.07222 [hep-ph]}}.

\bibitem{Kawamura:2009gr}
Y.~Kawamura and T.~Miura, ``{Orbifold Family Unification in $SO(2N)$ Gauge
  Theory},'' \href{http://dx.doi.org/10.1103/PhysRevD.81.075011}{{ Phys. Rev.}
  {\bfseries D81} (2010) 075011},
\href{http://arxiv.org/abs/0912.0776}{{\ttfamily arXiv:0912.0776 [hep-ph]}}.

\bibitem{Kawamura:2013rj}
Y.~Kawamura and T.~Miura, ``{Classification of Standard Model Particles in
  $E_6$ Orbifold Grand Unified Theories},''
  \href{http://dx.doi.org/10.1142/S0217751X13500553}{{ Int. J. Mod. Phys.}
  {\bfseries A28} (2013) 1350055},
\href{http://arxiv.org/abs/1301.7469}{{\ttfamily arXiv:1301.7469 [hep-ph]}}.

\bibitem{Cremmer:1979up}
E.~Cremmer and B.~Julia, ``{The $SO(8)$ Supergravity},''
\href{http://dx.doi.org/10.1016/0550-3213(79)90331-6}{{ Nucl. Phys.} {\bfseries
  B159} (1979) 141}.

\bibitem{Kugo:1983ai}
T.~Kugo and T.~Yanagida, ``{Unification of Families Based on a Coset Space
  $E_7/SU(5)\times SU(3)\times U(1)$},''
\href{http://dx.doi.org/10.1016/0370-2693(84)90007-8}{{ Phys. Lett.} {\bfseries
  B134} (1984) 313}.

\bibitem{Dienes:1996du}
K.~R. Dienes, ``{String Theory and the Path to Unification: A Review of Recent
  Developments},'' \href{http://dx.doi.org/10.1016/S0370-1573(97)00009-4}{{
  Phys. Rept.} {\bfseries 287} (1997) 447--525},
\href{http://arxiv.org/abs/hep-th/9602045}{{\ttfamily arXiv:hep-th/9602045
  [hep-th]}}.

\bibitem{Dienes:1996yh}
K.~R. Dienes and J.~March-Russell, ``{Realizing Higher Level Gauge Symmetries
  in String Theory: New Embeddings for String GUTs},''
  \href{http://dx.doi.org/10.1016/0550-3213(96)00406-3}{{ Nucl. Phys.}
  {\bfseries B479} (1996) 113--172},
\href{http://arxiv.org/abs/hep-th/9604112}{{\ttfamily arXiv:hep-th/9604112
  [hep-th]}}.

\bibitem{Hebecker:2001jb}
A.~Hebecker and J.~March-Russell, ``{The Structure of GUT Breaking by
  Orbifolding},'' \href{http://dx.doi.org/10.1016/S0550-3213(02)00016-0}{{
  Nucl. Phys.} {\bfseries B625} (2002) 128--150},
\href{http://arxiv.org/abs/hep-ph/0107039}{{\ttfamily arXiv:hep-ph/0107039
  [hep-ph]}}.

\bibitem{Hebecker:2003jt}
A.~Hebecker and M.~Ratz, ``{Group Theoretical Aspects of Orbifold and Conifold
  GUTs},'' \href{http://dx.doi.org/10.1016/j.nuclphysb.2003.07.021}{{ Nucl.
  Phys.} {\bfseries B670} (2003) 3--26},
\href{http://arxiv.org/abs/hep-ph/0306049}{{\ttfamily arXiv:hep-ph/0306049
  [hep-ph]}}.

\bibitem{Yamatsu:2017sgu}
N.~Yamatsu, ``{Special Grand Unification},''
  \href{http://dx.doi.org/10.1093/ptep/ptx088}{{ Prog. Theor. Exp. Phys.}
  {\bfseries 2017} (2017) 061B01},
\href{http://arxiv.org/abs/1704.08827}{{\ttfamily arXiv:1704.08827 [hep-ph]}}.

\bibitem{Yamatsu:2017ssg}
N.~Yamatsu, ``{String-Inspired Special Grand Unification},''
  \href{http://dx.doi.org/10.1093/ptep/ptx135}{{ Prog. Theor. Exp. Phys.}
  {\bfseries 2017} (2017) 101B01},
\href{http://arxiv.org/abs/1708.02078}{{\ttfamily arXiv:1708.02078 [hep-ph]}}.

\bibitem{Yamatsu:2018fsg}
N.~Yamatsu, ``{Family Unification in Special Grand Unification},''
  \href{http://dx.doi.org/10.1093/ptep/pty100}{{ Prog. Theor. Exp. Phys.}
  {\bfseries 2018} (2018) 091B01},
\href{http://arxiv.org/abs/1807.10855}{{\ttfamily arXiv:1807.10855 [hep-ph]}}.

\bibitem{Yamatsu:2020usp}
N.~Yamatsu, ``{$USp(32)$ Special Grand Unification},''
  \href{http://arxiv.org/abs/2007.08067}{{\ttfamily arXiv:2007.08067
  [hep-ph]}}.

\bibitem{Li:1973mq}
L.-F. Li, ``{Group Theory of the Spontaneously Broken Gauge Symmetries},''
\href{http://dx.doi.org/10.1103/PhysRevD.9.1723}{{ Phys. Rev.} {\bfseries D9}
  (1974) 1723--1739}.

\bibitem{Meljanac:1982rc}
S.~Meljanac, M.~Milosevic, and S.~Pallua, ``{Extrema of Higgs Potential and
  Higher Representations},''
\href{http://dx.doi.org/10.1103/PhysRevD.26.2936}{{ Phys. Rev.} {\bfseries D26}
  (1982) 2936--2939}.

\bibitem{Abud:1984xn}
M.~Abud, G.~Anastaze, P.~Eckert, and H.~Ruegg, ``{Minima of Higgs Potentials
  Corresponding to Nonmaximal Isotropy Subgroups},''
\href{http://dx.doi.org/10.1016/0003-4916(85)90232-5}{{ Annals Phys.}
  {\bfseries 162} (1985) 155}.

\bibitem{Kugo:2019isl}
T.~Kugo and N.~Yamatsu, ``{Is Symmetry Breaking into Special Subgroup
  Special?},'' \href{http://dx.doi.org/10.1093/ptep/ptz063}{{ Prog. Theor. Exp.
  Phys.} {\bfseries 2019} (2019) 073B06},
\href{http://arxiv.org/abs/1904.06857}{{\ttfamily arXiv:1904.06857 [hep-ph]}}.

\bibitem{Nambu:1961tp}
Y.~Nambu and G.~Jona-Lasinio, ``{Dynamical Model of Elementary Particles Based
  on an Analogy with Superconductivity. I},''
\href{http://dx.doi.org/10.1103/PhysRev.122.345}{{ Phys. Rev.} {\bfseries 122}
  (1961) 345--358}.

\bibitem{Nambu:1961fr}
Y.~Nambu and G.~Jona-Lasinio, ``{Dynamical Model of Elementary Particles Based
  on an Analogy with Superconductivity. II},''
\href{http://dx.doi.org/10.1103/PhysRev.124.246}{{ Phys. Rev.} {\bfseries 124}
  (1961) 246--254}.

\bibitem{Kugo:1994qr}
T.~Kugo and J.~Sato, ``{Dynamical Symmetry Breaking in an E(6) GUT Model},''
  \href{http://dx.doi.org/10.1143/ptp/91.6.1217, 10.1143/PTP.91.1217}{{ Prog.
  Theor. Phys.} {\bfseries 91} (1994) 1217--1238},
\href{http://arxiv.org/abs/hep-ph/9402357}{{\ttfamily arXiv:hep-ph/9402357
  [hep-ph]}}.

\bibitem{Kugo:2019wge}
T.~Kugo and N.~Yamatsu, ``{Dynamical Breaking to Special or Regular Subgroups
  in the $SO(N)$ Nambu--Jona-Lasinio Model},''
  \href{http://dx.doi.org/10.1093/ptep/ptaa001}{{ Prog. Theor. Exp. Phys.}
  {\bfseries 2020} (2020) 023B09},
\href{http://arxiv.org/abs/1911.09834}{{\ttfamily arXiv:1911.09834 [hep-ph]}}.

\bibitem{Inoue:1994qz}
K.~Inoue, ``{Generations of Quarks and Leptons from Noncompact Horizontal
  Symmetry},'' \href{http://dx.doi.org/10.1143/PTP.93.403}{{ Prog. Theor.
  Phys.} {\bfseries 93} (1995) 403--416},
\href{http://arxiv.org/abs/hep-ph/9410220}{{\ttfamily arXiv:hep-ph/9410220}}.

\bibitem{Inoue:2000ia}
K.~Inoue and N.-a. Yamashita, ``{Mass Hierarchy from $SU(1,1)$ Horizontal
  Symmetry},'' \href{http://dx.doi.org/10.1143/PTP.104.677}{{ Prog. Theor.
  Phys.} {\bfseries 104} (2000) 677--689},
\href{http://arxiv.org/abs/hep-ph/0005178}{{\ttfamily arXiv:hep-ph/0005178}}.

\bibitem{Inoue:2003qi}
K.~Inoue and N.-a. Yamashita, ``{Neutrino Masses and Mixing Matrix from
  $SU(1,1)$ Horizontal Symmetry},''
  \href{http://dx.doi.org/10.1143/PTP.110.1087}{{ Prog. Theor. Phys.}
  {\bfseries 110} (2004) 1087--1094},
\href{http://arxiv.org/abs/hep-ph/0305297}{{\ttfamily arXiv:hep-ph/0305297}}.

\bibitem{Yamatsu:2007}
K.~Inoue and N.~Yamatsu, ``{Charged Lepton and Down-Type Quark Masses in
  $SU(1,1)$ Model and the Structure of Higgs Sector},''
  \href{http://dx.doi.org/10.1143/PTP.119.775}{{ Prog. Theor. Phys.} {\bfseries
  119} (2008) 775--796},
\href{http://arxiv.org/abs/0712.2938}{{\ttfamily arXiv:0712.2938 [hep-ph]}}.

\bibitem{Yamatsu:2008}
K.~Inoue and N.~Yamatsu, ``{Strong CP Problem and the Natural Hierarchy of
  Yukawa Couplings},'' \href{http://dx.doi.org/10.1143/PTP.120.1065}{{ Prog.
  Theor. Phys.} {\bfseries 120} (2008) 1065--1091},
\href{http://arxiv.org/abs/0806.0213}{{\ttfamily arXiv:0806.0213 [hep-ph]}}.

\bibitem{Yamatsu:2009}
K.~Inoue, H.~Kubo, and N.~Yamatsu, ``{Vacuum Structures of Supersymmetric
  Noncompact Gauge Theory},''
  \href{http://dx.doi.org/10.1016/j.nuclphysb.2010.03.004}{{ Nucl. Phys.}
  {\bfseries B833} (2010) 108--132},
\href{http://arxiv.org/abs/0909.4670}{{\ttfamily arXiv:0909.4670 [hep-th]}}.

\bibitem{Yamatsu:2012}
N.~Yamatsu, ``{New Mixing Structures of Chiral Generations in a Model with
  Noncompact Horizontal Symmetry},''
  \href{http://dx.doi.org/10.1093/ptep/pts079}{{ Prog. Theor. Exp. Phys.}
  {\bfseries 2013} (2013) 023B03},
\href{http://arxiv.org/abs/1209.6318}{{\ttfamily arXiv:1209.6318 [hep-ph]}}.

\bibitem{Yamatsu:2013}
N.~Yamatsu, ``{A Supersymmetric Grand Unified Model with Noncompact Horizontal
  Symmetry},'' \href{http://dx.doi.org/10.1093/ptep/ptt100}{{ Prog. Theor. Exp.
  Phys.} {\bfseries 2013} (2013) 123B01},
\href{http://arxiv.org/abs/1304.5215}{{\ttfamily arXiv:1304.5215 [hep-ph]}}.

\bibitem{Maki:1962mu}
Z.~Maki, M.~Nakagawa, and S.~Sakata, ``{Remarks on the Unified Model of
  Elementary Particles},''
\href{http://dx.doi.org/10.1143/PTP.28.870}{{ Prog. Theor. Phys.} {\bfseries
  28} (1962) 870--880}.

\bibitem{Ishimori:2010au}
H.~Ishimori, T.~Kobayashi, H.~Ohki, Y.~Shimizu, and H.~Okada, ``{Non-Abelian
  Discrete Symmetries in Particle Physics},''
  \href{http://dx.doi.org/10.1143/PTPS.183.1}{{ Prog.Theor.Phys.Suppl.}
  {\bfseries 183} (2010) 1--163},
\href{http://arxiv.org/abs/1003.3552}{{\ttfamily arXiv:1003.3552 [hep-th]}}.

\end{thebibliography}\endgroup
